\documentclass{emulateapj}

\usepackage{amsmath}
\usepackage{amssymb}
\usepackage{amsfonts}
\usepackage{amstext}
\usepackage{graphicx}
\usepackage{fancybox}
\usepackage[usenames,dvipsnames]{color}
\usepackage{pstricks}
\usepackage{bm}
\usepackage{soul, color}
\usepackage{latexsym}
\usepackage{pifont}
\usepackage{shorttoc}
\usepackage{subfigure}
\usepackage{textcomp} 
\usepackage{float}

\usepackage{hyperref}
\usepackage{bm}
\usepackage{lipsum}
\usepackage{threeparttable}

\newcommand{\lya}{Ly$\alpha$\ }
\newcommand{\kms}{\, {\rm km \, s}^{-1}}
\newcommand{\cm}{\, {\rm cm}}
\newcommand{\angs}{\, {\rm \AA}}

\newcommand{\no}[1]{}
\renewcommand{\exp}[1]{\mathrm{exp}\left(#1\right)}

\newcommand{\myemail}{l.m.ribas@astro.uio.no}
\def\lsim{~\rlap{$<$}{\lower 1.0ex\hbox{$\sim$}}}

\def\gsim{~\rlap{$>$}{\lower 1.0ex\hbox{$\sim$}}}

\shorttitle{The mean metal-line spectrum of DLAs in BOSS}
\shortauthors{Ll. Mas-Ribas et al.}

\begin{document}

\title{The mean metal-line
absorption spectrum of Damped Lyman Alpha Systems in BOSS}

\author{Llu\'is Mas-Ribas\altaffilmark{1,2}} 
\author{Jordi Miralda-Escud\'e\altaffilmark{2,3}} 
\author{Ignasi P\'erez-R\`afols\altaffilmark{2,4}}
\author{Andreu Arinyo-i-Prats\altaffilmark{2}} 
\author{Pasquier Noterdaeme\altaffilmark{5}} 
\author{Patrick Petitjean\altaffilmark{5}} 
\author{Donald P. Schneider\altaffilmark{6,7}}
\author{Donald G. York\altaffilmark{8}} 
\author{Jian Ge\altaffilmark{9}} 

\altaffiltext{1}{Institute of Theoretical Astrophysics, University of Oslo,
 Postboks 1029, 0315 Oslo, Norway. \myemail}
 \altaffiltext{2}{Institut de Ci\`encies del Cosmos, Universitat de Barcelona (UB-IEEC),
Barcelona 08028, Catalonia.}
 \altaffiltext{3}{Instituci\'o Catalana de Recerca i Estudis Avan\c{c}ats, Barcelona,
Catalonia.}
 \altaffiltext{4}{Departament de F\'isica Qu\`antica i Astrof\'isica, Universitat de Barcelona,
 08028, Catalonia.}
 \altaffiltext{5}{Institut d'Astrophysique de Paris, UPMC \& CNRS, UMR7095 98bis Boulevard Arago,
75014 - Paris, France.}
\altaffiltext{6}{Department of Astronomy and Astrophysics, The Pennsylvania State University,
   University Park, PA 16802, USA.}
\altaffiltext{7}{Institute for Gravitation and the Cosmos, The Pennsylvania State University,
   University Park, PA 16802, USA.}
 \altaffiltext{8}{Department of Astronomy and Astrophysics and the Enrico Fermi Institute, University
of Chicago, 5640 South Ellis Avenue, Chicago, IL 60637, USA.}
 \altaffiltext{9}{Department of Astronomy, University of Florida, Bryant Space Science Center,
Gainesville, FL 32611-2055, USA.}


\begin{abstract}

  We study the mean absorption spectrum of the Damped Lyman alpha population 
at $z\sim 2.6$ by stacking normalized, rest-frame 
shifted spectra of $\sim 27\,000$ DLAs from the DR12 of BOSS/SDSS-III. 
We measure the equivalent widths
of 50 individual metal absorption lines in 5 intervals of DLA hydrogen
column density, 5 intervals of DLA redshift, and overall mean 
equivalent widths for an
additional 13 absorption features from groups of strongly blended lines.
The mean equivalent width of low-ionization lines increases
with $N_{\rm HI}$, whereas for high-ionization lines the increase is much weaker. 
The mean metal line equivalent widths decrease by 
a factor $\sim 1.1-1.5$ from $z\sim2.1$ to $z \sim 3.5$, with 
small or no differences between low- and high-ionization species. 
We develop a theoretical model,
inspired by the presence of multiple absorption components observed 
in high-resolution spectra, to infer mean metal column densities from the 
equivalent widths of partially saturated metal lines. We apply this model to 
14 low-ionization species and to AlIII, SIII, SiIII,
CIV, SiIV, NV and OVI. We use an approximate derivation for
separating the equivalent width contributions of several lines to blended absorption
features, and infer mean equivalent widths and column densities
from lines of the additional species NI, ZnII, CII${}^{*}$, FeIII, and
SIV. 
Several of these mean column densities of metal lines in DLAs are
obtained for the first time; their values generally agree
with measurements of individual DLAs from high-resolution, high
signal-to-noise ratio spectra when they are available.
\end{abstract}
\keywords{intergalactic medium -- DLA -- cosmology -- abundances --
 quasar: absorption lines.}


\section{Introduction}

  The existence of luminous quasars at high redshift is a gift of
Nature.  It allows us to explore in an unbiased way any population of
gas clouds in the Universe by means of the absorption lines they produce
in the spectra of the background sources. Without luminous quasars, we would not
have sources at high redshift that are sufficiently bright to obtain
spectra of high resolution and signal-to-noise ratio (hereafter, S/N)
in which the \lya line, as well as
numerous ultraviolet metal absorption lines, are shifted to the optical
range and can easily be observed from the ground. Damped Lyman Alpha
systems \citep[DLAs, hereafter;][]{Wolfe1986} are generally defined to 
have hydrogen column densities
$\rm{N_{HI}} > 2\times 10^{20} \cm^{-2}$. Systems of this high column density
have two important characteristics: ({\textit i}) they are self-shielded of the external
cosmic ionizing background, implying that the hydrogen in these systems is
mostly in atomic form \citep{Vladilo2001}, and (\textit{ii}) the damped profile 
of their hydrogen \lya line is clearly visible even in low-resolution spectra, 
therefore the column density can be measured from the absorption profile  
\citep[see][for detailed reviews]{Wolfe2005,Barnes2014}. DLAs  
provide a reservoir of atomic gas clouds that were available at high
redshift for the formation of galaxies. The mean cosmic density of
baryons contained in these systems is directly obtained from the
measurements of the column density distribution, and accounts for a small 
fraction of the critical density $\Omega_{\rm DLA}\simeq 10^{-3}$ at 
redshifts $2 < z < 3.5$ \citep[e.g.,][]{Peroux2003,Peroux2005,Prochaska2005,
Noterdaeme2012,Zafar2013,Padmanabhan2015,Crighton2015,Sanchez2015}. 
This value corresponds to $\sim$ 2\% of all the baryons in the Universe,
comparable to the fraction of baryons that had turned into stars at
these redshifts \citep{Prochaska2005,Prochaska2009,Noterdaeme2009}.

  The metal absorption lines of damped \lya systems have been explored since the 
discovery of DLAs. High-resolution spectra reveal a diversity of velocity structures
of the absorbers, characterized by multiple components. Sometimes a
single component with a narrow velocity width close to the thermal value
for photoionized gas clouds is observed, but often several components
are seen over a typical velocity range $\sim 100-300 \,{\rm \kms}$
\citep{Prochaska1997,Zwaan2008}. 
The derived metallicities are generally low, distributed over a
broad range of $10^{-3}$ to $10^{-1} Z_{\odot}$
\citep{Prochaska2002,Prochaska2003c,Kulkarni2005,Ledoux2006},
and on average declining with redshift \citep{Kulkarni2002,Vladilo2002,
Prochaska2003,Calura2003,Khare2004,Akerman2005,Kulkarni2005,Rafelski2012,
Jorgenson2013,Neeleman2013,Moller2013,Rafelski2014,Quiret2016}. The complex velocity
profiles suggest a highly turbulent environment, and models of clouds
moving in random orbits in galactic halos or thick disks
can generally explain the observations
\citep[][but see also \citealt{York1986}]
{Haehnelt1998,Fumagalli2011,Cen2012,Rahmati2014,Barnes2014b,
Bird2015,Neeleman2015}. The fact that several absorbing components are
typically seen along a given line of sight, each corresponding to clouds
moving at different velocities within a larger halo, implies that these
clouds are colliding with each other about once every orbital period 
\citep{McDonald1999}.
For individual DLAs it can be difficult to model the column densities
and velocity structure of the metal species due to a complex variety of
gas phases arising from photoionisation, shock-heating and collisions
leading to a broad range of temperatures and densities
\citep[e.g.,][]{Fox2007a,Berry2014,Dutta2014,Lehner2014,Neeleman2015,Cooke2015,Rubin2015}.

  In the context of the Cold Dark Matter model of structure formation,
the non-linear collapse of structure leads to hierarchical merging of
dark matter halos. The cosmological theory, starting from a matter
power spectrum that is now accurately determined from observations of
the Cosmic Microwave Background \citep[][and references therein]{Planck2015},
predicts the number density of halos as a function of halo mass that
exist at any epoch, $n(M_h,z)$. The observed rate of DLAs per unit of redshift
in any random direction due to halos of mass $M_h$ in the range $dM_h$
is then $n(M_h,z)\, \Sigma(M_h,z)\, c\, dt/dz\, dM_h$,
where $\Sigma(M_h,z)$ is the mean cross section (or area) within which
a DLA is observed in a halo of mass $M_h$. Although it has been
generally believed that DLAs are associated with dwarf galaxies
\citep[e.g.,][]{York1986,Dessauges2004,Khare2007,
Fumagalli2014,Webster2015,Bland2015,Cooke2015},
observations of the large-scale cross-correlation amplitude of DLAs with
the \lya forest absorption have determined their mean bias factor
$b_{\rm DLA}\simeq 2$ \citep{FontRibera2012}, which is consistent with
DLAs being distributed over a broad range of halo masses
$10^9 M_{\odot} \lesssim M_h \lesssim 10^{13} M_{\odot}$, from dwarf 
galaxies to halos of massive galaxies and galaxy groups.

  The Baryon Oscillation Spectroscopic Survey \citep[BOSS;][]{Dawson2013} of
the Sloan Digital Sky Survey-III \citep[SDSS-III;][]{Eisenstein2011} 
has obtained
spectra for more than $160\,000$ quasars at $z>2$, providing an
unprecedentedly large sample of DLAs. Despite the relatively low
resolution (R$\sim2000$) and S/N of the BOSS spectra, 
the large number of observed DLAs allows an accurate measurement of 
the mean metal-line absorption strength by 
stacking many systems, and studying the dependence of the
equivalent widths of any line on the hydrogen column density. This
approach has the advantage of directly providing mean properties 
of the DLA population, rather than properties of individual systems which have
a large intrinsic random variation. Although the study of individual
systems in detail obviously results in invaluable additional information
that is lost in a stacked spectrum of the global DLA population,
even a single DLA has absorption that probes a mixture of
different gas phases and is difficult to model in practice. Moreover,
absorption lines that are located in the \lya forest region can be
accurately measured only after averaging over a large number of
absorption systems, and they can provide important information that is
not accessible from lines on the red side of \lya \citep[see][for   
analysis of composite DLA spectra from 
BOSS]{Rahmani2010,Khare2012,Noterdaeme2014}.

  This paper presents the mean absorption spectra of metal lines in
DLAs that is derived from the Data Release 12 \citep{Alam2015} of BOSS,
using the DLA catalog generated with the technique described in
\cite{Noterdaeme2009}. The two main results are (\textit{i}) the mean
dependence of the equivalent width of each metal species on the hydrogen
column density, and (\textit{ii}) an analytical model of the mean
equivalent widths of multiple absorption lines of the same metal species
to account for the effect of saturation and derive mean column densities
in our DLA sample. We also present for the first time the mean
equivalent widths in DLAs of various species that are usually difficult
to measure owing to the confusion with the Lyman forest, e.g., SIV,
SIII, FeIII and NII, as well as accurate determinations of the mean
equivalent width and inferred column densities of more commonly measured
species like OVI.

  In \S~\ref{sec:data}, we present the method to calculate the mean 
quasar continuum spectrum and the DLA stacked absorption spectrum. 
We also detail the corrections we apply to improve the mean quasar 
continuum which, in turn, results in a more reliable stacked spectrum.
In \S~\ref{sec:results} we compute the mean equivalent 
width of detected metal lines, their dependence on the hydrogen 
column density is assessed and presented in \S~\ref{sec:nhi}, and in 
\S~\ref{sec:z} we address the dependence on DLA redshift. 
In \S~\ref{sec:model}, a simple model is proposed to correct for
line saturation and is used to infer the mean column 
densities of several low- and high-ionization species for which the mean
equivalent width of absorption lines has been measured.  
We discuss our results in \S~\ref{sec:discussion}, before summarizing
and concluding in \S~\ref{sec:summary}.

  All the equivalent widths in this paper are rest-frame.


\section{Data Analysis}\label{sec:data}
      
   We use the spectra of quasars in the SDSS-III BOSS 
Data Release 12 Quasar Catalogue `DR12Q' \citep{Paris2016}.
The SDSS telescope and camera are described in detail in 
\cite{Gunn1998,Gunn2006,Ross2012}, and the 
SDSS/BOSS spectrographs in \cite{Smee2013}.

\begin{figure*}     
\includegraphics[width=0.496 \textwidth]{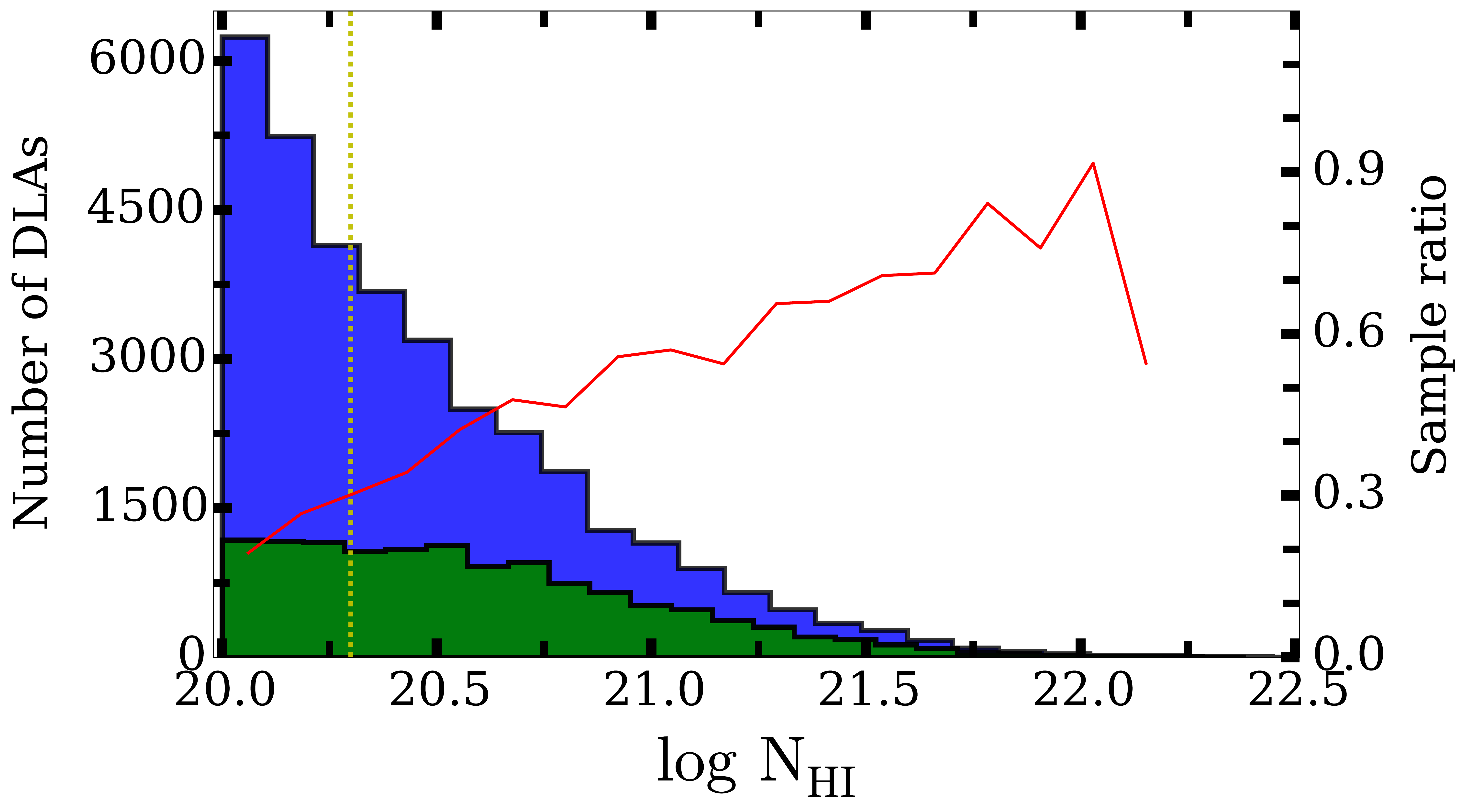}\hspace{5mm}\includegraphics[width=0.49 \textwidth]{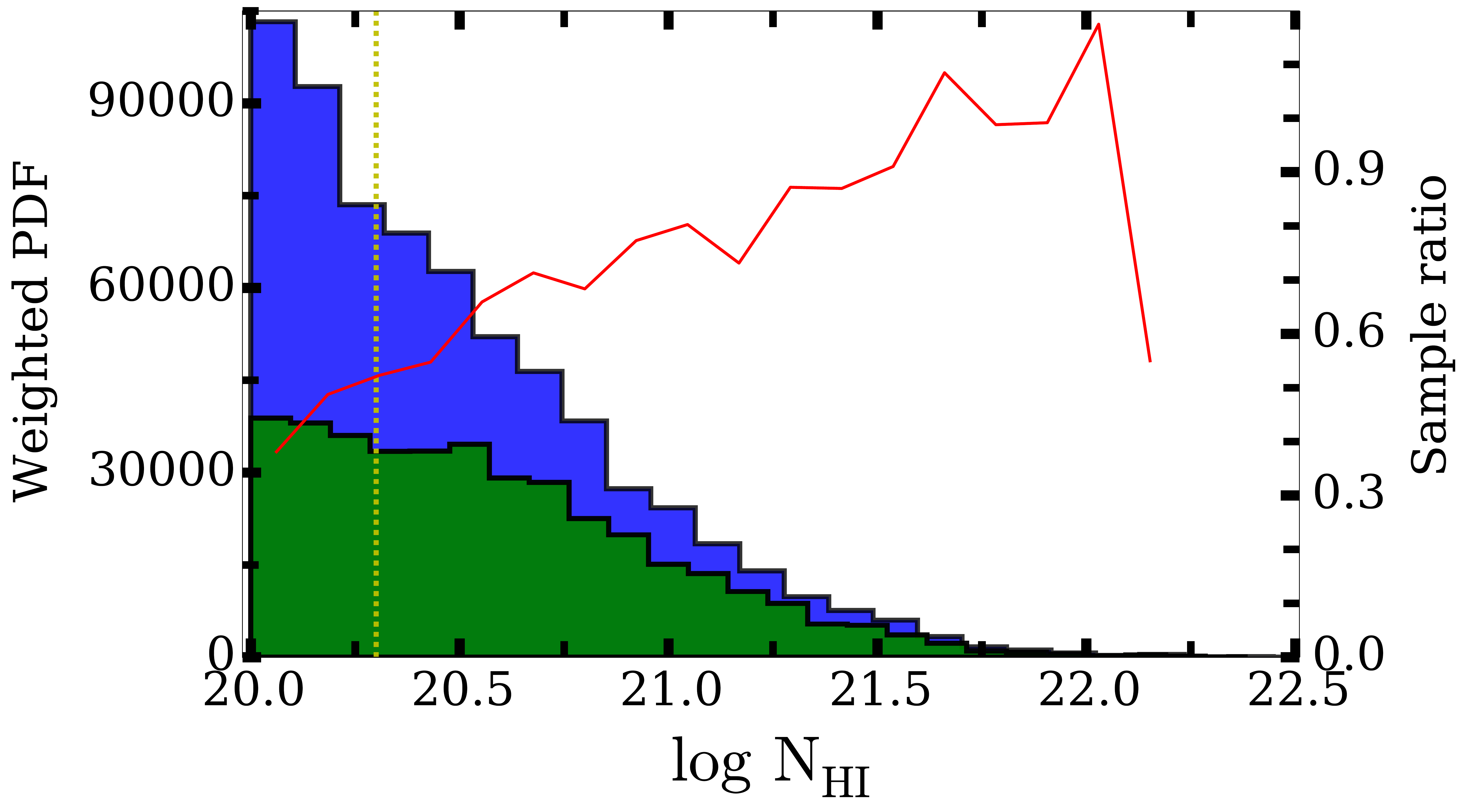}
\includegraphics[width=0.497 \textwidth]{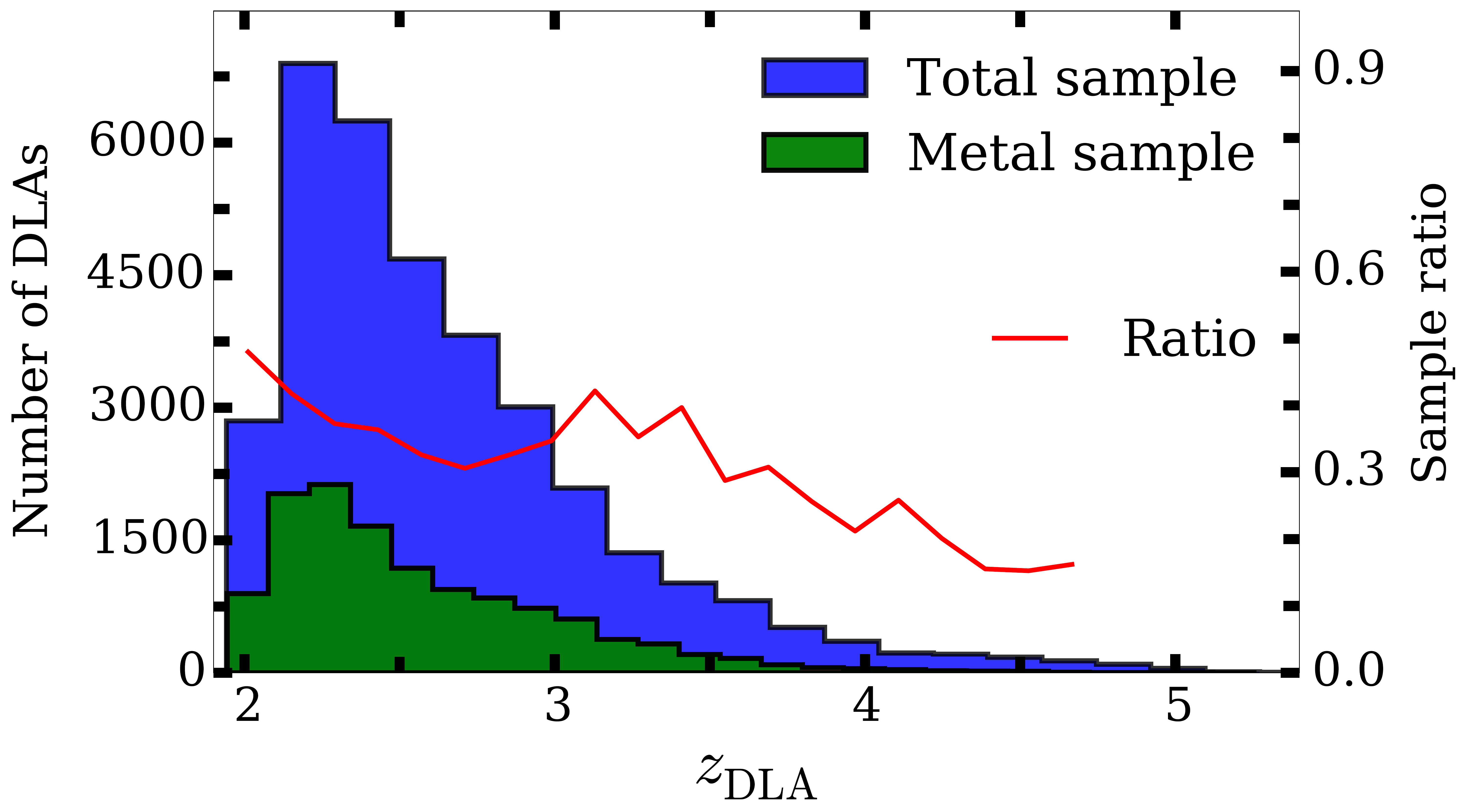}\hspace{5mm}\includegraphics[width=0.49 \textwidth]{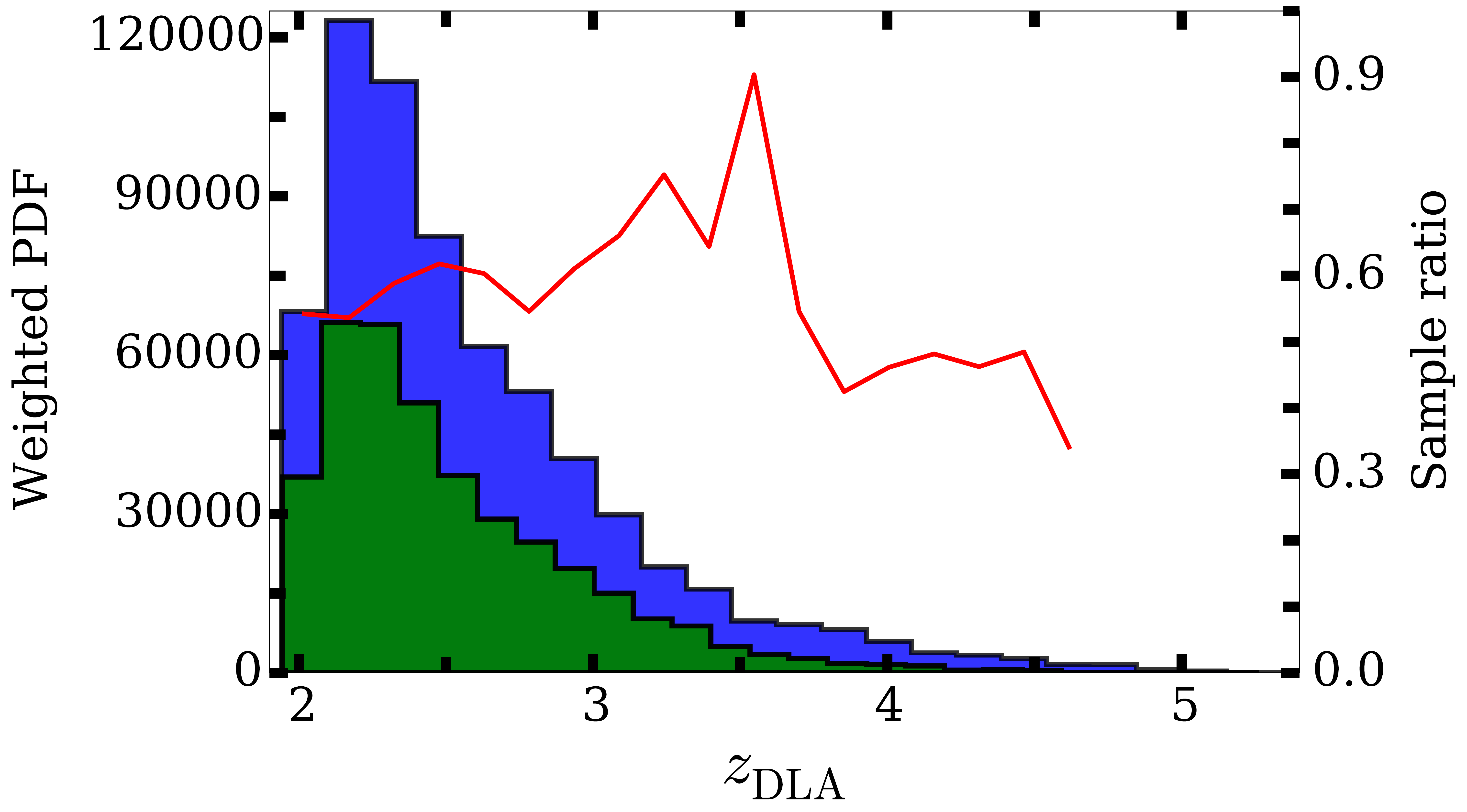}
 \caption{ {\it Left panels:} Neutral hydrogen column density (top panel) and redshift
(bottom panel) distributions for the total sample (blue historgram,
$34\,593$ DLAs) and the metal sample (green histogram,
$12\,420$ DLAs) of the entire DLA catalog. {\it Right panels:} Same distributions as in the left panel but 
considering only the systems that are used for the calculation of the composite spectra, after the 
application of our cuts, and weighting them according to the signal-to-noise calculation described 
in \S~\ref{sec:continuum}. These weighted total (blue histogram, $26\,931$ DLAs) and metal sample (green 
histogram, $10\,766$ DLAs) distributions are the true ones that give rise to the results presented in this work. 
In all panels, the red lines denote the ratio of DLAs between the metal and total samples, and  
the vertical dashed lines in the upper panels indicate the strict DLA column density lower limit, 
${ N_{\rm HI}=2\times10^{20}\,{\rm cm}^{-2}}$. }
\label{fig:distrib}
\end{figure*}

   We use the DR12 DLA catalog, which is the expanded version of the 
catalog presented in \cite{Noterdaeme2012} for the DRQ9 \citep{Paris2012},
and contains a total of $34\,593$ DLA candidates with a measured column 
density $\rm{N_{HI}} > 10^{20} \cm^{-2}$.
The detection of these systems is performed by means
of a fully-automatic procedure based on profile recognition using the
Spearman correlation analysis, as described in \cite{Noterdaeme2009}. 
Only $19\,376$ of the DLAs in the catalog have
$\log (N_{\rm HI}/{\rm cm}^{-2}) > 20.3$ and can therefore be designated DLAs if we strictly
use the standard definition of a DLA. This column density threshold was, however, set 
for observational purposes, and to ensure that the hydrogen gas in the inner regions of the 
DLAs is mostly neutral due to self-shielding from the background radiation \citep{Wolfe1986}. 
We include systems down to $\log (N_{\rm HI}/{\rm cm}^{-2})> 20$ because we find them 
to be also useful to characterize the mean properties of the population 
and their dependence on column density. We demonstrate below 
(\S~\ref{sec:curvenhi}) that the inclusion of systems with $\log 
(N_{\rm HI}/{\rm cm}^{-2}) < 20.3$ does not substantially change our results, although 
the mean equivalent widths do change with column density and redshift and, 
therefore, accurate comparisons with other stacked spectra in the future need to 
take into account our distribution of column densities and redshifts.
A relatively low minimum continuum-to-noise
ratio $C/N>2$ is required to include a DLA in the catalog, with the goal of
maximizing the size of the catalog without having a large number
of false DLA systems arising from spectral noise \citep{Noterdaeme2012}. 
This C/N threshold is specially important for systems with low column density since 
these have a higher probability to be false detections. 

  Most of the results presented in this paper are obtained from
stacks using the whole DLA catalog, which is designated here as  
{\bf total sample}.
This sample should be unbiased, in the sense that only the HI \lya
absorption line has been used to select the DLAs, and not the strength
of the metal lines. However, DLA samples from optically-selected quasars may be biased 
against systems containing substantial amounts of absorbing dust
\citep[e.g., ][]{Fall1993,Boisse1998,Ellison2001,Smette2005,Noterdaeme2015}, 
although the presence of dust in DLAs is expected to be 
small \citep[e.g.,][see also \citealt{Kulkarni2005,Fukugita2015,Krogager2016}]
{Pettini1997,Akerman2005,Vladilo2008,Khare2012,Murphy2016}. 
The presence of dust-biasing would have little impact on general HI 
studies \citep{Trenti2006,Ellison2008}, but it may significantly 
underestimate the metal content in DLAs \citep{Pontzen2009}. 
In addition, we also study a subsample of the
DLA catalog, which we call {\bf metal sample}, containing $12\,420$ DLA
candidates where metal lines can be individually detected by using templates 
(of these, $8\,699$ have $\log (N_{\rm HI}/{\rm cm}^{-2}) > 20.3$). The metal detection results from
a cross-correlation of the observed spectrum with an absorption template
containing the most prominent low-ionisation metal absorption lines,
which is done in addition 
to the previous Spearman correlation analysis for the \lya line.
 If one or more metal absorption lines are detected in the individual spectra, 
these are used to refine the absorption redshift of the DLA
\citep[see section 3.2 in][for further details]{Noterdaeme2009}.
The improved redshift of systems in the metal sample
gives rise to more sharply defined lines in the resulting stack, and
allows detection of  the weakest metal lines and measurement of 
 the effect of redshift uncertainties in the total
sample. However, the metal sample
is obviously biased in favor of DLA systems with strong metal lines 
and/or high S/N, and
therefore cannot be used to obtain mean properties of the true
population of DLAs. The absence of metal lines is never used to discard
candidate DLAs at a low signal-to-noise ratio from the total sample. 
The mean stacked spectra of the DLAs used
in each of these two samples are computed in a similar way, except 
for a few differences that are discussed below.

  The left panels in Figure \ref{fig:distrib} present the redshift and ${N_{\rm HI}}$
distributions of the total (blue histogram) and the metal sample (green
histogram). The ratio of the number of DLAs in the two samples is also
denoted as the red line, with the scale on the right axis. The metal
sample contains a greater fraction of high column density systems
because the strength of metal lines increases, on average, with ${N_{\rm HI}}$. 
The metal sample also has a smaller fraction of systems at high redshift,   
because of the declining mean metallicity of DLAs with redshift and  
the decreasing number of metal lines redwards of \lya that are 
observable with increasing redshift. In addition, the increased density of the 
\lya forest at high redshift may give rise to an increase of false positive DLA 
detections, specially in the total sample \citep[see, e.g.,][]{Rafelski2014}. 
The right panels display the same two distributions as before, but now considering 
only the systems that are finally used for the calculation of the composite spectra, after the 
application of cuts in the DLA sample and after weighting the spectra as described in the next 
subsection. The distributions on the right panels are the actual ones that give rise to 
the results of this paper, which should be used in order to precisely compare to any other 
future observational results or model predictions. The shape of the distributions in the right 
panels is similar to those in the left ones, but the metal sample has a higher contribution 
than in the left panels, in general. This difference is mostly because metal lines are more likely 
to be identified in high S/N spectra, which have higher weights, as explained in the next 
subsection, and because of the cuts that remove possible false DLAs in the catalog.
Additionally, the redshift 
distribution of the total sample narrows when applying the cuts and weighting the spectra.  
The mean values for the total sample in the right (left) panels are 
$\log { (\bar N_{\rm HI}/{\rm cm}^{-2})}=20.49$ (20.70) for column density, and ${\bar z=2.59}$ 
(2.65) for redshift.


\subsection{Continuum quasar spectrum calculation}\label{sec:continuum}

  A crucial part of computing a mean stacked spectrum of the
transmitted flux fraction for a sample of DLAs is the calculation of the
quasar continuum. We use a method that is similar to that in
\cite{PerezRafols2014}, who measured the mean absorption by Mg\,II
around the redshift of a galaxy near the line of sight to a quasar. Some
variations are necessary in our case, however, owing to the mean
absorption by the \lya forest and the presence of the DLA metal lines
themselves. We now describe in detail our procedure for estimating the
continuum.

  The method starts by computing a mean spectrum of the quasars 
used in both the total sample and the metal sample.
First, each quasar spectrum is shifted to the quasar rest-frame
wavelength, $\lambda_r = \lambda_{obs}/(1+z_q)$, where $\lambda_{obs}$
is the observed wavelength of every spectral bin, using the quasar
redshift $z_q$ provided in the DR12 DLA catalog
\citep[this is the visual inspection redshift of the
quasar catalog from][]{Paris2016}. The
values and errors of the flux are rebinned into new pixels of width
$1.0\, {\rm \AA}$ in the rest-frame by standard interpolation, averaging
the values in the original pixels as they are projected, partly or fully,
onto the new pixels. Any pixels affected by skylines, as reported in
\cite{Natalie2013}, are removed from the spectra and excluded from all
the analysis.

  The spectrum of each individual quasar is then normalized by
computing the mean flux in two fixed rest-frame wavelength intervals:
$1300 \, {\rm \AA} < \lambda_r < 1383\, {\rm \AA}$, and
$1408 \, {\rm \AA} < \lambda_r < 1500\, {\rm \AA}$. These intervals
are chosen to avoid the principal broad emission lines of quasars
and the region of the \lya forest absorption, and to be roughly
centered in the spectral range of interest for the DLA metal lines.
The normalization factor for each quasar $j$ is defined as
\begin{equation}
n_{j}= \sum_i{f_{ij} \over N_j} ~,
\end{equation}
where $f_{ij}$ is the flux per unit wavelength in the pixel $i$ of the quasar $j$, 
and the sum is done over all the $N_j$ pixels that are comprised within the two 
wavelength intervals for the normalization.
Some pixels in these two intervals are discarded because of the
skylines that are removed or because of additional corrections discussed
below (see \S~\ref{sec:correct}). Any quasar for which more
than 20\% of the pixels in the normalizing intervals are
discarded is removed from the sample. This results in the removal of
$1\,074$ quasars for the total sample and $298$ for the metal sample.

\begin{figure}       
\includegraphics[width=0.465 \textwidth]{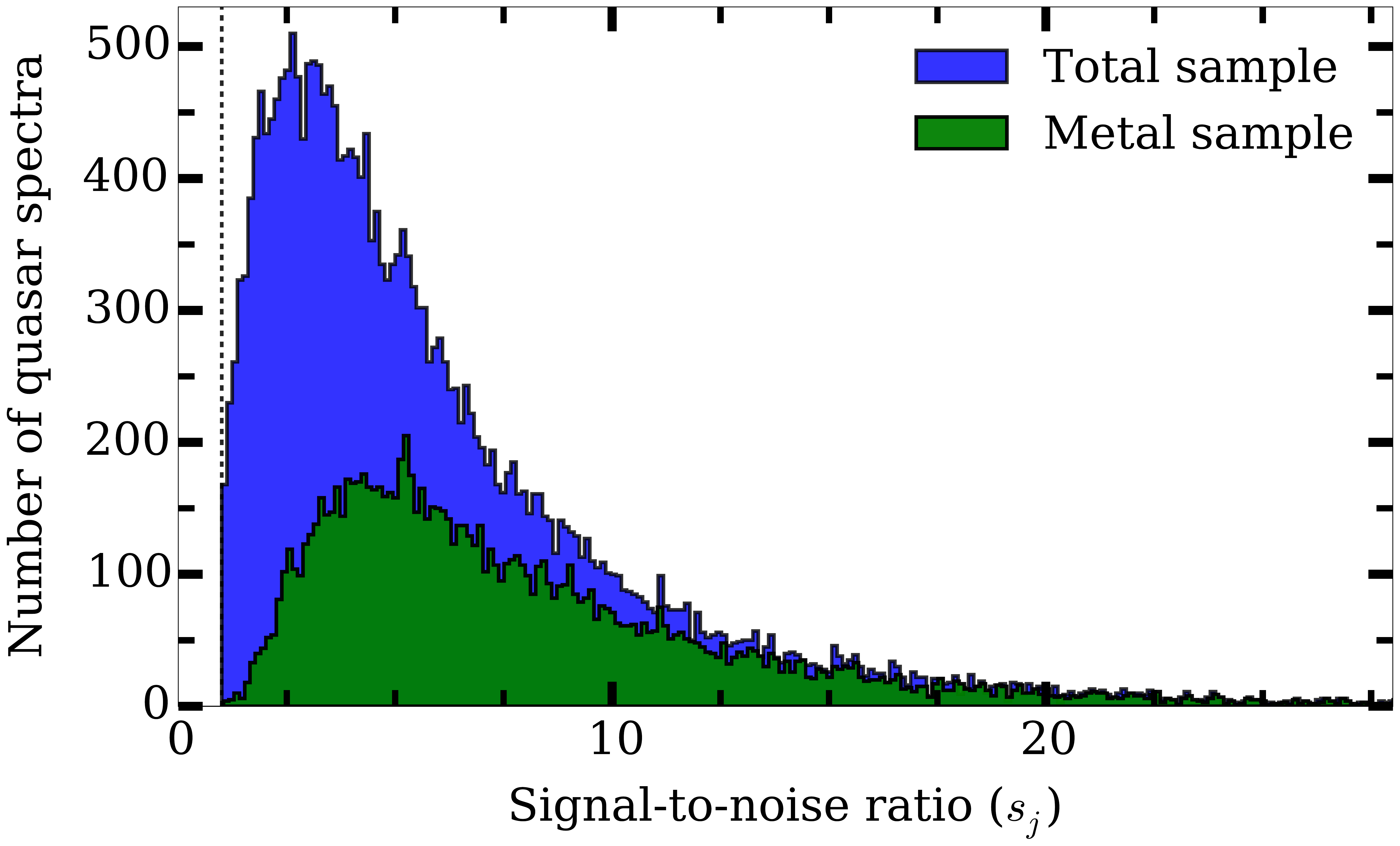}
\caption{Distribution of the mean S/N parameter $s_j$ for the total
sample (blue) and the metal sample (green). Spectra with $s_j < 1.0$
have been discarded. The vertical dashed line denotes  
the point where S/N reaches the threshold unity value.}\label{fig:snr}
\end{figure}

\begin{figure*}        
\includegraphics[width=1\textwidth]{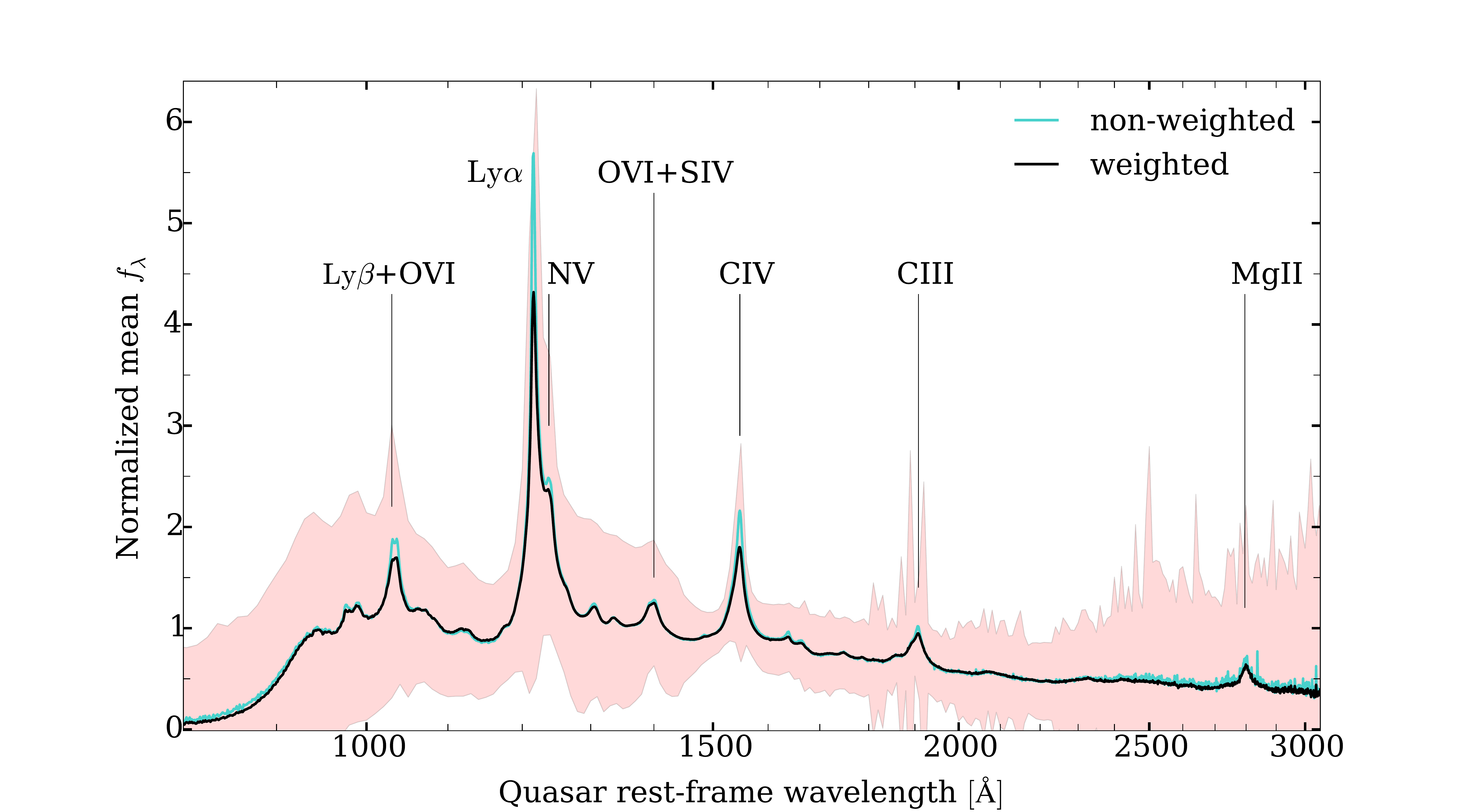}
\caption{Mean quasar normalized spectrum for the total sample. The {\it black 
line} represents our weighted mean spectrum. For comparison, the 
{\it cyan line} denotes the non-weighted mean, resulting in stronger emission lines. 
The {\it red region} illustrates the 68\% contours of the distribution of normalized 
spectra around the weighted mean. The wavelength scale is logarithmically spaced. 
We labelled the most prominent emission lines.}
\label{fig:mean}
\end{figure*}

  A mean S/N, $s_j$, is computed for each quasar spectrum using the
same two rest-frame wavelength intervals,
\begin{equation}
s_{j}=\frac{\sum_i{f_{ij}/N_j}}{\left(\sum_i{e_{ij}}^2/N_j\right)^{1/2}} ~,
\end{equation}
where $e_{ij}$ is the uncertainty for the flux $f_{ij}$. 
The resulting distribution of the $s_j$ parameter is presented in
Figure \ref{fig:snr} for the two samples. This distribution peaks at $s_j\simeq 2$ 
for the total sample, and at a higher value for the metal sample (as expected, 
because metal lines are more difficult to detect for low signal-to-noise). We discard 
from our sample any spectra with $s_j<1.0$, because of the very poor quality of these 
spectra, resulting in $264$ and $5$ spectra from the total and metal sample, respectively, 
not being further considered for our calculations. These discarded spectra are a small 
fraction of the total because of the independent constraint of a continuum-to-noise 
$C/N>2$ in the \lya forest region that was already applied to the DLA catalog with the 
method of \cite{Noterdaeme2009}.

  With the S/N values we assign a weight $w_j$ to each quasar spectrum,
defined as
\begin{equation}
\label{eq:weight}
w_{j}=\frac{1}{{s_j}^{-2}+\sigma^2} ~,
\end{equation}
where $\sigma$ is a constant that is introduced to prevent the quasars
with highest $s_j$ contributing excessively to the mean in the
presence of an intrinsic variability of quasar spectra, in addition to
observational noise. We choose, somewhat arbitrarily, a value
$\sigma = 0.1$, which represents our estimate that the typical 
intrinsic variability of quasar spectra is $10\%$. We ran the stacking using 
$\sigma = 0.2$ and saw that this difference do not produce 
any relevant effect in our resulting spectrum.

\begin{figure*}              
\includegraphics[width=1\textwidth]{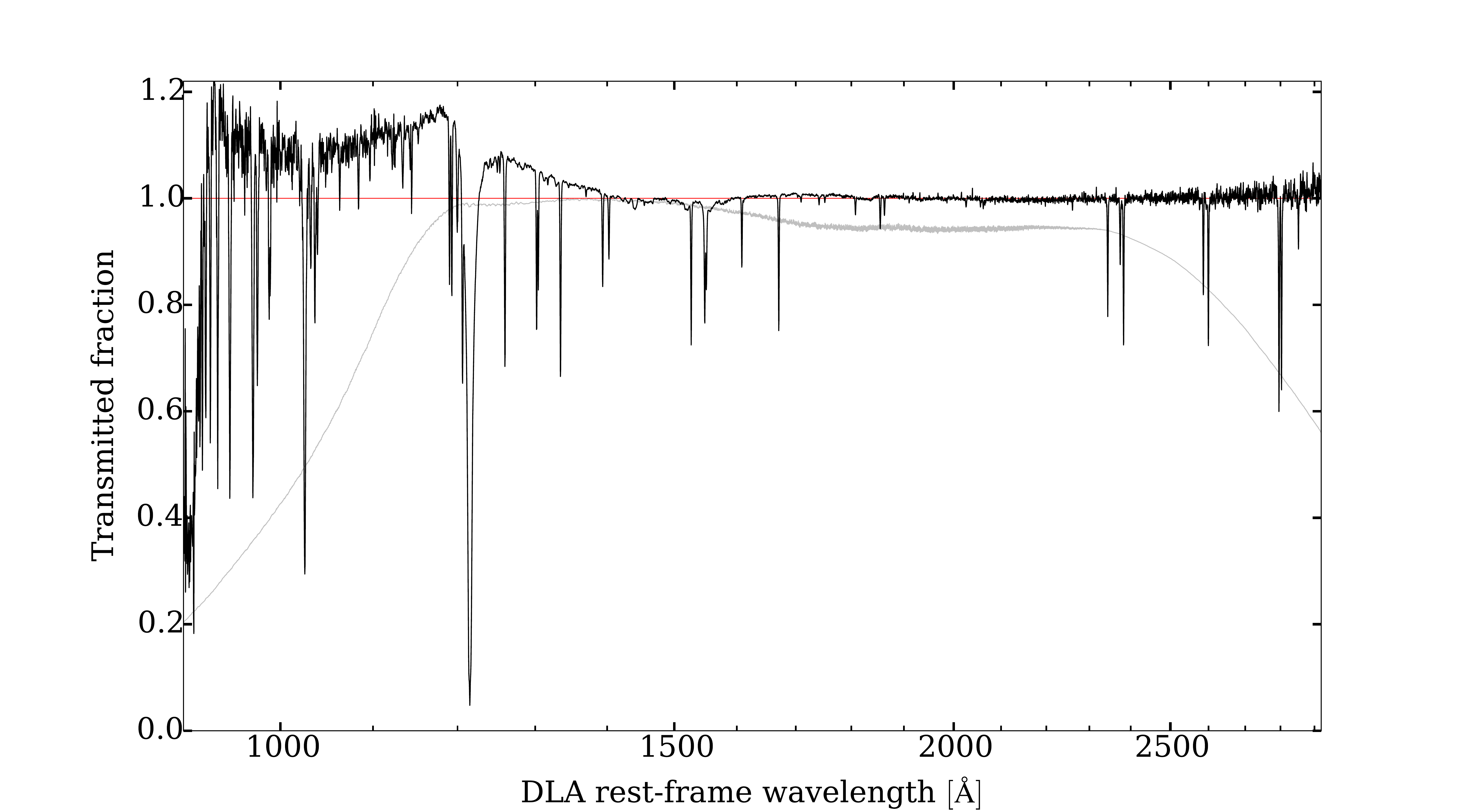}
\caption{Composite DLA transmission spectrum computed for the total
sample. 
The grey line indicates the fraction of the spectra used from the total sample 
contributing to the stacking at each wavelength pixel. The horizontal red line
denotes the value unity for the transmission. The wavelength scale is logarithmically 
spaced.}\label{fig:stack}
\end{figure*}

  Finally, the resulting mean normalized quasar spectrum is computed as a
weighted mean,
\begin{equation}
\bar{f_i}=\frac{\sum_jw_j(f_{ij}/n_j)}{\sum_jw_j} ~,
\label{eq:mnf}
\end{equation}
where $\overline{f}_i$ is the normalized flux per unit wavelength of the mean quasar
spectrum at the quasar rest-frame wavelength pixel $i$.
This mean quasar spectrum is displayed as the {\it black line} in Figure \ref{fig:mean}, 
with the most prominent emission lines labelled. This is the mean spectrum of quasars 
that have (at least) a DLA absorption system. Therefore, the mean spectrum incorporates any mean
modification that the DLA lines have produced. Some quasars ($\sim 20\% $ for both samples) have 
more than one DLA system in the catalog; in this case the same quasar contributes to the mean as 
many times as the number of DLAs it contains. The {\it red region} in Figure \ref{fig:mean} 
illustrates the 68\% contours of the distribution of normalized spectra around the mean. For 
comparison, we also plot the non-weighted mean, represented by the {\it cyan line}. This spectrum  
shows stronger emission lines compared to the weighted mean, because of the Baldwin effect: 
emission line equivalent widths decrease with quasar luminosity \citep{Baldwin1977}. A small 
difference is also present in the \lya forest region (barely visible in this plot), which is likely due to 
small variations of the distribution of S/N with redshift. Apart from this, we see that the {\it black line} 
is less noisy than the {\it cyan} one, which is our main reason to use the weights for computing the mean 
quasar spectrum.


\subsection{Composite DLA spectrum calculation}\label{sec:stack}

  To compute a stacked DLA absorption spectrum, we start by
shifting each quasar spectrum to the DLA rest-frame, rebinning 
now into a pixel width of $0.3\, {\rm \AA}$  to obtain better 
sampling. We also discard pixels affected by skylines, and
then divide by the previously computed mean quasar spectrum 
shifted to the same DLA
rest-frame, obtaining the transmission at the rebinned pixel $i$ of
the quasar spectrum $k$,
\begin{equation}
F_{ik}=\frac{f_{ik}/n_k}{\overline{f}_{ik}} ~,
\label{eq:trnorm}
\end{equation}
where $\overline{f}_{ik}$ now has the subindex $k$ labeling each DLA
only because the mean continuum $\bar f_i$ has been shifted and rebinned
on the DLA rest-frame. By means of error propagation, uncertainties are 
normalized and rebinned in the same
way to obtain the error, $E_{ik}$, of the transmission
spectrum of each DLA, $F_{ik}$. The final composite spectrum and its error
is again obtained from a weighted mean,
\begin{equation}
\bar{F_i}=\frac{\sum_kF_{ik}w_k}{\sum_kw_k} ~, \qquad
\bar{E_i}^{-2}=\frac{\sum_kE_{ik}^{-2}w_k}{\sum_kw_k} ~.
\end{equation} 
We set the weight $w_k$ to be the same as in Eq.~\ref{eq:weight},
with the same value of $\sigma=0.1$. This value does not
really have to be the same for computing the mean quasar continuum and the
mean DLA transmission spectrum; in general, we could choose a higher
$\sigma$ for stacking the DLA transmission because we need to take into
account the intrinsic variability of the DLA metal lines as well. As we did for 
the case of the continuum, we
have tested that increasing to $\sigma=0.2$ does not substantially alter the
results presented below; the two spectra have visually the same appearance. 

  The composite DLA transmission spectrum obtained after these 
calculations for the total sample is denoted by the black line in Figure 
\ref{fig:stack}. The grey line in this figure indicates the fraction of DLA
systems contributing to the estimated stacked spectrum at each
rest-frame wavelength bin. This fraction is less than unity 
on both sides of the range considered because a fraction of
the quasar spectra, depending on their redshift, do not extend along the
entire observed wavelength.
The mean of the Lyman series lines due to hydrogen and many metal
lines of our sample of DLAs are clearly evident. However, there are also
broad regions where the mean transmission deviates from unity, which are
clearly not associated with the narrow DLA absorption lines.
This deviation may be due to several effects: the spectra that 
contribute to a given wavelength for the quasar composite are not 
the same that contribute to a given wavelength in the DLA composite, thus 
the quasar continuum is not entirely cancelled out. In addition,
the \lya forest causes an absorption both in the mean quasar spectrum
and in the estimated transmission spectrum of each DLA, which does not
exactly cancel when dividing by the mean quasar spectrum due to the
redshift evolution of the \lya forest. 
In the next subsection, we assess and apply several corrections for these
effects in order to obtain a better quasar continuum and composite
spectrum for the two samples.  

   The mean metal line equivalent widths that we will obtain in this paper, which vary with 
column density and redshift, depend on the selection of our sample.  In addition, some 
fraction of the DLAs may be false and result from a concentration of \lya forest lines that, 
in noisy spectra, may look like a DLA, while others may have large errors in redshift. This should 
cause a reduction of the mean equivalent widths measured in our stacked spectra, implying that 
our results may be sensitive to our adopted cut in the $C/N$ ratio and the way we choose to 
weight the contribution of DLAs depending on the signal-to-noise ratio from Eq.~\ref{eq:weight}. 
We have tested the effect of eliminating the weights when calculating the composite spectrum 
(keeping the same sample of DLAs that we use after our cuts), which results in a very similar 
continuum, but substantially weaker metal absorption lines, with equivalent widths reduced 
typically by $\sim 30\%$. We believe most of this reduction is due to the fraction of false DLAs in 
the lowest S/N spectra, together with increased errors in redshift which wash out the metal lines 
in the stacked spectrum. Some of this reduction may also be due to a lower mean column density 
and higher mean redshift of the unweighted sample: The mean column density drops from 
from $\log (\bar N_{\rm HI}/{\rm cm}^{-2})= 20.49$ to 20.47, and the mean redshift increases from 
$\bar z = 2.59$ to 2.68, when eliminating the weights, but as we shall see in our results (Figures 
\ref{fig:nhi} and \ref{fig:z}, this accounts for only a small part of the reduction in mean
equivalent widths when removing the weights.

     However, by maintaining the weights, the contribution of false DLAs 
in low signal-to-noise spectra should be greatly reduced, and our 
systematic underestimate of the metal lines mean equivalent widths 
should be much less than 30\%. We have tested this by examining 
variations with the minimum threshold in $C/N$ to accept DLAs in our 
sample. We find that the median relative increase of equivalent widths 
of metal lines analysed in this paper is $2.5\%$, $3.7\%$ and $5\%$ as 
the minimum $C/N$ is raised to 3, 4, and 5, respectively. These 
fractional variations are, we believe, a fair estimate of our systematic 
errors caused by impurity and large redshift errors in our DLA sample. 
Again, part of this increasing mean equivalent width with $C/N$ may be 
caused by an increasing mean column density and decreasing redshift with 
$C/N$. In any case, this suggests that our weighting scheme is useful to 
reduce the effect of impurity in the DLA catalog, and that the remaining 
systematic reduction of mean equivalent widths is at the level of 
$\sim$ 5\%. 

   For illustrative purposes, we have created two movies displaying the evolution of the 
mean quasar and composite DLA spectra as the number of stacked objects is increased. 
The two movies are publicly available and can 
be found, together with a brief description of the calculations, in the url 
\url{https://github.com/lluism}.


\subsection{Corrections on the continuum spectrum}\label{sec:correct}

  We now present the corrections that we apply to improve our first
version of the transmission spectrum in Figure \ref{fig:stack}. 
Briefly, these corrections consist of detecting and removing bad pixels,
correcting for the mean \lya forest absorption, and correcting for the
average effect of the DLA lines on the continuum spectrum. In addition,
we describe the procedure applied to spectra where Ly$\beta$ absorptions
can be mistaken for Ly$\alpha$. All these corrections, described 
in detail below, are applied equally for the total and metal sample.


\subsubsection{Detection and removal of outliers}\label{sec:outliers}

  A variety of effects, e.g., cosmic rays, may cause large deviations
of the flux in a few pixels from the correct values that clearly set
them as outliers from the normal noise distribution.
We prefer to eliminate these outliers, rather that working with median 
values which reduce the sensitivity to outliers, because we want to obtain 
mean equivalent widths in the end, and a relation of median to mean values 
would be model-dependent. 
Outliers cannot be eliminated by simply setting a maximum noise deviation 
of the transmission from the expected range of zero to unity, 
because the intrinsic variability of the quasar spectra can be large, implying 
that a more generous transmission range should be allowed.

  Therefore, for the purpose of eliminating outliers, we first obtain
a fitted continuum, $C_{ik}$, to the transmission $F_{ik}$ computed previously for
each quasar. The detailed method we use to
fit this continuum is described below in \S~\ref{sec:wcalc};
essentially, $C_{ik}$ is a smoothed version of $F_{ik}$ computed once the
regions of the expected metal lines of the DLA have been excluded.
We then eliminate all pixels with a transmission $F_{ik}$ that is
outside the interval $[C_{ik}-2-3E_{ik}$ ,\, $C_{ik}+2+3E_{ik}]$, where
$E_{ik}$ is the transmission uncertainty in each pixel defined after Eq.~
\ref{eq:trnorm}. This is a generously broad range, which allows for an
uncertainty in the pixel flux of three times the estimated standard deviation
$E_{ik}$, and adds an additional variation of twice the normalized mean
quasar spectrum. Despite this broad range, it still excludes the most
important outliers without eliminating any pixels that are not obviously
bad. After the outliers are eliminated, we recalculate a new mean quasar
continuum and a new stacked DLA absorption spectrum, which we 
adopt as the new composite spectrum in our analysis. 


\subsubsection{Mean absorption of the \lya forest}\label{sec:lyaforest}

  The \lya forest causes a systematic, redshift-dependent flux 
decrement bluewards of the \lya emission line of the quasar. 
The redshift evolution of this
decrement means that if it is not corrected, its mean value at a
certain wavelength in the DLA rest-frame in our composite absorption
spectrum is generally not equal to the mean value of the decrement in
the mean quasar spectrum that was used to obtain the transmission from
the observed flux, leaving a residual effect. This residual effect consists  
in an increase of the transmission in the DLA composite spectrum bluewards 
of the \lya feature, more important for longer wavelengths.

  We use the fit obtained by \cite{Faucher2008} for the mean fractional
transmission as a function of redshift,
\begin{equation}\label{Fz}
       F_{\alpha}(z) = {\rm exp} \left[ -0.0018(1+z)^{3.92} \right] ~.
\end{equation} 
We divide the normalized flux, $f_{ij}/n_j$ in Eq.~\ref{eq:mnf}, in
the spectrum of each quasar by $F_{\alpha}(z)$ at the
redshift $z=\lambda_{obs}/\lambda_\alpha - 1$ (where $\lambda_{\alpha}=
1215.67 \, {\rm \AA}$), before stacking to obtain the mean quasar
spectrum. We then divide again each spectrum containing
a DLA by $F_{\alpha}$ at the same redshift, before dividing by the mean
quasar spectrum in Eq.~\ref{eq:trnorm}. The net correction does
not completely cancel, and depends on the probability distribution of
the DLA and quasar redshifts in our catalog, which is sensitive to
selection effects reflecting the DLA detection probability.

\begin{figure*} \center             
\includegraphics[width=0.9\textwidth]{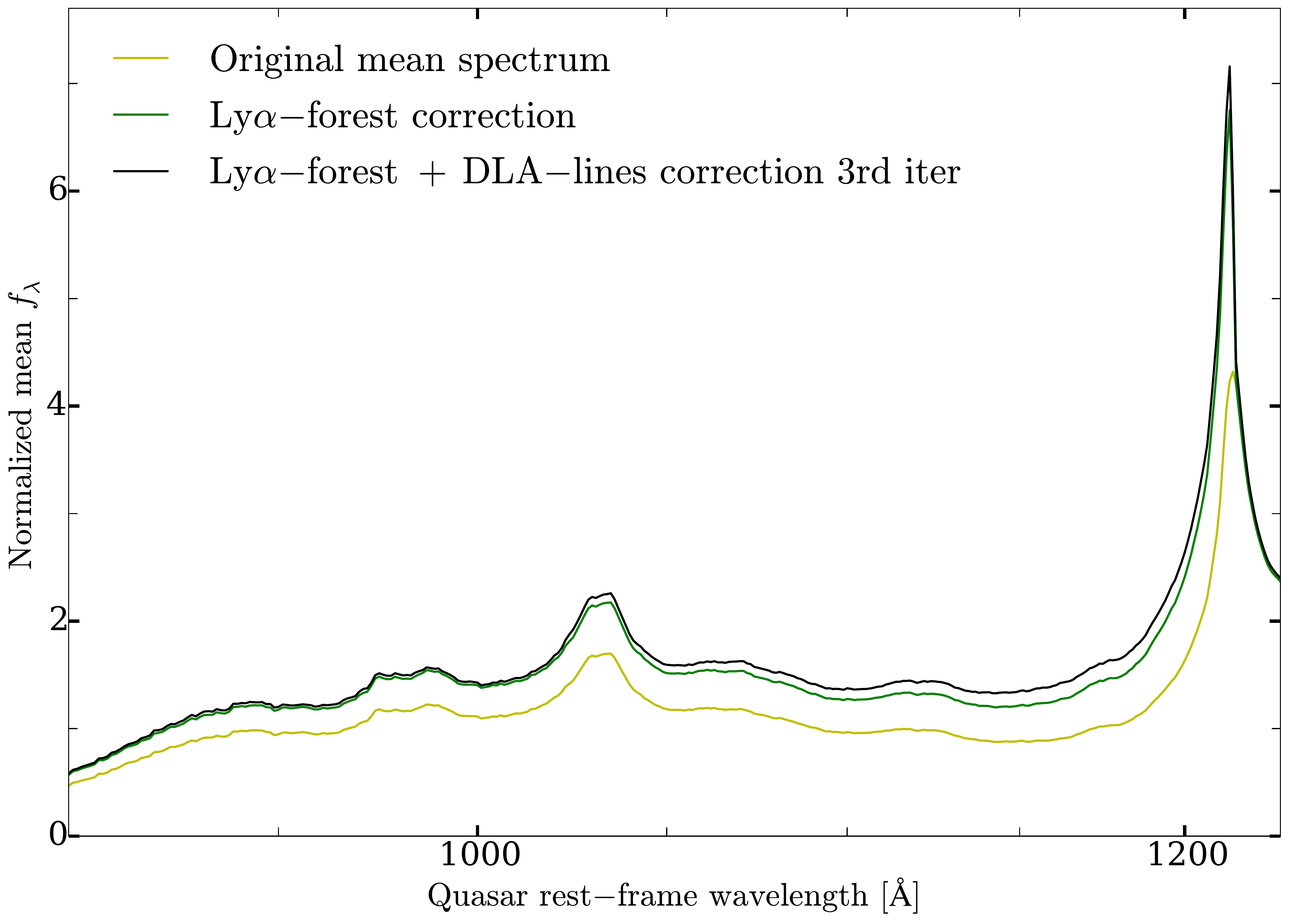}
\caption{Original mean quasar spectrum before corrections (yellow), after
correcting for the \lya forest absorption (dark green), and after the
third iteration of the correction for DLA absorption lines (black).
Modifications redwards of the \lya emission line are small
so only the \lya forest region is shown here.}
\label{fig:meancorr}
\end{figure*}

\begin{figure*}\center                  
\includegraphics[width=0.9\textwidth]{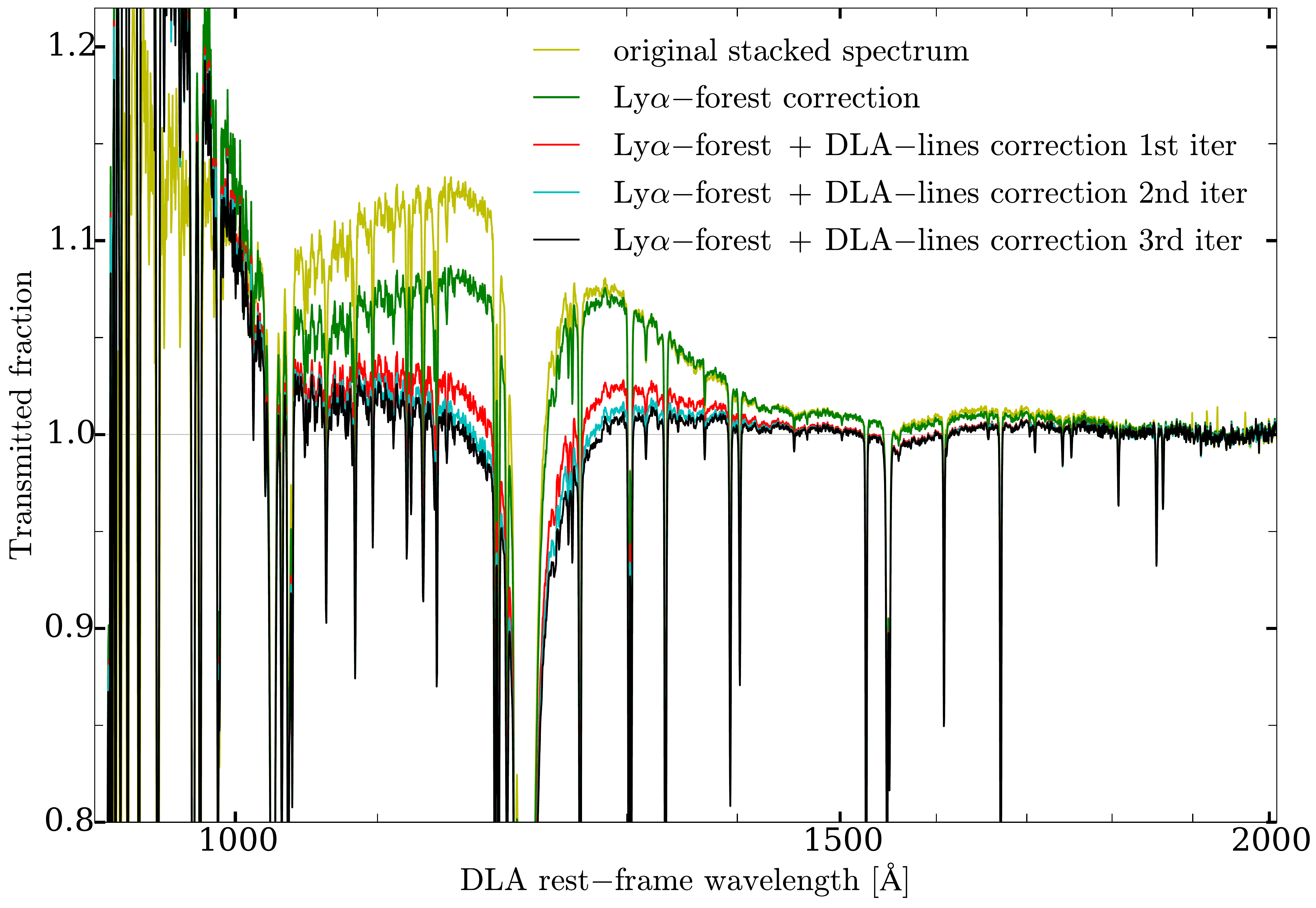}
\caption{Stacked DLA spectrum before and after corrections.
The original stacked spectrum, with some outliers causing deviations in
a few pixels near $1\,900$ \AA$\,$ and in the \lya forest region,
is shown in yellow. The spectrum after correcting for the \lya forest
mean decrement and the removal of outliers is in dark green. The DLA
lines correction is included in the red, blue and black lines, after the
first, second and third iteration, respectively.}
\label{fig:stackcorr}
\end{figure*}

  Figure \ref{fig:meancorr} displays the impact of the correction for the 
\lya forest transmission on the mean quasar continuum. The original
spectrum (yellow line) is raised by $\sim$ 20\% after this correction
(green line),
reflecting the mean decrement at the mean redshift of our quasar
sample. The change of the composite DLA spectrum after applying this
correction is presented in Figure \ref{fig:stackcorr} (yellow and green
lines for the original and corrected spectra, respectively). Both figures 
display the result for the total sample.

  Eq.~\ref{Fz} includes the effect of the \lya forest only. Metal
lines are believed to increase the decrement by $\sim 5\%$ of that due
to \lya \citep{Faucher2008}. We neglect these metal lines, and we also
do not include the additional corrections for the forests of Ly$\beta$
and higher-order hydrogen lines, although we do correct for the effect
of all the lines associated with the DLAs themselves
that are stacked in the final composite spectrum, as we describe below.


\subsubsection{Effect of the DLA absorption lines}\label{sec:lines}

  When we compute the mean quasar continuum, there is a mean 
incidence rate of hydrogen and metal lines of DLAs that are present in
each quasar spectrum and contribute to lower the measured flux. This
average flux decrement caused by the DLA lines is larger than in a random
sample of quasars because we use only the quasars in the DLA catalog to
obtain our mean quasar spectrum, where by construction, each quasar
spectrum contains at least one DLA. We now describe the procedure to
correct for the presence of these DLA lines, which is applied both to
the total and metal sample.

  For each DLA in the catalog, we redo the quasar continuum spectrum
calculation after modifying the flux values in pixels that are comprised
within predefined windows around each of the expected DLA absorption
lines. We also remove some of these pixels where the DLA absorption is
strongest. We account for all the DLA metal absorption lines listed in
Appendix \ref{sec:ew}, in Tables \ref{ta:low} and \ref{ta:high}, as well as for
the blended lines listed in Table \ref{ta:blended}. We include in
addition the hydrogen Lyman transitions from \lya up to $n=9$,
covering the range between $920$ \AA \,and $3\,000$ \AA.
The wavelength windows around each line 
center used for this correction are the same as those used 
for the computation of the equivalent width of the metal lines,
described in detail in \S~\ref{sec:wcalc} below. 

  For each DLA absorption line, the flux in each pixel within its
window is corrected as 
\begin{equation}\label{line}
       {f_{L}} = \frac{{f_{QSO}}}{{\overline{F}_D}},
\end{equation} 
where $f_{L}$ is the corrected flux, $f_{QSO}$ is the observed flux
before correction in the quasar rest-frame at redshift $z_q$, and
${\overline{F}_D}$ is the transmission in the mean DLA composite
spectrum, after shifting to the quasar rest-frame by multiplying the
wavelength by $(1+z_{DLA})/(1+z_q)$. 
The DLA composite spectrum is rebinned in order to match 
the quasar rest-frame, in the same way as in \S~\ref{sec:stack}
when computing the mean quasar spectrum.
This correction is applied only within each DLA absorption line window,
when $1.0 > \overline{F}_D \ge 0.4$. For the strongest DLA lines, 
pixels where the DLA mean transmission is $\overline{F}_D < 0.4$ in the
composite spectrum are removed instead of being corrected, and not taken
into account for calculating the improved quasar continuum. We adopted this
approach to avoid excessive noise from the regions that are highly absorbed.
Small variations for the threshold value $0.4$ do not significantly change 
our results. This correction is applied without considering  
detecting any DLA line in the individual spectra, to correct for their
mean expected absorption.

  The new mean quasar continuum is used to recalculate 
the DLA composite spectrum. We can now iterate the same procedure,
since the DLA composite spectrum is needed to compute the correction
to each quasar continuum, until there is no significant improvement.
This convergence is reached after three iterations. The black line in Figure 
\ref{fig:meancorr} shows how the mean quasar continuum is further
modified by these DLA lines in the region of the \lya forest (this
is mostly the effect of the DLA \lya line), and Figure
\ref{fig:stackcorr} indicates how the DLA composite spectrum is improved
after the first, second and third iterations (red, blue and black
lines, respectively; the spectrum is displayed only at $\lambda < 2000$ \AA, at longer
wavelengths the corrections to the continuum are very small).
A clear improvement is seen after the first iteration (red
line), in the sense that the continuum between the DLA metal lines
moves closer to unity over most of the regions, but a smaller
improvement is achieved with subsequent iterations.

  Despite the improvements that result from these corrections, 
Figure \ref{fig:stackcorr} demonstrates that the continuum still
deviates slightly from unity due to other uncorrected
systematics. One contribution is probably the proximity effect,
which accounts for the fact that the \lya forest near the quasar redshift has
a lower mean decrement than far from the quasar redshift. 
There is also a small rise of the continuum above unity in the region
longwards of the CIV line, which may be partly caused by the forest of
CIV lines associated with the \lya forest and Lyman limit systems.
Further improving our continuum model would clearly require more
detailed work to correct for these effects and other systematics that
are likely present. We have decided to stop here and to use a
simple method to flexibly fit the continuum between the DLA lines in
the next section.

  The correction involving the effect of the DLA absorption
lines assumes that there is only one DLA in each quasar spectrum. For each
DLA in the catalog, the quasar spectrum is corrected for the presence
of only that DLA, ignoring the possible presence of other DLAs in the
same spectrum. A more accurate procedure would take into account all the
detected DLAs in each quasar spectrum. As mentioned above, 
$\sim 80\%$ of the spectra in both samples contain only one DLA, so we
expect spectra with more than one DLA to produce a small
effect on the mean quasar continuum. 

   Finally, in order to compare our method for correcting the effect of the 
DLA lines, we adopt a different approach. We calculate the mean quasar 
spectrum, now using all the spectra in the DR12 quasar catalog from BOSS 
that do not contain DLAs. After our cuts, this results in the use of $\sim210\,000$ 
quasar spectra. The continuum of this mean quasar spectrum overlaps with our 
previously computed mean spectrum, except in the Ly$\alpha$ forest region. The 
difference between the two corresponds to a $\sim10\%$ increase of the flux in the 
forest for the sample without DLAs. Our correction ({\it black line} in Figure 
\ref{fig:meancorr}) increases the flux depending on wavelength, from $\sim5\%$ 
at $\sim1050$ \angs\, to $\sim10\%$ at $\sim1150$ \angs. Therefore, our correction 
might simply account for a fraction of the total DLA effect but, because of other 
possible contributions to the observed difference (e.g., the strength of the emission lines 
are different in the two samples), we consider our original approach for further 
calculations. This will not affect our results because of the use of the additional 
continuum fit to the final composite spectrum.

\subsubsection{Contamination by DLA Ly$\beta$ lines mistaken for \lya}
\label{sec:lyb}

  When we first obtained the DLA composite spectrum,
we noticed the presence of a few regions with unexplained anomalous
absorption features. These can be seen in Figure \ref{fig:stack}, where
the spectrum for the total sample has strange spectral features, for
example, near $1440\,{\rm \AA}$ or $1520\,{\rm \AA}$ (just to the left
of the SiII line), which do not appear in the spectrum of the metal
sample. The source of these features
is that some Ly$\beta$ absorption lines are
incorrectly identified as DLA \lya lines in the DLA catalog. This error 
produces spurious absorption features in the stacked spectrum and other
undesired effects (the features are not present in the metal sample
because the DLAs identified at a wrong redshift obviously do not
yield any metal line detections). To avoid this problem, we have
removed all the DLA spectra with a redshift
${(1+z_{DLA})\le 27/32\,(1+z_q)}$, ensuring that the detected \lya lines
cannot possibly be a Ly$\beta$ line of a higher redshift DLA. The
amplitude of the fictitious features caused by these DLAs indicates that
only $\sim 10\%$ of them are incorrect identifications of a Ly$\beta$
line, but we have not further attempted to separate these mistaken
detections in order to avoid any other selection effects in our total
sample. This criterion reduces the number of DLAs
in our total sample to $26\,931$, and to
$10\,766$ in our metal sample, to which the whole analysis described
above has been applied. In the rest of the paper, we analyze the results
obtained for these restricted catalogs.

\section{Analysis of the Composite DLA Spectrum}\label{sec:results}

   We measure the mean equivalent width of metal lines 
detected in the stacked DLA absorption spectra below.
We also divide the two DLA samples into five column density bins
to assess the dependence of these mean equivalent widths on
the hydrogen column density, ${{ N_{\rm HI}}}$, in \S~\ref{sec:nhi}.
The results are tabulated in
Table \ref{ta:blended}, and Tables \ref{ta:low}
to \ref{ta:highnhi76} in Appendix \ref{sec:ew}.

\begin{figure*}                  
\center
\includegraphics[width=0.75\textwidth]{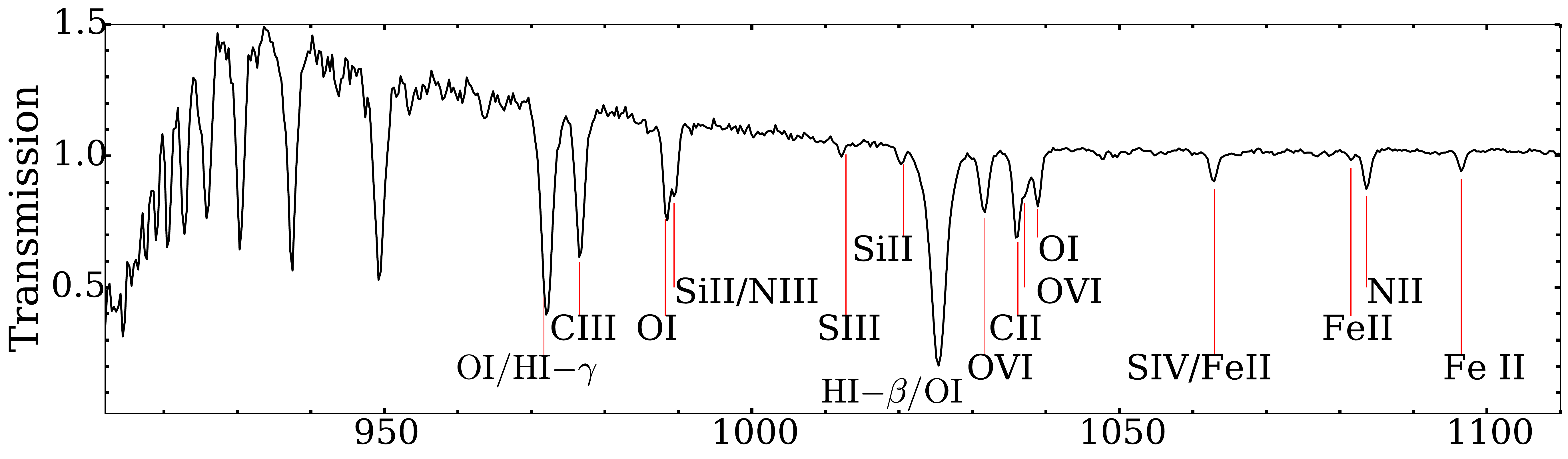}
\includegraphics[width=0.75\textwidth]{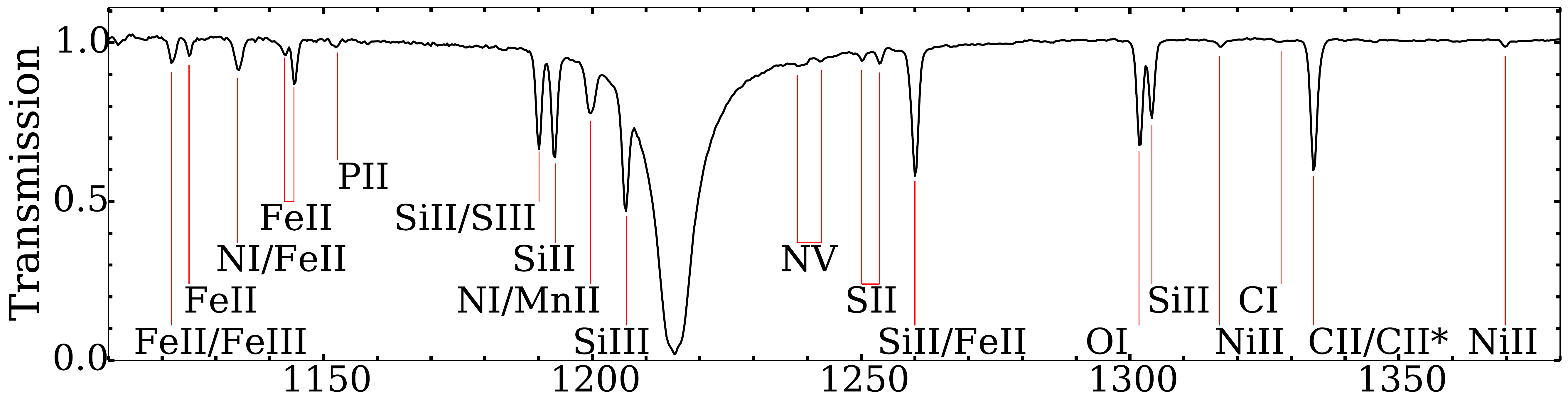}
\includegraphics[width=0.75\textwidth]{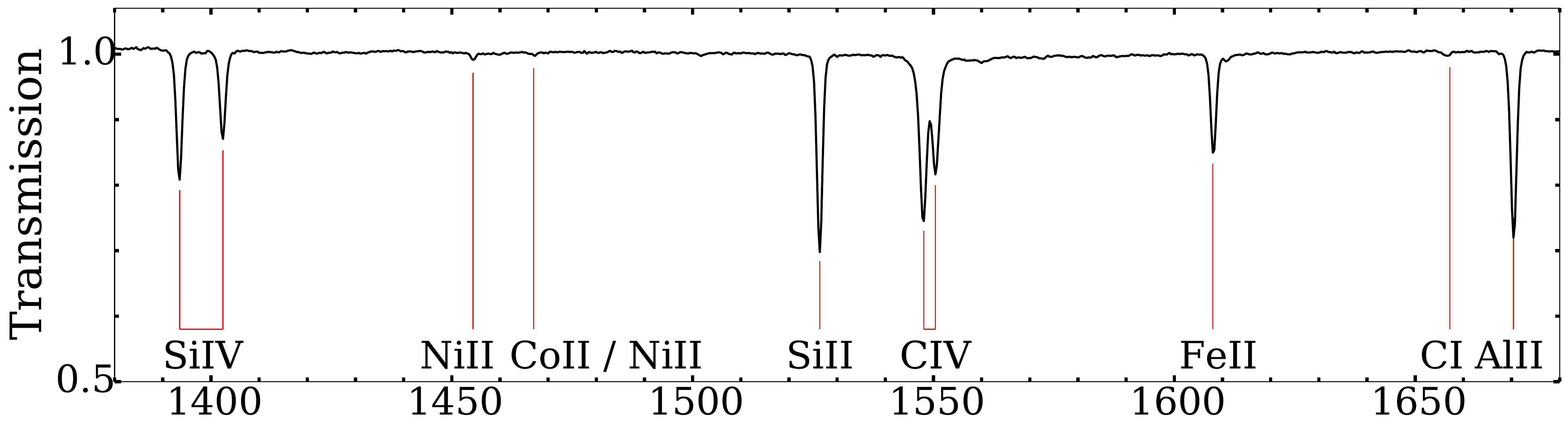}
\includegraphics[width=0.75\textwidth]{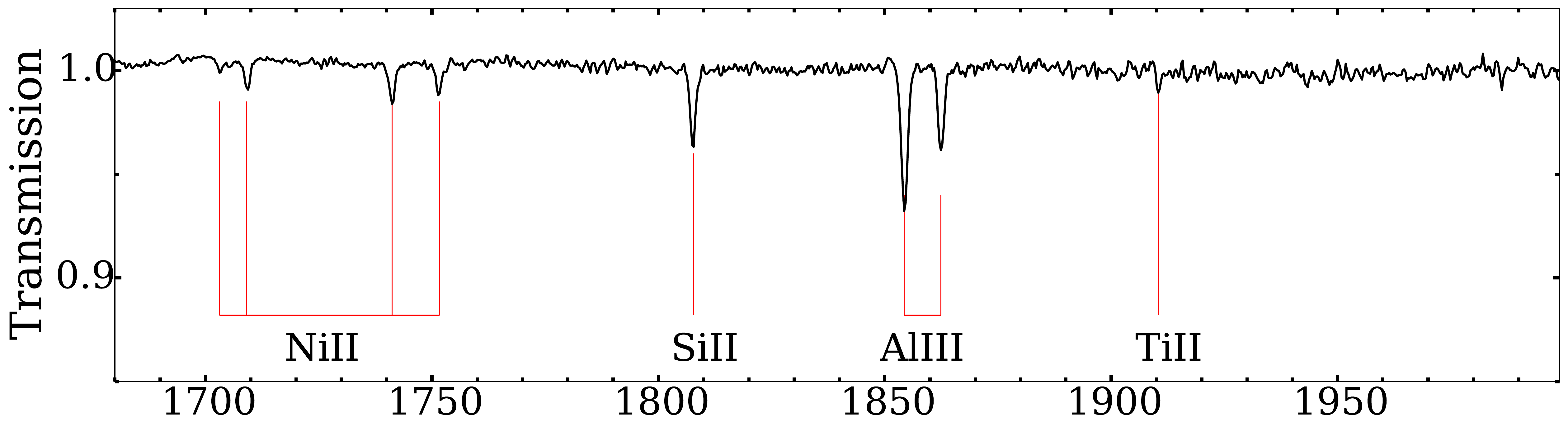}
\includegraphics[width=0.75\textwidth]{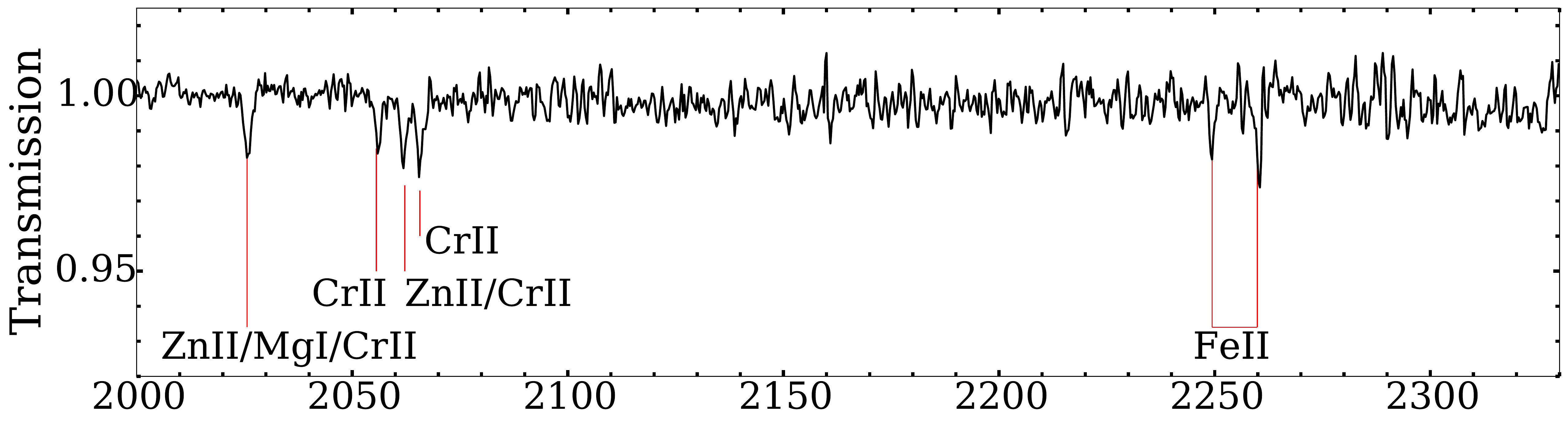}
\includegraphics[width=0.75\textwidth]{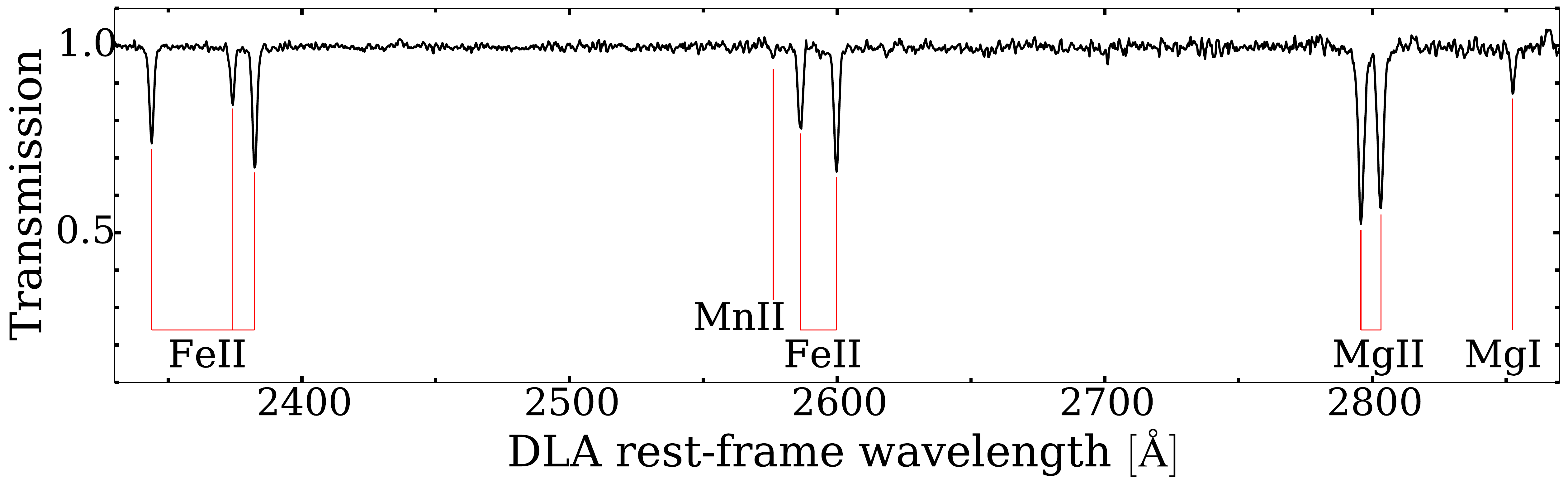}
\caption{Expanded version of the final composite DLA transmission
spectrum for the \textbf{total sample}. Labels under the red vertical
lines denote the species. Only lines with wavelengths longer than
$970 \, {\rm \AA}$ are shown. Lines at shorter wavelengths are severely blended 
with the \lya forest,
and the only features that are clearly visible are the higher-order Lyman
series of hydrogen. Each panel has a different scaling for the transmission, selected
to enhance the visibility of the lines.}
\label{fig:lines24}
\end{figure*}

\begin{figure*}                  
\center
\includegraphics[width=0.75\textwidth]{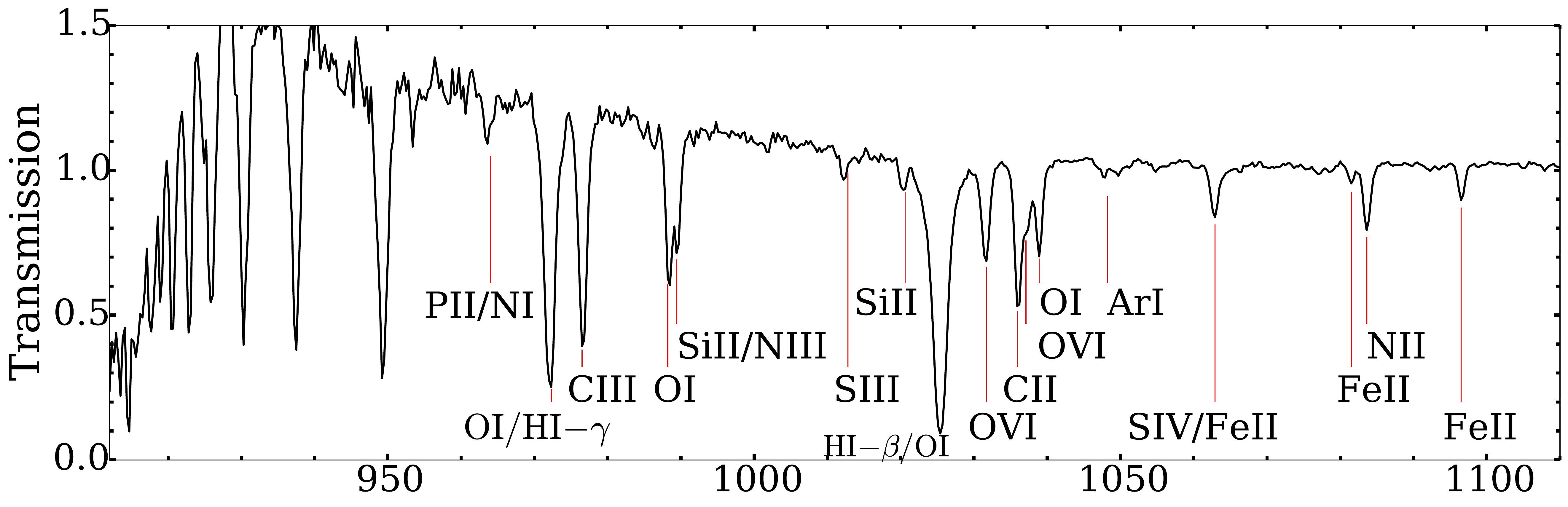}
\includegraphics[width=0.75\textwidth]{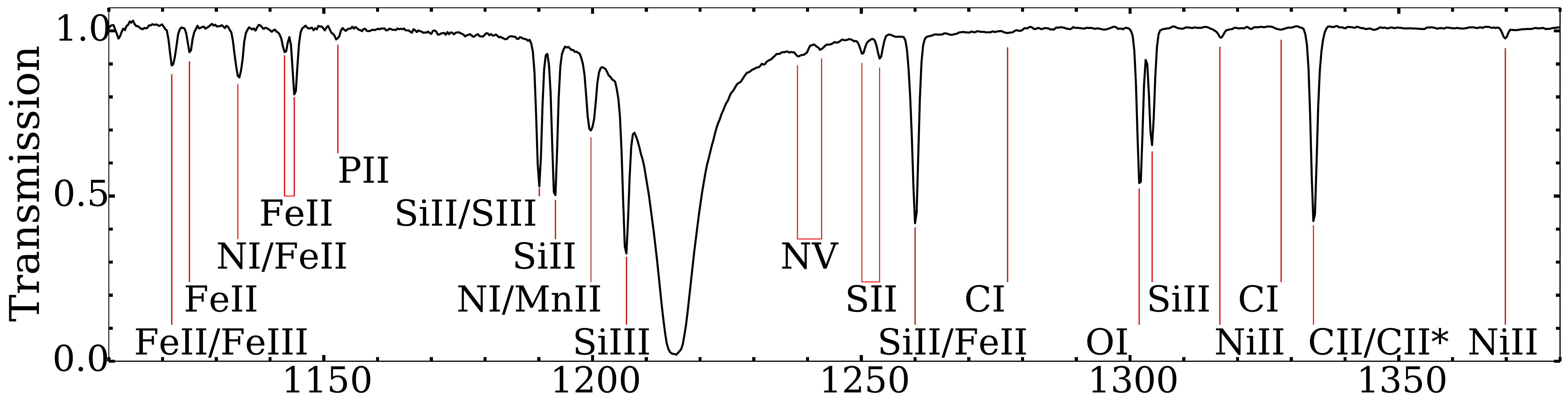}
\includegraphics[width=0.75\textwidth]{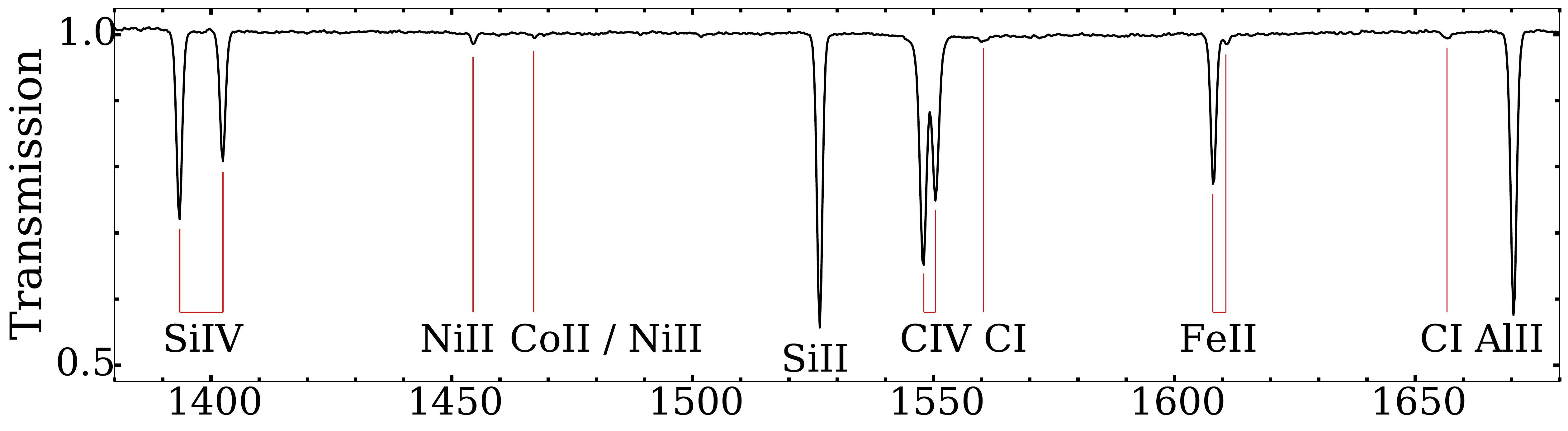}
\includegraphics[width=0.75\textwidth]{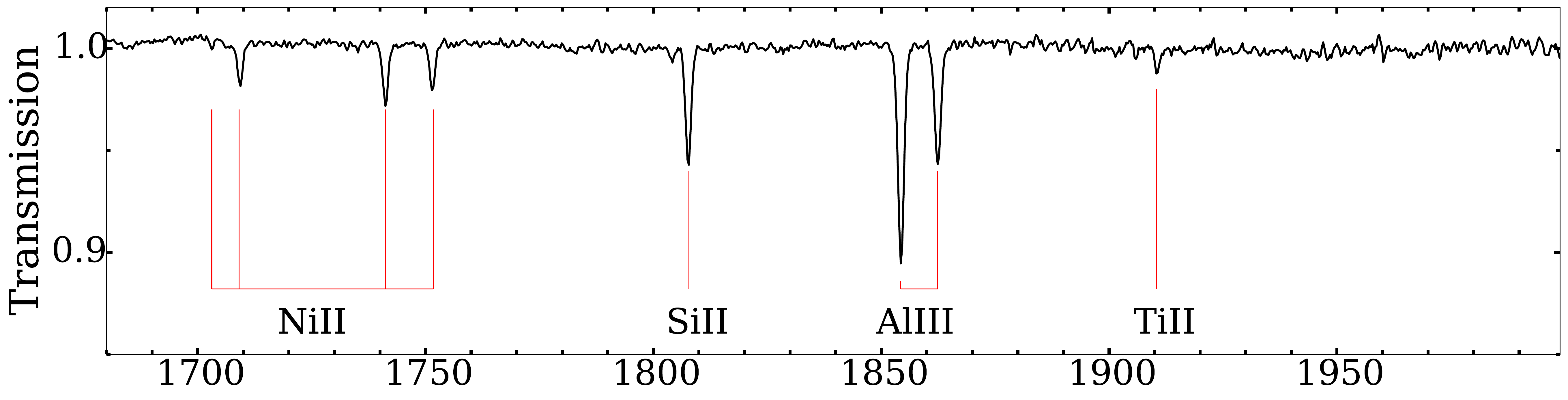}
\includegraphics[width=0.75\textwidth]{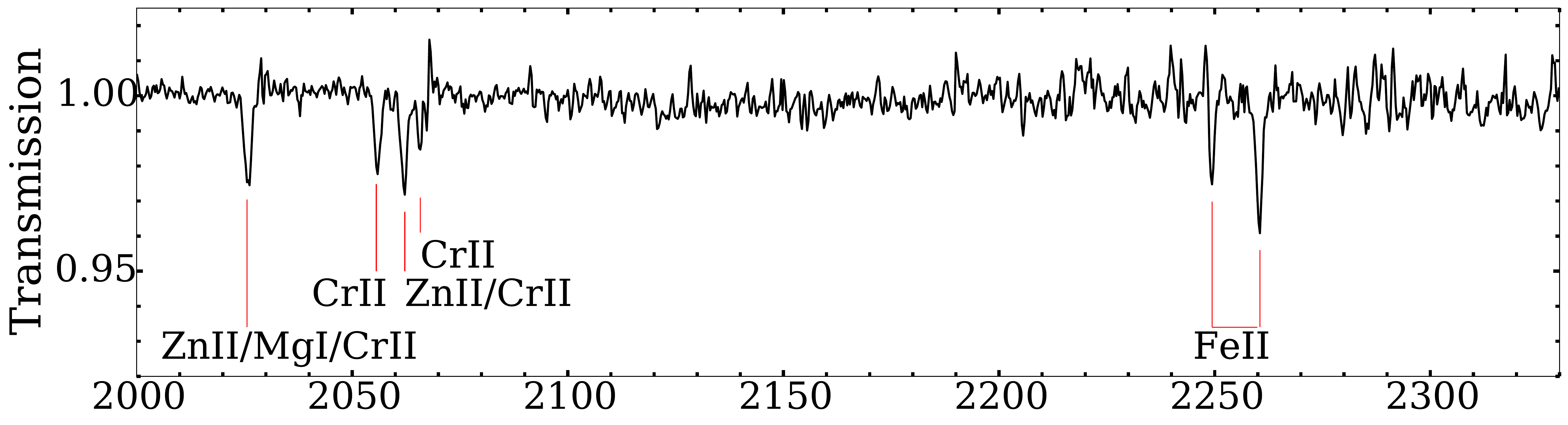}
\includegraphics[width=0.75\textwidth]{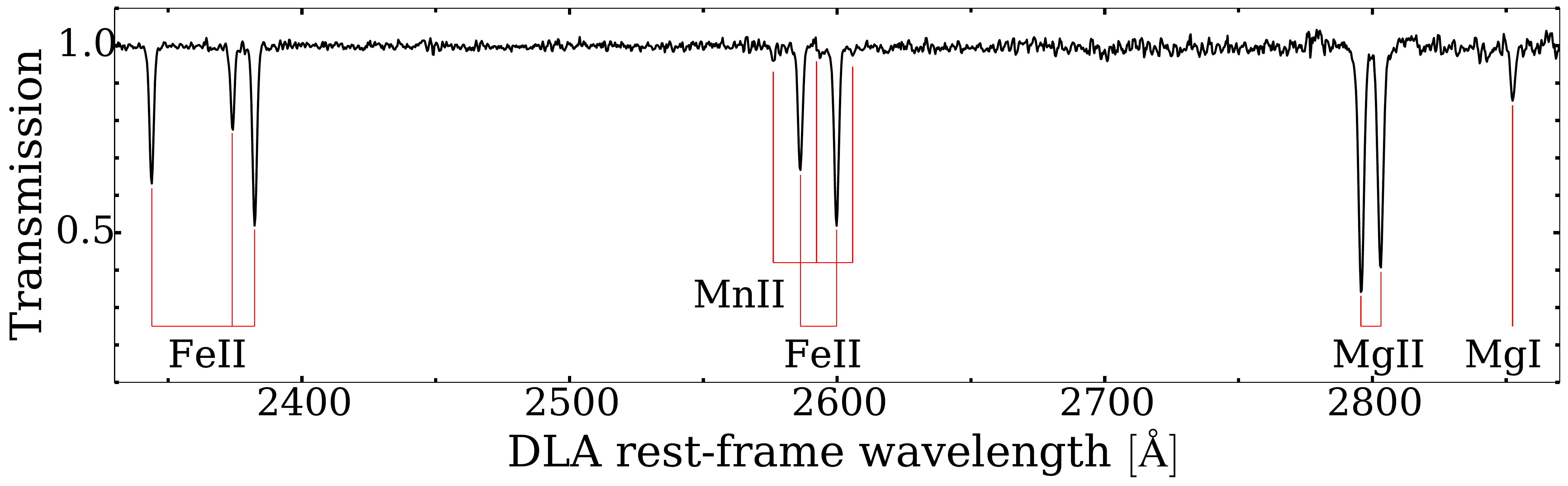}
\caption{Same as Figure \ref{fig:lines24}, here for the
\textbf{metal sample}.}
\label{fig:lines76}
\end{figure*}

\begin{figure*}              
\includegraphics[width=1\textwidth]{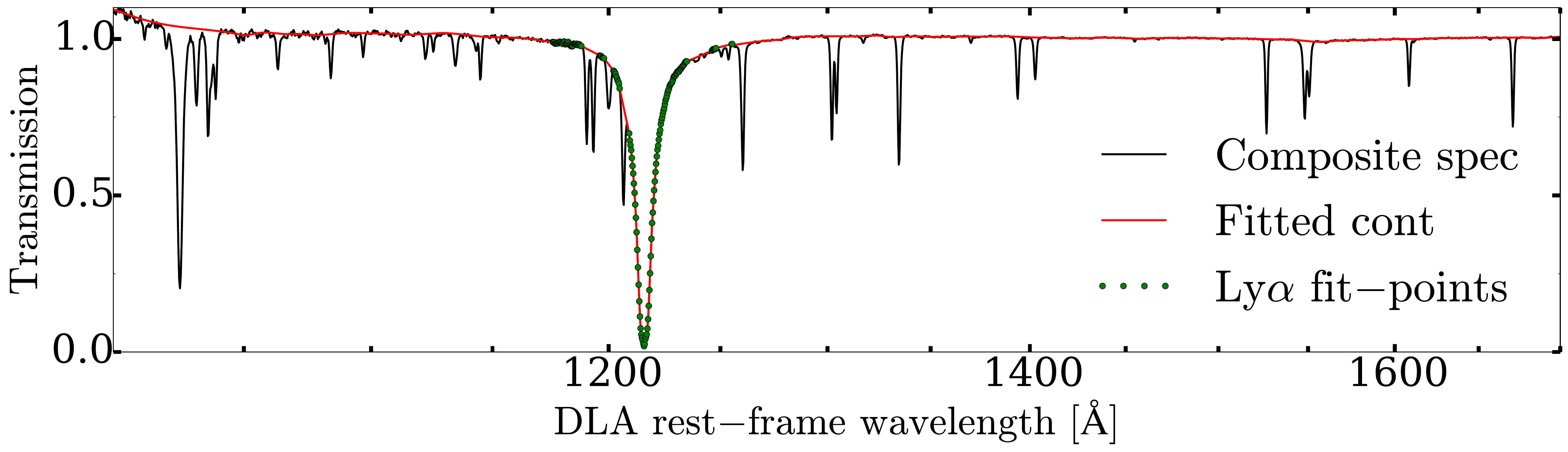}
\caption{Fitted continuum (red line) used for the calculation of the
rest-wavelength equivalent width of metal lines and detection of outliers 
over the stacked spectrum of the total sample (black line), shown over
the range $1000-1700\angs$. The pixels used for the linear fit in the
\lya line region are denoted as green circles.}
\label{fig:lyafit}
\end{figure*}

  Figures \ref{fig:lines24} and \ref{fig:lines76} present an expanded
version of the final composite absorption spectrum of DLAs, for the
total and the metal sample, respectively, with labels denoting the
detected absorption lines. 
Absorption lines in the metal sample (Figure \ref{fig:lines76}) are
usually stronger and present a sharper profile than those in the total
sample, allowing for more detections of weak absorption features 
despite the noise. This is simply because the metal sample is selected 
to include only DLAs with individually detected metal lines, so DLAs of 
low metallicity and/or low $C/N$ spectra are usually excluded, reducing 
the noise in the final stacked spectrum. 

  The \lya absorption feature in Figure \ref{fig:lines24} does not present 
a broad flat region with null flux at the position of the line center as observed 
in single systems. This effect is due to two reasons: First, there may be some 
zero-flux error with a mean transmission of $\sim 0.01$ that
is systematically added to all the 
spectra. Second, the DLA sample is not totally pure, and there may be a few
percent of the absorbers catalogued as DLAs that are actually arising as a
combination of lower column density absorbers and spectral noise, contributing
also a residual flux at the \lya line profile of the stacked spectrum. For the metal 
sample, where the percentage of false DLAs should be much
smaller, there is a flatter bottom of the DLA mean profile that is consistent
with the width of the damped profile expected for the lowest column densities
present in the catalog. This may also affect the metal lines: as mentioned 
earlier, our equivalent widths may be systematically underestimated because of a 
fraction of DLAs in the catalog that are not real or have large redshift errors.
The zero-flux error may also cause an additional underestimation of metal line 
equivalent widths, but the relative error should be similar to the fraction of zero flux 
present at the bottom of the \lya line in Figures \ref{fig:lines24} and \ref{fig:lines76}, 
which is only a few percent. We ignore these systematic errors in this paper; future 
improvements on this work should use mocks of the \lya spectra in BOSS to correct 
for the fraction of false DLAs in our catalog, as well as the zero-flux error \citep[see, 
e.g.,][]{Bautista2015}.

   We detect transitions of elements rarely seen in DLAs, such as TiII,
CII*, PII, CoII, ArI, and several lines of CI, a species that is
associated with ${\rm H_2}$ \citep{Srianand2005,Ledoux2015}. We also observe 
several high-ionization lines, including NV, OVI and SIV, which are extremely 
difficult to detect in individual spectra; in fact, NV and SIV have been
detected only in a few DLAs, and when they are detected they are hard to
separate unambiguously from the Lyman forest lines, particularly for
SIV and OVI lines which are always blended with the forest
\citep{Fox2007a,Lehner2008,Lehner2014}. 
Here, we will analyse a total of 42 low-ionization lines, 8
high-ionization lines, and 13 absorption features that are the result of
blends of several metal lines that are unresolved 
at the BOSS resolution. These lines are listed in Tables \ref{ta:low} and 
\ref{ta:high} in Appendix \ref{sec:ew} and in Table \ref{ta:blended}. The 
uncertainties in their equivalent widths indicate that not all of them are 
detected at a high confidence level, particularly in the total sample. We 
now describe how the equivalent widths and uncertainties are evaluated.


\subsection{Line windows and fitted continuum}
\label{sec:wcalc}

  We select the set of lines described above to measure their equivalent
widths. These features were chosen simply from their visual appearance 
to have a detection in our composite spectrum of the metal sample. These
lines, in addition to the first 8 hydrogen Lyman transitions (all of
them with wavelengths within
the range $920\, {\rm \AA} < \lambda < 3\,000\, {\rm \AA}$), are the 
ones used for the computation of the fitted continuum described below, as 
well as for the quasar continuum correction presented in Section
\ref{sec:lines}.
 
  Before equivalent widths can be measured, a set of windows around each
line need to be defined over which the absorption fraction is
to be computed. We generally choose a total window width of
$7\, {\rm \AA}$ centered in the DLA rest-frame line center,
which is wide enough to include all appreciable absorption for any
unblended line. The line profiles have all nearly the same widths
because they are unresolved, thus 
the equivalent width is, in practice, the only information that can be
obtained from these line profiles. We have tested that the measured equivalent
widths do not vary significantly under small variations of the
window widths. The Ly$\alpha$ and Ly$\beta$ transitions are treated
differently because they have a clearly resolved mean absorption
profile. We use halfwidths of $40$ and $5\, {\rm \AA}$ for their
windows, respectively. In addition, whenever several lines have
overlapping windows, we define broader windows which include all the
individual windows of the lines in the blend
(this is described in detail below, in \S~\ref{sec:blended}).

  Before measuring the equivalent widths, we perform a final continuum
fitting of the DLA composite spectrum to remove the residual variations
left after the corrections discussed in the previous section. This method 
is similar to that applied by, e.g., \cite{Pieri2014,Sanchez2015,Berg2016}; we proceed
in the following manner: for each pixel outside any of the line windows
described above, we compute the mean value of the transmission within a
$10\, {\rm \AA}$ width window centered on the pixel,
excluding any pixels that are inside the absorption line windows 
(pixels belonging to skylines or outliers have already been removed
and do not contribute to the fitting calculation). We then compute a
standard cubic-spline fit over the range $900\,-\, 3100 {\rm \AA}$ in
the DLA rest-frame, using only one (starting from the first) out
of every $15$ of these mean values for the transmission 
(pixels in the DLA stacked spectrum have widths of $0.3\angs$, 
therefore the averaging of the transmission is done over about 33 pixels, so
all pixels contribute to the determination of this final continuum).
The 15 pixels separation between successive points used for the spline
fitting corresponds to a distance of $4.5\, {\rm \AA}$ when there are no
absorption lines or other effects which can discard pixels in between. 
We have checked that using 20 or 10 pixels instead of 15 makes
no substantial difference. This approach produces a smoother continuum 
compared to using the averaged flux in all pixels, which
produces undesired `waves' over the regions of the absorption
lines. 

  Despite averaging the continuum in $10\, {\rm \AA}$ width windows,
this new fitted continuum is still affected by the pixel noise near the
window edges. The statistical error that this effect introduces is
accounted for with the bootstrap method described in Section
\ref{sec:bootstrap}, but any other possible systematic effects on
equivalent widths introduced by our method are not included in our
errors. The resulting cubic spline fit is used as the new continuum 
to calculate equivalent widths and limits on the detection of outliers 
described in \S~\ref{sec:outliers}. 

  The equivalent width calculation for the case of lines within the
\lya window needs special attention. We determine a continuum that
includes the \lya absorption, with the goal of being able to measure
the equivalent widths of metal lines that are blended with the \lya line. 
For the purpose of computing the continuum in the \lya line region, 
we ignore the previously defined metal line windows, and select
instead the following windows (in units
of \AA) that appear to be free of absorption by metal lines in both
the total and metal samples stacked spectra:
$[1175.67-1188],\,[1196-1198],\,[1202-1205],\,[1209-1235],\,
[1246-1248],\,[1255-1255.67]$. These intervals are selected as a
compromise for maximizing the number of points used for the continuum
and minimizing points near the metal absorption lines.
We then use linear interpolation to connect all the pixels throughout
this region, connecting also linearly the points at the window edges.
This approach yields our fitted
continuum over the region $[1175.67-1255.67\,{\rm \AA}]$.
Outside this range, we use the previously described cubic spline fitted
continuum.

  The final fitted continuum is displayed as the red line in Figure
\ref{fig:lyafit} over the $1000-1700\angs$ range. Green points are 
used to fit the linear continuum in the \lya line window. 

  We stress that our stacked \lya line does not have the shape of a single Voigt 
profile because it arises from a superposition of DLAs with different column densities. 
We have not attempted to fit the observed profile by modeling it with a column density 
distribution for our sample, in view of the zero flux error and other complicated effects 
(e.g., the cross-correlation of DLAs and the \lya forest). Ignoring this statement and 
forcing a fit to a single Voigt profile, we obtain a column density ${\rm \log (N_{HI}/cm^{-2})=
20.49}$, which coincides with the mean column density of our sample (with the weights applied 
in our stacked spectrum) as measured from 
the individual systems. In future work, it should be interesting to fit the mean profiles of all 
the Lyman series lines, which should contain valuable information on the distribution of 
velocity dispersions in the DLA systems. 

\subsection{Equivalent width estimator}
\label{sec:wcalct}

  We next fit a Gaussian optical depth line profile to each metal line
within its window,
\begin{equation} \label{eq:fitting}
   F = C_f\, {\rm exp} \left[ -b\, \exp{\frac{-(\lambda-\lambda_{0})^2}{2a^2}}
\right] ~,
\end{equation} 
where $b$ and $a$ are two free parameters, $\lambda$ is the pixel wavelength
in the DLA rest-frame, $\lambda_0$ is fixed to the known central
wavelength of the line, and $C_f$ is the value of the fitted continuum in each 
pixel. We perform a standard least-squares fit to the measured $F$ in the pixels of 
each line window with the two free parameters $a$ and $b$.

\begin{figure}              
\includegraphics[width=0.48\textwidth]{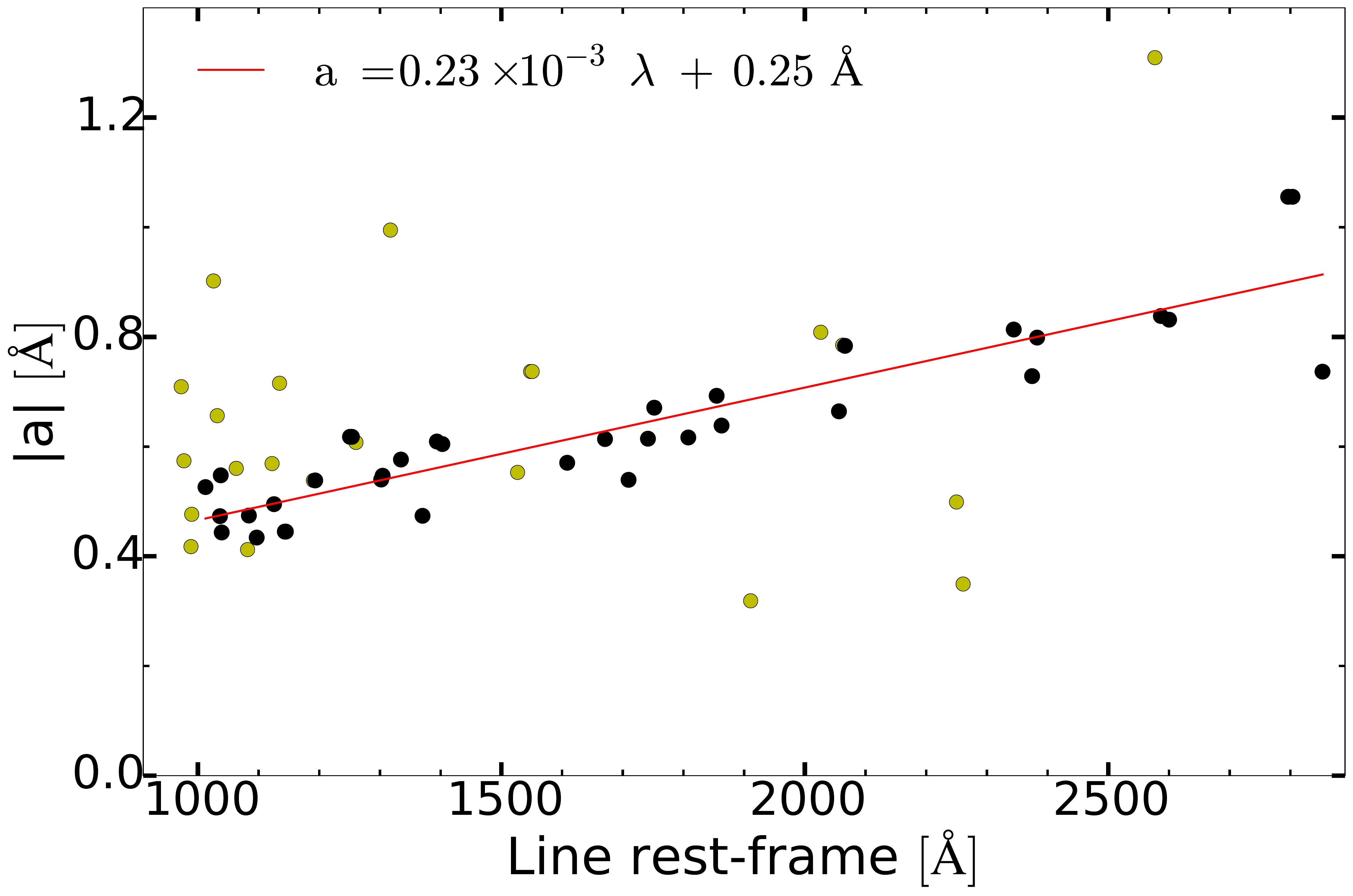}
\caption{Fitted values of the $a$ parameter for each absorption line in
the stacked spectrum for the total sample. 
Black data points are used to compute the linear regression, shown as
the red line. Yellow points are considered outliers, affected by blended
lines or low signal-to-noise ratio, and not used. The values for the $a$ 
parameter can be considered as an estimation of the spectrograph 
resolution assuming that the lines are unresolved. }
\label{fig:param}
\end{figure}

  In practice, the parameter $a$ in Eq.~\ref{eq:fitting},
reflecting the width of the lines,
is essentially determined by the spectrograph resolution of BOSS, except
in a few cases of blended lines. The BOSS resolution depends smoothly
on the observed wavelength, but once this smooth variation is taken into
account, the resolution should not vary among different metal lines. Therefore, the
accuracy of the fit to the line equivalent widths should improve if we
impose a fixed width parameter $a$ on the lines, assuming that the width
is not affected by variable levels of saturation of the absorption
lines. To examine the variation of $a$
with wavelength, Figure \ref{fig:param} displays the values of $a$
obtained for all the metal lines, as a function of their rest-frame
wavelength, for the case of the total sample. Yellow dots indicate 
blended lines, or lines that are
apparently very weak and strongly affected by noise, so that they are
deemed likely to present deviations of their width from any smooth
dependence. In cases of lines forming part of an atomic doublet,
the lines are jointly fitted and are required to have the same value
of $a$, but different values of $b$.

  It is apparent from Figure \ref{fig:param} that there is a smooth
increase of the width parameter $a$ with wavelength, with a small
scatter for the black points corresponding to lines that are not affected
by blends or a very weak signal-to-noise ratio. There is no evidence for any
difference in the width between low-ionization and high-ionization lines,
which might have indicated a different contribution from a physical
velocity dispersion to the measured widths. We fit a linear regression
to the values of $a$ as a function of wavelength using the black points
only (with each point weighted equally), and we obtain the result
\begin{equation}
 a = 0.23\times 10^{-3} \lambda +0.25 \, {\rm \AA} ~,
\label{eq:afit}
\end{equation}
shown as the red line in Figure \ref{fig:param}. The same
procedure for the metal sample yields a linear regression of the form 
\begin{equation}
 a = 0.25\times 10^{-3} \lambda +0.18 \, {\rm \AA} ~.
 \label{eq:afit76}
\end{equation}
These equations can be considered as 
an estimation of the spectrograph resolution assuming that the 
lines are unresolved.
The difference between the two samples is mostly due to the better
accuracy of the redshifts in the metal sample, which makes the lines
appear slightly narrower. 
Having tested that the deviations of the $a$ parameter from this linear
regression arise from noise and not from physical differences among
the lines, we now repeat the fits to each line with Eq.~
\ref{eq:fitting}, but keeping $a$ fixed to these linear regressions
and fitting only the $b$ parameter. We then integrate the area below the 
continuum represented by the function in
Eq.~\ref{eq:fitting} to compute the corresponding equivalent width
for each line. 

 In some cases, a group of absorption lines are close enough for their
wings to overlap, but the absorption maxima are still well separated. In
this case, we use a common window including the windows of all the
blended lines and we measure their equivalent widths in a single joint
fit. The optical
depths modeled as Gaussians are added, or equivalently, the
transmissions from each line are multiplied, to obtain the total
profile. The equivalent widths of these lines, marked with a superscript
denoting overlap, are listed together with all the other lines for the
total and metal samples in Tables \ref{ta:low} and \ref{ta:high} in
Appendix \ref{sec:ew}. Their errors are computed using the bootstrap
method described in \S~\ref{sec:bootstrap}.

  Figures \ref{fig:fit} and \ref{fig:fit2} in Appendix \ref{sec:linefit} 
present the fit for all the absorption features analysed in
the total sample. These figures display the line window, the continuum and 
the fitted absorption profile. In cases of overlap, each individual line
is indicated separately in addition to the total absorption profile.
Below, we discuss the case for lines that are strongly blended.


\subsection{Strongly blended absorption lines}
\label{sec:blended}

\begin{table*}
	\centering
	\caption{Mean rest equivalent widths of strongly blended
metal absorption lines in the total sample.} 
	\label{ta:blended}
	\begin{threeparttable}
	\begin{tabular}{cccccc}\\
		 		Total W		&Lines	        &$f$      	&Fitted $W$		&Model $W$							&Inferred $W$			\\ \hline  
   				$0.393\pm0.019$  		&					        &  		&		&									&$0.524\pm0.009$	\\  
 						&OI\,$\lambda988.58$			&0.0005		&$0.000\pm0.118$	&$0.034\pm0.002$						&  \\
 						&OI\,$\lambda988.65$			&0.008		&$0.283\pm0.135$	&$0.200\pm0.004$						&  \\
						&OI\,$\lambda988.77$			&0.047		&$0.131\pm0.073$	&$0.330\pm0.004$						& \\
				$0.271\pm0.021$ 		&						&    		&				&							& \\  
 						&NIII\,$\lambda989.80$		&0.123			&$0.089\pm0.108$	&									& \noindent\rule{0.3cm}{0.4pt}	\\
 						&SiII\,$\lambda989.87$		&0.171			&$0.191\pm0.110$	&$0.275\pm0.002$						&		\\
				$0.157\pm0.019$		&						&        	&					&						& \\  
  						&FeII\,$\lambda1062.15$		&0.003			&$0.004\pm0.012$	& $0.0115\pm0.0004$					& 	\\
 						&SIV\,$\lambda1062.66$		&0.049			&$0.020\pm0.011$	&									&$0.024\pm0.012$	\\
 						&FeII\,$\lambda1063.18$		&0.055			&$0.122\pm0.010$	&$0.110\pm0.003$						& 	\\
  						&FeII\,$\lambda1063.97$		&0.005			&$0.013\pm0.010$	&$0.0113\pm0.0004$					& 	\\
				$0.115\pm0.009$		&						&        	&					&  						&	\\
  						&FeII\,$\lambda1121.97$		&0.029			&$0.065\pm0.007$	&$0.077\pm0.002$						& 	\\
 						&FeIII\,$\lambda1122.53$		&0.054			&$0.051\pm0.007$	&									&$0.042\pm0.008$	\\  
   				$0.178\pm0.014$		&						&         	&					&						&$0.133\pm0.010$	\\ 
   						&FeII\,$\lambda1133.67$		&0.006			&$0.031\pm0.011$	&$0.0203\pm0.0007$					& 	\\
						&NI\,$\lambda1134.17$		&0.015			&$0.013\pm0.020$	&$0.0214\pm0.0012$					& 	\\
						&NI\,$\lambda1134.41$		&0.029			&$0.075\pm0.020$	&$0.039\pm0.002$						& 	\\
						&NI\,$\lambda1134.98$		&0.042			&$0.063\pm0.011$	&$0.054\pm0.003$						& 	\\
				$0.071\pm0.010$		&  						&   		&					& 						&$0.071\pm0.004$    \\
						&FeII\,$\lambda1142.36$		&0.004			&$0.023\pm0.006$	&$0.0140\pm0.0005$					& 	\\
						&FeII\,$\lambda1143.23$		&0.019			&$0.048\pm0.005$	&$0.057\pm0.002$						& 	\\
    				$0.411\pm0.006$		&						&        	&					& 						&$0.436\pm0.006$	\\   
						&SIII\,$\lambda1190.21$		&0.024			&$0.079\pm0.015$	&$0.056\pm0.013$					& 	\\
						&SiII\,$\lambda1190.42$		&0.292			&$0.345\pm0.012$	&$0.393\pm0.002$						& 	\\
				 $0.296\pm0.007$ 		&						&       	&					&						&$0.293\pm0.003$	\\ 
						&NI\,$\lambda1199.55$		&0.133			&$0.135\pm0.006$	&$0.137\pm0.005$						& 	\\  
						&NI\,$\lambda1200.22$		&0.087			&$0.104\pm0.008$	&$0.104\pm0.004$						& 	\\  
						&NI\,$\lambda1200.71$		&0.043			&$0.063\pm0.007$	&$0.061\pm0.003$						& 	\\  				
 				$0.623\pm0.008$		&						&        	&					&						&$0.688\pm0.008$ 	\\  
						&SII\,$\lambda1259.52$		&0.017			&$0.083\pm0.006$	&$0.095\pm0.005$						&     \\
						&SiII\,$\lambda1260.42$		&1.180			&$0.553\pm0.006$	&$0.550\pm0.003$						& 	\\
						&FeII\,$\lambda1260.53$		&0.024			&$-$				&$0.082\pm0.002$						& 	\\	
 				$0.012\pm0.007$		&						&        	&					&						&	\\  
						&TiII\,$\lambda1910.61$		&0.104			&$0.000\pm0.004$	&$0.006\pm0.003$						& 	\\
						&TiII\,$\lambda1910.95$		&0.098			&$0.012\pm0.006$	&$0.006\pm0.003$						& 	\\
 				$0.630\pm0.007$		&						&        	&					&						&	\\  
						&CII\,$\lambda1334.53$		&0.128			&$0.597\pm0.006$	&$0.597\pm0.014$						& 	\\
						&CII*\,$\lambda1335.71$		&0.115			&$0.037\pm0.005$	&									&$0.037\pm0.005$ 	\\
				$0.035\pm0.008$		&						&        	&					&						&	\\  
						&ZnII\,$\lambda2026.14$		&0.501			&$0.033\pm0.009$	&									&$0.025\pm0.007$	\\ 
						&CrII\,$\lambda2026.27$		&0.001			&$0.002\pm0.009$	&$0.00030\pm0.00009$					& 	\\   
						&MgI\,$\lambda2026.48$		&0.113			&$0.000\pm0.003$	&$0.010\pm0.002$						& 	\\  
				$0.034\pm0.007$	  	&						&       	&					&						&  \\  	
						&CrII\,$\lambda2062.23$		&0.076			&$0.031\pm0.008$	&$0.022\pm0.007$						& 	\\  
						&ZnII\,$\lambda2062.66$		&0.246			&$0.003\pm0.006$	&									&$0.011\pm0.007$	\\  \hline	
	\end{tabular} 
	\end{threeparttable}
	\tablecomments{First column: Total measured equivalent width (in $\angs$)
	for the absorption feature. Second column: Lines contributing to each blend.
        Third column: Oscillator strength of the contributing lines from
        \cite{Morton2003}. 
	Fourth column: Individual fitted equivalent widths.
	Fifth column: Equivalent widths predicted from our model, using
 other absorption lines of the same species in Table \ref{ta:low}.
        Sixth column: Inferred total and individual equivalent widths fixing the modeled lines
 to the predicted equivalent widths from our model. }

\end{table*}

  There are several blended absorption lines that do not present separate 
absorption maxima. We designate these groups 
`strongly blended lines'. These groups and their individual blended lines 
are listed in Table \ref{ta:blended}.

  We use the same procedure described above to perform a joint fit to
all the lines belonging to a blend: we fix all central wavelengths and
Gaussian widths, and only allow the amplitude of each line to vary. Any
lines that may be part of a blend with an equivalent width expected to 
contribute less than $1\%$ to the total equivalent width are ignored.
In these fits, the only reliable measurement is usually the total
equivalent width of each blended group, which is listed in the first
column with its bootstrap uncertainty. The individual equivalent widths
of each line (with wavelengths and oscillator strengths listed in the
second and third columns, respectively), with highly correlated and
larger errors, are listed in the fourth column.
The fifth column reports a modeled estimate of individual
equivalent widths, according to a theoretical model, 
described below in \S~\ref{sec:model}, that uses measurements of other
absorption lines of the same species. Finally, the sixth column gives
an inferred equivalent width for one of the blended lines once the
modeled lines are taken into account, and in cases where all the lines in
a blend are modeled, it lists the inferred total equivalent width of
the blend. These results will be discussed in detail in Section
\ref{sec:blend}. 

  We wish to warn here, however, that the results for the fitted
equivalent widths of blended lines in the fourth column are subject to
an important systematic error: we are simply assuming that the
transmission of the individual lines modeled with the profile in
Eq.~\ref{eq:fitting} can be multiplied to fit the entire blend.
This situation is actually not true because the absorbing components in each
individual line are not independent, but they have a highly correlated
velocity structure in each DLA. When the blended lines are very close
together (for example, for the first blend in Table \ref{ta:blended} for
OI lines), the combined equivalent widths of the various lines may
hardly increase in DLAs with very narrow lines, which are highly
saturated even if the blended line at the BOSS resolution appears far
from saturation, and may increase to some extent in DLAs with broader
velocity profiles. It is impossible to make a reliable estimate of how
the `fitted' equivalent widths of individual lines should be added to yield
the total blend equivalent width without a complete model of the
absorbing subcomponents width distribution and velocity correlations.
This effort would be much more ambitious than the simple model presented below
in \S~\ref{sec:curve} for modeling single line equivalent widths
in terms of mean column densities. We therefore choose to use the
simple assumption of multiplying transmissions of all the individual
lines. This approach produces approximately correct results only for blends of
lines that are sufficiently separated or weak that they do not have
frequently overlapping absorbing components that are saturated at full
resolution.

  The fits to the blended absorption profile in these cases are also
presented in Figures \ref{fig:fit} and \ref{fig:fit2}. Due to the
systematic errors mentioned above, individual line profiles are not
separately shown for these strongly blended lines.


\subsection{Bootstrap error computation}\label{sec:bootstrap}

  The errors on the line equivalent widths cannot be obtained from the
errors on the observed flux in individual pixels because they are often
dominated by continuum uncertainties. We therefore use a bootstrap
method: we randomly split our DLA sample (both the total and the metal
one) into $100$ subsets containing roughly the same number of DLA
candidates. We then generate 1000 new samples by randomly selecting
$100$ of these subsets, allowing for repetition, and we recompute the
composite spectrum, the continuum fit and the equivalent width
calculations for each new sample. We use the same values of $a(\lambda)$
that were obtained in equations \ref{eq:afit} and \ref{eq:afit76}, and
we obtain the errors from the standard deviation of each metal line
equivalent width among the bootstrap samples.


\section{Dependence of equivalent widths on $\rm{N_{HI}}$}
\label{sec:nhi}

\begin{figure*}                  
\includegraphics[width=0.5\textwidth]{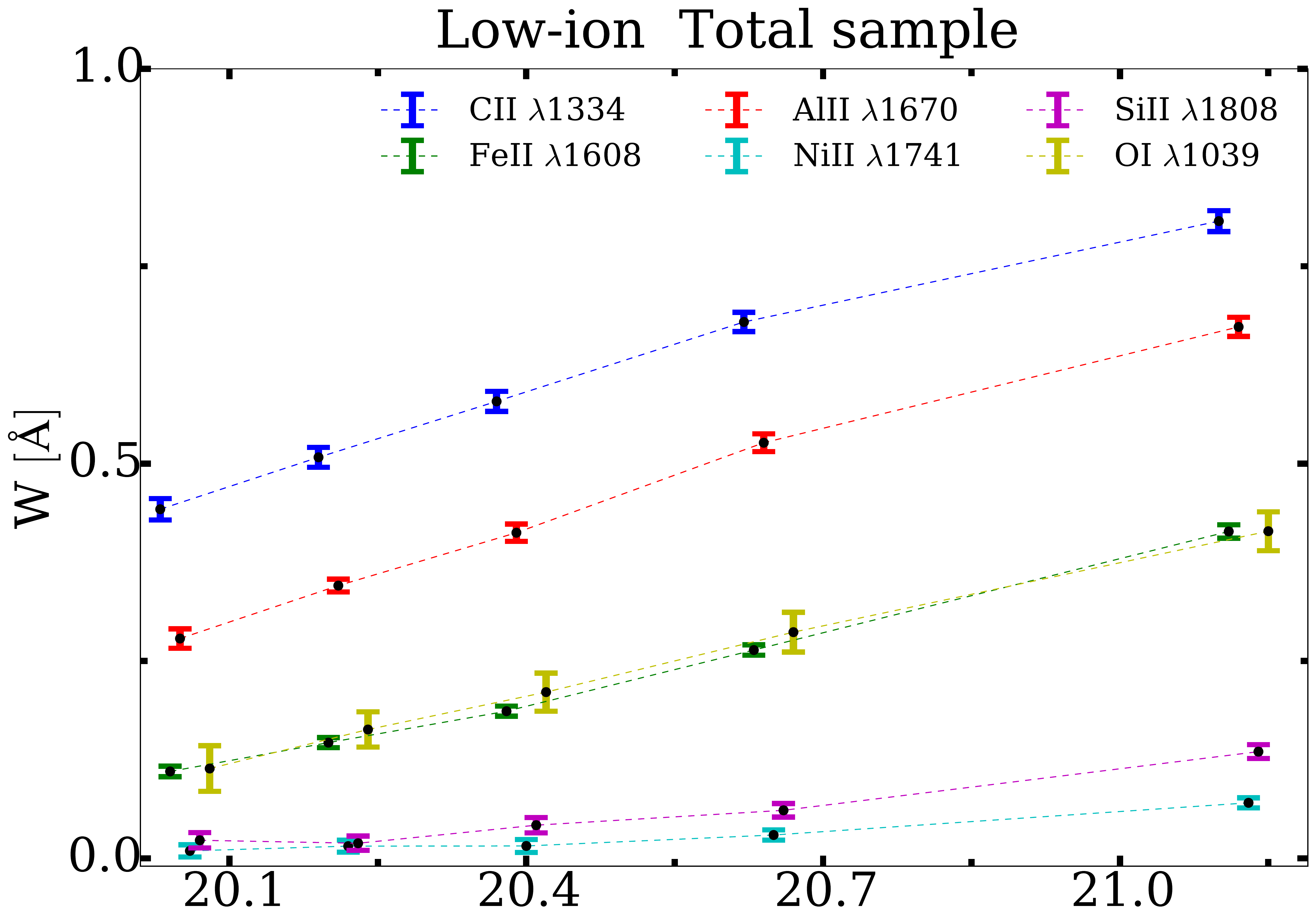}\includegraphics[width=0.475\textwidth]{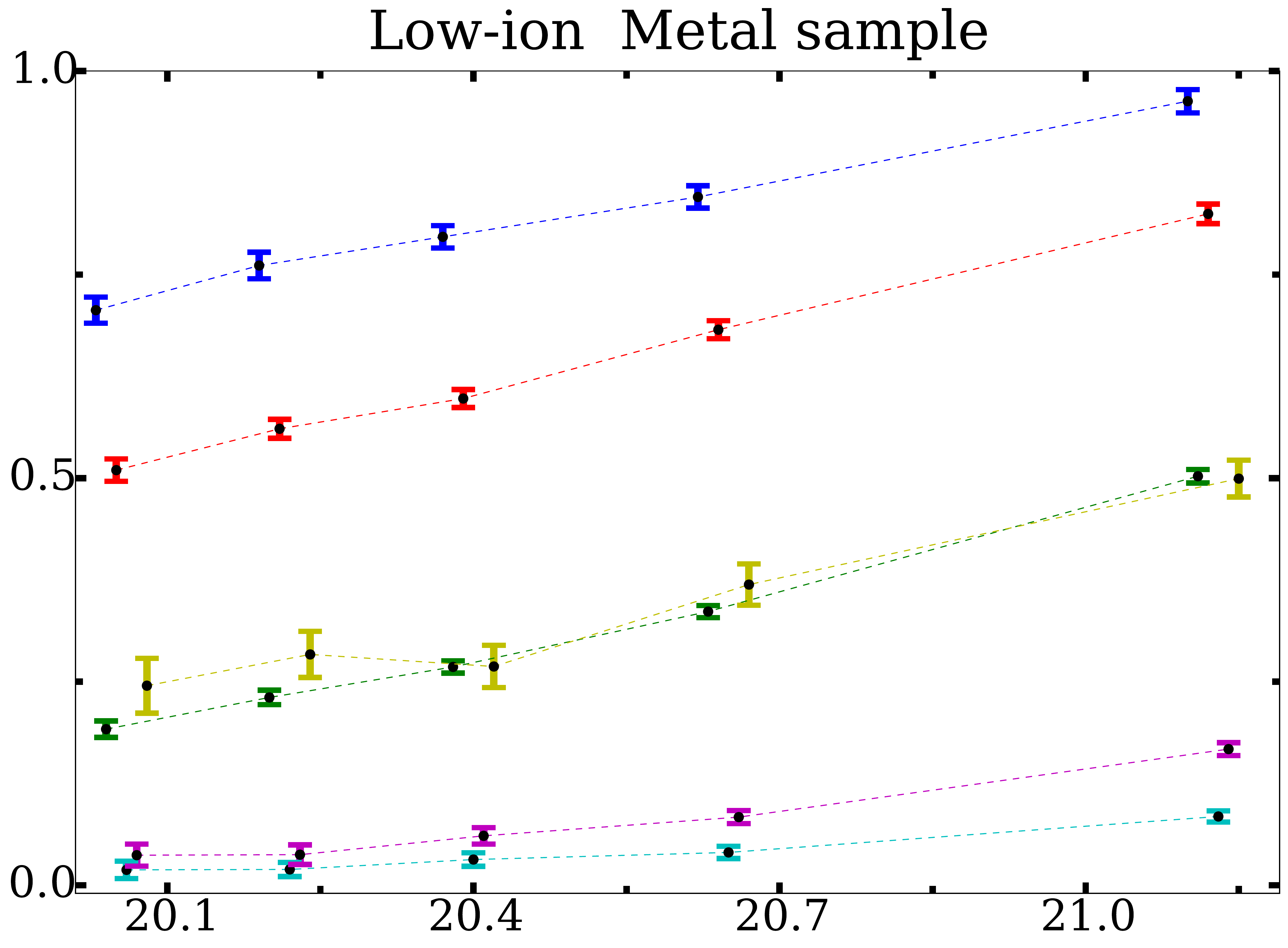}
\includegraphics[width=0.5\textwidth]{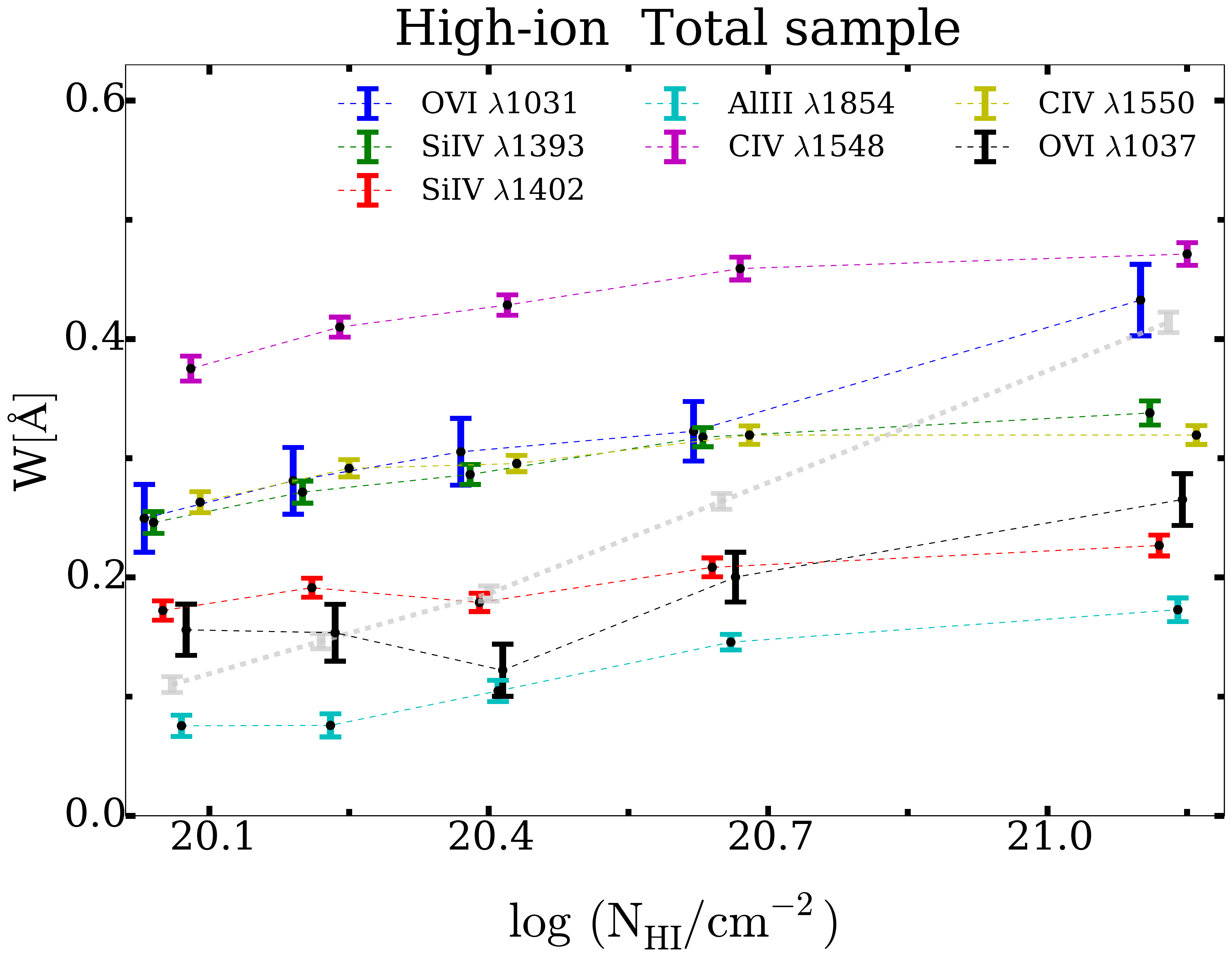}\includegraphics[width=0.475\textwidth]{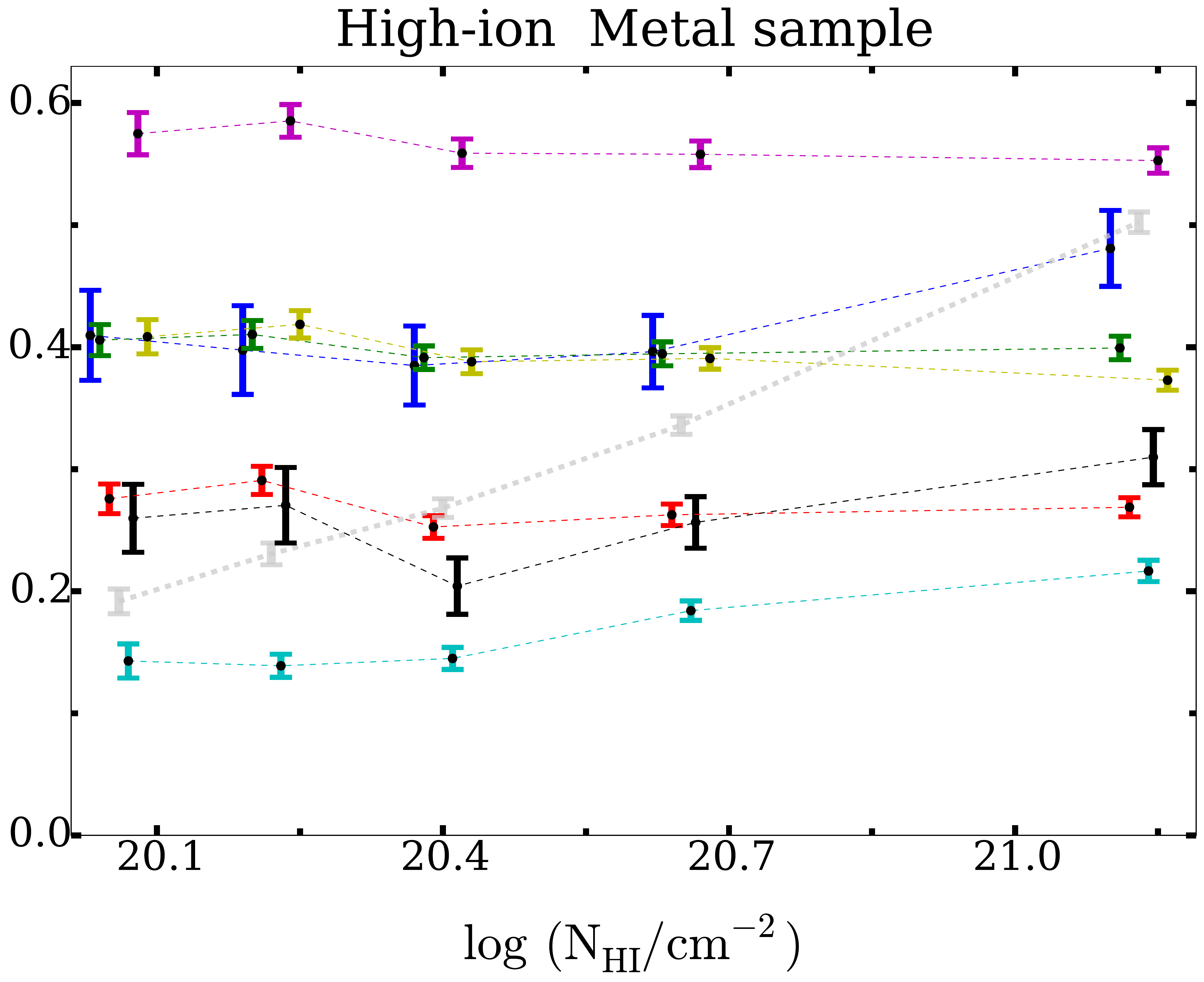}
\caption{\textit{Top:} Mean equivalent width of low-ionization metal lines
as a function of DLA HI column density, for the total (left panel) and 
metal sample (right panel). \textit{Bottom:} Same for the high-ionization
doublet lines of CIV, SiIV and OVI and the intermediate ionization line
of AlIII, for the total (left panel) and metal samples (right panel). 
The values of the  FeII $\lambda1608$ line are also indicated in light grey
for comparison between low- and high-ionisation behaviours.
Dotted lines connect the measured points to help visualize the dependence
on $N_{\rm HI}$. Horizontal positions of points with overlapping error bars
have been slightly shifted for visual aid, but they are all based on a single
set of 5 bins in $N_{\rm HI}$. The large error bars of OVI lines reflect the 
effect of blending with the \lya forest and other lines.}\label{fig:nhi}
\end{figure*}

  We now examine the dependence of the mean equivalent width of the
metal lines on the property that can be best measured for each individual
DLA: the hydrogen column density ${ N_{\rm HI}}$, which determines the damped
profile of the hydrogen \lya line.
We divide the total and the metal samples into five intervals in
$\log {\rm (N_{\rm HI}/cm^{-2})}$ chosen to contain a similar number of DLAs:
$[20.0-20.13],(20.13-20.30],(20.30-20.50],(20.50-20.80],(20.80-22.50]$.
The logarithm of the mean values of the column density in these ranges
are $20.06$, $20.22$, $20.40$, $20.65$ and $21.13$, respectively. 
For each interval, we redo the previous calculations: a composite
spectrum of transmission fraction is obtained for the DLAs that have a
column density in the specified range, the continuum is fitted, and
equivalent widths of our list of metal lines with bootstrap errors are
obtained. The $a$ parameter in Eq.~\ref{eq:fitting} is kept
fixed to the same values used for the entire samples, assuming that it
is independent of $N_{\rm HI}$. The bootstrap errors are computed as
before, using only the DLAs in every bootstrap subset which are in the
$N_{\rm HI}$ bin of interest when doing the stacking.
The bootstrap subsets are large enough to contain similar numbers of
DLAs in each column density bin.
Results for the mean equivalent widths are tabulated in Tables
\ref{ta:lownhi24} - \ref{ta:highnhi76} in Appendix \ref{sec:ew}, 
separately for low-ionization and high-ionization species, and for the 
total and metal samples. The intermediate species CIII, SiIII, AlIII 
and SIII are included in the table of low-ionization species owing to the
fact that the kinematic properties for these elements resemble those of
low-ionization species \citep{Wolfe2005}.

  The dependence of the mean equivalent widths on column density is
presented in Figure \ref{fig:nhi}. Several low-ionization species are shown
in the top panels (total sample on the left, metal sample on the right).
Points are connected with dotted lines to help visualize the trend
with ${N_{\rm HI}}$ for each species (they are slightly shifted
horizontally from the true mean column density values specified
previously to avoid overlapping errorbars). A general trend of
increasing $W$ with increasing $N_{\rm HI}$ is apparent, as expected.
If the mean
metallicity and dust depletion in DLAs do not vary substantially with
column density, hydrogen is mostly atomic, and the low-ionization
species displayed in Figure \ref{fig:nhi} are the dominant ionization
stage of the respective elements, then we expect the metal column densities
to increase linearly with $N_{\rm HI}$. This prediction, however, cannot be tested
directly because a large fraction of lines may be saturated, and the degree of 
saturation depends on a complex distribution of velocity dispersion, metallicity
and multi-component structure of the absorption systems.

  The bottom panels show the dependence of the mean equivalent widths 
on column density for three high-ionization species, SiIV, OVI and CIV, in the
total and metal samples (left and right panels, respectively), and for an 
intermediate species, AlIII. We also present the behaviour of the 
FeII$\,\lambda1608$ line for comparison (in soft grey). The total sample 
demonstrates that there is also an increase in
the mean equivalent width with column density, in all cases. However,
the increase is much smaller than for low-ionization lines.
Over the column density range $20 < \log ({ N_{\rm HI}/{\rm cm}^{-2}}) < 20.7$, the mean
equivalent width of the CIV and SiIV lines increases by a factor
$\sim 1.25$, whereas the increase for relatively weak lines of
low-ionization species in the top panels over the same range is a factor
$\sim 3$, and even the strongest low-ionization line, CII (which is
most saturated), increases by a factor $\sim 1.8$. The AlIII line
increases by a factor that is intermediate between the low-ionization
and high-ionization cases.

  The metal sample presents a different behavior: the mean equivalent width
is essentially independent of column density for the high-ionization
lines, and has a weaker increase with $N_{\rm HI}$ than in the total
sample for the low-ionization ones.
There are two possible reasons for this difference. First, the
DLAs in the metal sample are selected to be the ones with measured
redshifts from the metal lines, and therefore where the metal lines
have been individually detected in the BOSS spectra. This procedure selects
DLAs with strong metal lines, and also spectra of high signal-to-noise ratio.
The low column density DLAs are included in the metal sample with a
lower frequency than the high column density ones, with an important
selection in favor of systems with strong metal lines (either because
of high metallicity, or because of high velocity dispersion which
reduces the degree of line saturation). The second reason is that some
fraction of the DLAs at low column density may be false detections,
where a combination of noise and the presence of a cluster of
Ly$\alpha$ forest lines may be confused with a DLA in low
signal-to-noise ratio spectra. The low column density DLAs should have a
higher impurity fraction, or fraction of false systems, and this
impurity fraction is significantly reduced when selecting systems that have
associated metal lines detected. Therefore, the $\sim$ 25\% increase
in the high-ionization mean equivalent width with $N_{\rm HI}$ may in
part be real for the total sample (not suffering from metal selection
effects), but may also be due to a worsening sample purity at low
column density for the total sample. These effects illustrate how special care
must be taken in the sample selection for evaluating mean properties
of the DLA population.

  Even though the metal sample results in a composite spectrum where
weak lines can be detected at a greater statistical significance, it
will not be used in the rest of this paper because it does not produce
results that can be easily corrected for the sample selection effects. 
The question of the effect of the sample impurity (i.e., the rate of
false DLAs in the total sample) needs to be addressed with simulations
of DLA catalogs from mock spectra, and is left to be analyzed in future
work when these simulations are available.

\begin{figure*}                  
\includegraphics[width=0.5\textwidth]{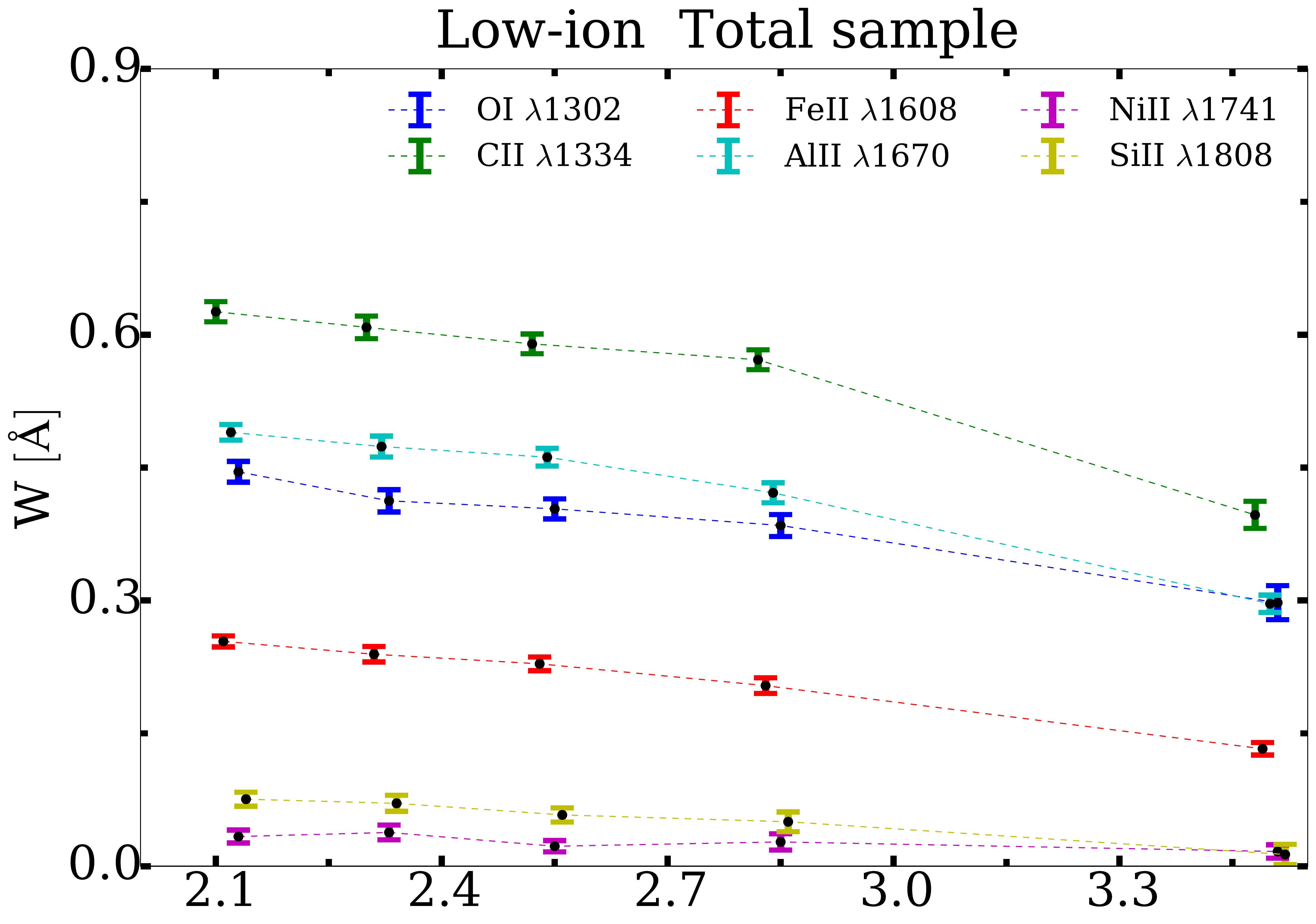}\includegraphics[width=0.475\textwidth]{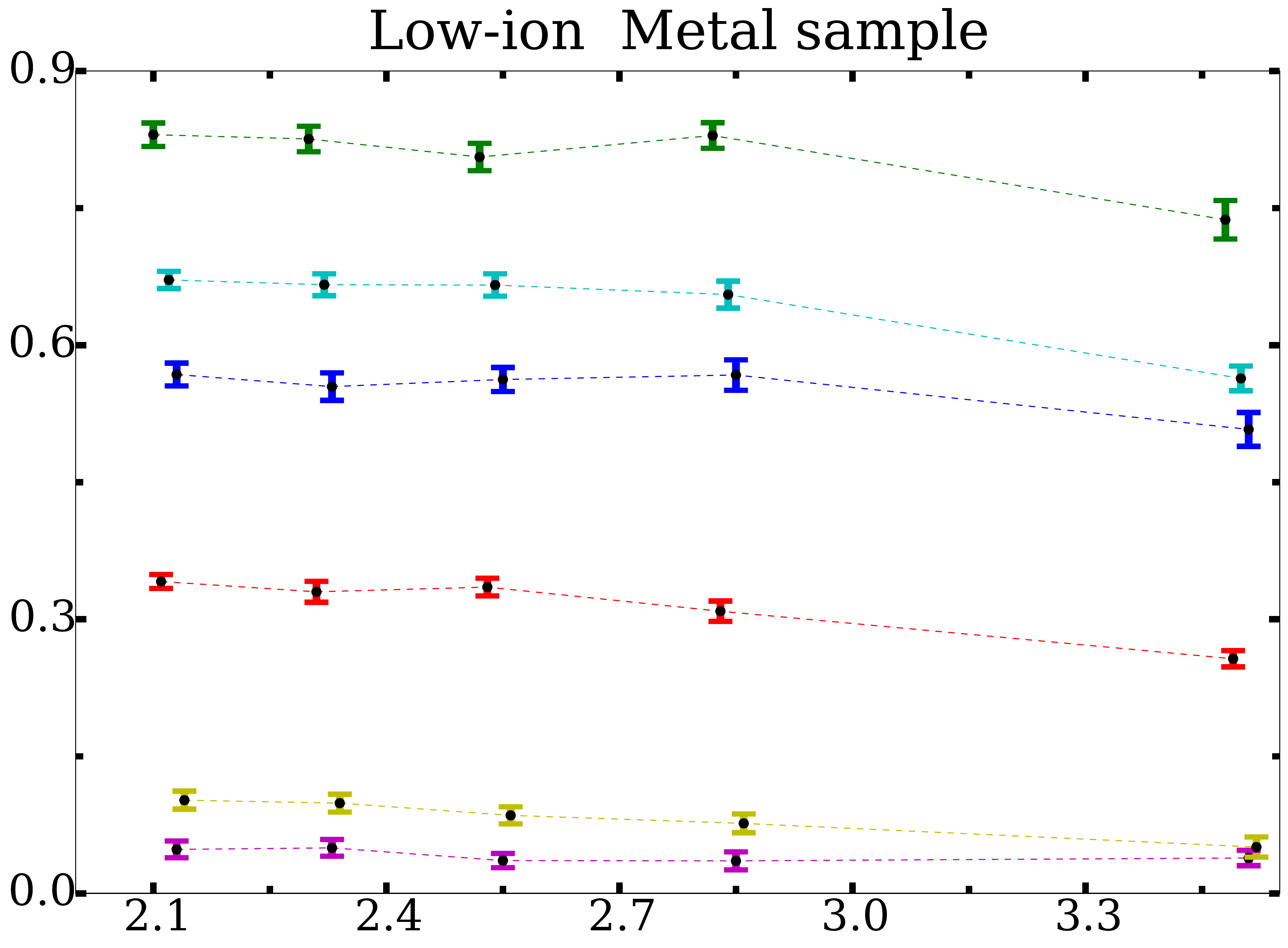}
\includegraphics[width=0.5\textwidth]{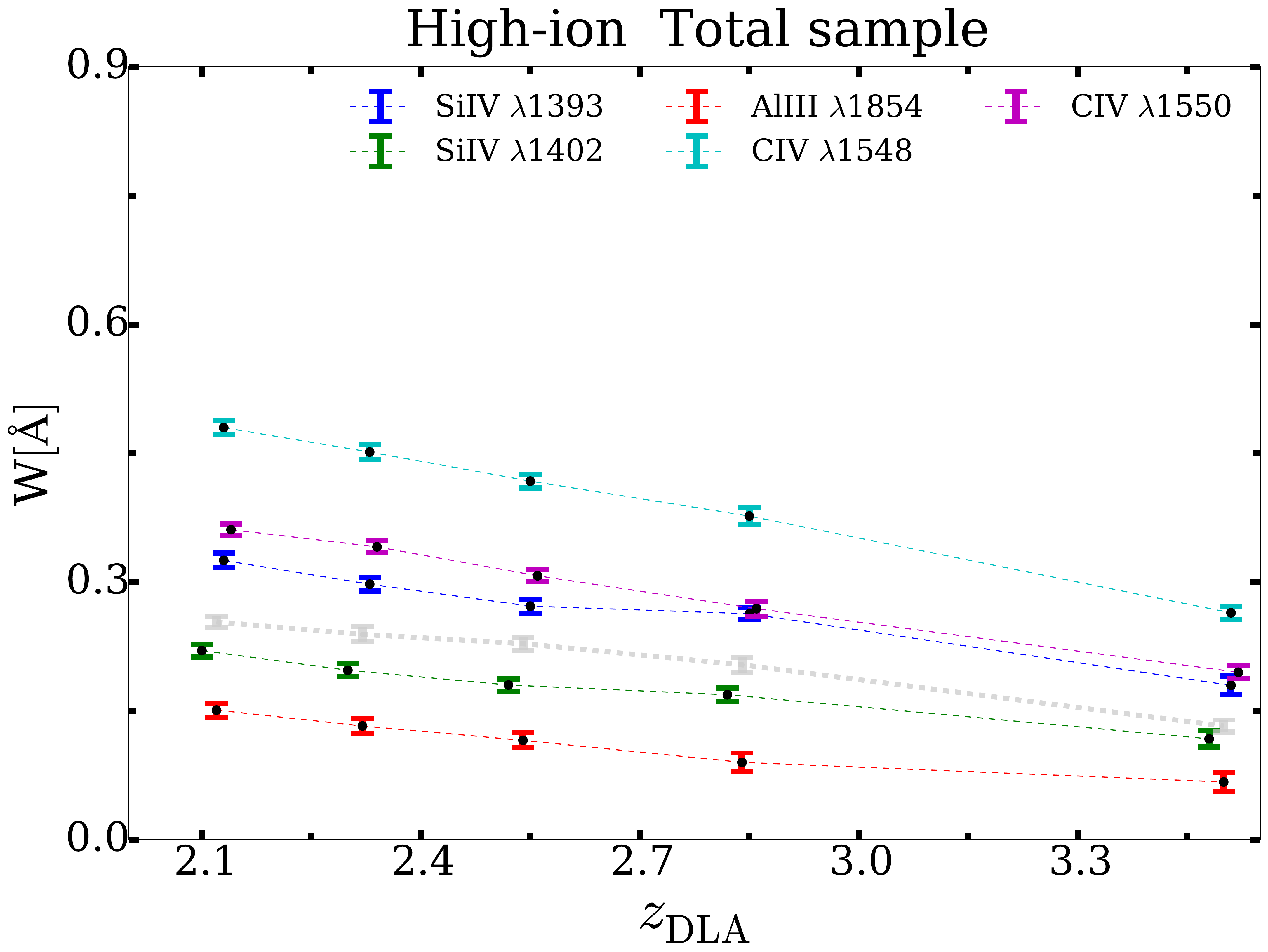}\includegraphics[width=0.475\textwidth]{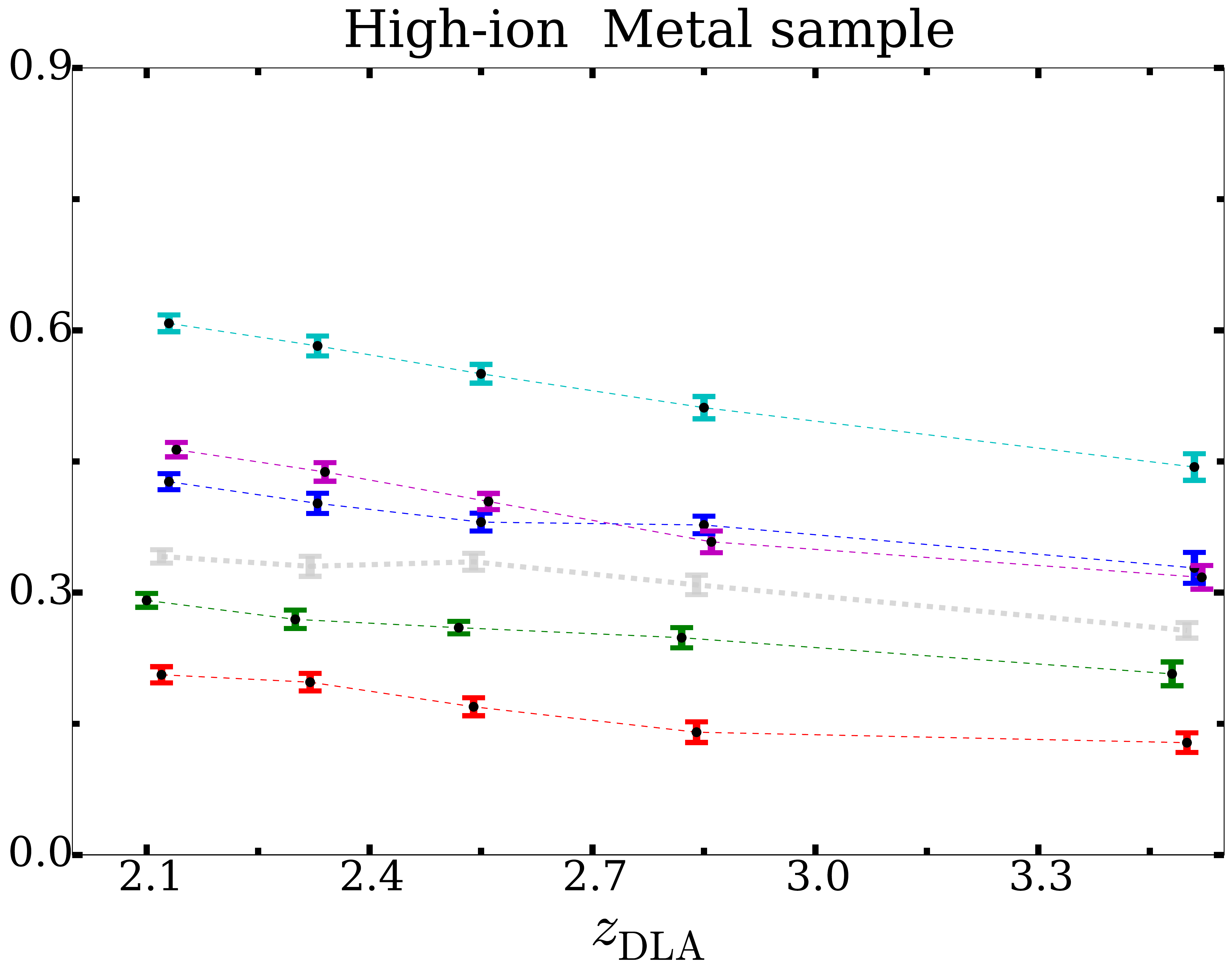}
\caption{\textit{Top:} Mean equivalent width of low-ionization metal lines
as a function of DLA redshift, for the total (left panel) and 
metal sample (right panel). \textit{Bottom:} Same for the high-ionization
doublet lines of CIV and SiIV, and the intermediate ionization line of AlIII. 
The values of the  FeII $\lambda1608$ line are also indicated in light grey
for comparison between low- and high-ionisation behaviours.
Dotted lines connect the measured points to help visualize the dependence
on redshift. Horizontal positions of points with overlapping error bars
have been slightly shifted for visual aid, but they are all based on a single
set of 5 $z_{\rm DLA}$ bins.}\label{fig:z}
\end{figure*}

  Additional interesting information can be obtained from the mean
equivalent width ratio of the doublets of CIV and SiIV. The equivalent
width ratio stays constant with column density near a value of $1.4$
to $1.5$ for
both CIV and SiIV. This factor is between the values of $2$ and $1$,
corresponding to the two extreme cases of completely optically thin
and completely saturated lines, respectively. The fact that this
ratio remains constant as the mean equivalent width increases suggests that
the velocity dispersion, or 
the mean number of subcomponents, should be increasing with column
density, although our uncertainties are still too large to place much confidence
on this interpretation. The ratio of $\sim 1.5$ indicates that the absorbers have a
mixture of weak, unsaturated components, and strong saturated ones, that
contribute about equally to the total equivalent widths.
The doublet equivalent width ratio of OVI 
appears to be larger, but the errors here are also larger and systematics
due to the blending of the weaker line of the doublet with CII and OI
lines are likely present.


\section{Dependence of equivalent widths on $z$}
\label{sec:z}

  We here explore the evolution of the mean metal line equivalent widths with redshift 
of the DLAs. We split the total and the metal samples into five DLA redshift intervals, 
$[1.9-2.24],(2.24-2.4],(2.4-2.7],(2.7-3.0],(3.0-6.4]$, with mean redshift values $2.12$, 
$2.32$, $2.54$, $2.84$ and $3.50$, respectively. The mean values of the column densities 
for each bin are $20.50$, $20.50$, $20.50$, $20.49$ and $20.46$, respectively, showing little 
differences between them. We repeat the equivalent width and uncertainty calculations we did 
for the case of column densities, but this time for each redshift bin. The $a$ parameter in Eq.~
\ref{eq:fitting} is kept fixed to the same values used for the entire samples, assuming 
that it is independent of DLA redshift. The mean equivalent widths in every redshift bin are tabulated 
in Tables \ref{ta:lowz24} - \ref{ta:highz76} in Appendix \ref{sec:ew}, separately for low-ionization and 
high-ionization species, and for the total and metal samples. 

  Figure \ref{fig:z} displays the evolution of the mean equivalent widths with redshift. The top panels 
show several low-ionization species. The points are slightly shifted horizontally from their original 
positions to clarify the visualization. A general decrease of equivalent width by a factor $\sim1.3-1.5$ 
from $z\sim2.1$ to $z\sim3.5$ is apparent for the total sample ({\it left panel}). The nearly 
equal values of the mean column density at every redshift bin confirm that the observed trend is 
not driven by changes in $N_{\rm HI}$. However, the effect of false positive DLA detections, which 
result in lower equivalent widths, might be present, specially at the highest $z$ bins. The evolution 
of the metal sample, which should be less affected by false positives, presents a slightly smoother 
decrease, by a factor $\sim1.1-1.2$ in the same redshift range ({\it right panel}).

   The bottom panels show the dependence of the mean equivalent widths 
on $z$ for two high-ionization species, SiIV and CIV, and for an 
intermediate species, AlIII. We also present the behaviour of the 
FeII$\,\lambda1608$ line for comparison (in soft grey). The doublet 
of OVI is not shown as its rest-frame wavelength is only covered by 
the three upper redshift bins which, in turn, are those most affected by noise. 
Here, as for the low-ionization species in the upper panels, the metal sample ({\it right panel}) 
suggests a smoother decrease with redshift than the total sample ({\it left panel}). 
However, for both samples, the equivalent width decreases by a factor $\gtrsim 1.5$ within the 
redshift range $z\sim2.1$ to $z\sim3.5$. Whether a different evolution between low- and 
high-ionization species exists or is an artifact driven by noise and/or systematics is unclear, 
and we defer more detailed examinations to future work.  


\section{Theoretical Model for line saturation}\label{sec:model}

  As previously described, the absorption profiles of the metal lines 
are mostly determined by the instrumental spectrograph
resolution. The absorption system components are often much narrower
than the BOSS resolution, and can therefore be saturated even though
they appear to be weak \citep{York2000}. Although a fraction of DLAs are
known to have velocity dispersions that are comparable to the 
BOSS resolution \citep{Prochaska1997}, this
intrinsic line width likely contributes only to extend the wings of the
mean absorption profile that can be seen in the strongest lines
(see, e.g., the CIV profile in Figure \ref{fig:fit2}). Deconvolving
the BOSS spectral point spread function from the mean profile
is difficult because of its wavelength and fiber-to-fiber dependence, and 
inaccuracies in its determination. We therefore use only the mean equivalent 
widths of the lines as the quantity that can be determined from the BOSS 
stacked spectra we have created.

  The physical quantities that the equivalent width depends on are the
column density and the velocity distribution of each species. The
velocity distribution is known to be generally complex, because
several absorbing components are often resolved in the metal line
profiles of DLAs \citep{Prochaska1997,Zwaan2008}, which means that 
there is not a simple
relation between mean equivalent widths and column densities that can
be expressed in terms of a single velocity dispersion, as for a Voigt
profile. The scenario that is usually invoked to explain the observed
absorption profiles is that a number of gas clouds are orbiting inside
a halo or thick disk \citep{Haehnelt1998,Wolfe1998,McDonald1999}, 
which randomly intercept the
line of sight to the quasar. Every absorbing system is 
characterized by a halo velocity dispersion, which describes the random
motions among the clouds seen as absorbing components in the spectra,
and an internal cloud velocity dispersion, which can be partly thermal
and partly turbulent, determining the width of the individual absorbing
components. We shall use this scenario to construct a simple model for
the relation of the mean equivalent width and column density for a
large sample of DLAs similar to the one we are using here.

\begin{figure*}                  
\includegraphics[width=0.495\textwidth]{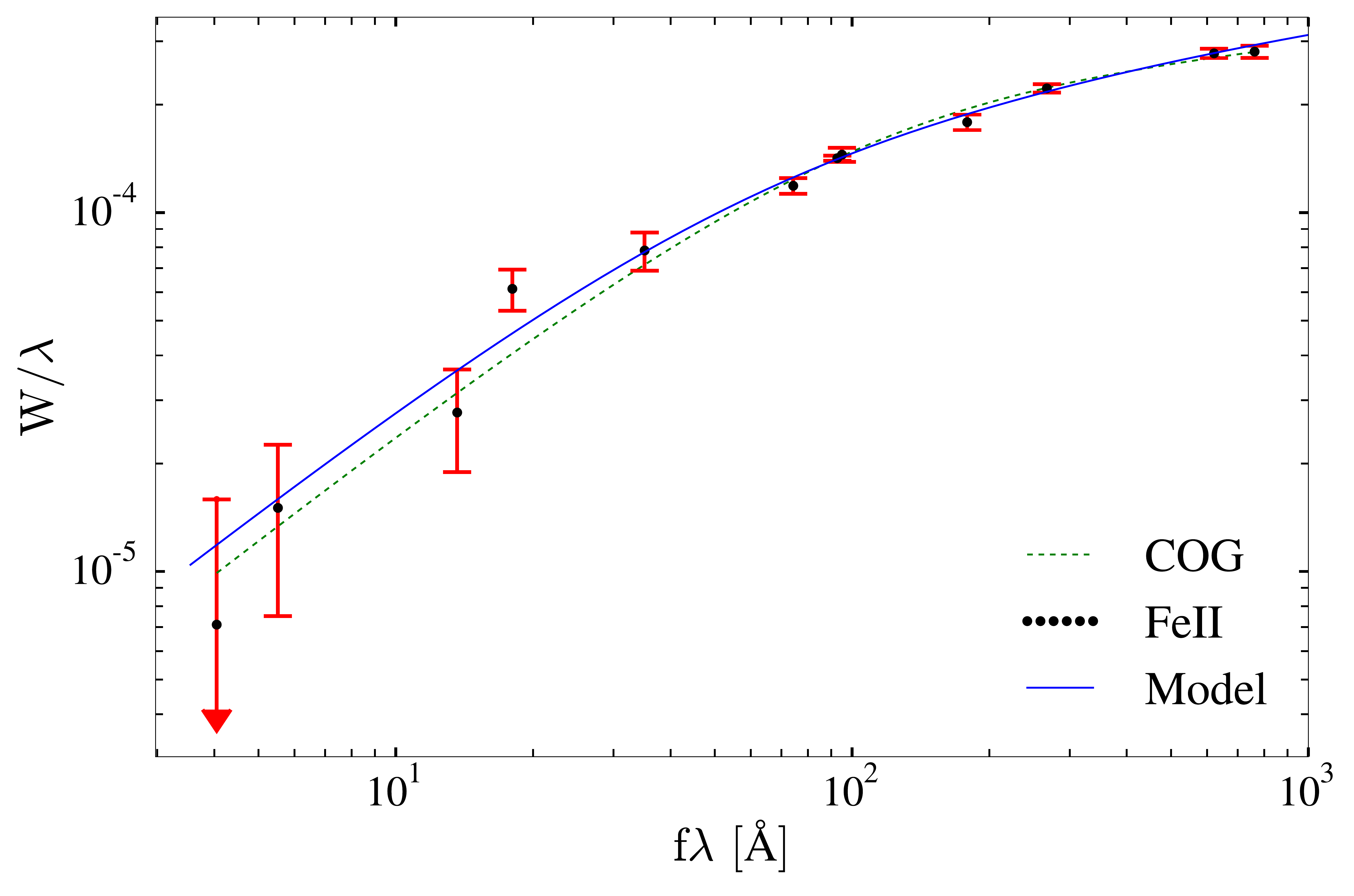}\includegraphics[width=0.49\textwidth]{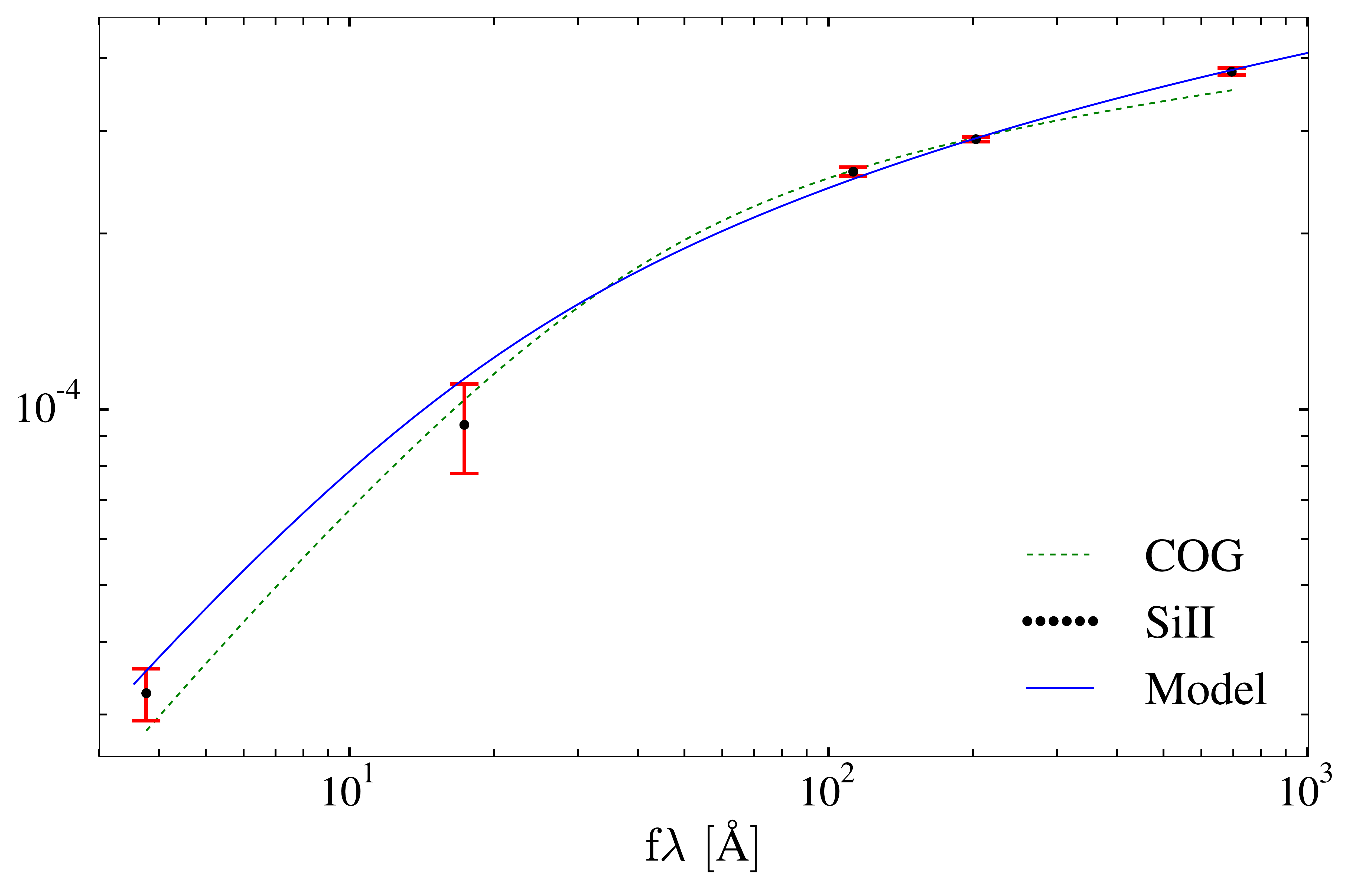}
\caption{Equivalent widths divided by the line wavelength measured for the FeII (\textit{left}) and
SiII (\textit{right}) transitions in the stacked spectrum of the total sample, and best fit curve obtained with 
our model as a function of $f\lambda$ ({\it solid blue lines}). For comparison, we show the best fit 
curves from the curve of growth (COG) method ({\it dashed green lines}). 
The red arrow in the left panel denotes that the error is larger than the expected value.}
\label{fig:curve}
\end{figure*}

  With this purpose in mind, we analyse in detail the transitions of
FeII and SiII for the total sample, which are the two species with the
largest number of measurable absorption lines, $12$ for FeII and $5$ for
SiII (except for NiII, which has 6 transitions, but they are weaker
lines with a narrower range of oscillator strengths), most of which
are strong enough to yield a reliable measurement of the mean equivalent
width, $W$. We plot in Figure \ref{fig:curve} the dimensionless ratio
$W/\lambda$ as a function of the product $f\lambda$ for each one of
these transitions, where $f$ is the oscillator strength of the line at
wavelength $\lambda$. For a completely optically thin absorption line,
the equivalent width is given by
\begin{equation}
 {W\over \lambda} = \pi f \lambda r_e N ~,
\end{equation}
where $r_e = e^2/(m_e c^2)$ is the classical electron radius, $e$ is the
electron charge and $m_e$ the electron mass. Therefore, in the optically
thin regime, we expect $W/\lambda \propto f\lambda$ for the mean of a
DLA sample with any distribution of column densities. As an increasing
fraction of the DLA components become optically thick with increasing
$f$, the mean relation should flatten until $W/\lambda$ becomes
nearly constant with $f\lambda$ at a value determined by the mean
maximum velocity range that is covered by the absorbing components in a
DLA. This is indeed the behavior seen in Figure \ref{fig:curve}. The
precise functional dependence of $W/\lambda$ on $f\lambda$ cannot be
reliably predicted from theory, because it depends on the detailed
velocity and column density distribution and the internal velocity
structure of the DLA absorbing components.

  However, we can argue that all the low-ionization species measured in
DLAs should follow the same functional form of $W/\lambda$ versus
$f\lambda$, except for a horizontal rescaling that reflects the element
abundance. This relation follows from assuming that all the low-ionization species
have the same distribution of velocities in the DLA sample we are
using, and that different ionization corrections do not lead to
significant differences, so that they all have the same distribution of
optical depths within the velocity range of the absorbing systems except
for the rescaling reflecting the abundance. In general, 
ionization corrections in DLAs are found to be $<0.2$ dex
\citep[e.g.,][]{Howk1999,Vladilo2001,Prochaska2002,Kisielius2015},
supporting the view that differences in the shape of the optical depth
distribution among metal species should be minimal. A model is proposed
below for the shape of $W/\lambda$ versus $f\lambda$, which is displayed 
as the blue curves in Figure \ref{fig:curve}. This simple model allows
us to infer mean column densities from the measured mean equivalent
widths, up to the systematic errors arising from the model assumptions. 
We shall generally assume that the model can be applied with the same
parameters to all low-ionization species. We will also apply the model
to high-ionization species allowing for different fit parameters, since
these absorbers are believed to reside in different regions and have different
velocity distributions than the low-ionization species
\citep[e.g.,][]{Wolfe2005}. We consider, however, this model to be less
reliable for high-ionization species because of the lack of a test with
many lines from the same species with different values of $f$, similar
to the one provided by FeII and SiII for low-ionization.

  Before describing this model, we should address the 
large differences in the error bars of different lines in Figure
\ref{fig:curve}. Large errors are sometimes caused by blends with other
absorption features that are difficult to separate (e.g., the
SiII$\,\lambda1020$ transition appearing at
$f\lambda=17.1\, {\rm \AA}$ in the right panel is blended
with the Ly$\beta$ transition), or when the lines are very weak.
The leftmost FeII line has an expected $W$ value smaller than its
uncertainty, so the lower edge of the error bar is replaced by an arrow.

\subsection{Modeling equivalent widths as a function of oscillator
strength}\label{sec:curve} 

  For any metal line centered at wavelength $\lambda_0$,
we define the rest-frame equivalent width in units of velocity as:
\begin{equation}
 W_v = {c\, W\over \lambda_0} = {c\over \lambda_0}
 \int d\lambda\, [1- {\rm exp}(-\tau(\lambda))] ~,
\end{equation}
where the integral is performed over a rest-frame wavelength interval
around $\lambda_0$ that contains the whole absorption line.
We define the {\it linearized equivalent width} as
\begin{equation}
\label{eq:wtau}
{ W_\tau\over c} =  \int {d\lambda\over\lambda_0}\, \tau(\lambda) 
 =  \pi f \lambda_0 r_e N ~,
\end{equation}
Although the mean of $W_\tau$ is the quantity that is directly
related to the mean column density, only the mean of $W_v$ is measured
from our BOSS stacked spectra.

  For absorption lines that contain a single absorbing component with
a Gaussian velocity dispersion (arising from thermal motions or from
a turbulent velocity dispersion that also has a Gaussian distribution),
the relation between $W_v$ and $W_\tau$ is the well known Voigt
curve of growth \citep{Goody1964}. For a velocity dispersion $\sigma$, when
$W_\tau \gg W_v$, we have
\begin{equation}
W_v \simeq 2\sigma \sqrt{2\log [W_\tau/(\sqrt{2\pi}\sigma)]} ~.
\end{equation}
The mean equivalent width resulting from the stacked metal lines of many
DLAs is not related in the same simple way to the mean column density
because each absorption system has a different distribution of optical
depths. We search for a simple, physically motivated fitting formula to
adjust the observed mean values of $W_v$ of the low-ionization lines as
a function of the values of $W_\tau$ that are theoretically predicted
from column densities.

  We assume the absorption lines are composed of subcomponents with an
internal velocity dispersion $\sigma$, given by
$\sigma^2 = \sigma_0^2 (m_H/m_i) + \sigma_h^2$, where $\sigma_h$ is the
turbulent (or hydrodynamic) velocity dispersion, assumed to be the same
for all the lines of all the species, and $\sigma_0$ is the thermal
velocity dispersion of hydrogen, which we shall fix to
$\sigma_0=10\kms$, corresponding to the typical equilibrium temperature
of photoionized gas $T\simeq 10^4\, {\rm K}$. The mass of the atom of
the species of each line is $m_i$, and $m_H$ is the hydrogen mass.

  We assume that the subcomponents of a certain absorption line are
contained within an interval of velocity width $V$, which is wide enough so that
any absorption outside this interval due to the line can be
neglected. Let the probability distribution of the optical depth at any
pixel in the spectrum within this interval of
width $V$ be $\psi(\tau)$. In other words, $\psi(\tau)\, d\tau$ is the
fraction of pixels in the intervals of width $V$ around the central
wavelength of the line in any DLA that have an optical depth $\tau$
within the range $d\tau$. The two measures of the equivalent widths are given by:
\begin{equation}
\label{eq:wprob}
 W_\tau = V \int d\tau\, \psi(\tau)\, \tau ~,
\end{equation}
and
\begin{equation}
 W_v = V \int d\tau\, \psi(\tau)\, \left[ 1 - e^{-\tau} \right] ~.
\end{equation}
As a model for the probability distribution of the optical depth, we
choose the function $\psi(\tau)=A/\tau$ within a certain range
$\tau_{min} < \tau < \tau_{max}$, and $\psi=0$ outside this range. The
normalization constant is $A=1/\log(\tau_{max}/\tau_{min})$. This definition is
chosen as a simple approximation that assumes that $\tau$ varies over a
broad range in a scale-invariant way. More realistically, there should
be a continuous distribution of $\tau$ reaching down to zero, but we
simply assume that $\tau_{min}$ is small enough to neglect this detail. Let us
now consider absorbers with a fixed $W_\tau$, i.e., a fixed column
density of a certain species absorption
line. The maximum optical depth is obtained when only one Gaussian
absorbing subcomponent accounts for all the absorption, with
\begin{equation}
\label{eq:taumax}
 \tau_{max}= {W_\tau\over \sqrt{2\pi} \sigma} ~.
\end{equation}
When several subcomponents are present, each one must have a smaller
central optical depth to produce the same $W_\tau$, and our model
assumes the distribution is flat in $\log\tau$. We now use the fact that
the average number of absorbing subcomponents in DLAs is of order unity, to
require that the mean $W_\tau$, from Eq.~ \ref{eq:wprob}, is
equal to that due to the case of a single component reaching the maximum
optical depth $\tau_{max}$, i.e.,
\begin{equation}
 V A (\tau_{max} - \tau_{min})= W_\tau= \sqrt{2\pi} \sigma \tau_{max} ~.
\end{equation}
We assume that the ratio $\tau_{max}/\tau_{min}$ is constant,
independent of $W_\tau$. We then obtain the following relation between $V$,
$\tau_{max}$ and $\tau_{min}$:
\begin{equation}
\label{eq:sig1}
 V = { \sqrt{2\pi} \sigma\, \log(\tau_{max}/\tau_{min}) \over
 1-\tau_{min}/\tau_{max} } \equiv \sqrt{2\pi} \sigma_1 ~,
\end{equation}
where $\sigma_1$ is defined by the last equality to replace $V$, and can
be interpreted as a rough estimate of the velocity dispersion of the
subcomponent absorbers, or DLA clouds, within their host halo.

  Finally, we find for the equivalent width $W_v$ the expression
\begin{equation}
\label{eq:Wfinal}
{W_v\over W_\tau} = {\log(\tau_{max}/\tau_{min}) -
\int_{\tau_{min}}^{\tau_{max}} d\tau\,
 e^{-\tau} /\tau \over \tau_{max} - \tau_{min} } ~.
\end{equation}

  The model relating $W_v$ to $W_\tau$ has only two free parameters:
$\sigma_h$ and $\sigma_1$. The thermal dispersion is fixed to $\sigma_0
= 10\, {\rm km\,s^{-1}}$. The fitting formula provides a value of $W_v$ for
any input value of $W_\tau$: Eq.~\ref{eq:taumax} first reveals
the value of $\tau_{max}$, the value of $\tau_{min}$ is derived from
Eq.~\ref{eq:sig1}, and we then obtain $W_v$ from Eq.~\ref{eq:Wfinal}. 
These two free parameters can be fixed to be the same
for all the low-ionization species, if we believe that the subcomponent
kinematics are an invariant characteristic of the absorption profiles
that has little dependence on the column density. Under this assumption,
the dependence of $W_v$ on $W_\tau$ is unique, and represents both the
variation with absorption lines of different $f$ for fixed column
density, and the variation with column density for a fixed line.
In addition, once $\sigma_h$ and $\sigma_1$ are determined, the relation
for each new species depends on one additional parameter only: 
the ratio of its column density to the hydrogen one. 

\begin{figure}           
\includegraphics[width=0.48\textwidth]{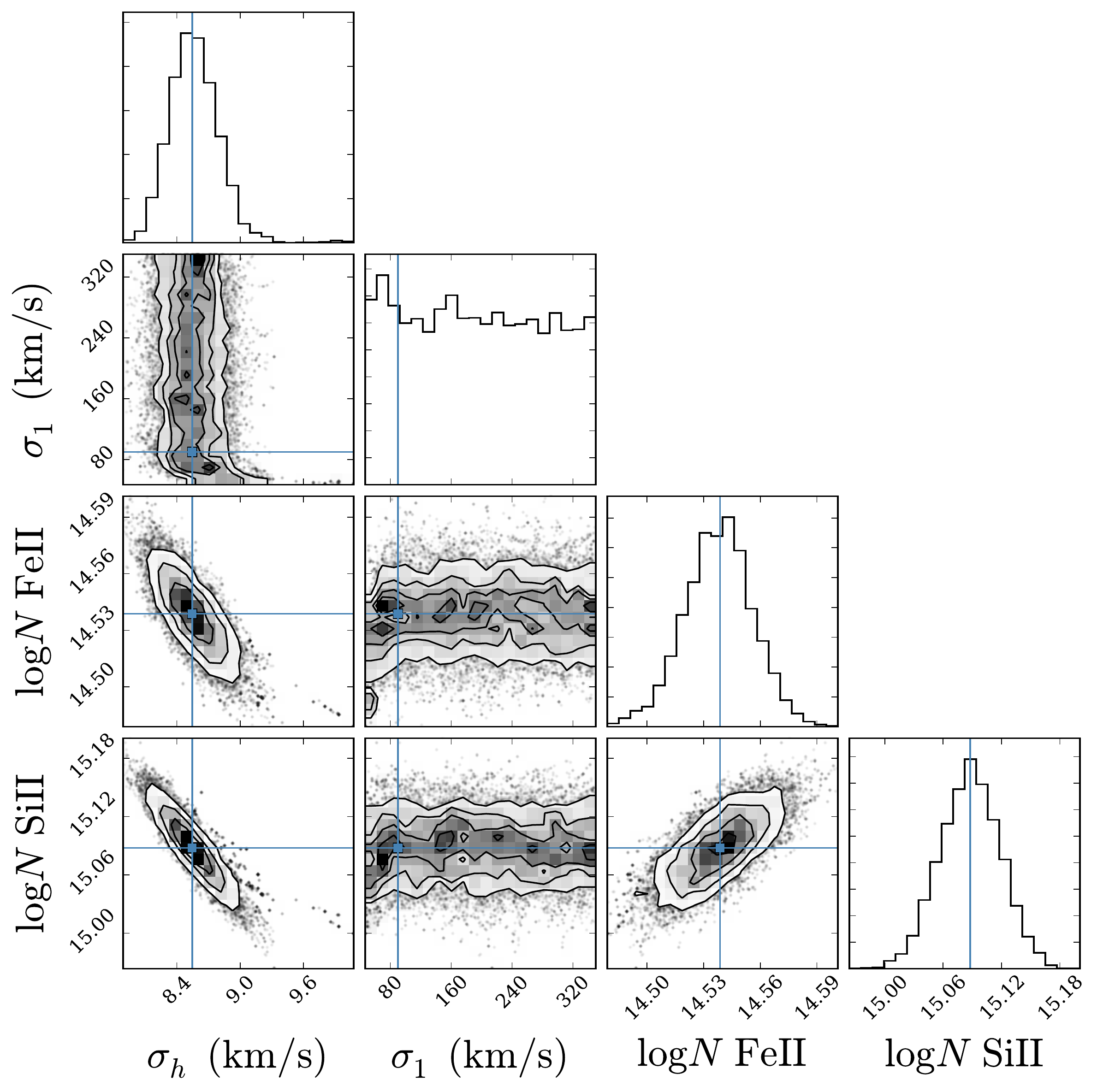}
\caption{Parameter posterior probabilities of the joint fit of FeII and
SiII transitions for the estimation of $\sigma_h$ and $\sigma_1$ and the
two mean column densities in the total sample. The $\chi^2$ contours
are at 0.5, 1, 1.5 and 2$\sigma$ levels (in some panels the 0.5$\sigma$
contour is too small to be visible). The blue dots and lines, plotted
over the contours, indicate the best fit. Note the large uncertainty of
the $\sigma_1$ parameter. This figure has been produced using the open
source code \textit{triangle.py} from \cite{triangle}.}
\label{fig:triangle}
\end{figure}

  We obtain the best fit parameters for low-ionization species 
by computing a joint fit to the
above mentioned FeII and SiII transitions, with the four free parameters
$N_{\rm FeII}$, $N_{\rm SiII}$, $\sigma_h$ and $\sigma_1$, which
minimizes the sum of the two $\chi{^2}$ values. The column densities
$N_{\rm FeII}$ and $N_{\rm SiII}$ are the mean values
for our DLA sample used directly in Eq.~\ref{eq:wtau} to
compute $W_\tau$ for any absorption line. The best fit and parameter
uncertainties are calculated with the open-source code \textit{EMCEE}
\citep{emcee2013}, using the Monte Carlo Markov Chain method (MCMC).
Figure \ref{fig:triangle} displays the one and two dimensional
projections of the posterior probability distributions of the
parameters; the contours are at 0.5, 1, 1.5 and 2$\sigma$ levels.

  The best fit values and uncertainties, indicating the
16th and 84th percentiles (the mean of the two) in the marginalized 
distribution for $\sigma_1$ (for $\sigma_h$ and $\log N$), are
$\sigma_h = 8.56\pm0.22\, {\rm km\,s^{-1}}$,  
$\sigma_1 = 90^{+205}_{-16}\, {\rm km\,s^{-1}}$, 
$\log N_{\rm FeII} = 14.538 \pm 0.011$ 
and ${\log N_{\rm SiII}} = 15.087 \pm 0.012$.
The fit reproduces the trend of the measured points, as seen in Figure
\ref{fig:curve} where our best fit model is indicated by blue lines.
The total ${\rm \chi^2}$ value of the fit is $\chi^2=14.98$, for an
expected value of $13$ (17 data points and four free parameters; 
some of the oscillator strengths have substantial uncertainties,
which are not taken into account and may contribute to some increase in
the value of $\chi^2$).
The halo velocity dispersion $\sigma_1$ is the parameter with the
largest uncertainty: any value in the range $\sim 70 - 300\, \rm{km\,s^{-1}}$ is
essentially equally good, with no correlation with any of the other
parameters. Our formula provides a good fit as long a
turbulent velocity dispersion of individual components provides 
rough equipartition of the thermal and turbulent kinetic energies
($\sigma_h\simeq 8.6\, {\rm km\,s^{-1}}$), and the well-determined normalizations
of the curves in Figure \ref{fig:curve} correspond to the values of
$N_{\rm FeII}$ and $N_{\rm SiII}$.

  For high-ionization species, we adopt the same assumptions as above and
perform a joint fit to the two doublets CIV and SiIV, which have
unblended, strong absorption lines. We use these species because they
have the strongest lines, and are believed to be produced mostly by
photoionization \cite[e.g.,][]{Fox2007a}. We assume the same fixed value
for $\sigma_0=10\,{\rm km\,s^{-1}}$ because these species are thought
to arise in warm gas regions with temperatures ${\rm T\sim10^4\, K}$
\citep{Lehner2008}. We obtain the values
$\sigma_h = 14.51\pm0.64\, {\rm km\,s^{-1}}$,  
$\sigma_1 = 185^{+112}_{-85}\, {\rm km\,s^{-1}}$, 
$\log N_{\rm CIV} = 14.396 \pm 0.019$ 
and ${\log N_{\rm SiIV}} = 13.783 \pm 0.016$.
The larger value for $\sigma_h$ compared to the low-ionization species
is indicative of a more violently turbulent environment for the
high-ionization species. Changing the thermal
velocity dispersion to $\sigma_0 = 30 \kms$ reduces $\sigma_h$ by
$\sim 15\%$ and increases $\sigma_1$ by $\sim 5\%$, with variations of
less than $1\%$ on the column densities. 

   As explained in the previous section, the use of a curve of growth for 
fitting the transitions of a given species and obtaining the column density 
is not accurate for our composite spectra, due to the large number of components 
and systems contributing to the mean absorption features. Although a detailed 
assessment of our theoretical model is beyond the scope of this work, we 
perform a simple comparison with the curve of growth approach, since the latter 
is broadly used in absorption line studies, e.g., in the recent work by 
\cite{Noterdaeme2014}. We use the publicly available package \texttt{linetools}\footnote{
\url{http://linetools.readthedocs.io/en/latest/index.html}} to find the best-fit parameters of 
the curve of growth, column density, $N\,{\rm (cm^{-2})}$, and velocity dispersion,  
$b\,{\rm (km/s)}$, applied to the 
species for which we have more transitions, i.e., FeII, SiII and NiII. The {\it dashed green 
lines} in Figure \ref{fig:curve} show the resulting curves for FeII and SiII. The best fit 
parameters for the three species are $\log N_{\rm FeII} = 14.452 \pm 0.010$, $\log N_{\rm SiII} = 
14.957 \pm 0.095$ and $\log N_{\rm NiII} = 13.843 \pm 0.072$, with Doppler parameters 
$b_{\rm FeII} = 24.04 \pm 0.73$, $b_{\rm SiII} = 26.18 \pm 1.39$ and $b_{\rm NiII} = 3.21 \pm 0.49$. 
These results are significantly different from the ones obtained with our model, which allows 
for a distribution of component structure and velocity dispersions.


\subsection{Mean column densities of low- and high-ionization species}
\label{sec:deduction}

  We now assume $\sigma_h$ and $\sigma_1$ to be constant for all the
low- and high-ionization species, fixing their values to the best fit obtained
for the FeII and SiII transitions, $\sigma_h = 8.56\, \rm{km\,s^{-1}}$ and
$\sigma_1 = 90\, \rm{km\,s^{-1}}$, and for CIV and SiIV, $\sigma_h = 14.51\, 
\rm{km\,s^{-1}}$ and $\sigma_1 = 185\, \rm{km\,s^{-1}}$, respectively.
For every species we fit the curve of this fixed model with the
mean column density $N$ as the only free parameter, i.e., we
assume the relation of $W_v$ and $W_\tau$ is fixed, and fit only the
constant of proportionality relating $W_\tau$ to $f\lambda$, which is 
proportional to the mean column density of the species in our DLA sample
through Eq.~\ref{eq:wtau}. This approach yields a ratio of the ion
abundance to that of HI considering the logarithm of the mean hydrogen column 
density of the total sample, $\log {\bar N}_{\rm HI}=20.49$. If the ionization corrections 
can be neglected (i.e., assuming that the fraction of the species in question is 
the same as the neutral hydrogen fraction), this result is equal to the element 
abundance compared to hydrogen.

  The results of this model are presented in the middle column in Table \ref{ta:N}.
Errors include equivalent width measurement uncertainties only, computed
earlier from our bootstrap analysis of the total DLA sample, and do not
include any systematic errors arising from our model assumptions.
We use all the lines reported in Table \ref{ta:low}. For species that
have only one line, the error of the column density directly
reflects the error of the equivalent width; when there are several
lines the error is reduced by obtaining the best fit to all the lines.

  We also report column densities for four intermediate ionization
species from Table \ref{ta:low}: Al\,III, C\,III, S\,III and Si\,III,
using the same parameters as for the low-ionization species since they
are believed to have similar velocity distributions \citep{Wolfe2005}.
For the high-ionization species we list the results obtained
for the fit to CIV and SiIV for the results in Table \ref{ta:high}
with their different parameters $\sigma_1$ and $\sigma_h$, and apply
them also to NV and OVI. These more highly ionized species
are believed to arise from higher temperature gas,
\citep[see][and \S~\ref{sec:discussion}]{Fox2007a,Lehner2008,Fox2011}, 
with velocity distributions that are likely broader than for CIV and SiIV,
implying perhaps an overestimate of the OVI column density, but not for
NV which produces lines that are mostly optically thin.

  The strongly blended lines in Table \ref{ta:blended} are in general
not used for the determination of column densities whenever we have
other lines of the same species in Table \ref{ta:low}, because blending
adds additional modeling uncertainties, as discussed in section
\ref{sec:blended}. However, there are several species that are
only measurable using these blended lines. In the case of TiII and NI,
we use the TiII blend at $ 1911 \angs$ and the NI blend at
$ 1200 \angs$, containing only lines from a single species, to
determine the mean column densities directly from the ``fitted W''
results of the lines. These lines are not very close together, so we
believe that the systematic effects discussed in section
\ref{sec:blended} are not important in this case. The TiII line is
in any case quite weak and has a large statistical error. The rest of the
lines with column densities listed in Table \ref{ta:N} derived from
blended groups are for ZnII, CII${}^{\*}$, FeIII and SIV, which
are inferred after correcting for lines of other species. We will
discuss these more complex cases further in \S~\ref{sec:blend}.

  The third column in Table \ref{ta:N} denotes the abundances obtained
using the expression
${\rm [X/HI]} = \log\left({N_{\rm X}}/{N_{\rm HI}}\right) -
 \log\left(N_{\rm X_e}/N_{\rm H}\right)_{\odot}$, where $X$ denotes the
ion of interest, $X_e$ the corresponding element, and $\odot$ denotes
solar values. We use solar abundances for elements from the photospheric
data in Table 1 of \cite{Asplund2009}.
These results for $\rm [X/HI]$ can be interpreted as element
abundances relative to solar values if ionization and dust depletion
corrections can be neglected. We give the abundances only for species
that are expected to have fractional columns compared to their elements
similar to the neutral hydrogen fraction (often the first ionized
species, but the neutral one for O).

\begin{table}
	\centering
	\caption{Derived mean column densities in the total sample.}
	\label{ta:N}
	\begin{tabular}{ccc}
Metal ion	&$\log \bar N$        		 & [X/HI] 	\\ \hline  

Al\,II		&$13.806 \pm 0.016$  		 &$-1.13 \pm 0.03$\\  
C\,II      	&$16.025 \pm 0.060$ 		&$-0.90 \pm 0.08$\\   
C\,II*      	&$13.351 \pm 0.060$ 		&\\ 
Cr\,II		&$12.908 \pm 0.117$ 		&$-1.22\pm0.12$\\  
Fe\,II		&$14.538 \pm 0.011$ 		&$-1.45\pm0.04$\\  
Mg\,I		&$12.387 \pm 0.073$  		 &\\  
Mg\,II      	&$14.943 \pm 0.041$ 		&$-1.15 \pm 0.06$\\  
Mn\,II	&$12.300\pm 0.188$ 		&$-1.62 \pm 0.19$\\  
N\,I		&$14.128 \pm 0.025$		&\\ 
N\,II		&$14.497 \pm 0.051$  		 &\\  
Ni\,II     	&$13.452 \pm 0.030$ 		&$-1.26\pm0.05$\\  
O\,I		&$15.959 \pm 0.025$  		 &$-1.22\pm0.06$\\  
P\,II		&$12.890\pm 0.141$  		 &$-1.01\pm0.14$\\ 
S\,II		&$14.731 \pm 0.032$ 		&$-0.88 \pm 0.04$\\  
Si\,II     	&$15.087 \pm 0.012$		 &$-0.91 \pm 0.03$\\  
Ti\,II     	&$12.264\pm 0.193$		 	&$-1.19\pm0.20$\\  
Zn\,II     	&$12.157\pm 0.111$			 &$-0.89 \pm 0.12$\\  
				       \hline
Al\,III     	&$12.946 \pm 0.023$ 		 & \\ 
C\,III		&$16.696 \pm 0.121$		&\\   
S\,III      	&$14.345 \pm 0.123$  		 &\\
Si\,III		&$15.107 \pm 0.036$		&\\   
Fe\,III	&$13.905\pm 0.095$			&\\   
					\hline
C\,IV 	&$14.396 \pm 0.019$		&\\  
N\,V 		&$13.074 \pm 0.090$		&\\     
O\,VI 	&$14.771 \pm 0.030$		&\\     
S\,IV 	&$13.707 \pm 0.184$		&\\      
Si\,IV 	&$13.783 \pm 0.016$		&\\      \hline

\end{tabular}
\tablecomments{
Mean column densities derived from the measured equivalent widths and our 
model relating $W_v$ to $W_\tau$, for low, intermediate and high
ionization species are listed in the central column.
Abundances, relative to solar values from \cite{Asplund2009} and derived with no 
ionization or dust depletion correction, are in the third column, using a mean
$N_{\rm HI}=10^{20.49}\cm^{-2}$. Intermediate ionization species, 
fitted using the low-ionization model parameters $\sigma_1$ and $\sigma_h$,
are presented in the middle rows, while high-ionization species,
using their own separate parameters, are listed at the bottom.
Uncertainties include only the equivalent width measurement errors, and
not modeling systematic errors, which are particularly severe for CIII
(derived from the line with the highest value of $W/\lambda$).  }

\end{table}

  Doubly ionized species generally have column densities
that are not much lower than the singly-ionized ones. In particular, the
SiIII and SiII column densities are equal within the
measurement error, and that of CIII is substantially higher than for CII. 
In the case of CIII, the column density we obtain is subject to a large
systematic error due to our modeling assumptions. The CIII column
density is derived only from the line at $977 \angs$, which has
the highest value of $W/\lambda\simeq 6.6\times 10^{-4}$ of any of our
lines, and involves a large extrapolation of our model relating
$W/\lambda$ to $f\lambda$ from the curve in Figure \ref{fig:curve} which
is not tested from the FeII and SiII lines in the figure. For the
cases of SiIII, SIII and AlIII, there is no extrapolation in the values
of $W/\lambda$; the velocity structure may still be different from
the singly ionized species, but the overestimate of the column density
this may induce is probably not large. These cases demonstrate that the
doubly ionized species can have column densities close to the singly
ionized species, implying a substantial ionization correction for the 
abundances. The ratios of double to single ionization species can be
used to constrain photoionization models of DLAs
\cite[e.g.,][]{Vladilo2001}.
The column densities of more highly ionized species are, as expected,
substantially lower than those of their low-ionization counterparts.
 
  In the case of neutral species, the mean column density of NI is about
half the mean column density of NII. Nitrogen, with an ionization
potential of $14.5$ eV, should be completely neutral in the deeply
self-shielded inner parts of DLAs, so
in this case the ionization correction is quite important. The fraction
of oxygen in the form of OI is probably even smaller than for nitrogen,
owing to its lower ionization potential, so our derived nitrogen and
oxygen abundances are highly dependent on these uncertain ionization
corrections.

  Without ionization and dust depletion corrections, the abundances for
iron and silicon would be $\rm{[FeII/HI]}= -1.45 \pm 0.04$ and
$\rm{[SiII/HI]}= -0.91 \pm 0.03$, respectively. In this case, the
ionization corrections are likely to be similar for silicon and iron
because of the similar ionization potentials. The measured ratio 
$\rm{[SiII/FeII]}\simeq 0.54 \pm 0.05$, significantly higher than the 
typical intrinsic $\alpha$-element enhancement in DLAs relative to the Sun 
\citep[${\rm [\alpha/Fe]\sim0.3}$, e.g.,][]{Prochaska2003b,Rafelski2012,Cooke2015}, 
reflects the effect of the large dust depletion of iron 
compared to silicon \citep[e.g.,][]{Pettini1997,Kulkarni1997,Akerman2005,Vladilo2002,
Vladilo2011,Cooke2011,Cooke2013}. 


\subsection{Modelling Blended Lines}\label{sec:blend}

  Many lines of high scientific interest are blended. As described in
\S~\ref{sec:blended}, the equivalent widths for the groups of
blended lines listed in Table \ref{ta:blended} are measured by jointly
fitting the parameter $b$ in Eq.~\ref{eq:fitting} for each line,
while the $a$ parameter is, as usual, kept fixed. This approach 
leads to large, correlated errors for the equivalent widths of
the individual lines contributing to the blend, as listed in the fourth
column of Table \ref{ta:blended}, and to systematic errors caused by
our assumption that the total transmission is simply the product of the
individual line transmissions. In several cases, we are particularly
interested in measuring the equivalent width of one line in one of these
groups, where all the other lines of the group can be modelled from other
lines of the same species measured independently. For example, the only
line available for SIV is at $1062 \angs$ and is blended with several
FeII lines, which can be modelled from the set of all lines used in
Figure \ref{fig:curve}. Using the model prediction for the FeII lines
can allow for a better estimate of the SIV line equivalent width.

  We use this approach for five of the line groups listed in Table
\ref{ta:blended}, in order to measure lines of SIV, FeIII, CII* and
ZnII (the latter in two different blended groups). The predictions for the
modelled lines are listed in the fifth column of Table \ref{ta:blended},
obtained from the column densities listed in Table \ref{ta:N} and the
fixed values of the $\sigma_1$ and $\sigma_h$ parameters of all the
low-ionization lines. The errors of these modelled equivalent widths
are small because they are derived from the error in the column density
of each element, without including model uncertainties. We repeat
the fit to the measured profile of
the blended group by fixing the $b$ parameter of all the modelled lines
to reproduce these predicted equivalent widths, and leaving as a single
free variable the $b$ parameter of the line that is inferred from this
model prediction (the $a$ parameter of all lines is fixed as usual by
Eq.~\ref{eq:afit}). The results for the equivalent widths of these
inferred lines are listed in the sixth column, for the five mentioned
groups. Errors are computed from bootstrap realizations that keep the
modelled lines fixed. They therefore include only the
observational noise and continuum modeling, but not the systematic
uncertainty of the modelled lines. This process is not applied in the
case of the NIII line at $990\angs$, for which the equivalent width
is strongly blended with a SiII line and partly blended also with the OI
group, and is too small to be measured from our stacked spectrum.

  As mentioned before, the TiII group at $1911 \angs$
and the NI group at $1200 \angs$ are used as the only
measurements to infer the TiII and NI column densities. The inferred
$W$ values in the fifth column are then derived from our model using
these inferred column densities.

  Finally, there are five absorption features in which all the lines
can be modelled from other measured lines: the OI group, the FeII-NI
group, the FeII group, the SIII-SiII group, and the SiII-FeII group.
These modelled
equivalent widths are also given in the fifth column of Table
\ref{ta:blended}. In addition, the table presents the total inferred equivalent
width from these modelled lines in the sixth column, for these five
blended groups. These values are often consistent with the measured total
equivalent widths of the group, supporting the reliability of our model,
with some exceptions. The main exception is the OI group, for which the
inferred $W$ is larger than the total $W$ measured for the blend. This
behavior can be attributed to the systematic error discussed in section
\ref{sec:blended}: the three OI lines are very close together, and
while we assume that their absorbing components are independent and
the transmission of the individual lines can simply be multiplied to
obtain the total transmission, in reality the absorbing components are
of course the same in all three lines in each individual DLA and are
often much narrower than the BOSS spectrograph resolution and almost 
certainly highly
saturated. Therefore the measured $W$ is smaller than the inferred $W$
when treating the three lines as independent. The other smaller discrepancies
in the remaining blend groups can be attributed to the same systematic
effect.

  Our treatment of the blend groups provides mean equivalent widths
for lines of three species that are in general extremely difficult to
measure: SIV\,$\lambda1062$, FeIII\,$\lambda1122$, and
CII*\,$\lambda1335$.
The first two are in the \lya forest, and blended with FeII and SiII
lines. The lines are therefore doubly difficult to measure from
individual DLA systems, even in spectra of high resolution and
signal-to-noise ratio.
The \lya forest contamination is automatically removed in our
technique, where we measure only the mean equivalent width of a
large sample of DLAs. The FeII or SiII contamination is reliably
predicted by our model, as illustrated by the small
scatter of the measured points in Figure \ref{fig:curve} and the
$\chi^2$ value of our three-parameter fit to these points. The
systematic effects of correlated components are in this case less
serious, because the line separations are comparable to the
BOSS resolution so that overlap of narrow components is not frequent.
Still, the errors we obtain in the end for these species are relatively
high, especially for SIV because it accounts for only $\sim$ 15\% of
the total equivalent width of the blended line group.

  Finally, we can derive the equivalent width of the ZnII lines from the
blends at $2026\angs$ and $2062\angs$. The most reliable measurement
originates from the $2026\angs$ blend, for which the CrII and MgI
contributions are a relatively small correction, so we use only this
group to derive the ZnII column density included in Table \ref{ta:N}.
This result yields a mean abundance with no ionization correction
of $\rm{[ZnII/HI]}= -0.89 \pm 0.12$. This element is a particularly good 
tracer of metallicity because it is one of the least affected elements by dust
depletion \citep[e.g.,][]{Pettini1990,Pettini1997,Prochaska2002,Jenkins2009,
Cooke2011,Cooke2013}. We discuss and compare 
these abundance values in \S~\ref{sec:discussion}.


\subsection{Behaviour of the model parameters with ${ N_{\rm HI}}$}
\label{sec:curvenhi}

\begin{figure*}                  
\includegraphics[width=0.5\textwidth]{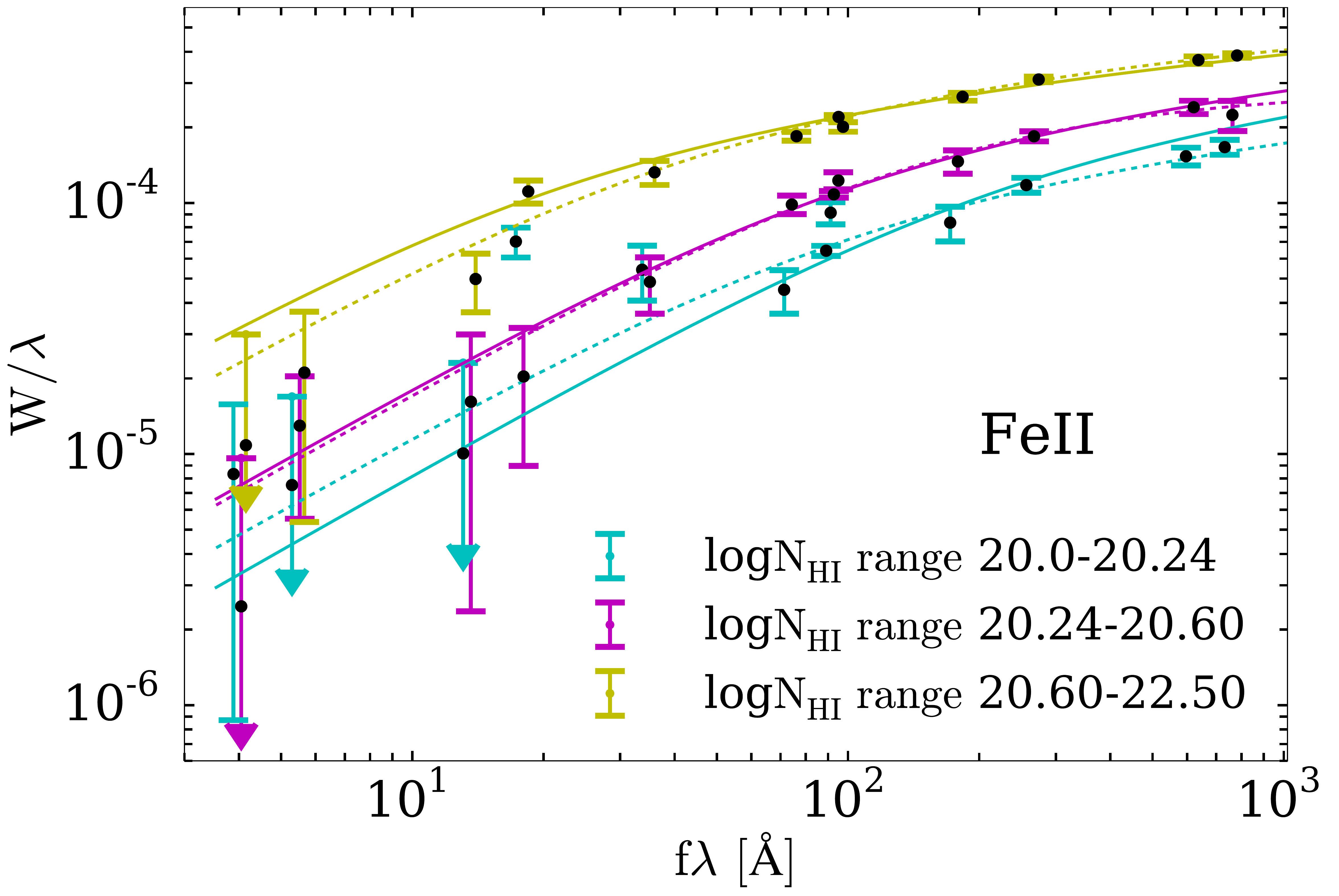}\includegraphics[width=0.48\textwidth]{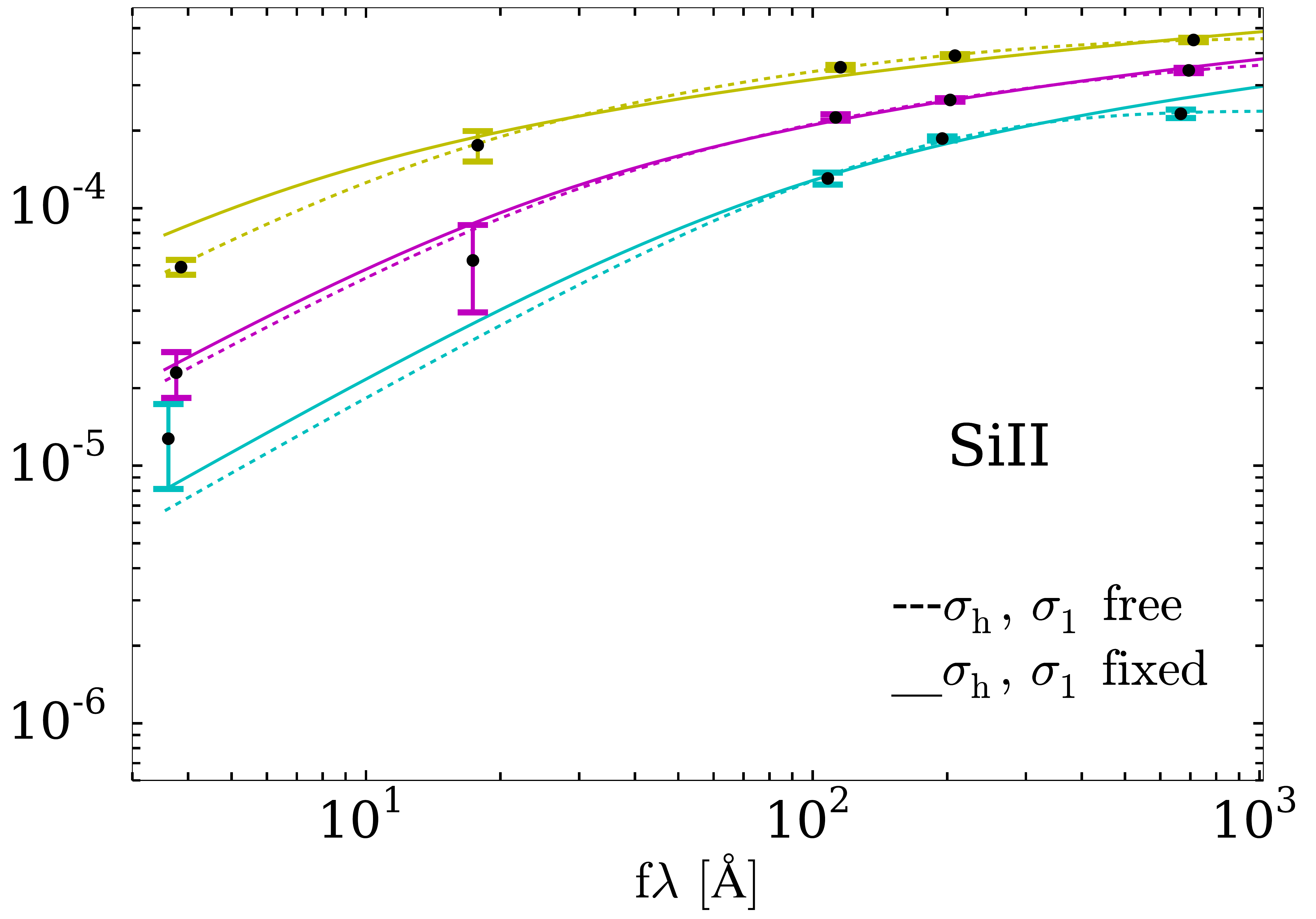}
\caption{Measurements of $W/\lambda$ versus $f\lambda$ in
3 ${ N_{\rm HI}}$ intervals. Arrows at the bottom end of an error bar
indicate that the error in W is larger than its expected value. The 
{\it solid lines} represent the modeled curves when the parameters $\sigma_{\rm h}=8.56\,{\rm km/s}$ 
and $\sigma_1=90\,{km/s}$ are fixed to the values found in \S~\ref{sec:curve}, and only the column 
density of the species, $N$, is set free. The {\it dashed lines} denote the model best-fit results when 
all three parameters are allowed to vary.  \textit{Left panel}: Fit to the FeII transitions. Only eleven 
lines are used for the fit, after excluding the line at $f\lambda \simeq 18 \angs$.
\textit{Right panel}: Fit to the SiII transitions. We consider 5 lines, except in the lowest column 
density range, where the effect of noise does not allow for reliably fitting one of the transitions. 
Data points are slightly shifted from their actual position in $f\lambda$ to avoid cluttering of error bars.}
\label{fig:cognhi}
\end{figure*}

  We now explore the behaviour of our three model parameters 
($N$, $\sigma_1$ and $\sigma_h$) for the FeII and SiII transition lines
with ${ N_{\rm HI}}$. We make use of the total sample, now dividing it
into the following three ${ N_{\rm HI}}$ intervals chosen to contain a
similar number of DLAs: $[20.0-20.24], (20.24-20.60], (20.6-22.50]$, with
mean values of ${\rm \log (N_{HI}/cm^{-2})}$ of $20.11$, $20.41$ and $20.97$,
respectively. We compute the three 
composite spectra following the procedure explained in the above sections 
and calculate the equivalent widths of the metal lines. The {\it left panel} of 
Figure \ref{fig:cognhi} presents the measurements of $W/\lambda$ versus $f\lambda$ of the FeII
transition lines in these three $N_{\rm HI}$ intervals. When modeling these results, we ignore the 
FeII\,$\lambda1125$ line, with points at $f\lambda \simeq 18\,{\rm \AA}$ in this panel of Figure 
\ref{fig:cognhi}. This line is, particularly for the lowest column density interval, strongly affected by the 
noise and continuum systematic errors, and is a clear outlier from the relation followed by
the other lines; for some reason that we have not been able to determine, this is not properly captured 
by our bootstrap calculation of the uncertainties. The {\it right panel} displays the 5 transitions of SiII. 
For the lowest column density range we have considered 4 points since in one case we have not been 
able to obtain a measurement for the equivalent width, owing to excessive noise in this very weak 
transition line.
 
  We obtain first a fit to the data setting $\sigma_h$ and $\sigma_1$ 
to the fixed values found in \S~\ref{sec:curve}, with the FeII and
SiII column densities as the only free parameter. The result is presented
in Table \ref{ta:cognhi}, and the fitted curves are displayed as the solid lines 
in the left and right panels of Figure \ref{fig:cognhi} for FeII and SiII, respectively. 
The expected value of $\chi^2$ is in this case equal to 10 for FeII and 4 for 
SiII, with 11 and 5 lines (4 in the lowest column density range due to noise effects) used 
for FeII and SiII, respectively, and only one free parameter for each species. The
values of the fits in Table \ref{ta:cognhi} demonstrate that the fits are not
good except for the middle $N_{\rm HI}$ interval: the slope of the
$W/\lambda$ versus $f\lambda$ curve is too shallow (steep) for the
high (low) $N_{\rm HI}$ interval, compared to the data, as clearly seen in the
left panel of Figure \ref{fig:cognhi} for the case of FeII. A similar trend is observed for 
the SiII lines in the right panel.

  Good fits are obtained when all three parameters are allowed to vary in each
$N_{\rm HI}$ interval, as represented by the values of $\chi^2$ in Table
\ref{ta:cognhi} for the free cases. We have not been able to fit one of the 
SiII transitions, that at $f\lambda\sim18\,\angs$, due to the effect of noise and 
for this reason we do not present the values for this free case. The general improvement 
is mainly related to an increase of $\sigma_h$ with $N_{\rm HI}$, which allows adjusting the
slope of the curve to that shown by the data, as visible in 
Figure \ref{fig:cognhi}. The value of $\sigma_1$ is lower than for
the fixed case, but generally also with a large uncertainty for the allowed upper
range. Surprisingly, for the upper density range of SiII,
$\sigma_1$ is tightly constrained, contrary to the results in other
cases. This behaviour is produced by the three rightmost data points, which have the
smallest uncertainties and are well fitted by the model. Therefore, small departures from the 
best fit $\sigma_1$ value produce large changes in $\chi^2$, thus constraining the 
range of allowed values in the parameter space.  
The ratio of the metal column density to $N_{\rm HI}$ now
decreases slightly with ${N_{\rm HI}}$.

\begin{table*}
	\centering
	\caption{Best fit parameters for the three ${ N_{\rm HI}}$ ranges}
	\label{ta:cognhi}
	\begin{tabular}{cccccc}
				&			&${\rm \hspace{3.5cm}	FeII}$		&						&${\rm \hspace{3.5cm}		SiII}$		&\\ \hline
	$\log {\rm (N_{\rm HI}/cm^{-2})}$	&			&Fixed 			&Free                                        &Fixed 			&Free \\  \hline 										
			&${\rm logN_{X}}$	&$13.967 \pm 0.020$	&$14.098 \pm 0.072$	&$14.423 \pm 0.017$ 	&\\
$20.11$		&${\rm \sigma_h}$    &$8.56$       			&$5.59\pm0.61$                &$8.56$				&\\     
			&${\rm \sigma_1}$	&$90$				&$45^{+211}_{-23}$		&$90$				&\\  
			&${\rm \chi^2}$		&$33.36$				&$13.20$				&$23.55$				&\\ \hline
			&${\rm logN_{X}}$	&$14.334 \pm 0.017$	&$14.318 \pm 0.053$	&$14.913 \pm 0.019$ 	&$14.863 \pm 0.086$\\     
$20.41$            	&${\rm \sigma_h}$ 	&$8.56$				&$9.92\pm1.01$		&$8.56$				&$9.56\pm0.61$\\       
			&${\rm \sigma_1}$	&$90$				&$31^{+191}_{-12} $		&$90$				&$46^{+202}_{-23} $\\  
			&${\rm \chi^2}$		&$4.92$				&$3.96$				&$3.49$				&$1.25$\\ \hline
			&${\rm logN_{X}}$	&$14.994\pm 0.015$		&$14.773 \pm 0.076$	&$15.540 \pm 0.017$ 	&$15.320 \pm 0.036$\\   
$20.97$		&${\rm \sigma_h}$ 	&$8.56$				&$12.03\pm0.88$  	        &$8.56$				&$12.15\pm0.66$ \\ 
			&${\rm \sigma_1}$	&$90$				&$51^{+198}_{-9}$		&$90$				&$54^{+2}_{-2}$	\\  
			&${\rm \chi^2}$		&$31.02$				&$9.43$				&$68.26$				&$0.35$\\ \hline
	\end{tabular}
	\tablecomments{Best fit values and errors for the three fit
parameters, and values of $\chi^2$ for the fits, in the 3 column density
intervals. The mean hydrogen column density is denoted in the first
column. Parameter values for the fixed case (with the $\sigma_h$ and
$\sigma_1$ values found in \S~\ref{sec:curve}) are given in the
third and fifth columns for FeII and SiII, and for the free case with
three parameters allowed to vary in the fourth and sixth columns. Values
of $\sigma_h$ and $\sigma_1$ are in units of ${\rm km\, s^{-1}}$.
The expected ${\rm \chi^2}$ values are ${\chi^2=9}$ and $7$ for FeII,
and $\chi^2=4$ and $2$ for SiII, for the \textit{fixed} and
\textit{free} cases, respectively. No fit is obtained for the lowest
$N_{\rm HI}$ interval in the \textit{free} case for SiII, because two
of the SiII lines do not have significant detections.}
\end{table*}

  We conclude that although our model works well as a fitting formula
for our observational results of the equivalent width dependence on
$f$ displayed in Figure \ref{fig:curve}, it should not be interpreted as a
physical model in a straightforward way, because the values of
$\sigma_h$ and $\sigma_1$ need to vary when we consider DLAs in
different $N_{\rm HI}$ intervals. The increase of $\sigma_h$ with
$N_{\rm HI}$ is probably not directly related to a
real increase of the internal dispersion from turbulence of absorbing
components, but to the need to fit a variable shape of the $W/\lambda$
versus $f\lambda$ curves that is caused by other effects. A more
accurate model that is calibrated to precisely account for the
properties of metal lines in high-resolution spectra would be required
to improve the physical interpretation of these fits. In the meantime,
the validity of the column densities in Table \ref{ta:N} are subject to
our assumption that the curve presented in Figure \ref{fig:curve} has a
unique shape for all the species we consider, and are furthermore
not affected by any photoionization and dust depletion corrections.

  The trend of the declining ratio $N_{\rm FeII}/N_{\rm HI}$ and
$N_{\rm SiII}/N_{\rm HI}$ with column density in the results of 
Table \ref{ta:cognhi} appears only when all three parameters are allowed
to vary.  Several works have suggested a decrease of metallicity with hydrogen
column density in DLAs \citep[e.g.,][]{Boisse1998,Kulkarni2002,Khare2004,
Akerman2005, Meiring2006}, which could also extend to sub-DLAs \citep[][]
{Peroux2003b,Dessauges2003,York2006,Kulkarni2007,Khare2007,Som2013,
Som2015,Kulkarni2015}. As seen in Figure
\ref{fig:triangle}, the value of $\sigma_h$ and the derived metal
column densities are anticorrelated. The trend of an increasing
$\sigma_h$ with $N_{\rm HI}$ is therefore correlated with the decrease
of the ratio of metal columns to the hydrogen column. In fact, when we
fix the parameters $\sigma_1$ and $\sigma_h$, the column density ratios
(see Table \ref{ta:cognhi}) remain constant. We have further checked this
point by dividing our sample of DLAs in only two groups: systems that are
usually classified as sub-DLAs in the literature, with $N_{\rm HI} <
10^{20.3} \cm^{-2}$ (and a mean column density in our total sample of
$10^{20.14} \cm^{-2}$), and standard DLAs, with $N_{\rm HI} > 10^{20.3}
\cm^{-2}$ (with a mean column density of $10^{20.72}\cm^{-2}$ in our
total sample). We find results very similar to those in Table
\ref{ta:cognhi}: the ratios of $N_{\rm FeII}$ and $N_{\rm SiII}$ to
$N_{\rm HI}$ are higher in sub-DLAs by $\sim 0.2$ dex compared to
standard DLAs, but the value of $\sigma_h$ is $7\kms$ for sub-DLAs
compared to $10 \kms$ for standard DLAs. These variations are consistent
with the expected correlation of errors seen in Figure
\ref{fig:triangle}, and we therefore conclude that the decline of column
density ratios with $N_{\rm HI}$ is not necessarily a real effect. 

   We do not assess the evolution of our model parameters with redshift, 
because of the unknown systematic uncertainties and dependencies on the model, 
and the small redshift range covered by our DLA sample.


\section{Discussion}\label{sec:discussion}

  This paper has presented a new technique to study mean properties
of the metal lines of a sample of DLAs. After evaluating a continuum for
the mean quasar spectrum in our sample of $34\,593$ detected DLAs, the
composite DLA absorption spectrum presented in Figures \ref{fig:lines24} and
\ref{fig:lines76} is obtained for the total and metal samples,
respectively, from which we obtain mean equivalent widths of all the
detectable metal lines. This is the largest sample of DLAs ever analyzed
for this purpose. Previously, similar stacking techniques were applied
by \cite{Khare2012} for the purpose of measuring the effect of dust
reddening and determining mean equivalent widths. Here, we have focused on
completing a more extensive analysis of mean equivalent widths of all
the metal lines we can detect with our much larger sample, for which we
are presenting detailed tables with bootstrap errors that include the
uncertainty in the continuum determination. Specifically, we have
analyzed the dependence of the mean equivalent width on $N_{\rm HI}$ for
low- and high-ionization lines, developed a model for the effects
of line saturation to relate equivalent widths to mean column densities,
and used the model to separate the contributions of lines
contributing to several blended groups listed in Table \ref{ta:blended}.

  The advantage of using this stacking technique with a very large
sample is that the superposition of the forest of Ly$\alpha$ and higher
order Lyman series lines from the intergalactic medium with the metal
lines associated with DLAs is automatically removed. We can therefore
measure several lines at wavelengths which have never been previously 
measured without the ambiguity due to the contamination by the
forests. Of course, the BOSS spectra are missing all the
information on the rich velocity structure of the DLA metal lines that
is observed in high-resolution, high signal-to-noise ratio spectra, but we
must bear in mind that even in the highest quality spectra, the
individual components in metal lines can be highly saturated and arise
from gas at different densities and temperatures, with a variable degree 
of turbulence, a situation which is not
essentially different from the situation we face when trying to model the mean
equivalent widths we measure here.


\subsection{Dependence of mean equivalent widths on $N_{\rm HI}$}

  As described in \S~\ref{sec:nhi}, whereas there is a clear increase
of $W$ with $N_{\rm HI}$ for low-ionization lines, the equivalent widths
of high-ionization lines stay practically constant as $N_{\rm HI}$
increases. This result is clear observational evidence in favor of the 
widely-believed picture that low-ionization metal lines arise in self-shielded
gas that is centrally concentrated in the absorption systems in a
similar way as the atomic hydrogen
\citep[e.g.,][and references therein]{Wolfe2005}. 
The high-ionization lines must, on
the other hand, arise in a more extended gas distribution around the
self-shielded gas to explain their weak dependence on $N_{\rm HI}$.
This picture has also been supported in the past by the different
velocity profiles of low-ionization and high-ionization species,
indicating that they arise from different gaseous structures
\citep{WolfeProchaska2000,Prochaska2002,Fox2007a}, but our results
uniquely demonstrate that the high-ionization gas arises in a
more extended spatial region that surrounds the low-ionization gas, as
this is the only way to understand the nearly uniform properties of
high-ionization lines over DLAs with highly variable low-ionization
column densities, combined with independent observations of the higher
incidence rate of absorption systems selected from CIV and other
high-ionization lines compared to DLAs \citep{Shull2012}.

  As explained in \S~\ref{sec:nhi}, it is not clear from our observations
if the high-ionization lines have a weak increase of $W$ with
$N_{\rm HI}$, owing to selection effects in the DLA samples we use.
Our metal sample, which includes only DLAs with a detected presence of
metal lines at the individual basis, is consistent with a constant $W$
with $N_{\rm HI}$, but this result may be affected by a stronger selection in
favor of metal-rich systems at low $N_{\rm HI}$, compared to the
mean. This interpretation is in fact consistent with the histograms in the top
panel of Figure \ref{fig:distrib}, demonstrating a clear increase of the
fraction of DLAs in the metal sample with $N_{\rm HI}$. However, the
effect may also be due in part to a fraction of false DLAs in our
sample that is higher at the low end of our column density range. The
relative importance of these two effects can only be modelled with
detailed mocks that simulate the entire process of DLA detection. 
Undertaking such a task is difficult because false DLAs will often be the result of
clusters of \lya forest lines of high column density (not reaching DLA
values) that may also have metals associated with them.

  At present, we can only conclude that any correlation of the
high-ionization lines with $N_{\rm HI}$ in DLAs cannot be stronger than
the result we have found for our total sample. The general picture
described above postulates that most of the atomic hydrogen in DLAs,
together with low-ionization metal lines, should arise from relatively
dense, self-shielded clouds, whereas the high-ionization species should
mostly arise in lower density, more extended, unshielded gas. Therefore, in 
this picture, the column densities of the two types of species should present little
correlation in any individual halo. However, a correlation may be induced 
by a dependence of the $N_{\rm HI}$ radial profile in the self-shielded clouds 
of DLAs on the metallicity or velocity dispersion of the halo (note that the 
velocity dispersion increases the metal equivalent widths, even at fixed 
metallicity, because of line saturation effects). Indeed, a correlation between 
the equivalent widths of the lines CIV$\,\lambda$1548 and SiII$\,\lambda$1526 
was reported by \citealt{Prochaska2008} (their figure 8). If absorption systems
with weak metal lines have broad cores and rarely reach the highest
hydrogen column densities in our sample, whereas systems with stronger
metal lines have more cuspy $N_{\rm HI}$ profiles with higher central
values, that would induce an increase of the high-ionization lines
$W$ with $N_{\rm HI}$ even if the high-ionization gas
has a constant $W$ unrelated to $N_{\rm HI}$ for individual systems.
This topic can motivate further work on correcting for any sample selection
effects in the future.

  We have also measured the dependence of $W$ on $N_{\rm HI}$ for the
intermediate ionization species Al\,III, C\,III, S\,III and Si\,III.
The results, given in Tables \ref{ta:lownhi24} and \ref{ta:lownhi76},
and in Figure \ref{fig:nhi} for the Al\,III case, tend to show an
increase of $W$ with $N_{\rm HI}$ that lies between the behavior of
low- and high-ionization lines, although
with fairly large uncertainties. These intermediate species can be
formed via X-ray photoionization in self-shielded regions, and may arise
in the same gas region as low-ionization species \citep{Wolfe2005}, but
may also have an important contribution from the more extended regions
of the high-ionization gas. The results suggest that S\,III and Si\,III
behave more similarly to the high-ionization lines, and Al\,III and
C\,III behave more similarly to the low-ionization lines, but the
uncertainties are too large to reach a clear conclusion at this point,
and further modeling of the sample selections is required.
The point we wish to stress here is that these observations have a high
potential to constrain models for the distribution of gas in halos and
their photoionization state, and to test theoretical results from
simulations of galaxy formation.

\subsection{Mean abundances of low-ionization species}\label{sec:abundances}

  In \S~\ref{sec:model}, we have presented our model for line
saturation used to derive the column densities listed in
Table \ref{ta:N}. How reliable are these column densities? Our basic
argument for believing they are approximately correct is that the curves
presented in Figure \ref{fig:curve} have reached the unsaturated regime at
low $f\lambda$, i.e., they follow a linear relation
$W/\lambda \propto f\lambda$ from which the column
densities are derived. The problem is, of course, that the lines we can
measure at low $f\lambda$ have large equivalent width errors, and the
column densities directly derived from the weakest lines are also
subject to these errors. Our model is used to reduce the error by
fitting all the other lines at higher $f\lambda$ with a three-parameter
curve that approaches the linear regime at low $f\lambda$ according to
reasonable assumptions on the distribution and velocity dispersions of
the individual absorption components. For internal bulk velocity
dispersions of these components $\sigma_h\simeq 9 \kms$, metal lines
should reach the unsaturated regime roughly at $W/\lambda \sim
\sigma_h/c \simeq 3\times 10^{-5}$, which is consistent with our
measurements shown in Figure \ref{fig:curve}. For ions other than FeII
and SiII, we assume that their lines follow the same curve, with variations 
only in the mean column density. If the lines for which our column
densities are based on are very weak (i.e., in the unsaturated regime),
they are not subject to any model dependence, but if they are saturated
then our model assumption of a single curve for all the low-ionization
ions is used to extrapolate to low values of $f\lambda$ and correct for
the saturation effects.

  In general, our derived abundances are slightly higher, although in  
broad agreement, than previous determinations.
Our results for $\rm {[FeII/HI]}=-1.45 \pm 0.04$, 
$ \rm {[SiII/HI]}= -0.91 \pm 0.03$, 
$\rm {[CrII/HI]}=-1.22 \pm 0.12$, $\rm{[MnII/HI]}= -1.62 \pm 0.19$, 
$\rm {[NiII/HI]}=-1.26 \pm 0.05$ and $\rm{[ZnII/HI]}= -0.89\pm 0.11$ 
are within $0.5$ dex of those reported by \cite{Khare2012} (their table 5). 
We can also
compare our results with high-resolution studies of smaller numbers of
DLAs. Our zinc result is within $3\sigma$ of the mean value reported by
\cite{Pettini2006}, ${\rm [\langle Zn/H \rangle]=-1.2}$, at $1.8<z<3.5$.
To compare with the results of \cite{Rafelski2012}, we use our mean
DLA redshift $z=2.59$ in their expression for the evolution of the mean
DLA metallicity with redshift, $\langle Z\rangle \,=\,(-0.22\pm0.03)\,
z_{DLA}\,-\, (0.65\pm0.09)$, to find $\langle Z\rangle= -1.22\pm0.10$
for their result, which is in agreement within 3$\sigma$ of our
$\rm{[ZnII/HI]}$ value and also with our result for sulfur,
$\rm{[SII/HI]}= -0.88\pm 0.04$, both species being commonly used as
metallicity indicators \citep{Wolfe2005}. Considering equation 4 in 
\cite{Quiret2016}, $\langle Z \rangle_{\rm DLAs}= (-0.15\pm0.03)z_{\rm DLA} 
-(0.60\pm0.13)$, we obtain a HI-weighted mean DLA metallicity of 
$\langle Z \rangle_{\rm DLAs}=-0.99\pm0.15$, consistent within 1$\sigma$ 
with our $\rm{[ZnII/HI]}$, $\rm{[SII/HI]}$ and $\rm{[SiII/HI]}$ results.
Our mean abundances of titanium and phosphorus from Table \ref{ta:N},
$\rm{[TiII/HI]}=-1.19 \pm 0.20$ and $\rm{[PII/HI]}= -1.01 \pm 0.14$,
are within 2$\sigma$ and 3$\sigma$, respectively, from a small number 
of detections reported by \cite{Prochaska2001} (their table 41), with mean 
values $\rm {[Ti/H]}\sim-1.45$ (13 data points) and $\rm {[P/H]} \sim -1.42$
(5 data points).

\begin{figure}       
\includegraphics[width=0.465 \textwidth]{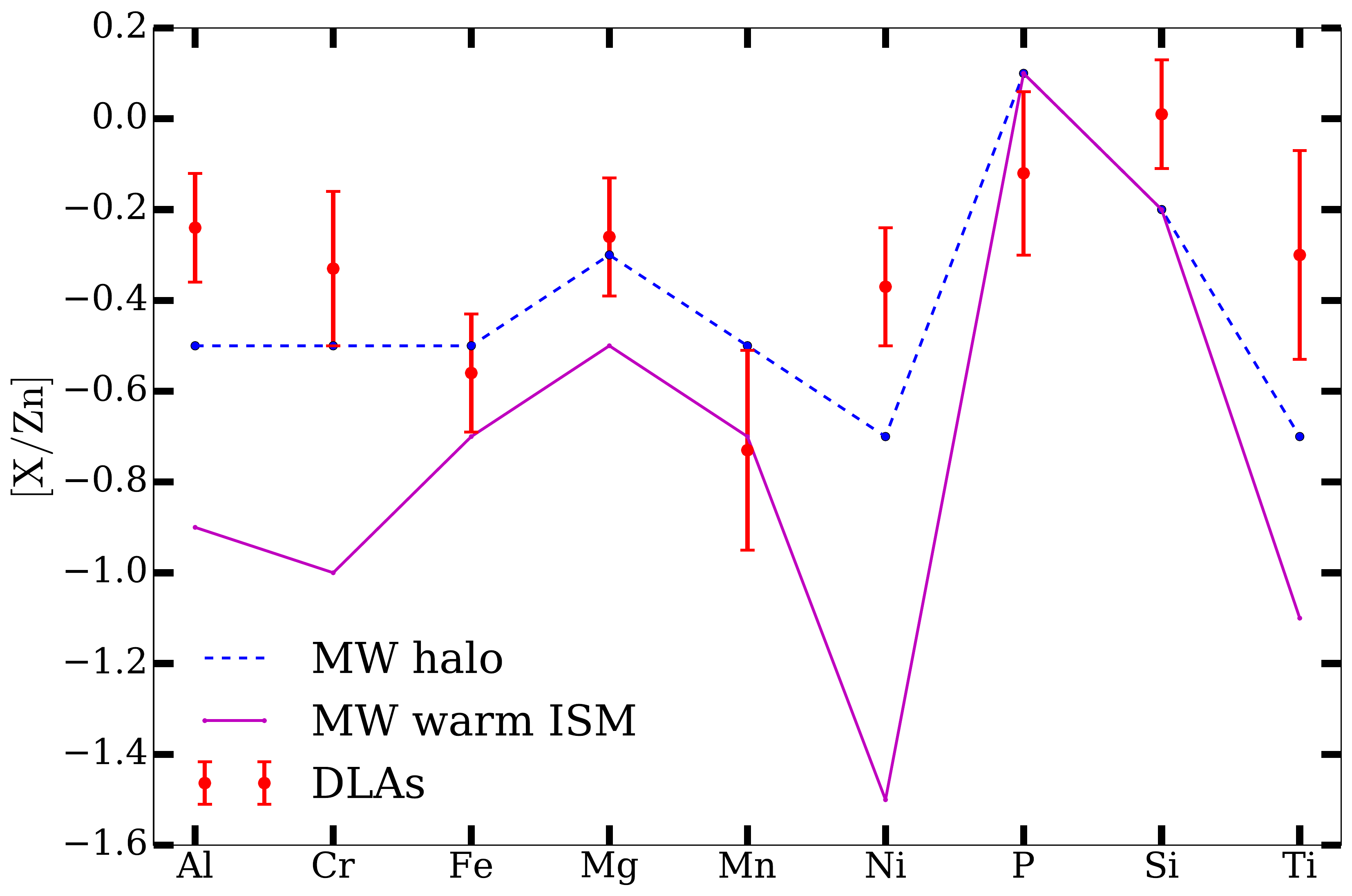}
\caption{Measurements of relative abundances, [X/Zn], and their uncertainties,    
for several species compared with the depletion patterns of the Milky Way halo 
({\it dashed blue line}) and warm ISM ({\it magenta line}) gas from \cite{Welty1999}.
}\label{fig:dp}
\end{figure}

  Our measurement of $\rm{[CrII/ZnII]}= -0.33 \pm 0.17$, which is the ratio
normally used to estimate the dust content in DLAs, is in
agreement with high-resolution studies of DLAs with abundances 
$\rm{[ZnII/HI]}\sim-1$ \citep{Pettini1997, Prochaska2002,Akerman2005}.
We also find $\rm{[FeII/ZnII]}= -0.56 \pm 0.13$ and $\rm{[NiII/ZnII]}=
-0.37 \pm 0.14$, which agree with the values by, e.g.,
\cite{Prochaska2001,Pettini2004,Khare2004,Rafelski2012}. 
Figure \ref{fig:dp} illustrates that, in general, our [X/Zn] measurements 
are in broad agreement with the gas depletion pattern in the halo of the Milky Way  
reported by \cite{Welty1999}, who analysed the absorption signatures of the 
Milky Way and LMC gas in the spectra of the supernova SN1987A. The 
value for [FeII/ZnII] indicates a modest dust depletion effect
in DLAs which yields a correction for the abundances
\citep[e.g.,][]{Vladilo2002,Vladilo2011}. 
Silicon is more weakly depleted, and sulfur is extremely weakly depleted,
and the values $\rm{[SiII/FeII]}=0.54\pm0.05$ and
$\rm{[SII/FeII]}=0.57\pm0.06$ are in agrement, for example, with
\cite{Prochaska2002,Dessauges2006,Rafelski2012,Berg2015}, when we use
our measured value of ${\rm [ZnII/FeII]} \simeq 0.6$.
Ratios of strongly depleted elements, such as 
${\rm [CrII/FeII]} = 0.23 \pm 0.12$ and
${\rm [MnII/FeII]} = -0.17 \pm 0.19$, are also in agreement with
various high-resolution studies \citep[e.g.,][]{Dessauges2006}.

   Adding the column densities of neutral and singly ionized nitrogen,
our measured ratios of $\rm{[N/SII]}= -0.79 \pm 0.08$ and
$\rm{[N/SiII]}= -0.76 \pm 0.08$ agree with the results
found for evolved DLAs, ${\rm [N/{\alpha}] \sim -0.7}$ 
\citep{Prochaska2002,Pettini2002,Centurion2003,Petitjean2008,Zafar2014}.

\subsection{Doubly-ionized species}\label{sec:highlow}

  On the intermediate ionization species, our mean measured ratio for
${\rm [AlIII/AlII]} = -0.86\pm 0.03$ is in good agreement with the
results of \cite{Vladilo2001}, who reported this ratio to vary from $-0.2$
to $-1.2$ as $\log (N_{\rm HI}/{\rm cm}^{-2})$ varies from $20.3$ to $21.0$. We find
higher values for the ratios ${\rm [SiIII/SiII]} = 0.02\pm 0.04$,
${\rm [SIII/SII]}= -0.39 \pm 0.12$, ${\rm [FeIII/FeII]} = -0.63\pm 0.10$,
and ${\rm [CIII/CII]} = 0.67 \pm 0.12$, which indicate that doubly-ionized 
species often have comparable column densities to the singly
ionized ones. As explained in \S~\ref{sec:deduction}, we do not
consider the unusually high ratio obtained for the case of carbon to be
reliable, because the CIII column density depends on an extrapolation of
our model to a large value of $W/\lambda$ that is untested in our data. The other
ratios, however, are probably not affected so severely by the systematic
modeling uncertainty of saturated lines, and they illustrate the
potential of our method to derive mean column densities for various
species in DLAs. These results can be further developed and analyzed in
the future to probe the photoionization state of DLAs.

\subsection{High-ionization species}\label{sec:highlow}

  We also report in this paper measurements of the mean column density
of several lines of high-ionization species. The strongest features are
the doublets of CIV and SiIV. The doublet ratios for the total sample
are, from Table \ref{ta:high}, equal to $1.44\pm 0.02$ for CIV and
$1.50 \pm 0.04$ for SiIV. The doublet ratio for both species 
denotes that the absorbers have a mixture of weak, unsaturated, and 
strong, saturated components, contributing almost equally to the total 
absorption feature. The fact that this ratio is constant with DLA column 
density implies that the velocity dispersion, or the mean number of 
subcomponents, should increase with column density. 

  Our result for the mean column density of CIV in our DLA sample is
${\rm \log N_{\rm CIV}=14.40\pm 0.02}$. We can compare that to the results
on CIV absorption of DLAs reported by \cite{Fox2007a}, who
studied a sample of 63 DLAs and found a relation of CIV column density
to the DLA metallicity of ${\rm \log N_{\rm CIV}=15.8 + 1.2 [Z/H]}$. Using
our mean metallicity derived from the zinc abundance, ${\rm [ZnII/HI]=-0.89}$
(from our Table \ref{ta:N}), there is a remarkably good agreement with
our measured mean CIV column density.
Our measured SiIV column density  is lower than that of CIV by a factor
$\sim 4$, also in agreement with the results of \cite{Fox2007b}.
We also obtain for the first time a mean column density of SIV in DLAs,
which is difficult to obtain because of the blend with FeII lines, and
we find a value similar to the column density of SiIV.
   
  The OVI lines at $1031$ and $1037 \angs$ are clearly detected, thanks
to the elimination in our stacked spectrum of the usual difficulty due
to blending with Lyman forest lines. We also detect the much weaker
lines of NV at $1238$ and $1242 \angs$. It has been demonstrated that
OVI is commonly present in DLAs \citep{Rahmani2010,Fox2007a}. The mean
column density we obtain, $N_{\rm OVI}=14.77 \pm 0.03$, should therefore
be a typical value in halos hosting DLAs, which, as we have seen in
Figure \ref{fig:nhi}, does not strongly depend on the HI column density.
For NV, which
has nearly optically thin lines, the mean column density we obtain
is $N_{\rm NV}=13.07 \pm 0.09$, a remarkable measurement given the small
number of cases where NV has been detected in individual DLAs
\citep{Pettini1995,Prochaska2002,Centurion2003,Henry2007,Fox2007a,Lehner2008,
Petitjean2008,Fox2009}. This result confirms the presence of phases of highly
ionized gas around DLAs, and provides observational constraints for
clarifying the photoionized or collisionally ionized state of this gas.

\subsection{Atomic and excited ionized carbon}
\label{sec:lownhi}

  We report a detection of the average column of excited CII, or CII*,
from the absorption line at $1335.71 \angs$. We are able to
deblend this line from the CII line at $1334.53\angs$, but, as can be
seen in Figure \ref{fig:fit}, the detection is only made from a
slight asymetry in the blended absorption profile and is therefore
subject to systematic errors. Nevertheless, our mean inferred column
density of $\log N_{\rm CII^{*}} = 13.35 \pm 0.06$ is in rough agreement
with the results of \cite{Wolfe2003}. Measuring the ${\rm CII}^{*}$ column
density yields estimates of the cooling rate in the DLA gas
\citep{Wolfe2003,Wolfe2004}.

  There is clear evidence for the presence of CI absorption
lines in our stacked spectrum, a species that is usually a good tracer
of molecular gas \citep{Srianand2005,Ledoux2015}. These absorption lines
are indicated 
in Figure \ref{fig:lines76} at $\lambda \simeq 1278$, 1329, 1560 and $1657 \angs$, 
but they are extremely weak and we have not attempted to measure them.

\subsection{Possible contamination by broad absorption line systems}
\label{sec:lownhi}

   Some of the DLAs in the catalog we use might be broad absorption
line systems that have been incorrectly identified as DLAs in the
low signal-to-noise ratio BOSS spectra. To test possible influence of
this potential contamination, we remove from our total 
sample all DLAs at a velocity separation $v<5000\, {\rm \kms}$ 
from the quasar. There are $3\,295$ such objects, which is $\sim10\%$ of the
total sample. After calculating the stacked spectrum for the remaining
systems, we visually confirm that there are no significant differences
between this and the total sample spectrum. 

\section{Summary and conclusions}\label{sec:summary}

  We have calculated DLA composite absorption spectra using the DLA
catalog of \cite{Noterdaeme2012} of Data Release 12 of BOSS. We have
measured the mean equivalent width of 50 absorption lines, 38 from
low-ionization species (neutral or singly-ionized), 4 from doubly
ionized species, and 8 from high-ionization species. In addition, we
have measured the total equivalent widths of 13 groups of strongly
blended lines. We have performed the same analysis with a subsample of
DLAs with individually identified metal lines, called the metal sample,
containing about a third of the total sample, which allows for the
detection of fainter absorption lines but is not representative of the
mean DLA properties. We have divided the two previous samples in 5
${ N_{\rm HI}}$ and 5 $z$ ranges and have analysed the dependence of the 
metal equivalent widths on the DLA hydrogen column density and redshift, in 
\S~\ref{sec:nhi} and \S~\ref{sec:z}, respectively.

  The increase of the mean $W$ with $N_{\rm HI}$ for the low-ionization
lines confirms that these species are closely associated with the
self-shielded atomic hydrogen in DLAs. The much weaker dependence of
the high-ionization lines on $N_{\rm HI}$, conversely, demonstrates 
that these species occur in a different gas phase at lower density that
is more extended and surrounds the low-ionization region, in view of
the fact that high-ionization lines are ubiquitous in DLAs.
 The equivalent widths decrease by a factor $\sim 1.1 - 1.5$ from 
redshift $z\sim2.1$ to $z\sim 3.5$, in general, with the high-ionization 
species showing a slightly steeper evolution. However, it is not clear 
whether this possible difference has a physical origin or it is simply 
driven by systematics. 

  We have presented a new simple model to correct for line saturation
and derive column densities for all the species we measure, described in
\S~\ref{sec:model}. Our model is quite successful in fitting the
available lines of FeII and SiII with different oscillator strengths,
as displayed in Figure \ref{fig:curve}. The inferred abundances for other
species generally agree with the determinations that have been made
from high-resolution spectra, when available. We have also been able
to measure the mean column density of OVI associated with
DLAs, which is otherwise difficult to do because of the superposition
with the \lya forest. For the first time, we obtain also a mean
equivalent width and inferred column density for NV and SIV for DLAs.
We obtained inferred column densities of several doubly ionized
species, like AlIII, SIII and SiIII, which we believe are generally
reliable (although not in the case of CIII owing to the required
extrapolation of our simple model for line saturation correction), and
can be used to test models of photoionization of the various layers
surrounding DLAs.

  In conclusion, the techniques we have developed here to use
stacked DLA absorption spectra demonstrate a promise to explore the
photoionization state and abundances of the various heavy element
species present in DLAs, with some advantages that can be exploited
with very large samples of DLAs even when the resolution and
signal-to-noise ratio of individual spectra are poor.
Further refinement of our simple model for deriving column densities
from partly saturated lines, measurement of correlations among
different absorption lines, and correlations of the large-scale bias
factor of DLAs \citep{FontRibera2012} with metal lines are all promising
avenues for future research.


\section*{Acknowledgements}

We thank the anonymous referee for a detailed and careful 
analysis which improved the quality of our work.
We are very grateful to Mat Pieri, H\'elion du Mas des Bourboux, 
Hadi Rahmani, Benjamine Racine, 
Pilar Gil-Pons, Signe Riemer-Sorensen, Xavier Prochaska, Sebasti\'an 
L\'opez, C\'eline P\'eroux, George Becker
and Max Pettini for their useful comments and suggestions during the
different stages of this work. 
This research was partially supported by the Munich Institute for Astro- and Particle
Physics (MIAPP) of the DFG cluster of excellence `Origin and Structure 
of the Universe'.
LM and JM have been partly supported by Spanish grant AYA2012-33938.
Funding for SDSS-III has been provided by the Alfred P. Sloan Foundation, 
the Participating Institutions, the National Science Foundation, and the U.S. 
Department of Energy Office of Science. The SDSS-III web site is 
\url{http://www.sdss3.org/}. SDSS-III is managed by the Astrophysical Research
Consortium for the Participating Institutions of the SDSS-III Collaboration 
including the University of Arizona, the Brazilian Participation Group, Brookhaven 
National Laboratory, Carnegie Mellon University, University of Florida, the French 
Participation Group, the German Participation Group, Harvard University, the 
Instituto de Astrofisica de Canarias, the Michigan State/Notre Dame/JINA 
Participation Group, Johns Hopkins University, Lawrence Berkeley National 
Laboratory, Max Planck Institute for Astrophysics, Max Planck Institute for 
Extraterrestrial Physics, New Mexico State University, New York University, 
Ohio State University, Pennsylvania State University, University of Portsmouth, 
Princeton University, the Spanish Participation Group, University of Tokyo, 
University of Utah, Vanderbilt University, University of Virginia,
University of Washington, and Yale University. 

\clearpage
\appendix
\section{Appendix A: Rest Equivalent Width Measurement}\label{sec:ew}

   This appendix presents the tables for the calculated rest 
equivalent widths and their uncertainties. Tables 
\ref{ta:low} and \ref{ta:high} present the values for the 
low and high-ionization species in  the metal and total samples,  
respectively. For the case of the $5$ $\rm{N_{HI}}$ 
ranges, we present
Tables \ref{ta:lownhi24}, \ref{ta:highnhi24}, \ref{ta:lownhi76}
and \ref{ta:highnhi76}, which list the measurements of low-  
and high-ionization species, in the total and metal samples, 
respectively. Whenever the convergence of the fitting method is not
reached or the measurements present negative values, the equivalent 
widths are set to zero while the uncertainties remain unchanged.
The values for $\rm{N_{HI}}$  displayed in the tables are those computed taking 
the mean column density within the corresponding interval. 
Tables \ref{ta:lowz24}, \ref{ta:highz24}, \ref{ta:lowz76} and \ref{ta:highz76} 
correspond to the same cases for the $5$ redshift bins, $\bar z_{\rm DLA}$. 
Positions with no values indicate the cases where the corresponding line wavelength 
is not covered by the spectrum in that redshift bin. 
The oscillator strengths, $f$, are those tabulated
by \cite{Morton2003} except for the NiII$\,\lambda$1317  
line, where we use the value measured by \cite{Dessauges2006}.

\begin{table}[H]
	\centering
	\caption{Rest equivalent widths of \textbf{low-ionization} metal absorption
	 lines for the two samples.}
         \label{ta:low}
	\begin{threeparttable}
	\begin{tabular}{cccc}\\
		&		&Total sample               &Metal sample   			\\ \hline        
Transition			&$f$		&$W$ ($\angs$)  	&$W$ ($\angs$)	\\ \hline
			
 AlII\,$\lambda1670$  &$1.740$  &$0.452\pm0.005$   &$0.656\pm0.005$   \\  \hline 
 
 AlIII\,$\lambda1854$  &$0.559$  &$0.117\pm0.006$   &$0.172\pm0.006$   \\ 
 AlIII\,$\lambda1862$  &$0.278$  &$0.067\pm0.006$   &$0.096\pm0.006$   \\  \hline 

 CII\,$\lambda1036^*$ &$0.118$  &$0.418\pm0.012$   &$0.579\pm0.020$   \\ \hline 

 CIII\,$\lambda977^*$  &$0.757$  &$0.646\pm0.019$   &$0.884\pm0.033$   \\  \hline
 
 CrII\,$\lambda2056$  &$0.103$  &$0.028\pm0.009$   &$0.043\pm0.008$   \\ 
 CrII\,$\lambda2066^*$  &$0.051$  &$0.034\pm0.020$   &$0.028\pm0.009$   \\  \hline
 
  FeII\,$\lambda1081^*$  &$0.013$  &$0.030\pm0.010$   &$0.058\pm0.013$   \\
 FeII\,$\lambda1096$  &$0.032$   &$0.086\pm0.010$   &$0.134\pm0.013$   \\   
 FeII\,$\lambda1125^*$ &$0.016$  &$0.069\pm0.009$   &$0.090\pm0.010$   \\  
 FeII\,$\lambda1144^*$  &$0.083$  &$0.166\pm0.007$   &$0.228\pm0.010$   \\  
 FeII\,$\lambda1608$  &$0.058$  &$0.228\pm0.004$   &$0.329\pm0.004$   \\  
 FeII\,$\lambda2249$  &$0.002$  &$0.016\pm0.020$   &$0.029\pm0.024$   \\  
 FeII\,$\lambda2260$  &$0.002$  &$0.034\pm0.017$   &$0.068\pm0.012$   \\  
 FeII\,$\lambda2344$  &$0.114$  &$0.520\pm0.014$   &$0.717\pm0.014$   \\  
 FeII\,$\lambda2374$  &$0.031$ &$0.282\pm0.014$   &$0.413\pm0.014$   \\ 
 FeII\,$\lambda2382$  &$0.320$  &$0.669\pm0.026$   &$0.973\pm0.013$   \\ 
 FeII\,$\lambda2586$  &$0.069$  &$0.462\pm0.023$   &$0.680\pm0.023$   \\  
 FeII\,$\lambda2600$  &$0.239$  &$0.722\pm0.021$   &$1.002\pm0.025$   \\  \hline
 
 MgI\,$\lambda2852$  &$1.830$  &$0.231\pm0.030$   &$0.277\pm0.031$   \\ \hline
 
 MgII\,$\lambda2796^*$	&$0.616$  &$1.149\pm0.034$   &$1.598\pm0.030$   \\  
 MgII\,$\lambda2803^*$    &$0.306$  &$1.067\pm0.025$   &$1.437\pm0.025$   \\  \hline
 
 MnII\,$\lambda2576$  &$0.361$  &$0.040\pm0.021$   &$0.059\pm0.019$   \\ \hline
 
 NII\,$\lambda1084^*$  &$$0.111$$  &$0.168\pm0.010$   &$0.250\pm0.011$   \\  \hline

 NiII\,$\lambda1317$  &$0.057$  &$0.032\pm0.005$   &$0.040\pm0.005$   \\ 
 NiII\,$\lambda1370$  &$0.077$  &$0.027\pm0.005$   &$0.037\pm0.005$   \\
 NiII\,$\lambda1454$  &$0.032$  &$0.016\pm0.002$   &$0.041\pm0.003$   \\
 NiII\,$\lambda1709$  &$0.032$  &$0.023\pm0.004$   &$0.032\pm0.004$   \\  
 NiII\,$\lambda1741$  &$0.043$  &$0.029\pm0.005$   &$0.042\pm0.005$   \\
 NiII\,$\lambda1751$  &$0.028$  &$0.024\pm0.005$   &$0.034\pm0.005$   \\  \hline  
 
 OI\,$\lambda1039^*$  &$0.009$  &$0.251\pm0.013$   &$0.343\pm0.016$   \\  
 OI\,$\lambda1302$     &$0.048$  &$0.460\pm0.006$   &$0.621\pm0.007$   \\ \hline
 
 PII\,$\lambda1152$     &$0.245$  &$0.021\pm0.008$   &$0.038\pm0.010$   \\ \hline
 
 SII\,$\lambda1250^*$  &$0.005$  &$0.044\pm0.005$   &$0.064\pm0.006$   \\  
 SII\,$\lambda1253^*$  &$0.011$  &$0.063\pm0.005$   &$0.086\pm0.005$   \\     \hline

 SIII\,$\lambda1012$  &$0.044$    &$0.067\pm0.017$   &$0.093\pm0.022$   \\ \hline 
 
 SiII\,$\lambda1020^*$  &$0.017$  &$0.096\pm0.017$   &$0.116\pm0.026$   \\  
 SiII\,$\lambda1193^*$  &$0.582$  &$0.452\pm0.006$   &$0.617\pm0.008$   \\  
 SiII\,$\lambda1304^*$  &$0.086$  &$0.333\pm0.006$   &$0.453\pm0.007$   \\  
 SiII\,$\lambda1526$  &$0.133$  &$0.443\pm0.004$   &$0.630\pm0.004$   \\  
 SiII\,$\lambda1808$  &$0.002$  &$0.059\pm0.006$   &$0.088\pm0.006$   \\   \hline
 
 SiIII\,$\lambda1206$  &$1.630$  &$0.555\pm0.008$   &$0.755\pm0.007$   \\  \hline
	\end{tabular}
	\end{threeparttable}
	 \tablecomments{Third and fourth columns give the measured rest 
equivalent widths in the total and metal samples and their bootstrap errors.
All low-ionization lines are included, except for the strongly blended
ones that are in Table \ref{ta:blended}.
The doubly-ionized aluminum and silicon lines are also included here.
The superscript $^*$ denotes weakly blended lines with
equivalent widths measured from a joint fit in a common window. 
Figures \ref{fig:fit} and \ref{fig:fit2} in Appendix \ref{sec:linefit}
display the lines included in each blend.}
\end{table}

\begin{table*}
	\centering
	\caption{Same as Table \ref{ta:low} for \textbf{high-ionization}
metal absorption lines.
	}\label{ta:high}
	\begin{tabular}{cccc}\\
		&		&Total sample               &Metal sample   			\\ \hline        
Transition			&$f$		&$W$ ($\angs$)  	&$W$ ($\angs$)	\\ \hline
 CIV\,$\lambda1548^*$  &$0.190$  &$0.429\pm0.004$   &$0.565\pm0.005$   \\  
 CIV\,$\lambda1550^*$  &$0.095$  &$0.298\pm0.003$   &$0.394\pm0.005$   \\  \hline
 
 NV\,$\lambda1238^*$  &$0.156$  &$0.023\pm0.006$   &$0.031\pm0.007$   \\ 
 NV\,$\lambda1242^*$  &$0.078$  &$0.015\pm0.006$   &$0.015\pm0.007$   \\  \hline
 
 OVI\,$\lambda1031$  &$0.133$  &$0.331\pm0.015$   &$0.411\pm0.025$   \\ 
 OVI\,$\lambda1037^*$  &$0.066$  &$0.191\pm0.011$   &$0.256\pm0.014$   \\  \hline
 
 SiIV\,$\lambda1393$  &$0.513$  &$0.292\pm0.004$   &$0.400\pm0.005$   \\ 
 SiIV\,$\lambda1402$  &$0.254$  &$0.195\pm0.004$   &$0.270\pm0.005$   \\  \hline

	\end{tabular}
\end{table*}

\begin{table*}
	\centering
	\caption{Rest equivalent widths  of absorption \textbf{low-ionization} lines with respect to the hydrogen column density for the \textbf{total sample}.} \label{ta:lownhi24}
	\begin{tabular}{ccccccc}\\
$\log({\bar N}_{\rm HI} / {\rm \cm^{-2}}) $& & 20.06 & 20.22 & 20.40 & 20.65 & 21.13 \\ \hline
Transition			&$f$			& $W$ ($\angs$) & $W$ ($\angs$) & $W$ ($\angs$) & $W$ ($\angs$) & $W$ ($\angs$) \\ \hline        
 AlII\,$\lambda1670$  &$1.740$  &$0.278\pm0.012$    &$0.345\pm0.008$  &$0.412\pm0.011$   &$0.526\pm0.011$   &$0.673\pm0.012$  \\  \hline
 
 AlIII\,$\lambda1854$  &$0.559$  &$0.076\pm0.009$    &$0.076\pm0.010$  &$0.105\pm0.009$   &$0.146\pm0.007$   &$0.173\pm0.010$  \\  
 AlIII\,$\lambda1862$  &$0.278$  &$0.039\pm0.011$    &$0.045\pm0.010$  &$0.060\pm0.010$   &$0.085\pm0.008$   &$0.097\pm0.008$  \\  \hline
 
 CII\,$\lambda1036$  &$0.118$  &$0.260\pm0.022$    &$0.318\pm0.022$  &$0.356\pm0.021$   &$0.481\pm0.022$   &$0.613\pm0.024$  \\  \hline 
 
 CIII\,$\lambda977$  &$0.757$  &$0.553\pm0.048$    &$0.699\pm0.042$  &$0.551\pm0.048$   &$0.701\pm0.043$   &$0.811\pm0.045$  \\  \hline
 
 CrII\,$\lambda2056$  &$0.103$  &$0.000\pm0.014$    &$0.000\pm0.014$  &$0.022\pm0.015$   &$0.043\pm0.013$   &$0.079\pm0.013$  \\  
 CrII\,$\lambda2066$  &$0.051$  &$0.011\pm0.010$    &$0.094\pm0.089$  &$0.016\pm0.011$   &$0.024\pm0.012$   &$0.042\pm0.012$  \\  \hline
 
 FeII\,$\lambda1081$  &$0.013$  &$0.030\pm0.018$    &$0.000\pm0.017$  &$0.005\pm0.018$   &$0.027\pm0.019$   &$0.087\pm0.015$  \\  
 FeII\,$\lambda1096$  &$0.032$  &$0.025\pm0.018$    &$0.057\pm0.018$  &$0.064\pm0.017$   &$0.096\pm0.017$   &$0.179\pm0.019$  \\  
 FeII\,$\lambda1125$  &$0.016$  &$0.056\pm0.015$    &$0.015\pm0.013$  &$0.048\pm0.017$   &$0.076\pm0.014$   &$0.154\pm0.015$  \\  
FeII\,$\lambda1144$  &$0.083$  &$0.096\pm0.015$    &$0.101\pm0.013$  &$0.150\pm0.013$   &$0.184\pm0.013$   &$0.286\pm0.012$  \\  
 FeII\,$\lambda1608$  &$0.058$  &$0.110\pm0.007$    &$0.147\pm0.006$  &$0.186\pm0.006$   &$0.264\pm0.007$   &$0.414\pm0.009$  \\  
 FeII\,$\lambda2249$  &$0.002$   &$0.033\pm0.020$    &$0.028\pm0.021$  &$0.000\pm0.020$   &$0.031\pm0.015$   &$0.003\pm0.068$  \\  
 FeII\,$\lambda2260$  &$0.002$  &$0.011\pm0.029$    &$0.018\pm0.021$  &$0.027\pm0.020$   &$0.050\pm0.017$   &$0.049\pm0.055$  \\  
 FeII\,$\lambda2344$  &$0.114$  &$0.279\pm0.027$    &$0.342\pm0.026$  &$0.463\pm0.026$   &$0.621\pm0.023$   &$0.816\pm0.024$  \\  
 FeII\,$\lambda2374$  &$0.031$  &$0.101\pm0.027$    &$0.158\pm0.028$  &$0.264\pm0.023$   &$0.315\pm0.020$   &$0.544\pm0.022$  \\  
 FeII\,$\lambda2382$  &$0.320$  &$0.441\pm0.024$    &$0.406\pm0.101$  &$0.616\pm0.072$   &$0.778\pm0.023$   &$1.041\pm0.023$  \\  
 FeII\,$\lambda2586$  &$0.069$  &$0.212\pm0.046$    &$0.300\pm0.035$  &$0.414\pm0.062$   &$0.484\pm0.031$   &$0.846\pm0.030$  \\  
 FeII\,$\lambda2600$  &$0.239$  &$0.451\pm0.035$    &$0.529\pm0.040$  &$0.644\pm0.060$   &$0.911\pm0.040$   &$0.997\pm0.042$  \\  \hline
 
 MgI\,$\lambda2852$  &$1.830$  &$0.185\pm0.053$    &$0.083\pm0.065$  &$0.100\pm0.048$   &$0.282\pm0.062$   &$0.364\pm0.036$  \\  \hline

 MgII\,$\lambda2796$  &$0.616$  &$0.700\pm0.080$    &$1.014\pm0.048$  &$1.256\pm0.078$   &$1.258\pm0.095$   &$1.500\pm0.050$  \\  
 MgII\,$\lambda2803$  &$0.306$  &$0.760\pm0.053$    &$0.793\pm0.063$  &$1.062\pm0.048$   &$1.215\pm0.042$   &$1.491\pm0.045$  \\  \hline
 
 MnII\,$\lambda2576$  &$0.361$  &$0.017\pm0.033$    &$0.001\pm0.039$  &$0.004\pm0.034$   &$0.051\pm0.056$   &$0.134\pm0.025$  \\  \hline
 
 NII\,$\lambda1084$   &$0.111$  &$0.103\pm0.018$    &$0.127\pm0.019$  &$0.142\pm0.020$   &$0.201\pm0.016$   &$0.258\pm0.020$  \\  \hline
 
 NiII\,$\lambda1317$  &$0.057$  &$0.029\pm0.011$    &$0.018\pm0.010$  &$0.029\pm0.011$   &$0.031\pm0.012$   &$0.055\pm0.011$  \\  
 NiII\,$\lambda1370$  &$0.077$  &$0.002\pm0.009$    &$0.015\pm0.009$  &$0.021\pm0.010$   &$0.021\pm0.009$   &$0.067\pm0.008$  \\  
  NiII\,$\lambda1454$  &$0.032$  &$0.000\pm0.004$    &$0.010\pm0.005$  &$0.011\pm0.004$   &$0.022\pm0.004$   &$0.043\pm0.004$  \\  
 NiII\,$\lambda1709$  &$0.032$  &$0.012\pm0.009$    &$0.008\pm0.007$  &$0.011\pm0.008$   &$0.023\pm0.006$   &$0.051\pm0.007$  \\  
 NiII\,$\lambda1741$  &$0.043$  &$0.009\pm0.008$    &$0.015\pm0.008$  &$0.016\pm0.008$   &$0.030\pm0.006$   &$0.070\pm0.006$  \\  
 NiII\,$\lambda1751$  &$0.028$  &$0.012\pm0.007$    &$0.016\pm0.008$  &$0.012\pm0.008$   &$0.029\pm0.006$   &$0.050\pm0.007$  \\  \hline
 
 OI\,$\lambda1039$  &$0.009$  &$0.114\pm0.029$    &$0.163\pm0.022$  &$0.211\pm0.024$   &$0.287\pm0.025$   &$0.414\pm0.025$  \\
  OI\,$\lambda1302$     &$0.048$ &$0.288\pm0.011$    &$0.362\pm0.010$  &$0.429\pm0.012$   &$0.543\pm0.012$   &$0.654\pm0.013$  \\  \hline
  
 PII\,$\lambda1152$     &$0.245$  &$0.010\pm0.015$    &$0.027\pm0.013$  &$0.025\pm0.015$   &$0.030\pm0.013$   &$0.000\pm0.016$  \\  \hline
 
 SII\,$\lambda1250$  &$0.005$  &$0.014\pm0.012$    &$0.023\pm0.010$  &$0.041\pm0.011$   &$0.042\pm0.011$   &$0.092\pm0.010$  \\  
 SII\,$\lambda1253$  &$0.011$  &$0.039\pm0.011$    &$0.022\pm0.010$  &$0.046\pm0.010$   &$0.081\pm0.009$   &$0.116\pm0.010$  \\  \hline
 
 SIII\,$\lambda1012$  &$0.044$  &$0.074\pm0.032$    &$0.020\pm0.031$  &$0.065\pm0.037$   &$0.029\pm0.037$   &$0.121\pm0.036$  \\  \hline
 
 SiII\,$\lambda1020$  &$0.017$  &$0.000\pm0.032$    &$0.093\pm0.031$  &$0.076\pm0.032$   &$0.048\pm0.033$   &$0.240\pm0.029$  \\  
 SiII\,$\lambda1193$  &$0.582$  &$0.333\pm0.018$    &$0.362\pm0.010$  &$0.433\pm0.014$   &$0.529\pm0.012$   &$0.605\pm0.014$  \\  
 SiII\,$\lambda1304$  &$0.086$  &$0.177\pm0.012$    &$0.251\pm0.010$  &$0.315\pm0.011$   &$0.390\pm0.010$   &$0.513\pm0.012$  \\  
 SiII\,$\lambda1526$  &$0.133$  &$0.275\pm0.008$    &$0.337\pm0.008$  &$0.407\pm0.008$   &$0.514\pm0.009$   &$0.656\pm0.010$  \\   
 SiII\,$\lambda1808$  &$0.002$  &$0.023\pm0.010$    &$0.019\pm0.009$  &$0.042\pm0.010$   &$0.061\pm0.009$   &$0.135\pm0.009$  \\  \hline
 
 SiIII\,$\lambda1206$  &$1.630$  &$0.472\pm0.016$    &$0.517\pm0.014$  &$0.586\pm0.015$   &$0.614\pm0.014$   &$0.612\pm0.017$  \\  \hline
 
	\end{tabular}
		\tablecomments{The values for $N_{\rm HI}$ are
the mean values of each range, for which values of the rest-frame
equivalent width W with its error are given for each line.}
\end{table*}

\begin{table*}
	\centering
	\caption{Rest equivalent widths  of absorption \textbf{high-ionization} lines with respect to the hydrogen column density for the \textbf{total sample}. }\label{ta:highnhi24}
	\begin{tabular}{ccccccc}\\
$\log({\bar N}_{\rm HI} / {\rm \cm^{-2}}) $& & 20.06 & 20.22 & 20.40 & 20.65 & 21.13 \\ \hline
Transition			&$f$			& $W$ ($\angs$) & $W$ ($\angs$) & $W$ ($\angs$) & $W$ ($\angs$) & $W$ ($\angs$) \\ \hline    

 CIV\,$\lambda1548$  &$0.190$  &$0.375\pm0.010$    &$0.410\pm0.008$  &$0.429\pm0.009$   &$0.459\pm0.010$   &$0.471\pm0.010$  \\  
 CIV\,$\lambda1550$  &$0.095$  &$0.263\pm0.009$    &$0.291\pm0.007$  &$0.296\pm0.007$   &$0.319\pm0.008$   &$0.319\pm0.008$  \\  \hline
 
 NV\,$\lambda1238$  &$0.156$ &$0.024\pm0.013$    &$0.061\pm0.013$  &$0.018\pm0.012$   &$0.022\pm0.012$   &$0.000\pm0.011$  \\  
 NV\,$\lambda1242$  &$0.078$  &$0.010\pm0.013$    &$0.025\pm0.014$  &$0.028\pm0.012$   &$0.018\pm0.013$   &$0.000\pm0.013$  \\  \hline
 
 OVI\,$\lambda1031$  &$0.133$  &$0.250\pm0.028$    &$0.281\pm0.028$  &$0.305\pm0.028$   &$0.323\pm0.025$   &$0.433\pm0.030$  \\  
 OVI\,$\lambda1037$  &$0.066$  &$0.156\pm0.021$    &$0.154\pm0.024$  &$0.122\pm0.022$   &$0.200\pm0.021$   &$0.265\pm0.022$  \\  \hline

 SiIV\,$\lambda1393$  &$0.513$  &$0.246\pm0.009$    &$0.272\pm0.009$  &$0.286\pm0.008$   &$0.318\pm0.008$   &$0.338\pm0.010$  \\  
 SiIV\,$\lambda1402$  &$0.254$  &$0.172\pm0.008$    &$0.191\pm0.008$  &$0.179\pm0.008$   &$0.208\pm0.008$   &$0.227\pm0.009$  \\  \hline

	\end{tabular}
\end{table*}

\begin{table*}
	\centering
	\caption{Rest equivalent widths  of absorption \textbf{low-ionization} lines with respect to the hydrogen column density for the \textbf{metal sample}.} \label{ta:lownhi76}
	\begin{tabular}{ccccccc}\\
$\log({\bar N}_{\rm HI} / {\rm \cm^{-2}}) $& & 20.06 & 20.22 & 20.40 & 20.65 & 21.13 \\ \hline
Transition			&$f$			& $W$ ($\angs$) & $W$ ($\angs$) & $W$ ($\angs$) & $W$ ($\angs$) & $W$ ($\angs$) \\ \hline

 AlII\,$\lambda1670$  &$1.740$  &$0.510\pm0.014$    &$0.561\pm0.012$  &$0.598\pm0.011$   &$0.682\pm0.011$   &$0.825\pm0.012$  \\  \hline

 AlIII\,$\lambda1854$  &$0.559$  &$0.143\pm0.014$    &$0.139\pm0.009$  &$0.145\pm0.009$   &$0.184\pm0.008$   &$0.217\pm0.009$  \\  
 AlIII\,$\lambda1862$  &$0.278$  &$0.082\pm0.014$    &$0.085\pm0.011$  &$0.079\pm0.010$   &$0.104\pm0.009$   &$0.114\pm0.009$  \\  \hline

 CII\,$\lambda1036$  &$0.118$  &$0.505\pm0.033$    &$0.513\pm0.030$  &$0.504\pm0.024$   &$0.595\pm0.022$   &$0.731\pm0.025$  \\  \hline 
 
 CIII\,$\lambda977$  &$0.757$  &$1.053\pm0.077$    &$1.229\pm0.074$  &$1.156\pm0.071$   &$1.300\pm0.082$   &$1.489\pm0.097$  \\  \hline

 CrII\,$\lambda2056$  &$0.103$ &$0.017\pm0.021$    &$0.015\pm0.017$  &$0.018\pm0.014$   &$0.043\pm0.014$   &$0.093\pm0.012$  \\  
 CrII\,$\lambda2066$  &$0.051$  &$0.011\pm0.007$    &$0.013\pm0.014$  &$0.011\pm0.018$   &$0.022\pm0.012$   &$0.057\pm0.016$  \\  \hline
 
 FeII\,$\lambda1081$  &$0.013$  &$0.024\pm0.022$    &$0.039\pm0.023$  &$0.023\pm0.021$   &$0.063\pm0.019$   &$0.097\pm0.015$  \\  
 FeII\,$\lambda1096$  &$0.032$  &$0.083\pm0.024$    &$0.098\pm0.023$  &$0.098\pm0.021$   &$0.121\pm0.019$   &$0.216\pm0.016$  \\  
 FeII\,$\lambda1125$  &$0.016$  &$0.040\pm0.022$    &$0.038\pm0.016$  &$0.075\pm0.016$   &$0.088\pm0.015$   &$0.169\pm0.016$  \\  
 FeII\,$\lambda1144$  &$0.083$  &$0.151\pm0.019$    &$0.156\pm0.016$  &$0.205\pm0.016$   &$0.220\pm0.015$   &$0.331\pm0.014$  \\  
 FeII\,$\lambda1608$  &$0.058$  &$0.192\pm0.010$    &$0.231\pm0.009$  &$0.268\pm0.008$   &$0.336\pm0.008$   &$0.502\pm0.008$  \\  
 FeII\,$\lambda2249$  &$0.002$  &$0.042\pm0.025$    &$0.042\pm0.022$  &$0.023\pm0.024$   &$0.043\pm0.018$   &$0.006\pm0.077$  \\  
 FeII\,$\lambda2260$  &$0.002$  &$0.029\pm0.028$    &$0.060\pm0.022$  &$0.040\pm0.019$   &$0.062\pm0.018$   &$0.120\pm0.017$  \\  
 FeII\,$\lambda2344$  &$0.114$  &$0.447\pm0.033$    &$0.519\pm0.037$  &$0.640\pm0.030$   &$0.783\pm0.020$   &$0.989\pm0.022$  \\  
 FeII\,$\lambda2374$  &$0.031$  &$0.168\pm0.032$    &$0.265\pm0.042$  &$0.364\pm0.026$   &$0.435\pm0.023$   &$0.657\pm0.020$  \\  
 FeII\,$\lambda2382$  &$0.320$  &$0.686\pm0.030$    &$0.794\pm0.029$  &$0.922\pm0.027$   &$1.016\pm0.024$   &$1.230\pm0.022$  \\  
 FeII\,$\lambda2586$  &$0.069$  &$0.433\pm0.048$    &$0.484\pm0.050$  &$0.601\pm0.067$   &$0.676\pm0.030$   &$0.984\pm0.031$  \\  
 FeII\,$\lambda2600$  &$0.239$  &$0.719\pm0.037$    &$0.818\pm0.047$  &$0.884\pm0.081$   &$1.108\pm0.051$   &$1.261\pm0.052$  \\  \hline
 
 MgI\,$\lambda2852$  &$1.830$  &$0.263\pm0.077$    &$0.202\pm0.051$  &$0.231\pm0.056$   &$0.306\pm0.069$   &$0.394\pm0.040$  \\  \hline
 
 MgII\,$\lambda2796$  &$0.616$  &$1.168\pm0.159$    &$1.519\pm0.057$  &$1.580\pm0.048$   &$1.746\pm0.050$   &$1.812\pm0.049$  \\  
 MgII\,$\lambda2803$  &$0.306$  &$1.142\pm0.088$    &$1.210\pm0.068$  &$1.450\pm0.053$   &$1.482\pm0.042$   &$1.733\pm0.047$  \\  \hline
 
 MnII\,$\lambda2576$  &$0.361$  &$0.003\pm0.042$    &$0.037\pm0.035$  &$0.000\pm0.043$   &$0.040\pm0.035$   &$0.147\pm0.027$  \\  \hline
 
 NII\,$\lambda1084$  &$0.111$   &$0.183\pm0.023$    &$0.224\pm0.025$  &$0.211\pm0.022$   &$0.268\pm0.019$   &$0.311\pm0.019$  \\  \hline
 
 NiII\,$\lambda1317$  &$0.057$  &$0.051\pm0.014$    &$0.028\pm0.012$  &$0.032\pm0.014$   &$0.032\pm0.012$   &$0.057\pm0.010$  \\  
 NiII\,$\lambda1370$  &$0.077$  &$0.017\pm0.012$    &$0.029\pm0.012$  &$0.035\pm0.009$   &$0.020\pm0.010$   &$0.071\pm0.008$  \\  
 NiII\,$\lambda1454$  &$0.032$  &$0.000\pm0.006$    &$0.020\pm0.005$  &$0.011\pm0.005$   &$0.021\pm0.004$   &$0.042\pm0.004$  \\  
 NiII\,$\lambda1709$  &$0.032$  &$0.012\pm0.013$    &$0.018\pm0.009$  &$0.017\pm0.008$   &$0.031\pm0.008$   &$0.059\pm0.006$  \\  
 NiII\,$\lambda1741$  &$0.043$  &$0.019\pm0.011$    &$0.019\pm0.009$  &$0.032\pm0.008$   &$0.040\pm0.008$   &$0.084\pm0.007$  \\  
 NiII\,$\lambda1751$  &$0.028$  &$0.023\pm0.011$    &$0.021\pm0.011$  &$0.020\pm0.009$   &$0.031\pm0.007$   &$0.063\pm0.007$  \\  \hline

 OI\,$\lambda1039$  &$0.009$  &$0.245\pm0.034$    &$0.283\pm0.028$  &$0.269\pm0.026$   &$0.369\pm0.025$   &$0.499\pm0.023$  \\ 
 OI\,$\lambda1302$     &$0.048$ &$0.471\pm0.017$    &$0.533\pm0.013$  &$0.573\pm0.013$   &$0.658\pm0.014$   &$0.771\pm0.013$  \\ \hline
 
  PII\,$\lambda1152$     &$0.245$  &$0.051\pm0.020$    &$0.031\pm0.018$  &$0.024\pm0.018$   &$0.028\pm0.015$   &$0.064\pm0.018$  \\  \hline
 
 SII\,$\lambda1250$  &$0.005$  &$0.037\pm0.014$    &$0.034\pm0.012$  &$0.063\pm0.014$   &$0.051\pm0.012$   &$0.109\pm0.013$  \\  
 SII\,$\lambda1253$  &$0.011$  &$0.038\pm0.013$    &$0.060\pm0.011$  &$0.070\pm0.014$   &$0.088\pm0.010$   &$0.140\pm0.011$  \\  \hline
 
 SIII\,$\lambda1012$  &$0.044$  &$0.131\pm0.060$    &$0.068\pm0.046$  &$0.090\pm0.043$   &$0.020\pm0.044$   &$0.149\pm0.042$  \\  \hline
 
 SiII\,$\lambda1020$  &$0.017$  &$0.059\pm0.050$    &$0.134\pm0.045$  &$0.095\pm0.040$   &$0.081\pm0.038$   &$0.200\pm0.033$  \\  
 SiII\,$\lambda1193$  &$0.582$  &$0.506\pm0.022$    &$0.546\pm0.014$  &$0.591\pm0.015$   &$0.645\pm0.014$   &$0.728\pm0.013$  \\  
 SiII\,$\lambda1304$  &$0.086$  &$0.316\pm0.016$    &$0.372\pm0.010$  &$0.414\pm0.013$   &$0.478\pm0.011$   &$0.595\pm0.012$  \\  
 SiII\,$\lambda1526$  &$0.133$  &$0.477\pm0.012$    &$0.532\pm0.010$  &$0.579\pm0.009$   &$0.655\pm0.009$   &$0.793\pm0.010$  \\  
 SiII\,$\lambda1808$  &$0.002$  &$0.037\pm0.014$    &$0.038\pm0.012$  &$0.061\pm0.010$   &$0.084\pm0.008$   &$0.167\pm0.008$  \\  \hline
 
 SiIII\,$\lambda1206$  &$1.630$  &$0.740\pm0.018$    &$0.757\pm0.018$  &$0.769\pm0.015$   &$0.751\pm0.015$   &$0.762\pm0.019$  \\  \hline

	\end{tabular}
		\tablecomments{The values for $N_{\rm HI}$ are
the mean values of each range, for which values of the rest-frame
equivalent width W with its error are given for each line.}
\end{table*}

\begin{table*}
	\centering
	\caption{Rest equivalent widths  of absorption \textbf{high-ionization} lines with respect to the hydrogen column density for the \textbf{metal sample}. }\label{ta:highnhi76}
	\begin{tabular}{ccccccc}\\
$\log({\bar N}_{\rm HI} / {\rm \cm^{-2}}) $& & 20.06 & 20.22 & 20.40 & 20.65 & 21.13 \\ \hline
Transition			&$f$			& $W$ ($\angs$) & $W$ ($\angs$) & $W$ ($\angs$) & $W$ ($\angs$) & $W$ ($\angs$) \\ \hline    

 CIV\,$\lambda1548$  &$0.190$  &$0.575\pm0.017$    &$0.585\pm0.013$  &$0.559\pm0.012$   &$0.558\pm0.011$   &$0.553\pm0.010$  \\  
 CIV\,$\lambda1550$  &$0.095$  &$0.409\pm0.014$    &$0.419\pm0.011$  &$0.388\pm0.010$   &$0.391\pm0.009$   &$0.373\pm0.008$  \\  \hline
 
 NV\,$\lambda1238$  &$0.156$  &$0.045\pm0.019$    &$0.077\pm0.015$  &$0.020\pm0.016$   &$0.044\pm0.012$   &$0.000\pm0.013$  \\  
 NV\,$\lambda1242$  &$0.078$  &$0.011\pm0.017$    &$0.023\pm0.016$  &$0.029\pm0.013$   &$0.021\pm0.014$   &$0.000\pm0.014$  \\  \hline
  
 OVI\,$\lambda1031$  &$0.133$  &$0.410\pm0.037$    &$0.398\pm0.036$  &$0.385\pm0.032$   &$0.396\pm0.030$   &$0.481\pm0.031$  \\  
 OVI\,$\lambda1037$  &$0.066$  &$0.260\pm0.028$    &$0.270\pm0.031$  &$0.204\pm0.023$   &$0.256\pm0.021$   &$0.310\pm0.023$  \\  \hline

 SiIV\,$\lambda1393$  &$0.513$  &$0.406\pm0.013$    &$0.410\pm0.011$  &$0.391\pm0.010$   &$0.395\pm0.010$   &$0.399\pm0.010$  \\  
 SiIV\,$\lambda1402$  &$0.254$  &$0.276\pm0.012$    &$0.291\pm0.012$  &$0.253\pm0.009$   &$0.263\pm0.009$   &$0.269\pm0.008$  \\  \hline
	\end{tabular}
\end{table*}

\begin{table*}
	\centering
	\caption{Rest equivalent widths  of absorption \textbf{low-ionization} lines with respect to the redshift  for the \textbf{total sample}.} \label{ta:lowz24}
	\begin{tabular}{ccccccc}\\
$ \bar z_{\rm DLA}$& & 2.12 & 2.32 & 2.50 & 2.49 & 2.46 \\ \hline
Transition			&$f$			& $W$ ($\angs$) & $W$ ($\angs$) & $W$ ($\angs$) & $W$ ($\angs$) & $W$ ($\angs$) \\ \hline        
 AlII\,$\lambda1670$  &$1.740$  &$0.490\pm0.009$    &$0.474\pm0.012$  &$0.462\pm0.010$   &$0.422\pm0.011$   &$0.296\pm0.010$  \\    \hline
 
 AlIII\,$\lambda1854$  &$0.559$   &$0.151\pm0.008$    &$0.133\pm0.009$  &$0.116\pm0.009$   &$0.090\pm0.011$   &$0.067\pm0.011$  \\   
 AlIII\,$\lambda1862$  &$0.278$  &$0.091\pm0.009$    &$0.077\pm0.009$  &$0.064\pm0.009$   &$0.054\pm0.012$   &$0.035\pm0.011$  \\    \hline
 
 CIII\,$\lambda977$  &$0.757$  &$-$    &$-$  &$0.809\pm0.118$   &$0.655\pm0.034$   &$0.607\pm0.022$  \\   \hline
 
 CrII\,$\lambda2056$  &$0.103$  &$0.028\pm0.012$    &$0.029\pm0.013$  &$0.030\pm0.012$   &$0.026\pm0.017$   &$0.023\pm0.022$  \\    \hline
 FeII\,$\lambda1081$  &$0.013$  &$-$    &$0.058\pm0.025$  &$0.043\pm0.016$   &$0.028\pm0.015$   &$0.010\pm0.015$  \\
 FeII\,$\lambda1096$  &$0.032$  &$-$    &$0.084\pm0.021$  &$0.100\pm0.016$   &$0.076\pm0.017$   &$0.047\pm0.015$  \\   
 FeII\,$\lambda1608$  &$0.058$  &$0.254\pm0.006$    &$0.239\pm0.009$  &$0.228\pm0.008$   &$0.204\pm0.009$   &$0.133\pm0.007$  \\  
 FeII\,$\lambda2249$  &$0.002$    &$0.038\pm0.017$    &$0.035\pm0.016$  &$0.040\pm0.016$   &$0.018\pm0.021$   &$-$  \\    
 FeII\,$\lambda2260$  &$0.002$  &$0.051\pm0.018$    &$0.039\pm0.016$  &$0.046\pm0.017$   &$0.053\pm0.021$   &$0.056\pm0.043$  \\    
 FeII\,$\lambda2344$  &$0.114$   &$0.553\pm0.019$    &$0.542\pm0.019$  &$0.517\pm0.019$   &$0.481\pm0.024$   &$0.392\pm0.070$  \\    
 FeII\,$\lambda2374$  &$0.031$  &$0.332\pm0.019$    &$0.315\pm0.019$  &$0.298\pm0.020$   &$0.233\pm0.030$   &$0.148\pm0.067$  \\    
 FeII\,$\lambda2382$  &$0.320$  &$0.756\pm0.020$    &$0.742\pm0.021$  &$0.717\pm0.019$   &$0.667\pm0.028$   &$0.516\pm0.075$  \\    
 FeII\,$\lambda2586$  &$0.069$   &$0.534\pm0.025$    &$0.505\pm0.025$  &$0.456\pm0.029$   &$0.333\pm0.048$   &$-$  \\  
 FeII\,$\lambda2600$   &$0.239$   &$0.768\pm0.024$    &$0.748\pm0.025$  &$0.682\pm0.027$   &$0.556\pm0.081$   &$-$  \\   \hline
 
 MgI\,$\lambda2852$  &$1.830$   &$0.228\pm0.029$    &$0.205\pm0.038$  &$0.329\pm0.059$   &$-$   &$-$  \\ \hline

 MgII\,$\lambda2796$  &$0.616$  &$1.267\pm0.028$    &$1.168\pm0.034$  &$1.117\pm0.032$   &$-$   &$-$  \\ 
 MgII\,$\lambda2803$  &$0.306$  &$1.117\pm0.027$    &$1.046\pm0.034$  &$1.127\pm0.042$   &$-$   &$-$  \\  \hline
 
 MnII\,$\lambda2576$  &$0.361$  &$0.056\pm0.025$    &$0.041\pm0.023$  &$0.035\pm0.025$   &$0.017\pm0.050$   &$-$  \\  \hline
 
 NII\,$\lambda1084$   &$0.111$  &$-$    &$0.207\pm0.025$  &$0.187\pm0.016$   &$0.174\pm0.013$   &$0.105\pm0.014$  \\  \hline
 
 NiII\,$\lambda1317$  &$0.057$  &$0.036\pm0.009$    &$0.041\pm0.010$  &$0.031\pm0.009$   &$0.019\pm0.011$   &$0.003\pm0.015$  \\  
 NiII\,$\lambda1370$  &$0.077$  &$0.019\pm0.008$    &$0.035\pm0.010$  &$0.023\pm0.009$   &$0.019\pm0.008$   &$0.018\pm0.010$  \\ 
 NiII\,$\lambda1709$  &$0.032$  &$0.020\pm0.007$    &$0.029\pm0.008$  &$0.019\pm0.007$   &$0.021\pm0.009$   &$0.015\pm0.007$  \\  
 NiII\,$\lambda1741$  &$0.043$   &$0.034\pm0.007$    &$0.038\pm0.008$  &$0.023\pm0.006$   &$0.027\pm0.009$   &$0.017\pm0.008$  \\  
 NiII\,$\lambda1751$  &$0.028$  &$0.026\pm0.007$    &$0.033\pm0.009$  &$0.027\pm0.008$   &$0.020\pm0.010$   &$0.010\pm0.009$  \\  \hline
 
  OI\,$\lambda1302$     &$0.048$  &$0.445\pm0.012$    &$0.412\pm0.013$  &$0.403\pm0.011$   &$0.385\pm0.012$   &$0.297\pm0.016$  \\  \hline
  
 SII\,$\lambda1250$  &$0.005$   &$0.031\pm0.010$    &$0.045\pm0.010$  &$0.045\pm0.010$   &$0.019\pm0.013$   &$0.037\pm0.015$  \\  
 SII\,$\lambda1253$  &$0.011$  &$0.067\pm0.007$    &$0.057\pm0.009$  &$0.055\pm0.010$   &$0.053\pm0.010$   &$0.040\pm0.016$  \\  \hline
 
 SIII\,$\lambda1012$  &$0.044$  &$-$    &$-$  &$0.060\pm0.042$   &$0.064\pm0.026$   &$0.045\pm0.021$  \\  \hline
 
 SiII\,$\lambda1020$  &$0.017$  &$-$    &$-$  &$0.109\pm0.030$   &$0.075\pm0.023$   &$0.088\pm0.021$  \\  
 SiII\,$\lambda1304$  &$0.086$  &$0.378\pm0.011$    &$0.352\pm0.012$  &$0.327\pm0.011$   &$0.322\pm0.011$   &$0.224\pm0.016$  \\  
 SiII\,$\lambda1526$  &$0.133$  &$0.469\pm0.007$    &$0.461\pm0.009$  &$0.440\pm0.008$   &$0.428\pm0.010$   &$0.289\pm0.008$  \\  
 SiII\,$\lambda1808$  &$0.002$  &$0.076\pm0.008$    &$0.071\pm0.009$  &$0.058\pm0.008$   &$0.050\pm0.011$   &$0.013\pm0.011$  \\  \hline
 
 SiIII\,$\lambda1206$  &$1.630$  &$0.576\pm0.013$    &$0.541\pm0.011$  &$0.540\pm0.011$   &$0.529\pm0.015$   &$0.371\pm0.019$  \\  \hline
 
	\end{tabular}
		\tablecomments{The values for $\bar z_{\rm DLA}$ are
the mean values of each range, for which values of the rest-frame
equivalent width W with its error are given for each line.}
\end{table*}

\begin{table*}
	\centering
	\caption{Rest equivalent widths  of absorption \textbf{high-ionization} lines with respect to the redshift  for the \textbf{total sample}. }\label{ta:highz24}
	\begin{tabular}{ccccccc}\\
$ \bar z_{\rm DLA}$& & 2.12 & 2.32 & 2.50 & 2.49 & 2.46 \\ \hline
Transition			&$f$			& $W$ ($\angs$) & $W$ ($\angs$) & $W$ ($\angs$) & $W$ ($\angs$) & $W$ ($\angs$) \\ \hline    

 CIV\,$\lambda1548$  &$0.190$  &$0.480\pm0.008$    &$0.452\pm0.009$  &$0.418\pm0.008$   &$0.377\pm0.009$   &$0.264\pm0.008$  \\  
 CIV\,$\lambda1550$  &$0.095$  &$0.361\pm0.007$    &$0.341\pm0.007$  &$0.307\pm0.007$   &$0.269\pm0.009$   &$0.195\pm0.008$  \\  \hline
 
 NV\,$\lambda1238$  &$0.156$  &$0.029\pm0.010$    &$0.008\pm0.012$  &$0.025\pm0.011$   &$0.030\pm0.013$   &$0.005\pm0.019$  \\  
 NV\,$\lambda1242$  &$0.078$  &$0.006\pm0.010$    &$0.011\pm0.011$  &$0.016\pm0.012$   &$0.011\pm0.015$   &$0.014\pm0.021$  \\  \hline

 SiIV\,$\lambda1393$  &$0.513$   &$0.325\pm0.008$    &$0.298\pm0.008$  &$0.272\pm0.008$   &$0.263\pm0.007$   &$0.180\pm0.011$  \\  
 SiIV\,$\lambda1402$  &$0.254$  &$0.220\pm0.008$    &$0.197\pm0.008$  &$0.180\pm0.007$   &$0.169\pm0.008$   &$0.118\pm0.010$  \\ \hline

	\end{tabular}
\end{table*}

\begin{table*}
	\centering
	\caption{Rest equivalent widths  of absorption \textbf{low-ionization} lines with respect to the redshift  for the \textbf{metal sample}.} \label{ta:lowz76}
	\begin{tabular}{ccccccc}\\
$\bar z_{\rm DLA}$& & 2.12 & 2.32 & 2.50 & 2.49 & 2.46 \\ \hline
Transition			&$f$			& $W$ ($\angs$) & $W$ ($\angs$) & $W$ ($\angs$) & $W$ ($\angs$) & $W$ ($\angs$) \\ \hline

 AlII\,$\lambda1670$  &$1.740$  &$0.671\pm0.009$    &$0.666\pm0.012$  &$0.666\pm0.012$   &$0.655\pm0.015$   &$0.564\pm0.014$  \\  \hline

 AlIII\,$\lambda1854$  &$0.559$  &$0.206\pm0.009$    &$0.198\pm0.010$  &$0.169\pm0.010$   &$0.140\pm0.012$   &$0.128\pm0.011$  \\  
 AlIII\,$\lambda1862$  &$0.278$  &$0.119\pm0.011$    &$0.109\pm0.011$  &$0.097\pm0.010$   &$0.080\pm0.012$   &$0.061\pm0.010$  \\  \hline
 
 CIII\,$\lambda977$  &$0.757$   &$-$    &$-$  &$1.198\pm0.159$   &$0.904\pm0.040$   &$0.896\pm0.031$  \\  \hline

 CrII\,$\lambda2056$  &$0.103$  &$0.031\pm0.015$    &$0.047\pm0.014$  &$0.047\pm0.013$   &$0.035\pm0.017$   &$0.056\pm0.021$  \\  \hline
 
 FeII\,$\lambda1081$  &$0.013$  &$-$    &$0.082\pm0.027$  &$0.071\pm0.017$   &$0.031\pm0.018$   &$0.029\pm0.016$  \\ 
 FeII\,$\lambda1096$  &$0.032$  &$-$    &$0.116\pm0.021$  &$0.136\pm0.018$   &$0.132\pm0.019$   &$0.090\pm0.021$  \\ 
 FeII\,$\lambda1608$  &$0.058$   &$0.341\pm0.008$    &$0.330\pm0.012$  &$0.335\pm0.010$   &$0.309\pm0.011$   &$0.257\pm0.009$  \\  
 FeII\,$\lambda2249$  &$0.002$  &$0.053\pm0.021$    &$0.056\pm0.017$  &$0.056\pm0.018$   &$0.042\pm0.022$   &$-$  \\  
 FeII\,$\lambda2260$  &$0.002$  &$0.073\pm0.021$    &$0.061\pm0.018$  &$0.068\pm0.017$   &$0.074\pm0.024$   &$0.039\pm0.034$  \\
 FeII\,$\lambda2344$  &$0.114$  &$0.750\pm0.020$    &$0.751\pm0.019$  &$0.737\pm0.021$   &$0.715\pm0.026$   &$0.583\pm0.086$  \\  
 FeII\,$\lambda2374$  &$0.031$  &$0.451\pm0.021$    &$0.438\pm0.019$  &$0.443\pm0.022$   &$0.382\pm0.030$   &$0.254\pm0.079$  \\  
 FeII\,$\lambda2382$  &$0.320$  &$1.016\pm0.021$    &$1.020\pm0.022$  &$1.004\pm0.021$   &$0.966\pm0.030$   &$0.786\pm0.098$  \\  
 FeII\,$\lambda2586$  &$0.069$  &$0.716\pm0.027$    &$0.713\pm0.026$  &$0.674\pm0.029$   &$0.619\pm0.058$   &$-$  \\
 FeII\,$\lambda2600$  &$0.239$  &$1.048\pm0.026$    &$1.049\pm0.024$  &$0.992\pm0.030$   &$0.844\pm0.129$   &$-$  \\  \hline
 
 MgI\,$\lambda2852$  &$1.830$  &$0.295\pm0.035$    &$0.282\pm0.040$  &$0.389\pm0.065$   &$-$   &$-$  \\  \hline
 
 MgII\,$\lambda2796$  &$0.616$  &$1.688\pm0.029$    &$1.656\pm0.037$  &$1.604\pm0.042$   &$-$   &$-$  \\  
 MgII\,$\lambda2803$  &$0.306$  &$1.501\pm0.029$    &$1.493\pm0.033$  &$1.550\pm0.045$   &$-$   &$-$  \\ \hline
 
 MnII\,$\lambda2576$  &$0.361$  &$0.073\pm0.029$    &$0.078\pm0.025$  &$0.043\pm0.028$   &$-$   &$-$  \\  \hline
 
 NII\,$\lambda1084$  &$0.111$   &$-$    &$0.281\pm0.030$  &$0.274\pm0.018$   &$0.255\pm0.016$   &$0.193\pm0.020$  \\ \hline
 
 NiII\,$\lambda1317$  &$0.057$  &$0.040\pm0.011$    &$0.059\pm0.012$  &$0.041\pm0.012$   &$0.036\pm0.016$   &$0.012\pm0.017$  \\ 
 NiII\,$\lambda1370$  &$0.077$  &$0.033\pm0.009$    &$0.045\pm0.012$  &$0.033\pm0.011$   &$0.033\pm0.013$   &$0.033\pm0.015$  \\  
 NiII\,$\lambda1709$  &$0.032$  &$0.035\pm0.007$    &$0.040\pm0.010$  &$0.029\pm0.009$   &$0.028\pm0.009$   &$0.027\pm0.009$  \\  
 NiII\,$\lambda1741$  &$0.043$  &$0.048\pm0.009$    &$0.050\pm0.009$  &$0.036\pm0.008$   &$0.036\pm0.010$   &$0.039\pm0.008$  \\ 
 NiII\,$\lambda1751$  &$0.028$  &$0.035\pm0.009$    &$0.042\pm0.011$  &$0.035\pm0.008$   &$0.028\pm0.010$   &$0.027\pm0.009$  \\  \hline

 OI\,$\lambda1302$     &$0.048$ &$0.568\pm0.012$    &$0.555\pm0.015$  &$0.562\pm0.013$   &$0.567\pm0.017$   &$0.508\pm0.019$  \\ \hline
 
 SII\,$\lambda1250$  &$0.005$  &$0.051\pm0.010$    &$0.075\pm0.013$  &$0.065\pm0.012$   &$0.035\pm0.017$   &$0.068\pm0.026$  \\  
 SII\,$\lambda1253$  &$0.011$   &$0.093\pm0.009$    &$0.076\pm0.012$  &$0.079\pm0.011$   &$0.094\pm0.014$   &$0.047\pm0.021$  \\   \hline
 
 SIII\,$\lambda1012$  &$0.044$  &$-$    &$-$  &$0.083\pm0.061$   &$0.104\pm0.032$   &$0.075\pm0.027$  \\  \hline
 
 SiII\,$\lambda1020$  &$0.017$  &$-$    &$-$  &$0.156\pm0.040$   &$0.128\pm0.027$   &$0.133\pm0.024$  \\  
 SiII\,$\lambda1304$  &$0.086$  &$0.492\pm0.012$    &$0.478\pm0.013$  &$0.457\pm0.012$   &$0.458\pm0.016$   &$0.391\pm0.019$  \\  
 SiII\,$\lambda1526$  &$0.133$  &$0.634\pm0.009$    &$0.634\pm0.010$  &$0.624\pm0.011$   &$0.634\pm0.012$   &$0.557\pm0.013$  \\  
 SiII\,$\lambda1808$  &$0.002$  &$0.102\pm0.010$    &$0.099\pm0.010$  &$0.085\pm0.009$   &$0.077\pm0.010$   &$0.051\pm0.011$  \\  \hline
 
 SiIII\,$\lambda1206$  &$1.630$   &$0.763\pm0.014$    &$0.730\pm0.017$  &$0.736\pm0.015$   &$0.763\pm0.019$   &$0.665\pm0.023$  \\   \hline

	\end{tabular}
		\tablecomments{The values for $\bar z_{\rm DLA}$ are
the mean values of each range, for which values of the rest-frame
equivalent width W with its error are given for each line.}
\end{table*}

\begin{table*}
	\centering
	\caption{Rest equivalent widths  of absorption \textbf{high-ionization} lines with respect to the redshift  for the \textbf{metal sample}. }\label{ta:highz76}
	\begin{tabular}{ccccccc}\\
$ \bar z_{\rm DLA}$& & 2.12 & 2.32 & 2.50 & 2.49 & 2.46 \\ \hline
Transition			&$f$			& $W$ ($\angs$) & $W$ ($\angs$) & $W$ ($\angs$) & $W$ ($\angs$) & $W$ ($\angs$) \\ \hline    

 CIV\,$\lambda1548$  &$0.190$  &$0.608\pm0.010$    &$0.582\pm0.011$  &$0.550\pm0.011$   &$0.512\pm0.013$   &$0.444\pm0.015$  \\  
 CIV\,$\lambda1550$  &$0.095$  &$0.463\pm0.008$    &$0.438\pm0.011$  &$0.404\pm0.009$   &$0.358\pm0.012$   &$0.317\pm0.013$  \\  \hline
 
 NV\,$\lambda1238$  &$0.156$  &$0.034\pm0.012$    &$0.021\pm0.013$  &$0.029\pm0.016$   &$0.035\pm0.018$   &$0.017\pm0.024$  \\  
 NV\,$\lambda1242$  &$0.078$ &$0.004\pm0.013$    &$0.004\pm0.013$  &$0.014\pm0.017$   &$0.007\pm0.018$   &$0.040\pm0.025$  \\ \hline

 SiIV\,$\lambda1393$  &$0.513$  &$0.427\pm0.009$    &$0.402\pm0.012$  &$0.381\pm0.010$   &$0.377\pm0.010$   &$0.328\pm0.018$  \\ 
 SiIV\,$\lambda1402$  &$0.254$  &$0.291\pm0.008$    &$0.269\pm0.011$  &$0.260\pm0.007$   &$0.248\pm0.012$   &$0.207\pm0.014$  \\  \hline
	\end{tabular}
\end{table*}

\clearpage
\section{Appendix B: Absorption line fitting}\label{sec:linefit}
Figures \ref{fig:fit} and \ref{fig:fit2} display the result of the profile fits
of every absorption feature analysed in this work.

\begin{figure}[H]               
\includegraphics[width=0.25\textwidth]{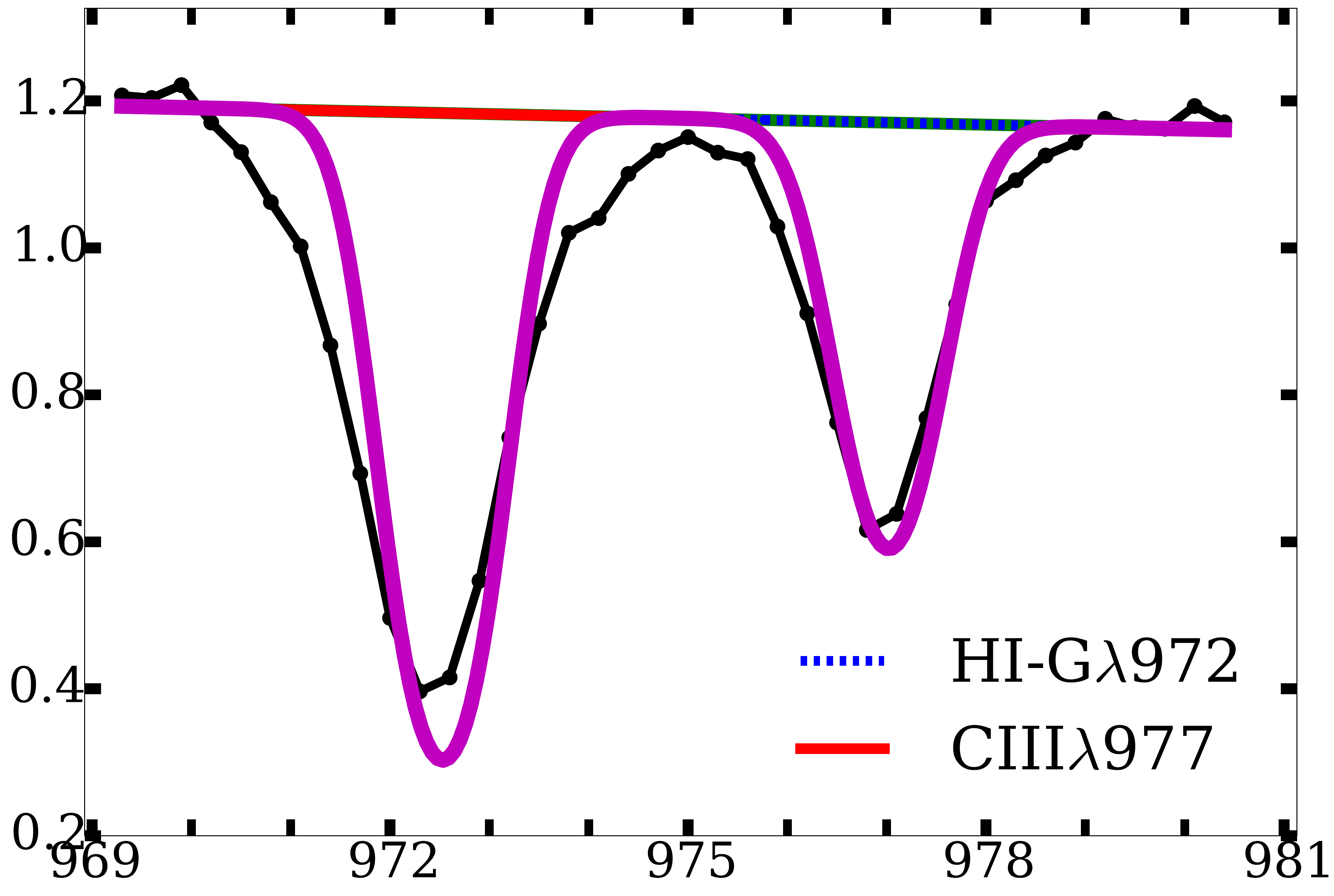}\includegraphics[width=0.25\textwidth]{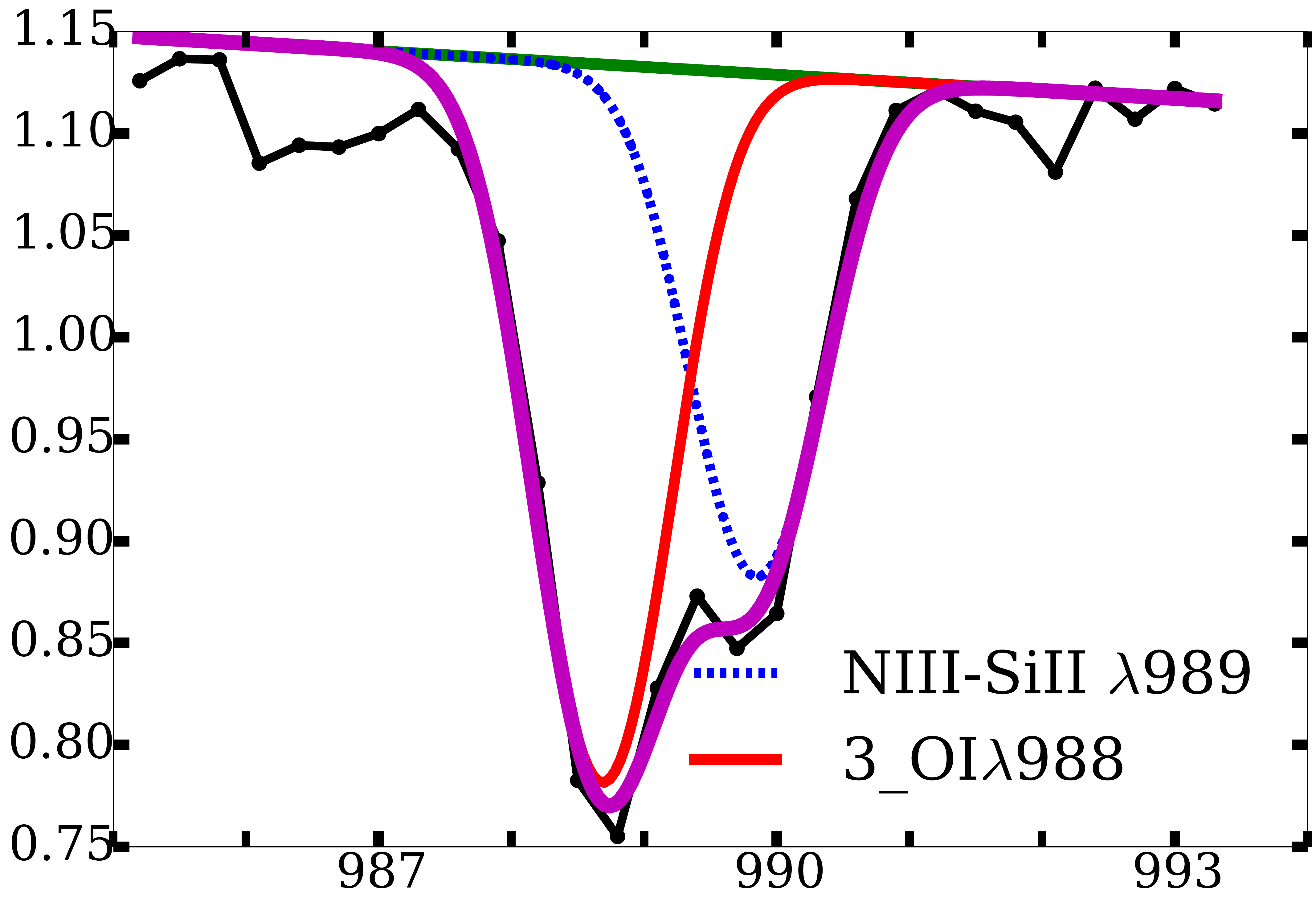}\includegraphics[width=0.25\textwidth]{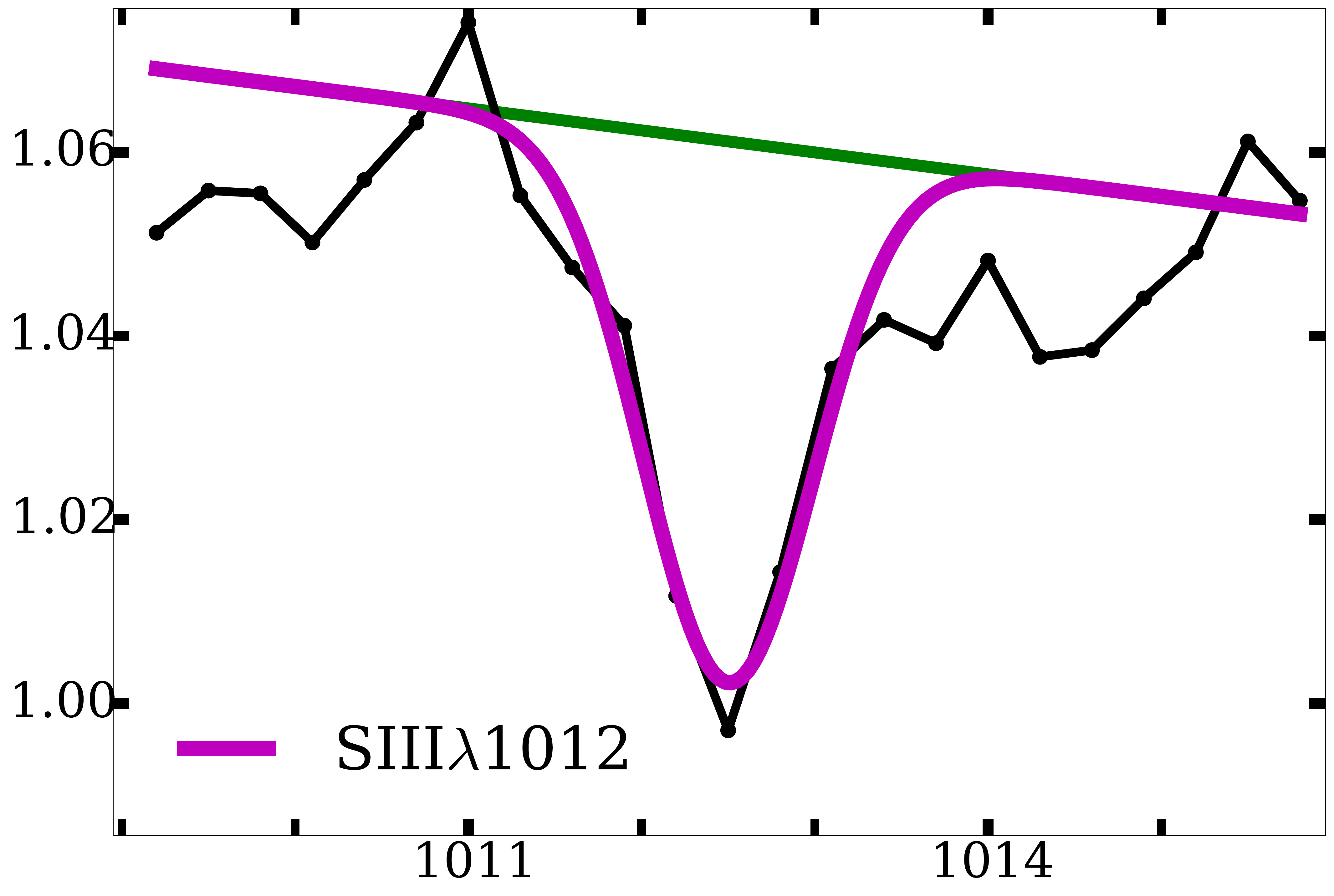}\includegraphics[width=0.25\textwidth]{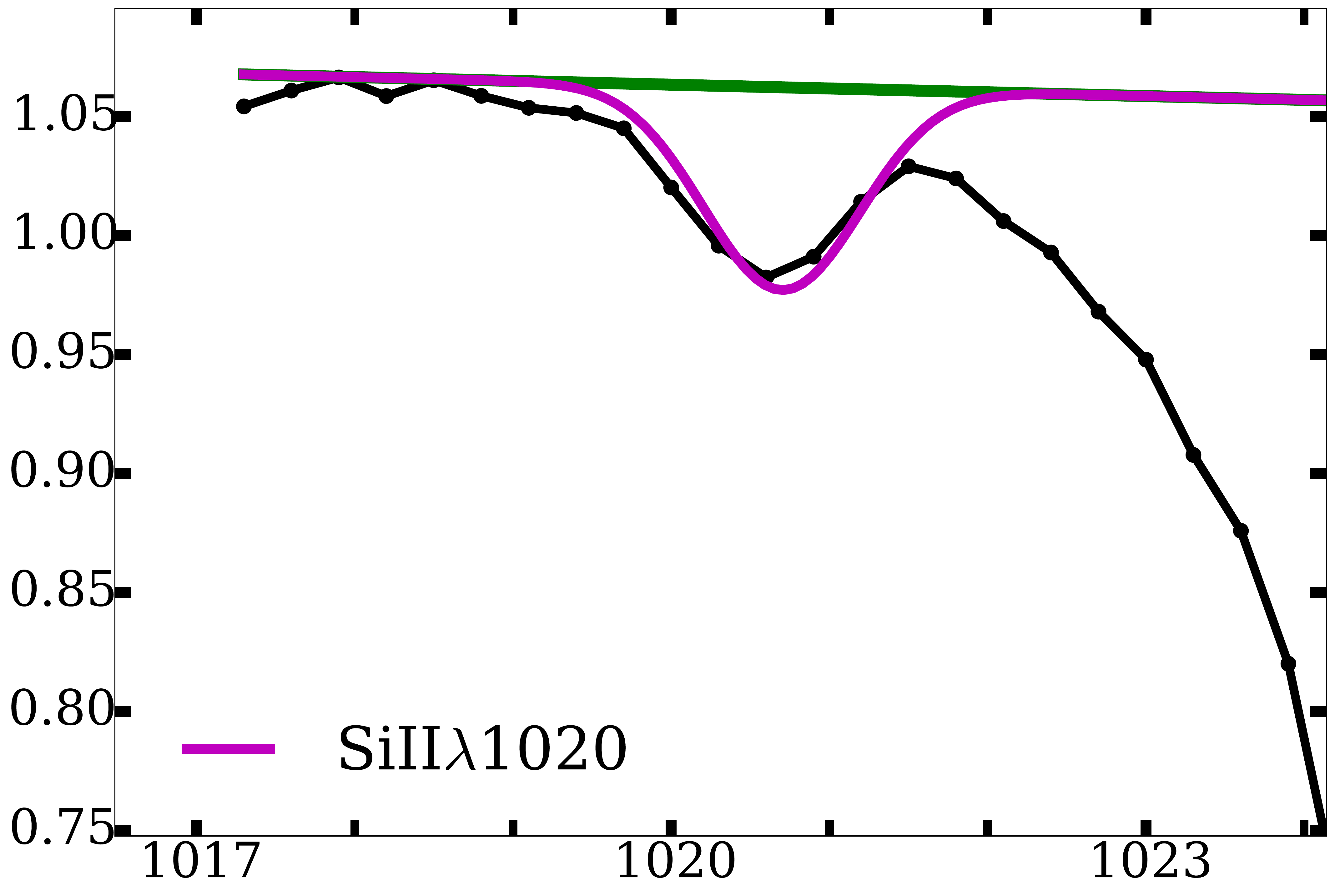}
\includegraphics[width=0.25\textwidth]{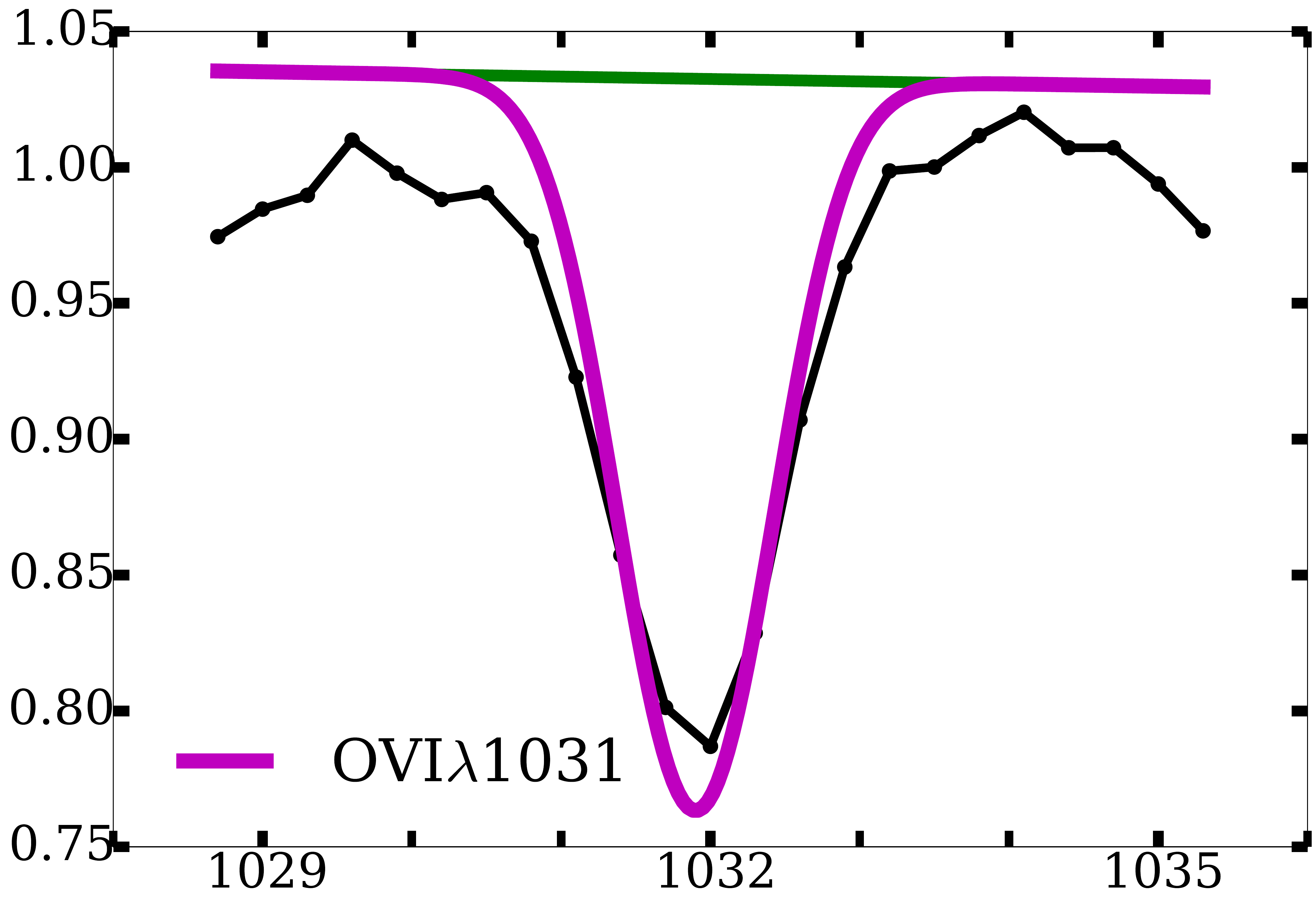}\includegraphics[width=0.25\textwidth]{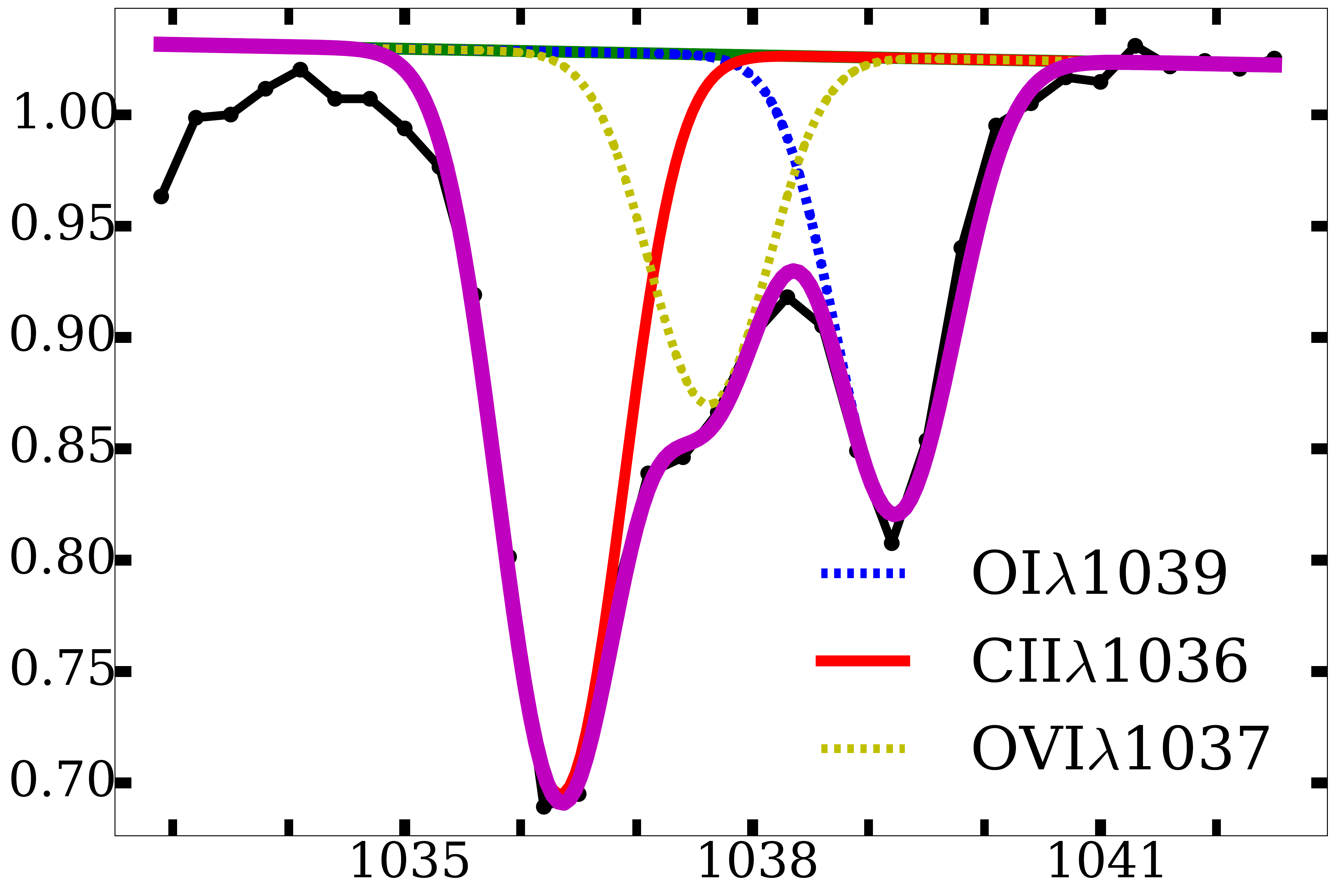}\includegraphics[width=0.25\textwidth]{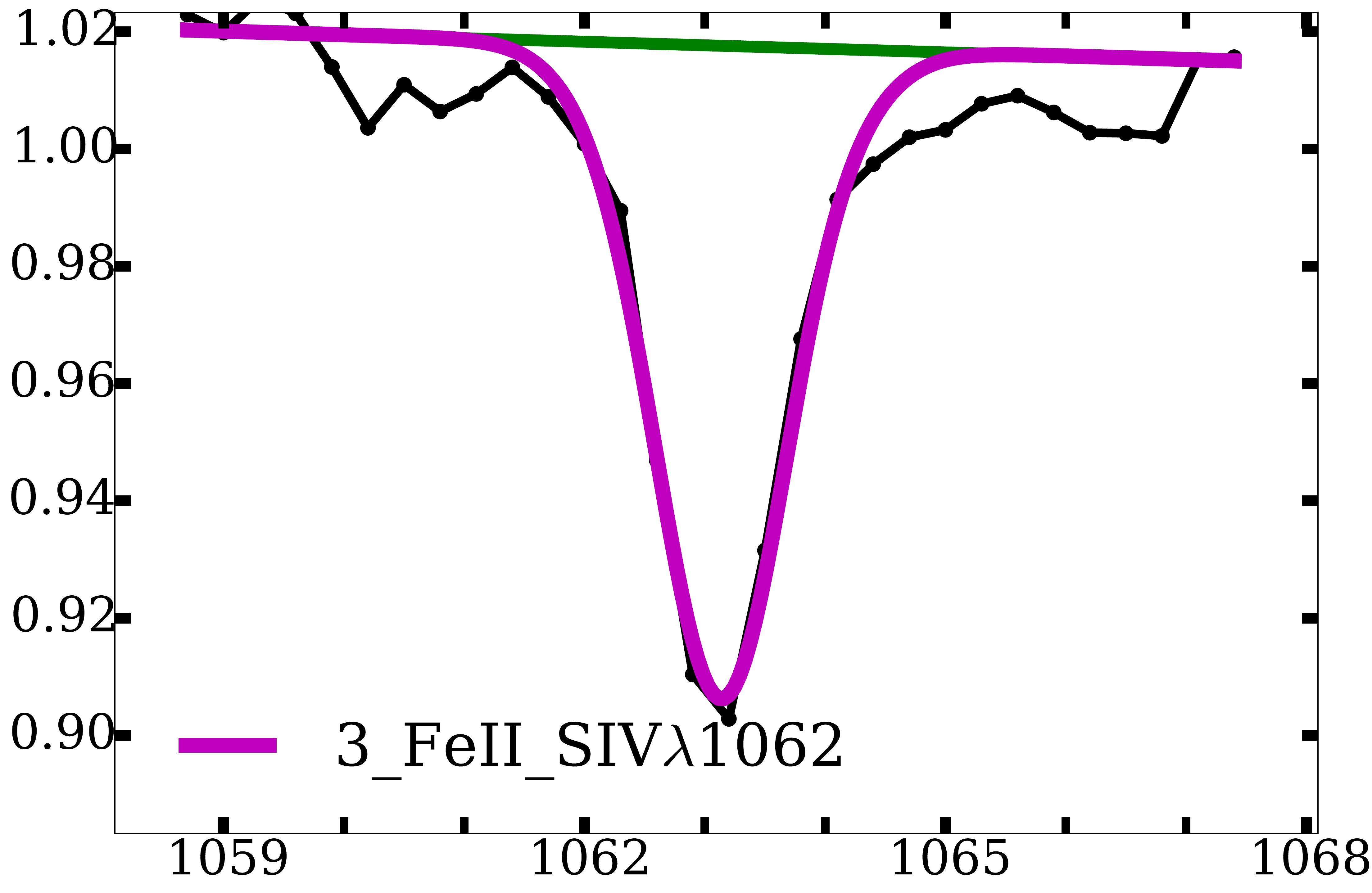}\includegraphics[width=0.25\textwidth]{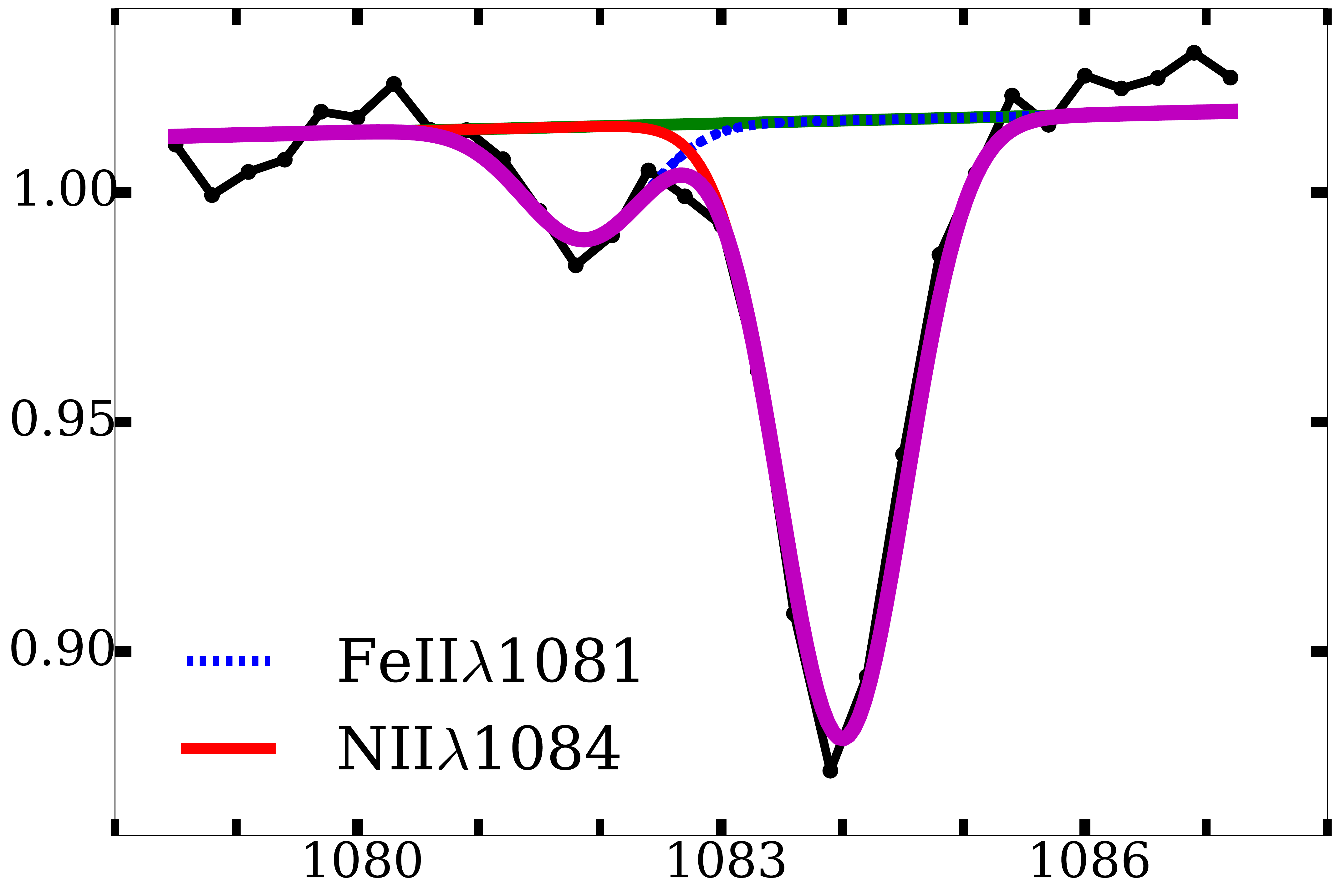}
\includegraphics[width=0.25\textwidth]{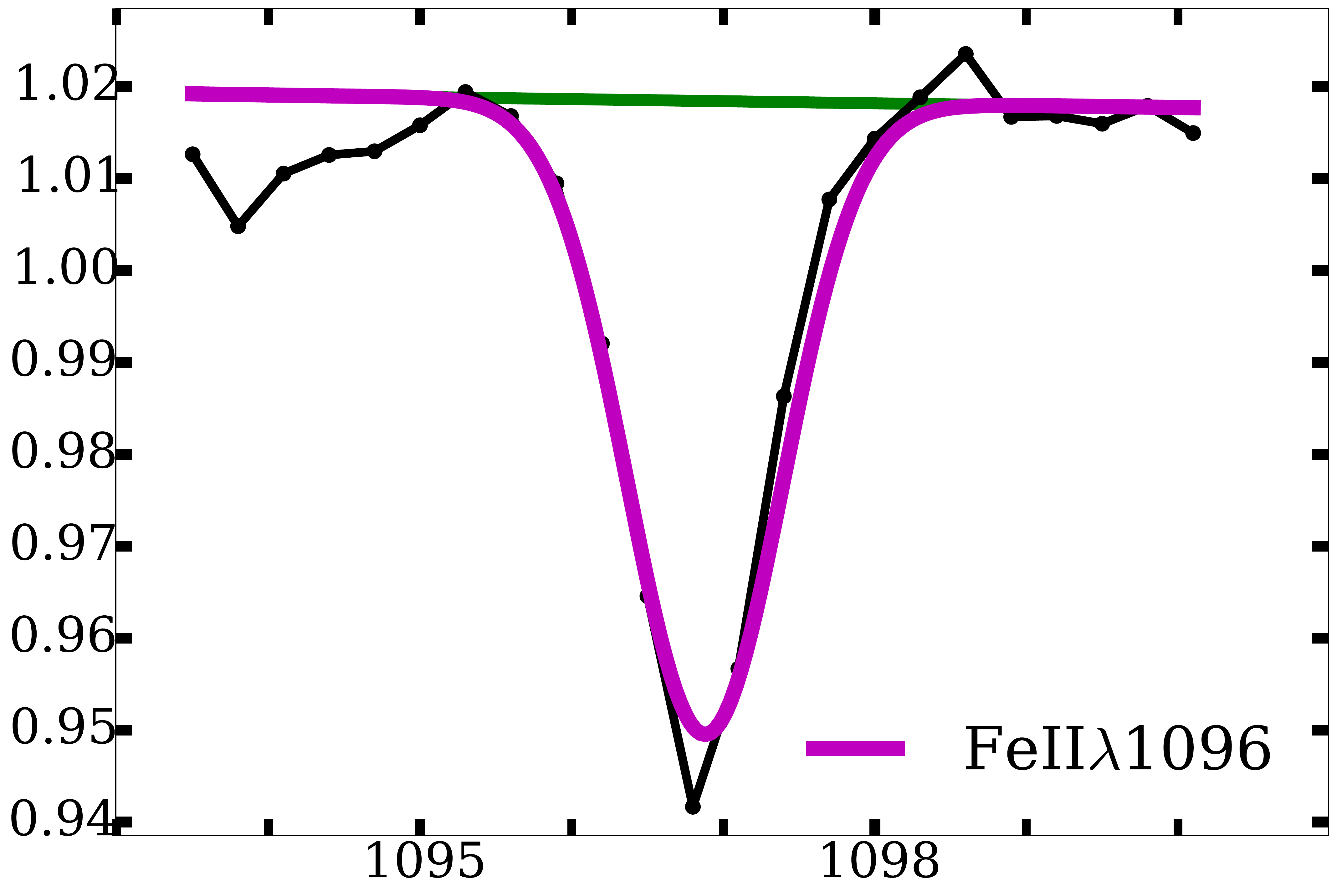}\includegraphics[width=0.25\textwidth]{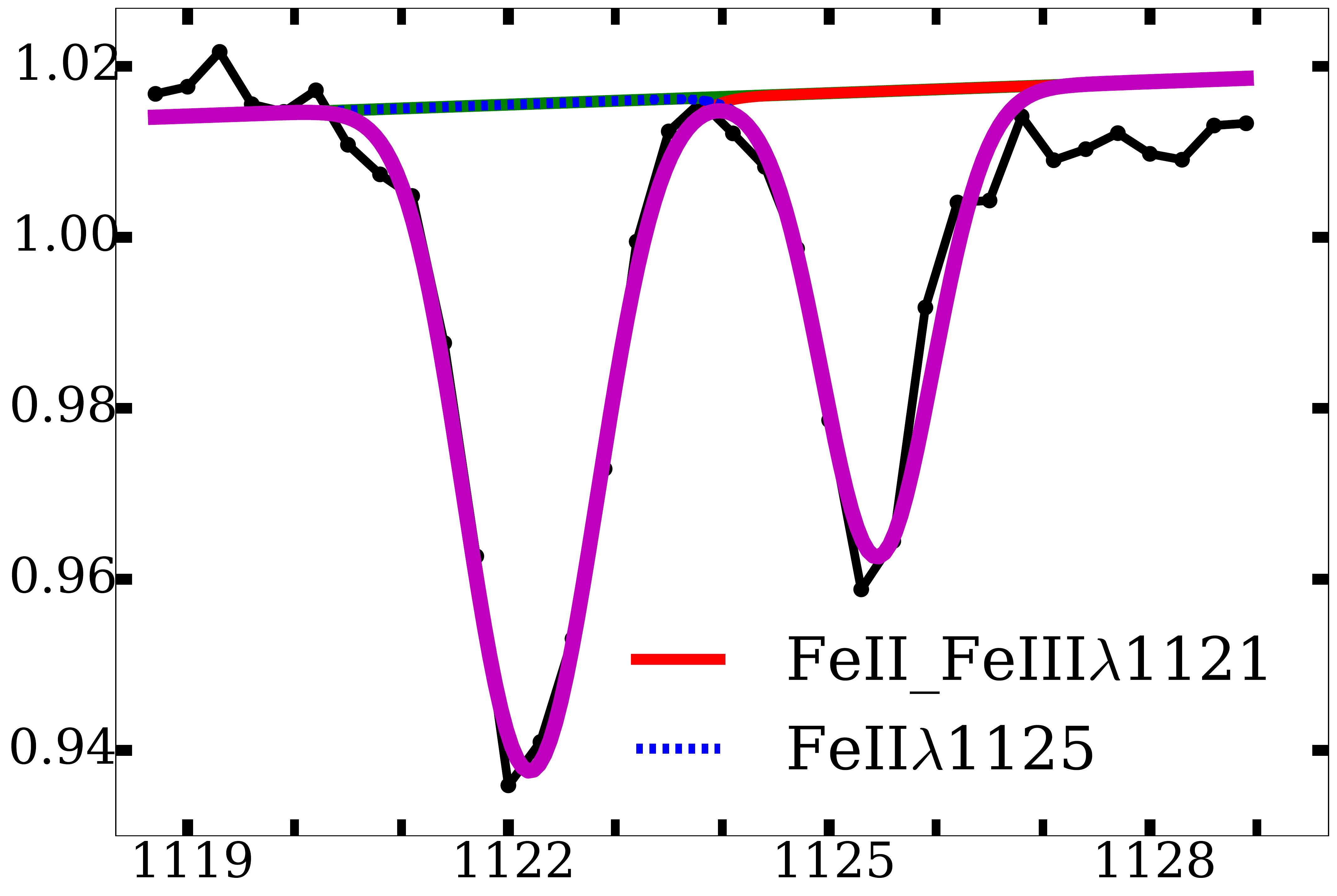}\includegraphics[width=0.25\textwidth]{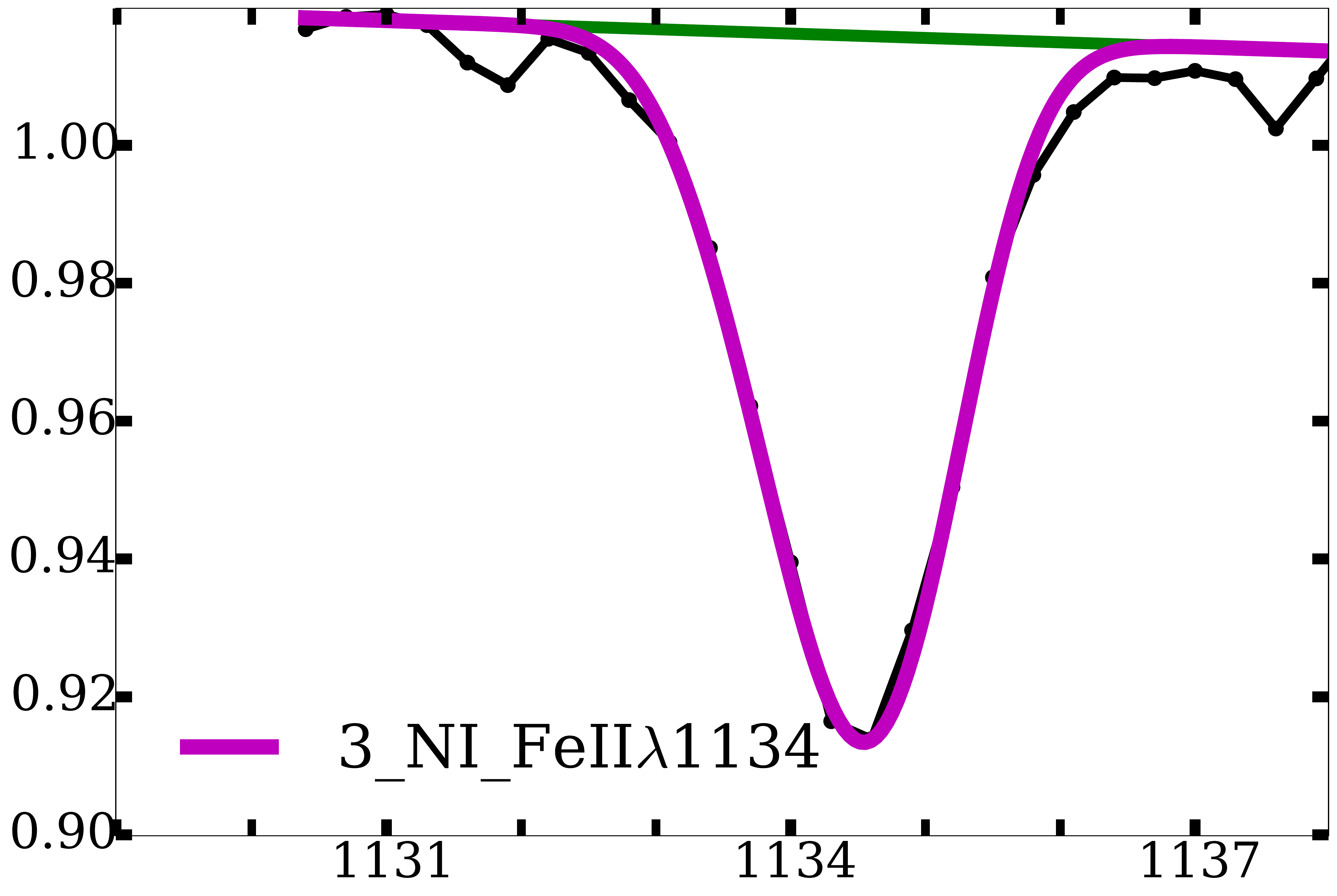}\includegraphics[width=0.25\textwidth]{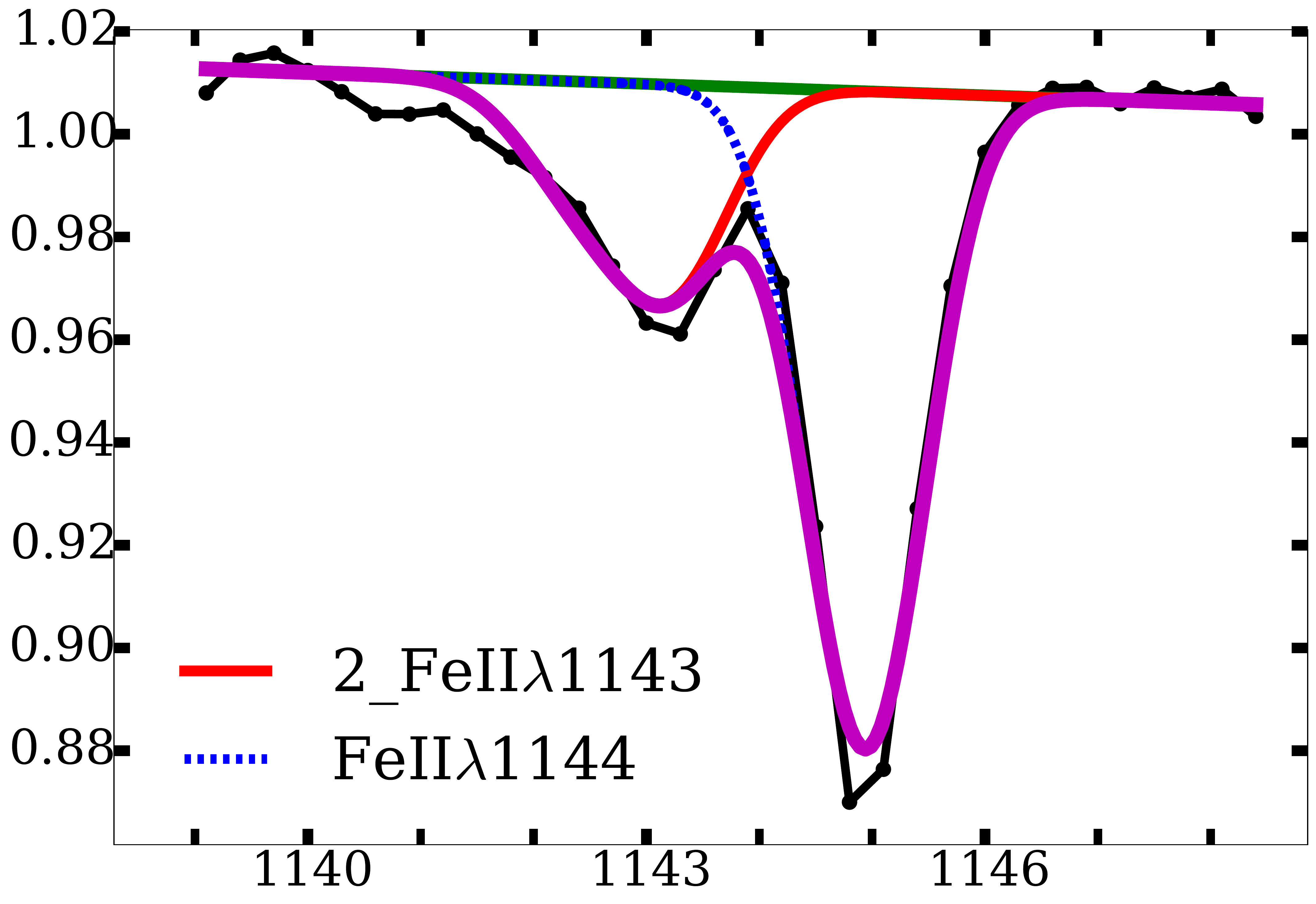}
\includegraphics[width=0.25\textwidth]{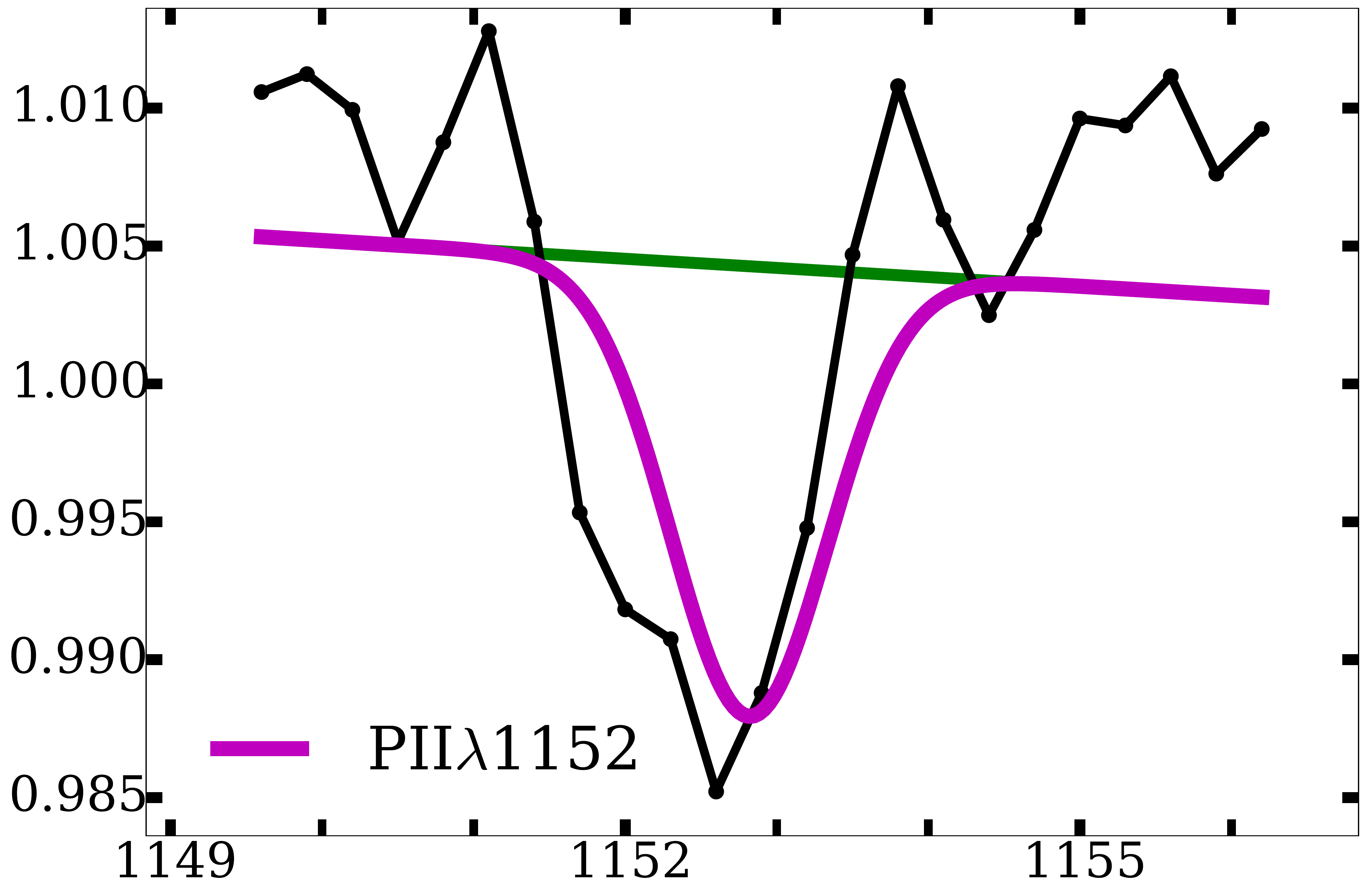}\includegraphics[width=0.25\textwidth]{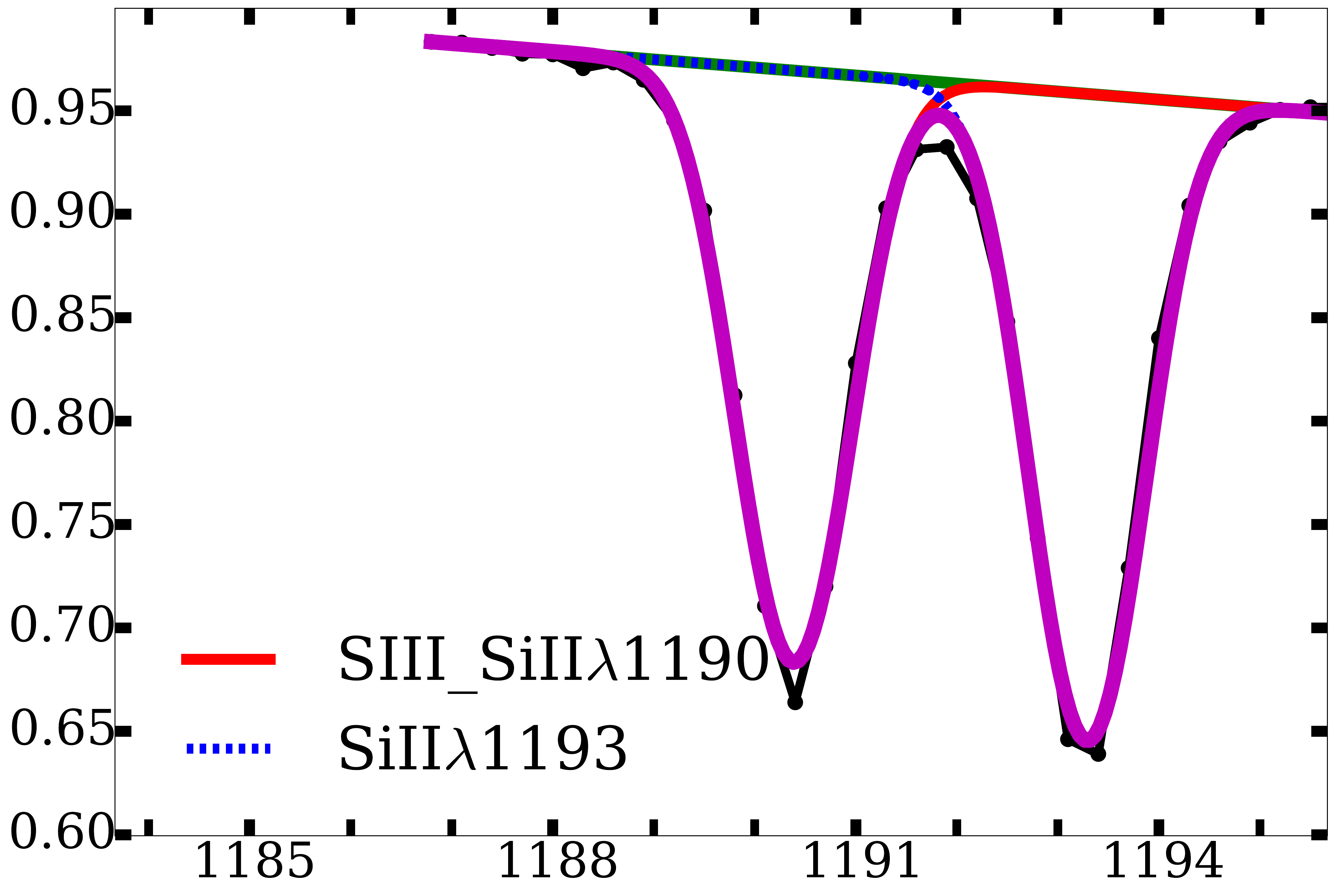}\includegraphics[width=0.25\textwidth]{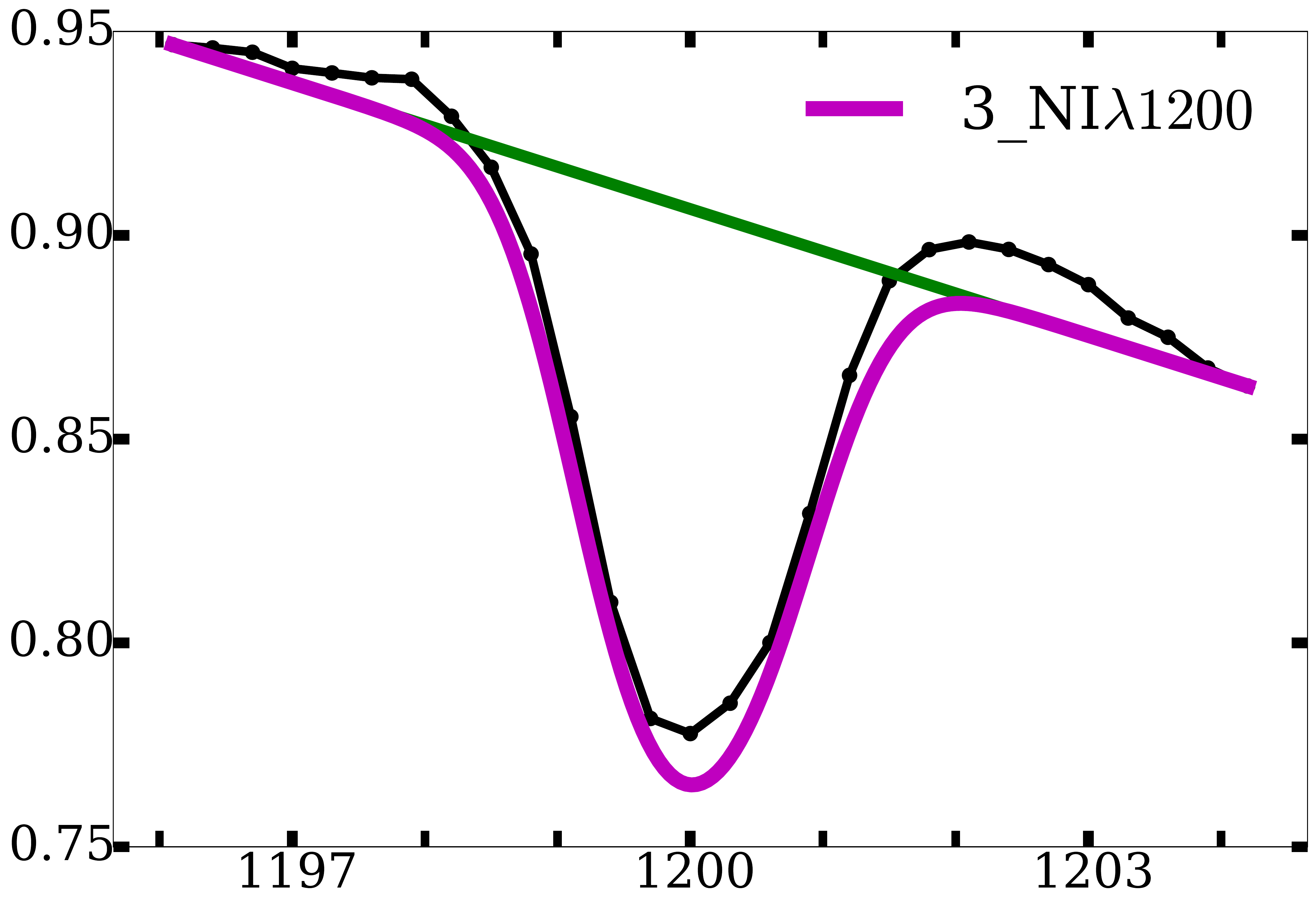}\includegraphics[width=0.25\textwidth]{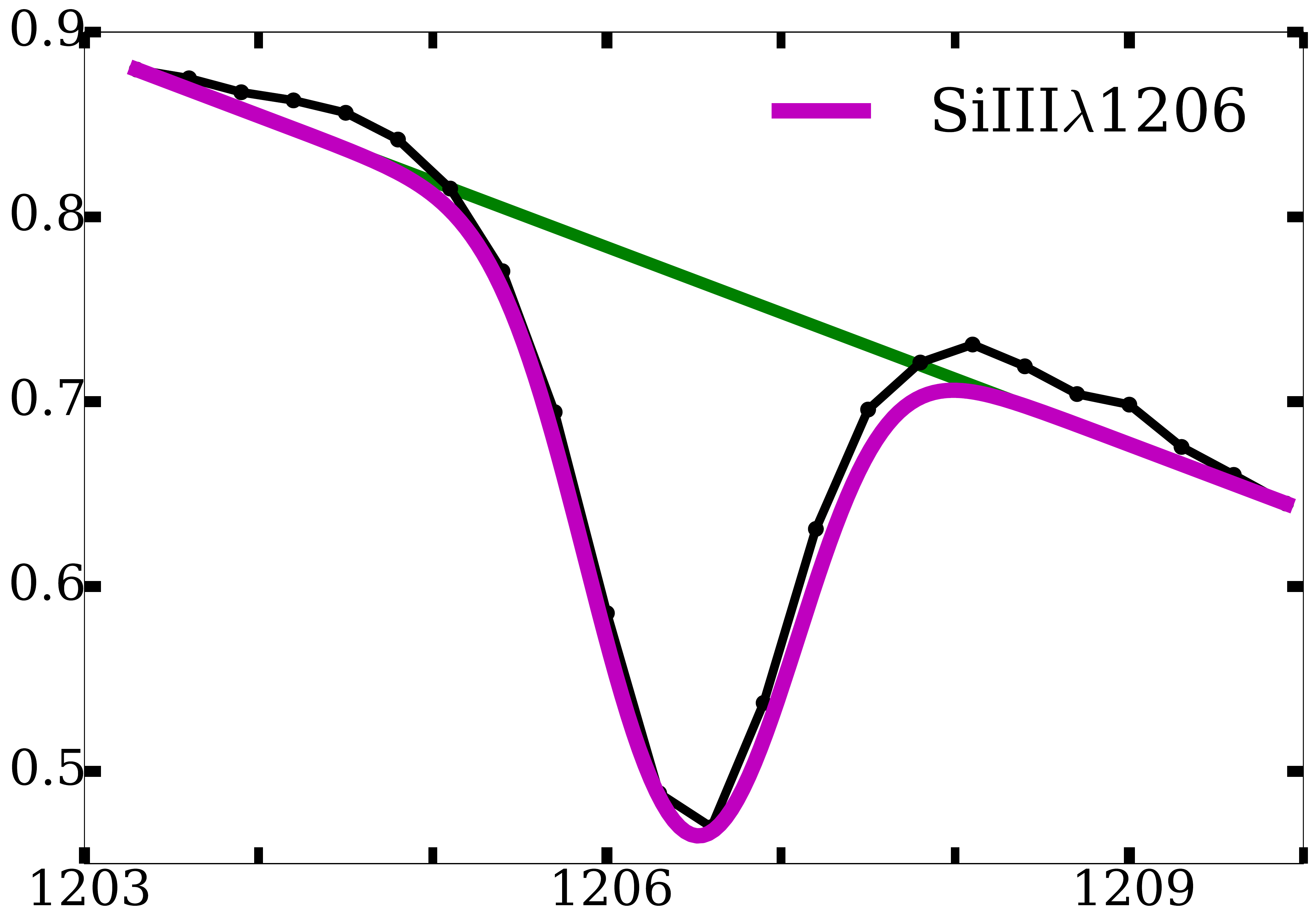}
\includegraphics[width=0.25\textwidth]{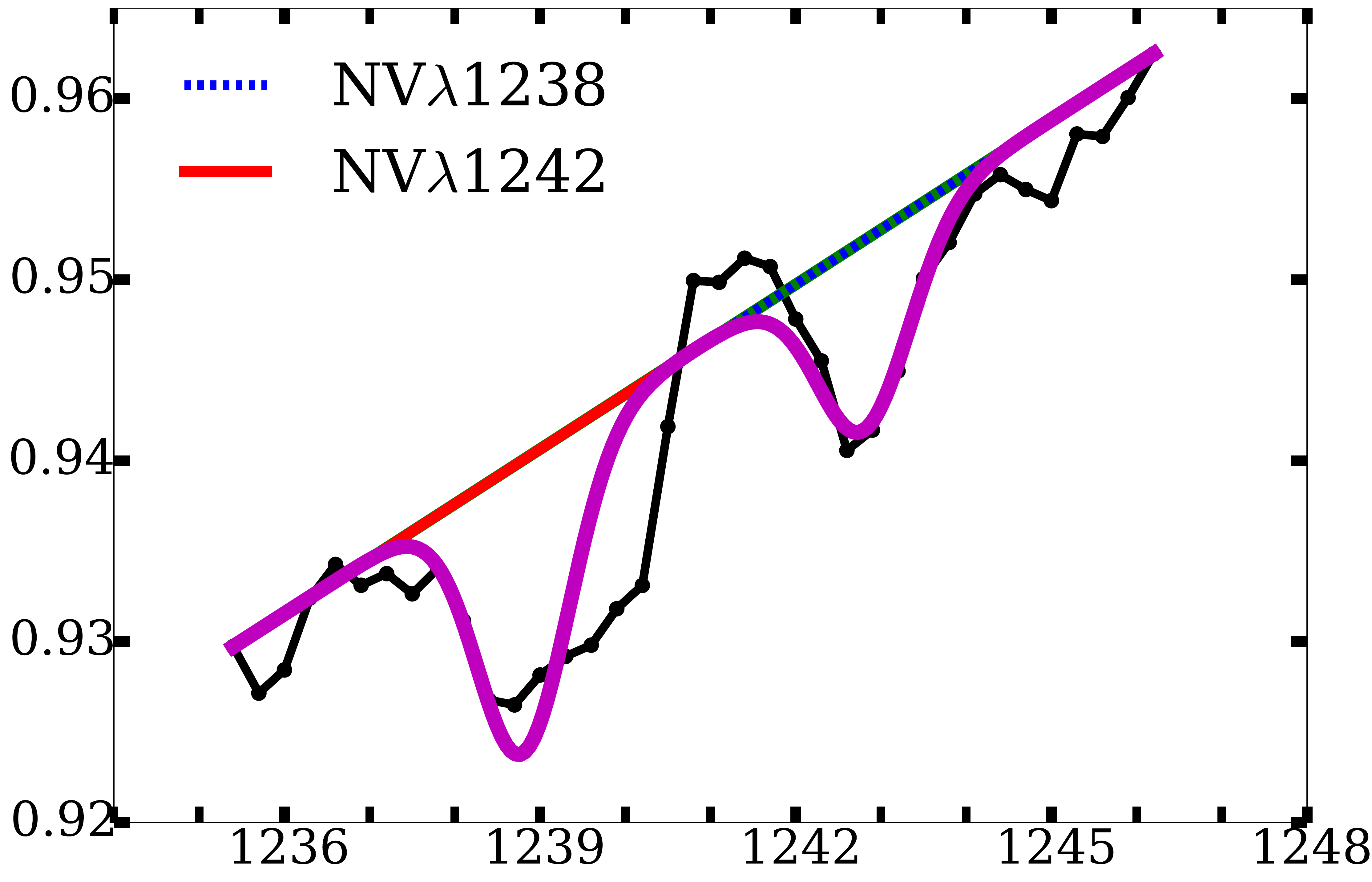}\includegraphics[width=0.25\textwidth]{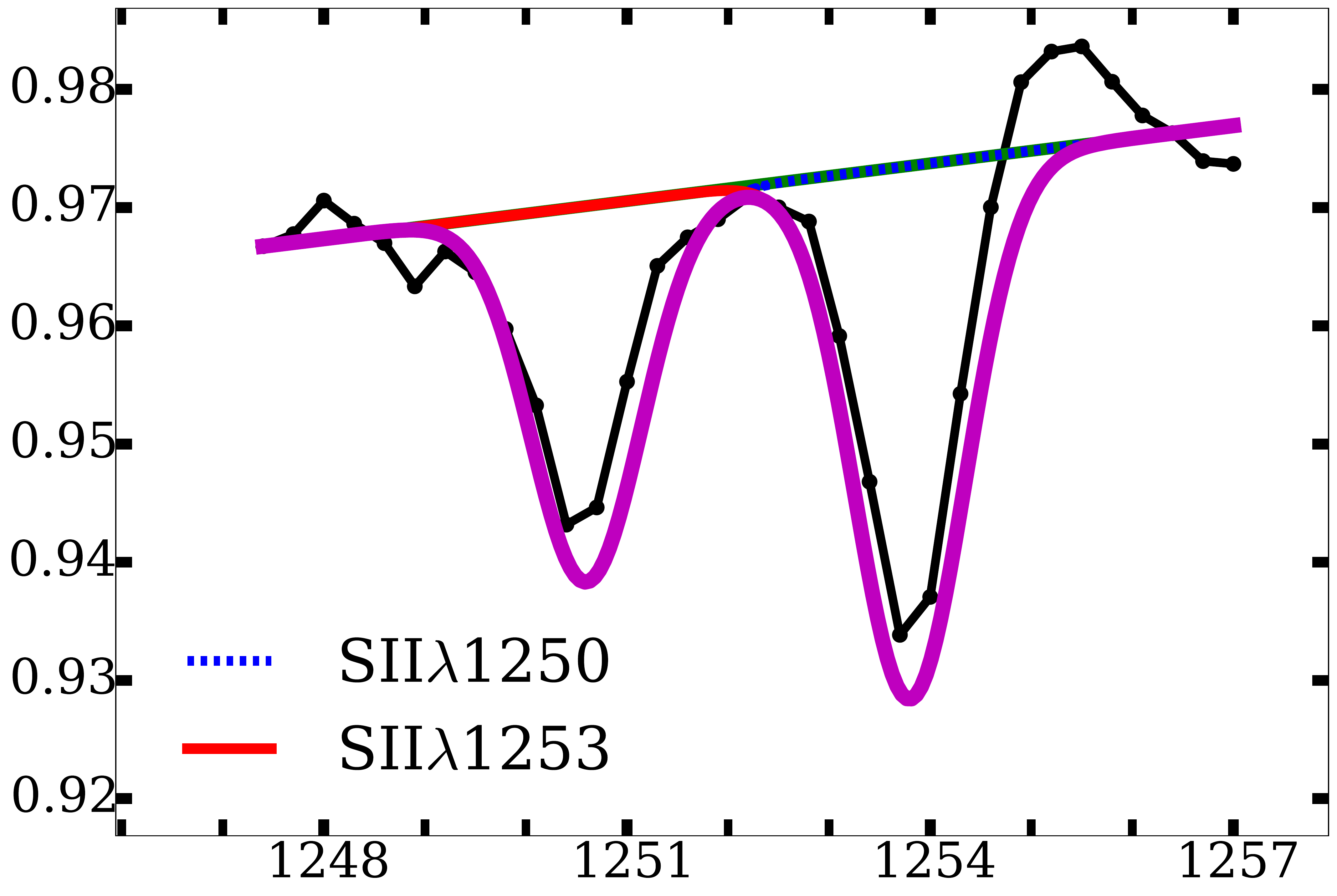}\includegraphics[width=0.25\textwidth]{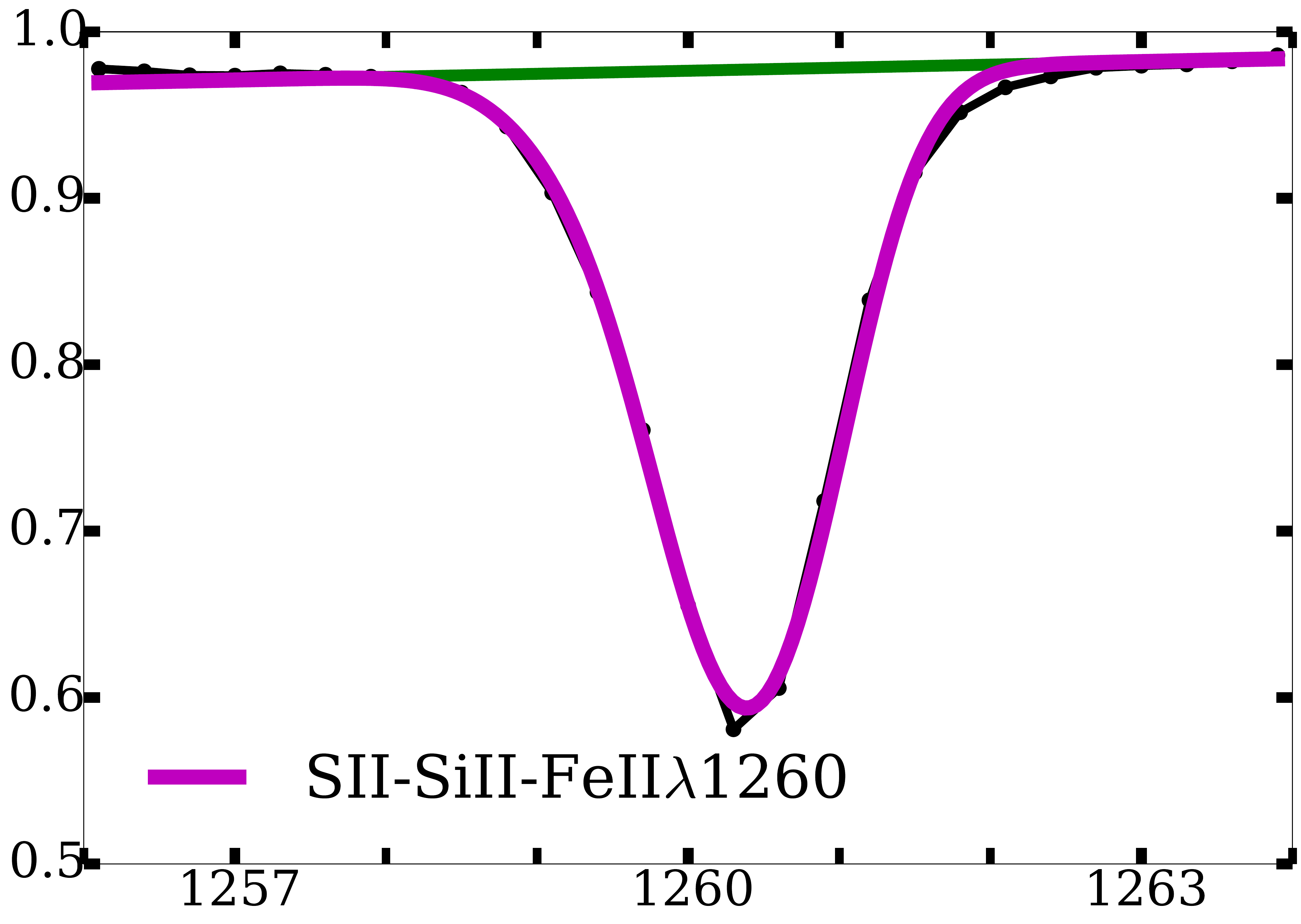}\includegraphics[width=0.25\textwidth]{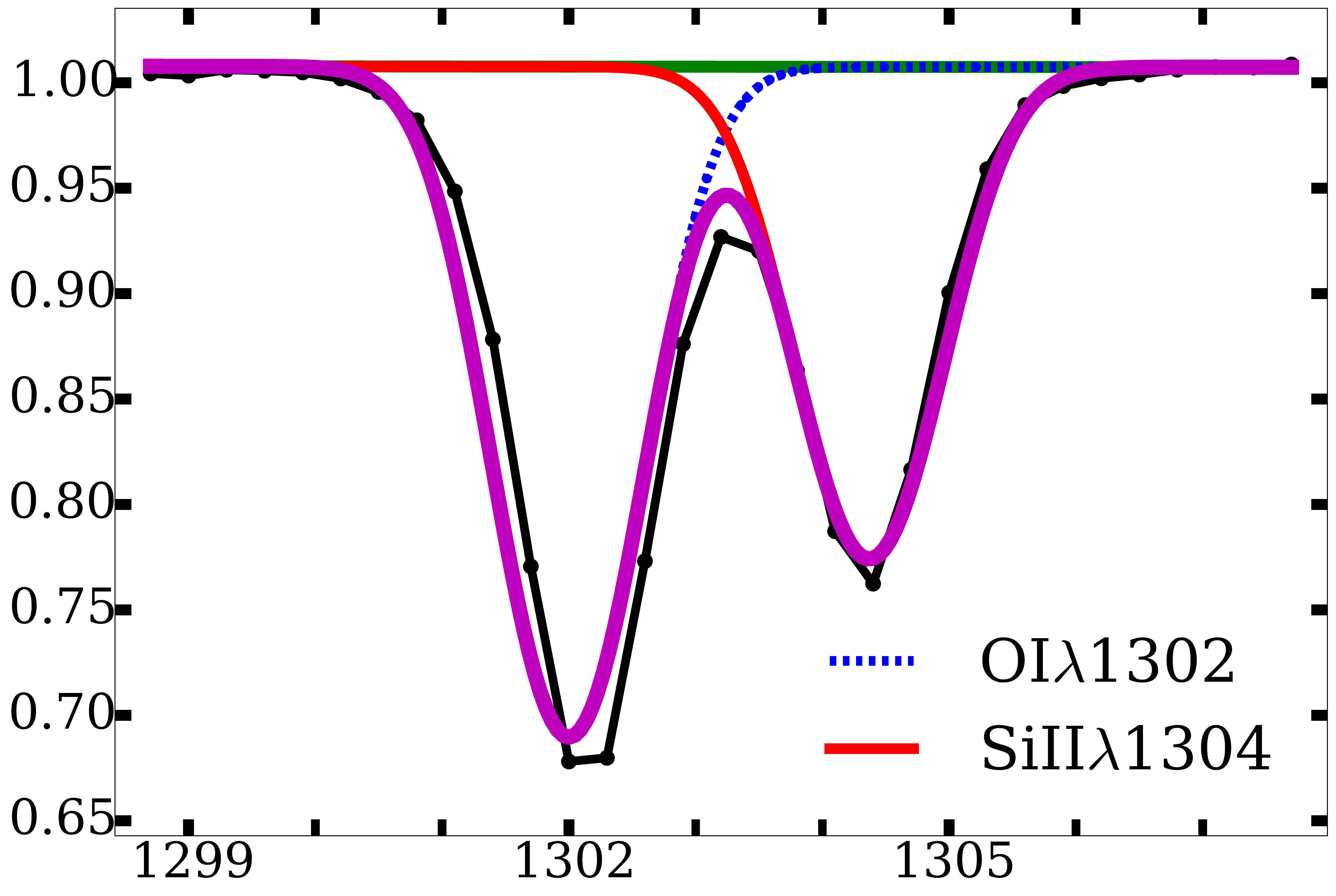}
\includegraphics[width=0.25\textwidth]{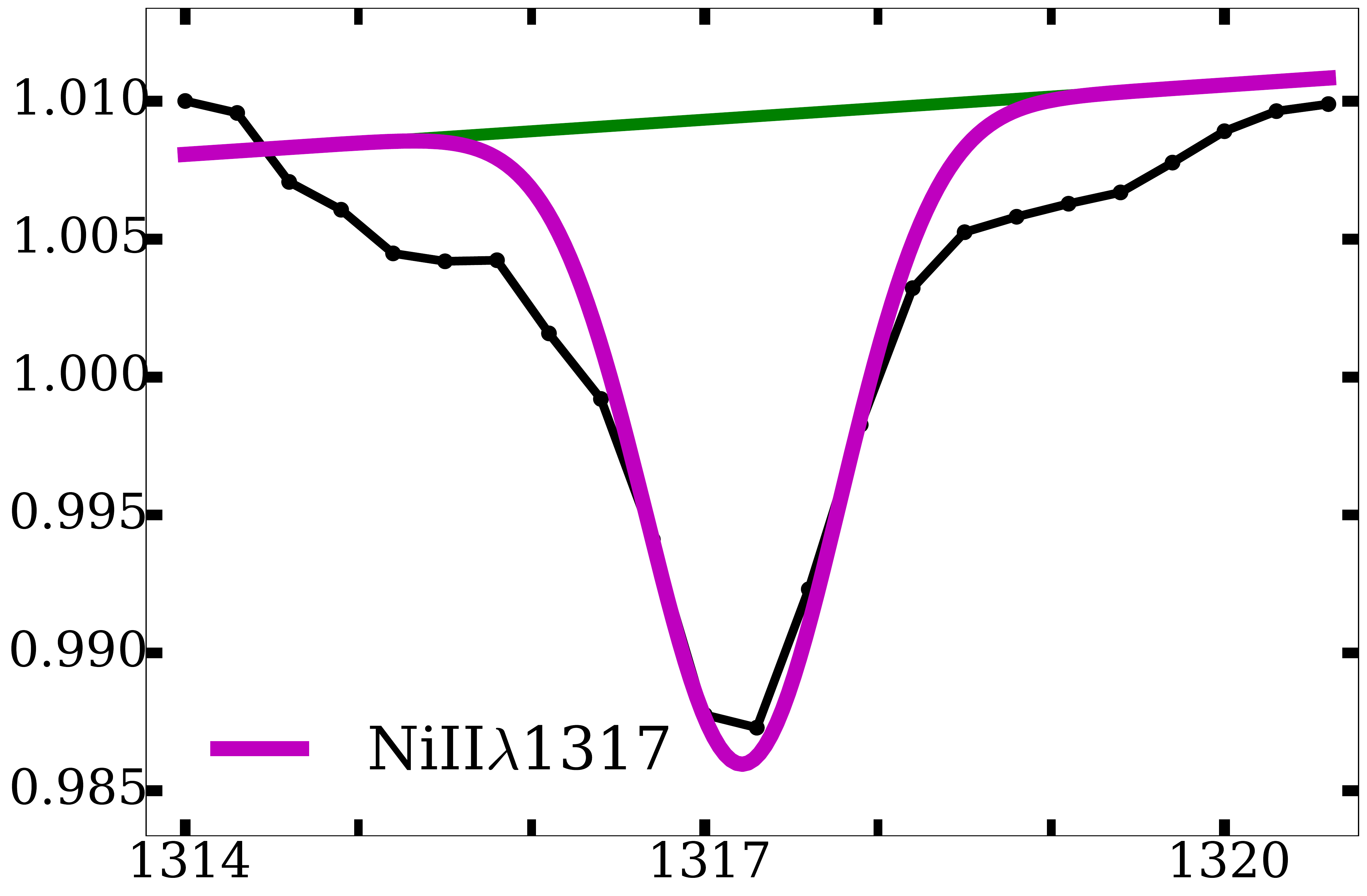}\includegraphics[width=0.25\textwidth]{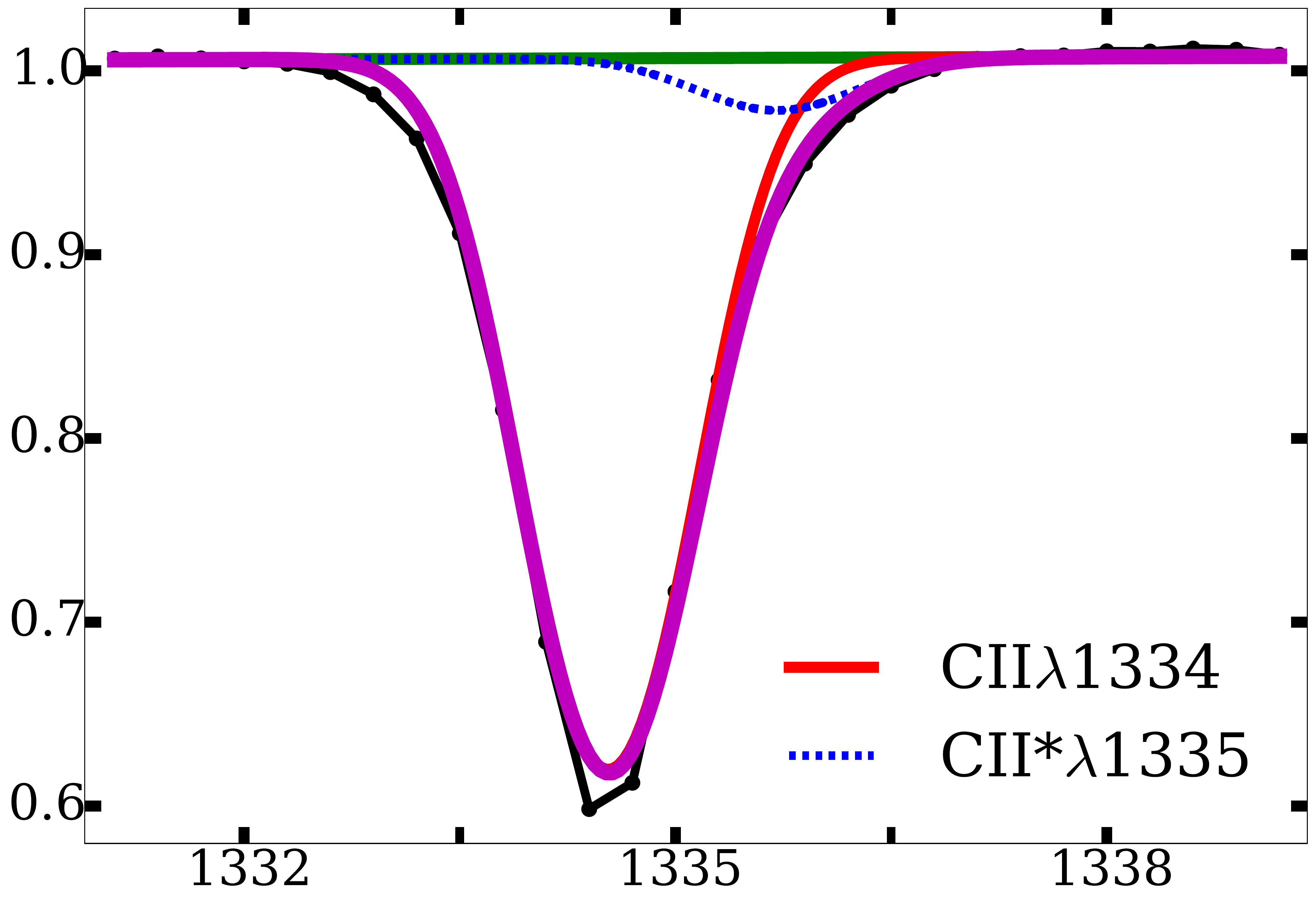}\includegraphics[width=0.25\textwidth]{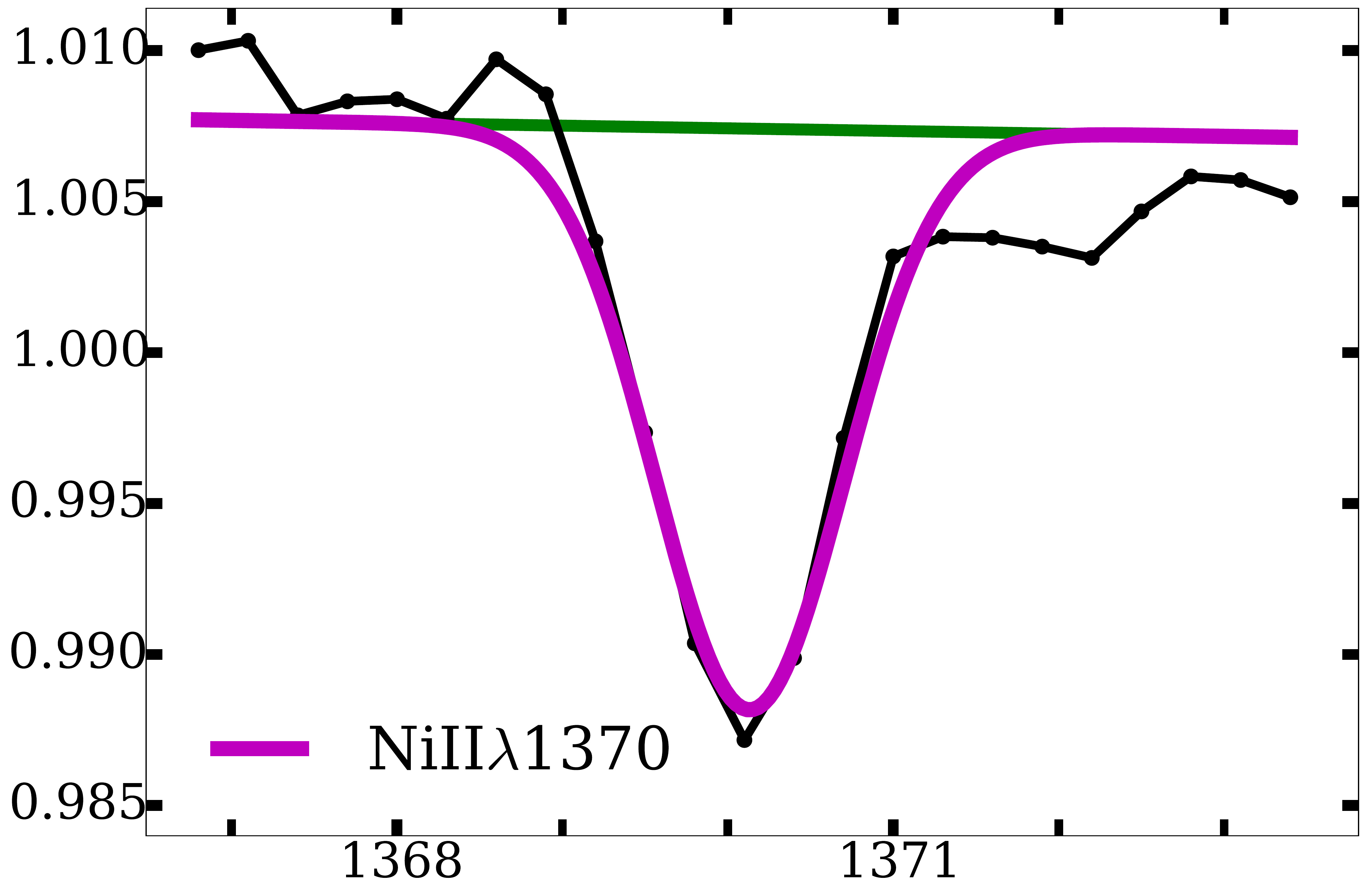}\includegraphics[width=0.25\textwidth]{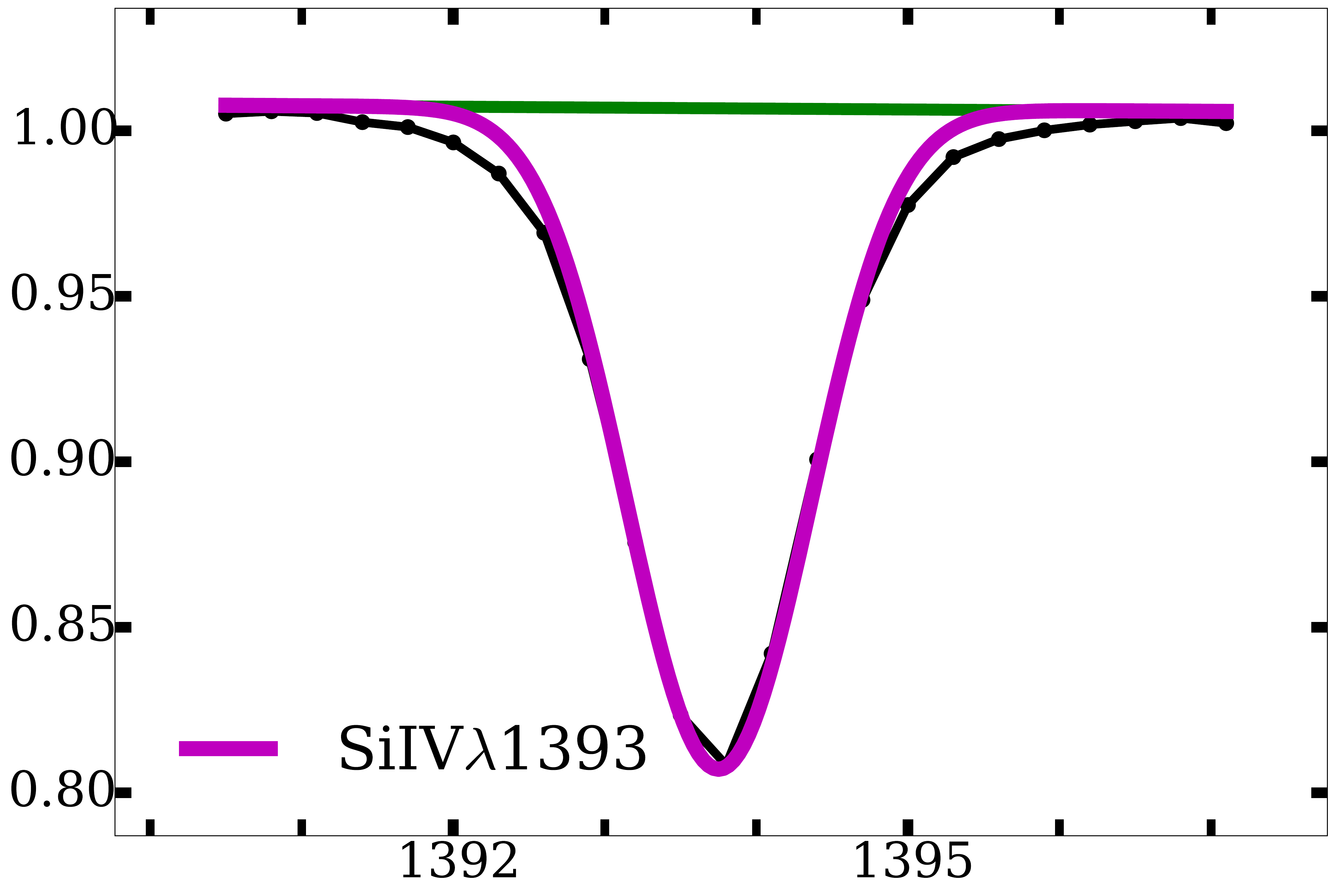}
\caption{ Absorption line profile fits for the total sample, denoting 
transmission as a function of DLA rest-frame wavelength (in ${\rm \AA}$).
\textit{Black thin line}: final stacked spectrum, with points indicating
the data pixels.
\textit{Green thick line}: fitted continuum.
\textit{Magenta thick line}: Total fitted profile. Other lines
(\textit{dotted thin blue and yellow, and solid red})
represent the fits of individual absorption features. Good fits are generally
obtained for the central parts of the lines. In some of the strong lines,
the observed tails are broader than the fits, possibly due to the intrinsic
velocity dispersions of the absorption systems. 
The Ly$\beta$ and Ly$\gamma$ lines are strong and have substantial damped
wings, which explains why the fit is substandard in these cases. The fit to the 
SiII\,$\lambda1020$ line does not account for the blend with the Ly$\beta$ line. 
The transmission scales vary with each panel to maximize the visibility of the lines.}
\label{fig:fit}
\end{figure}

\begin{figure*}                  
\includegraphics[width=0.25\textwidth]{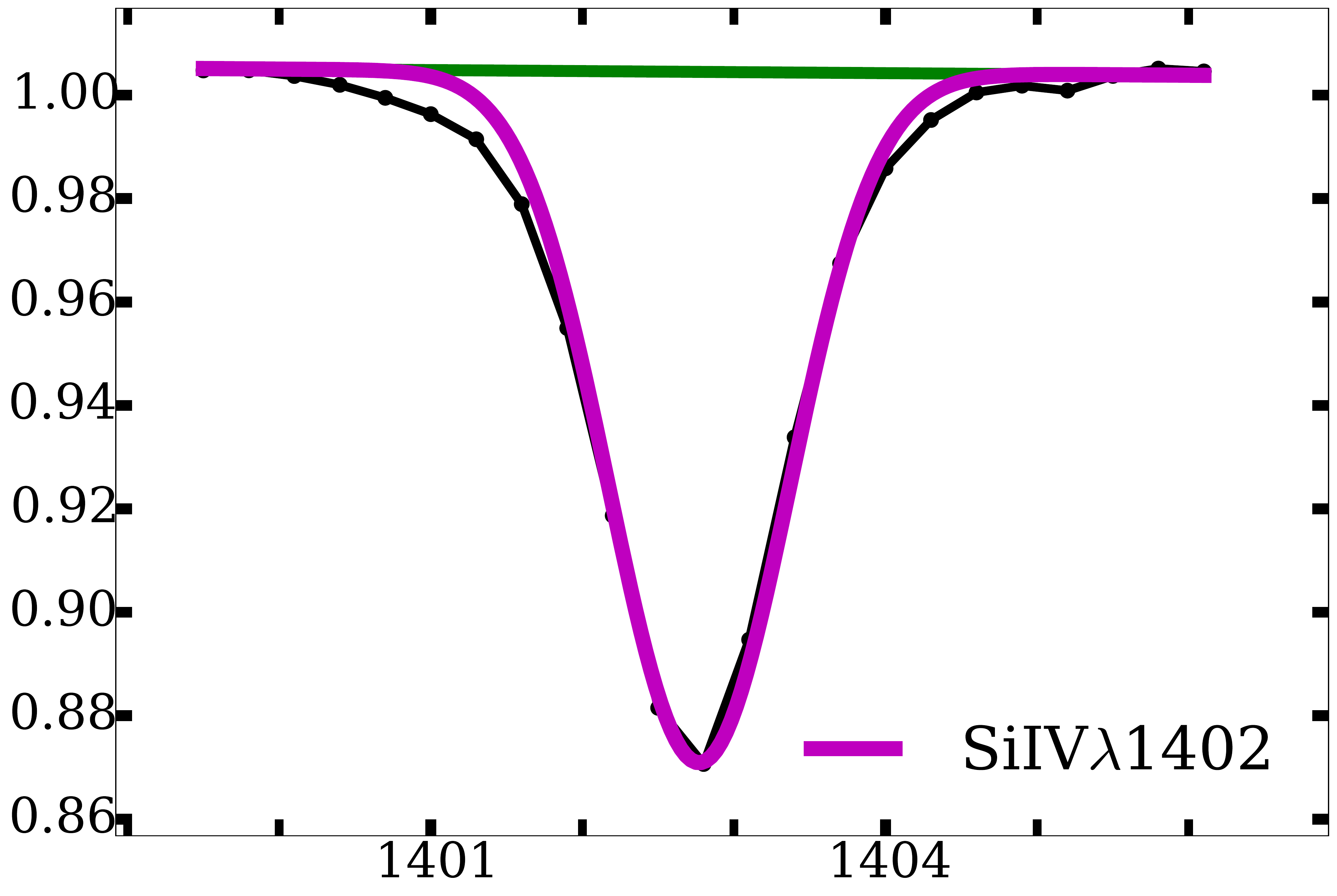}\includegraphics[width=0.25\textwidth]{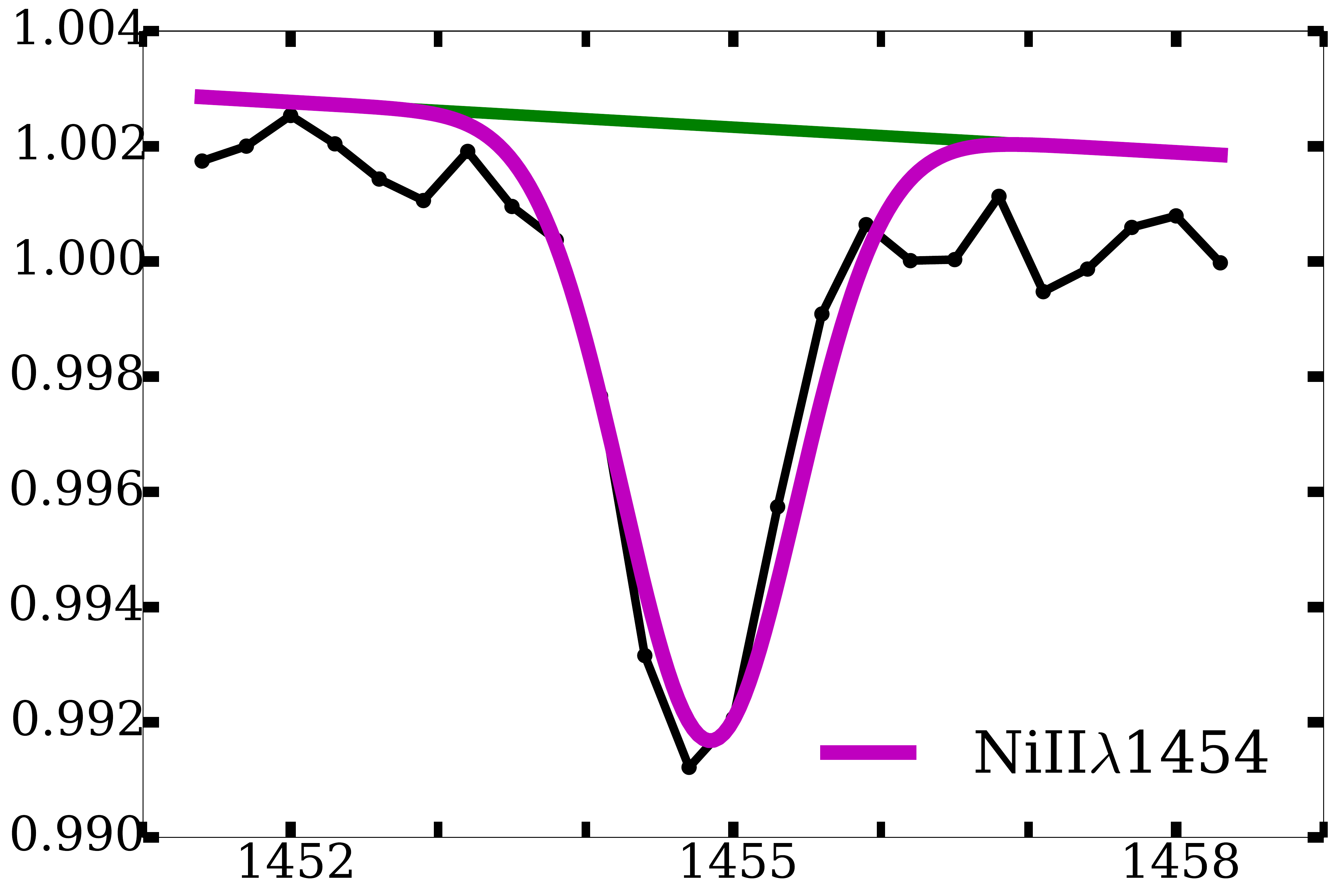}\includegraphics[width=0.25\textwidth]{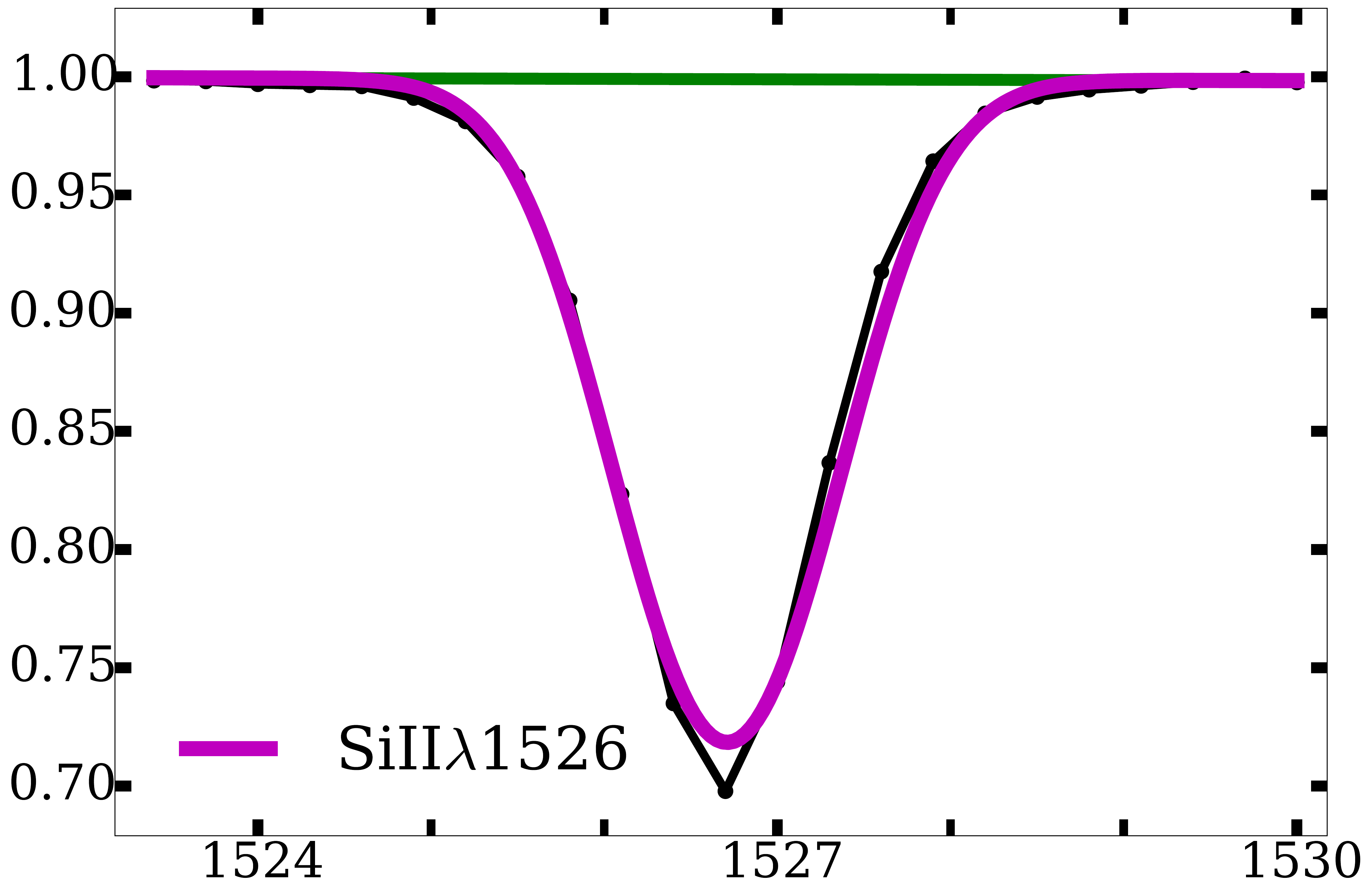}\includegraphics[width=0.25\textwidth]{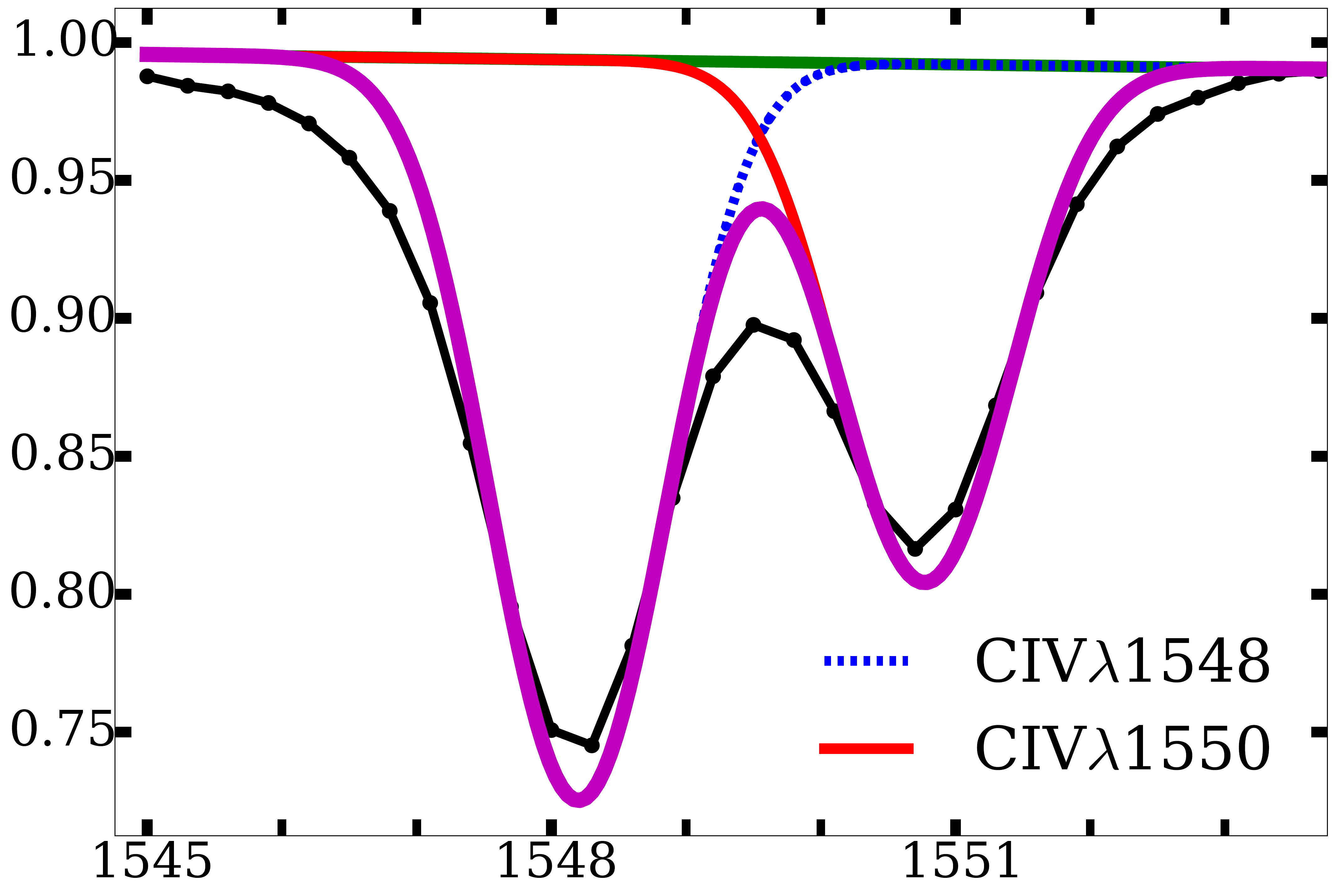}
\includegraphics[width=0.25\textwidth]{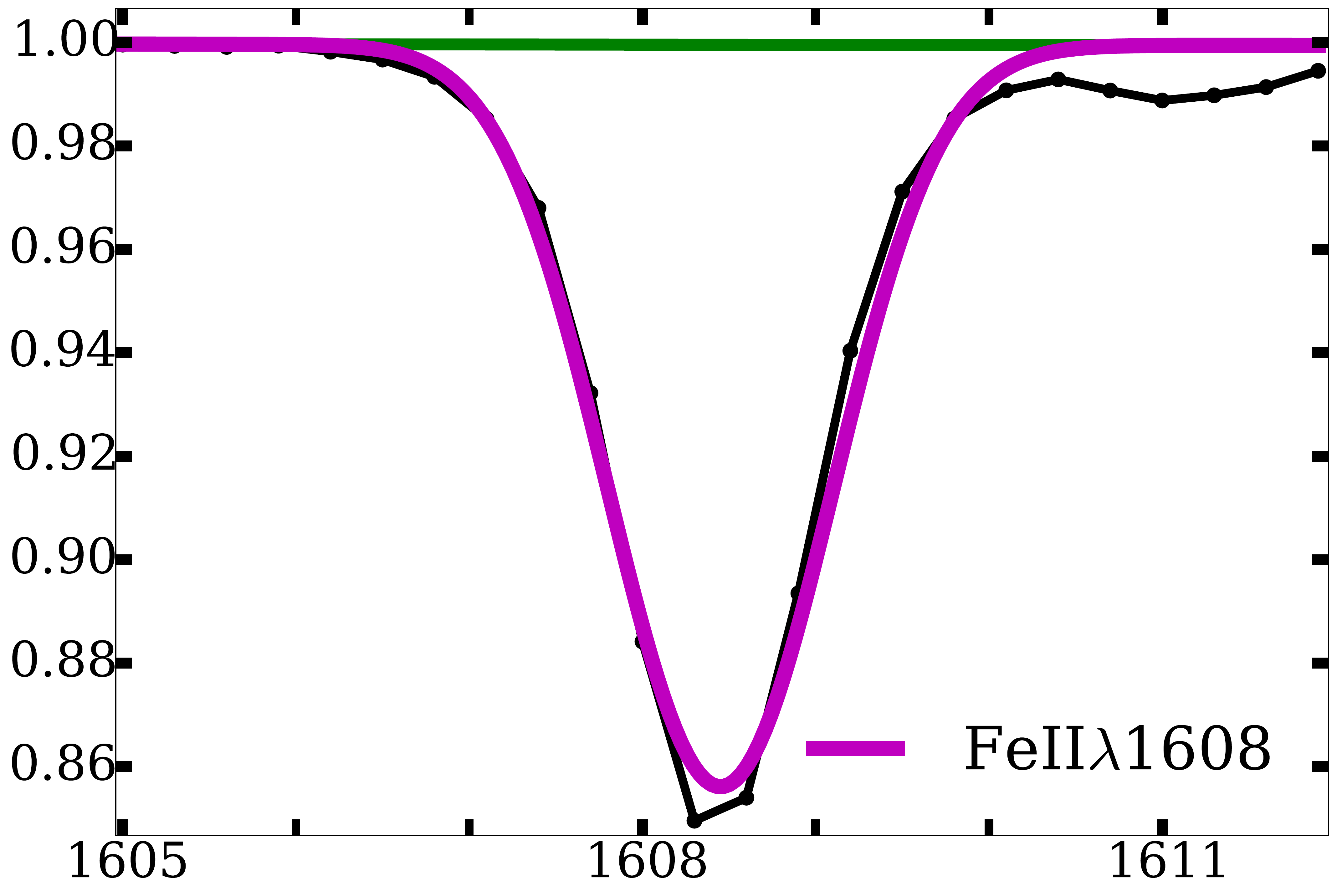}\includegraphics[width=0.25\textwidth]{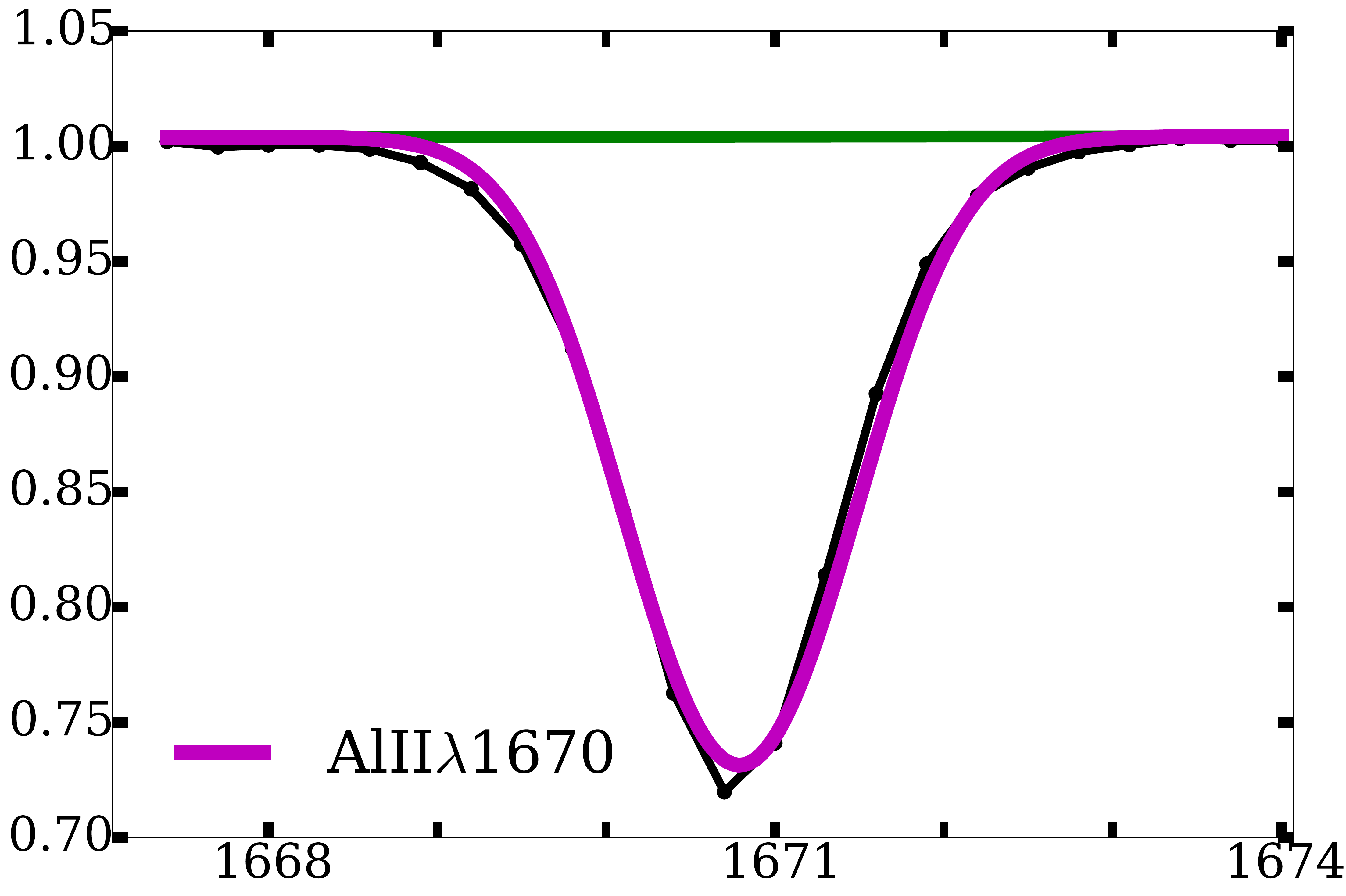}\includegraphics[width=0.25\textwidth]{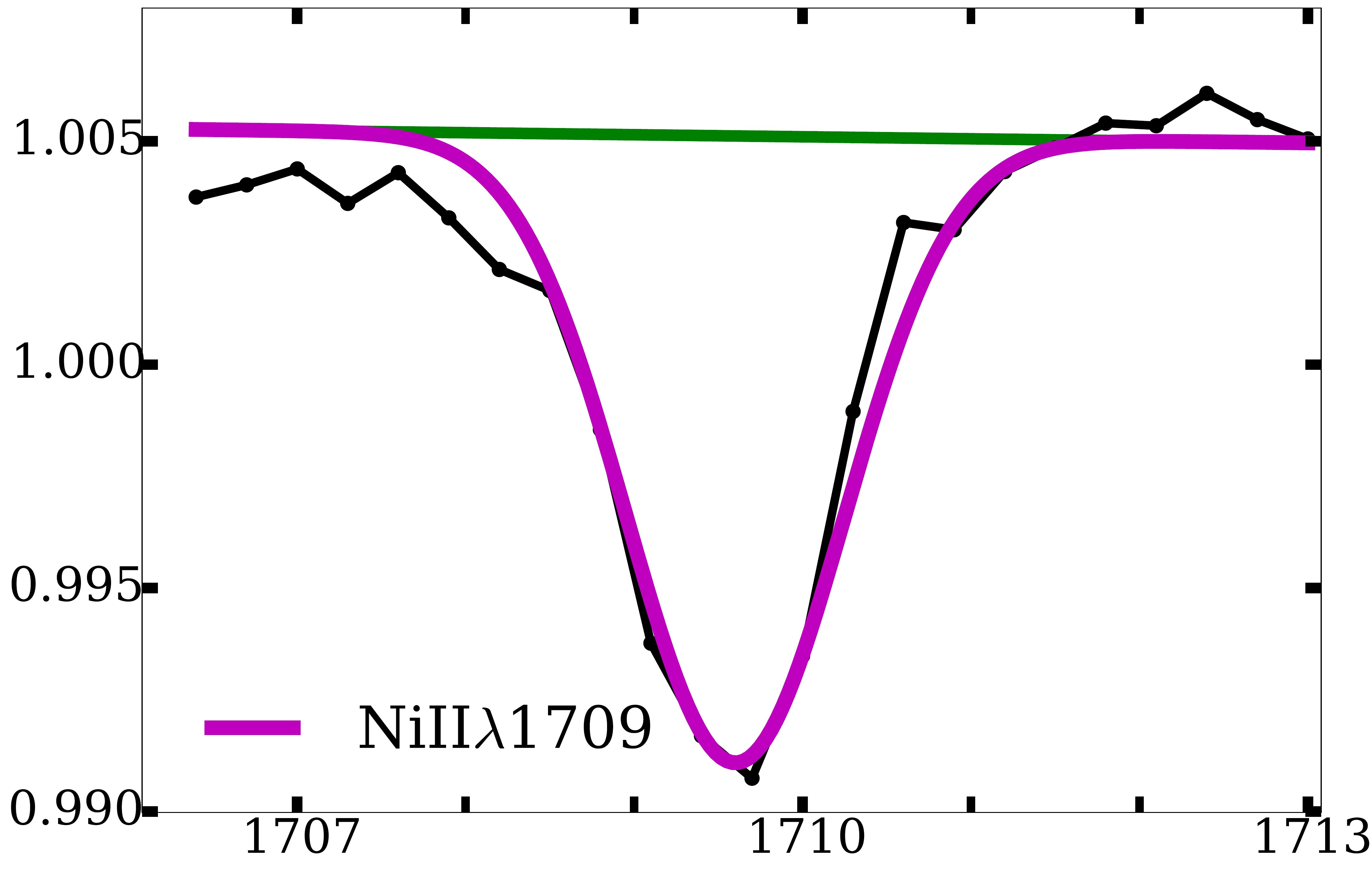}\includegraphics[width=0.25\textwidth]{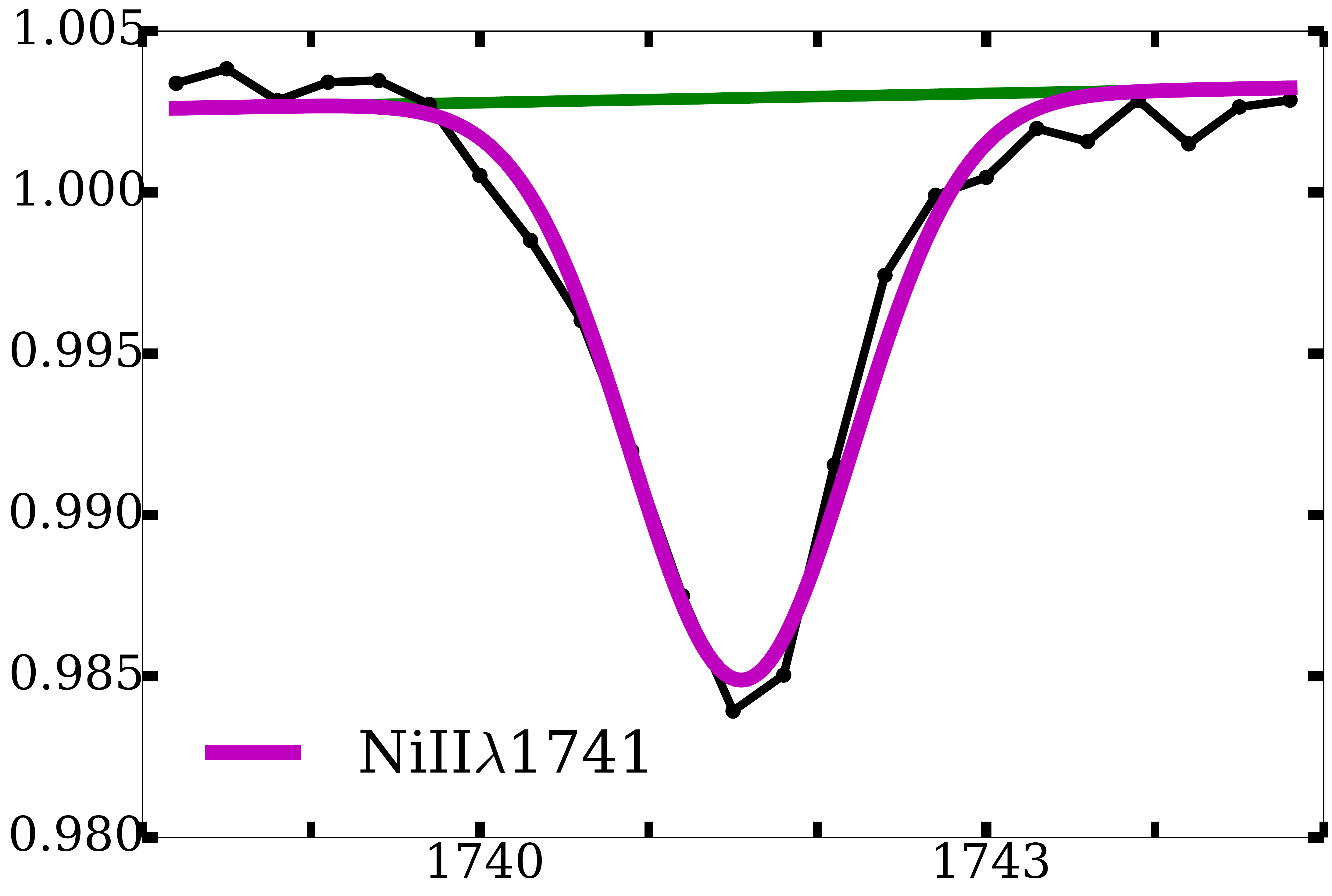}
\includegraphics[width=0.25\textwidth]{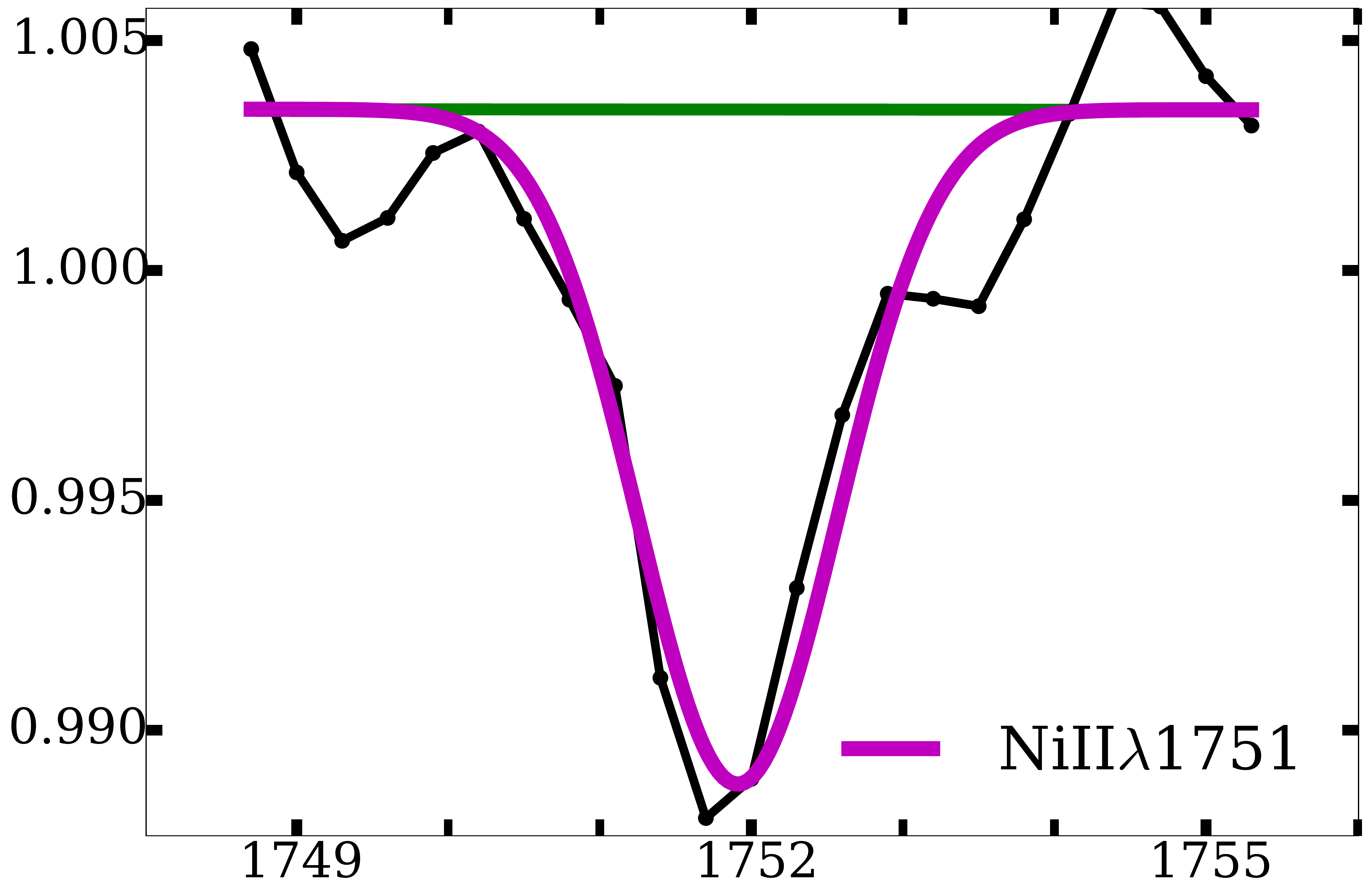}\includegraphics[width=0.25\textwidth]{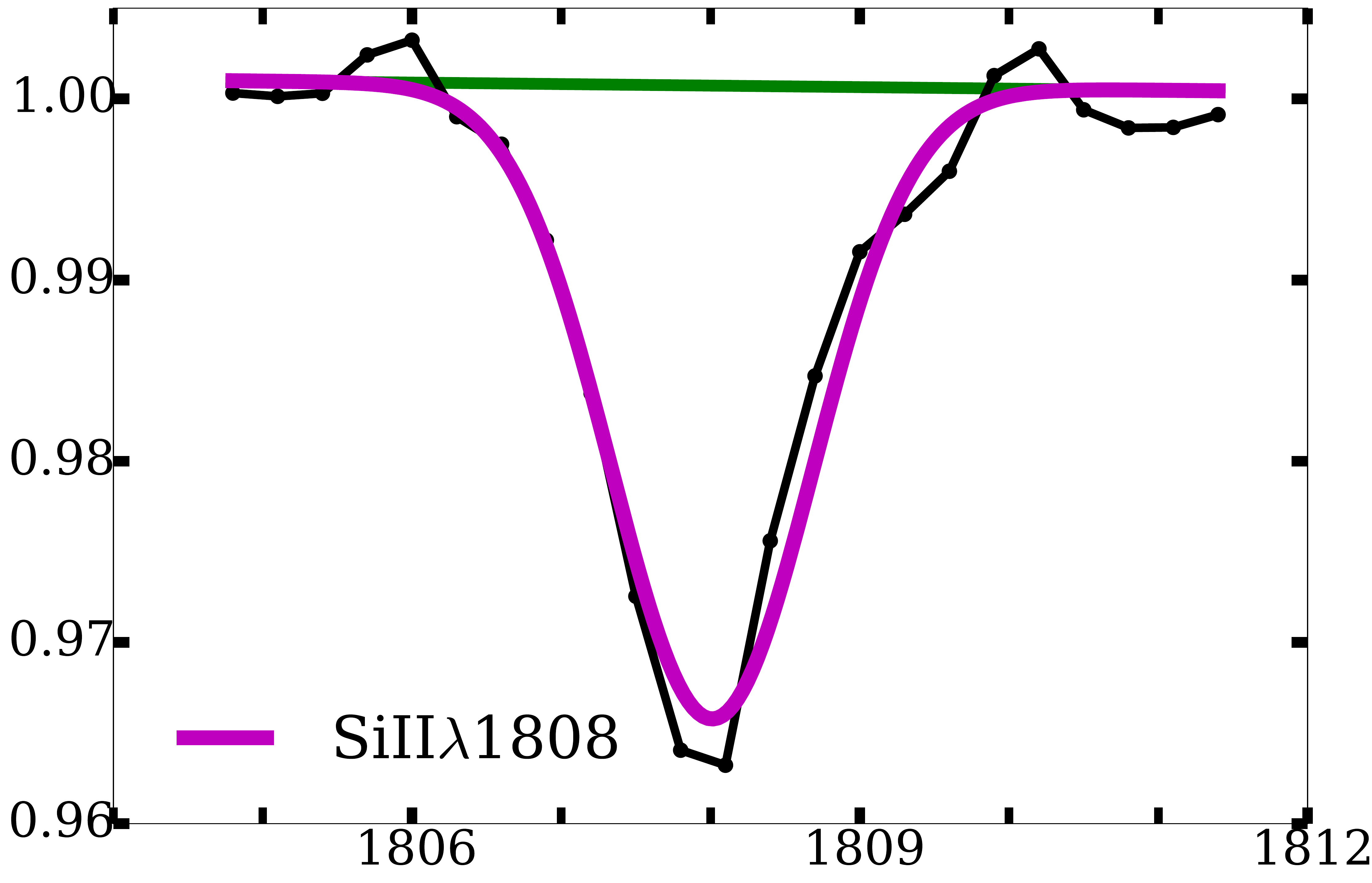}\includegraphics[width=0.25\textwidth]{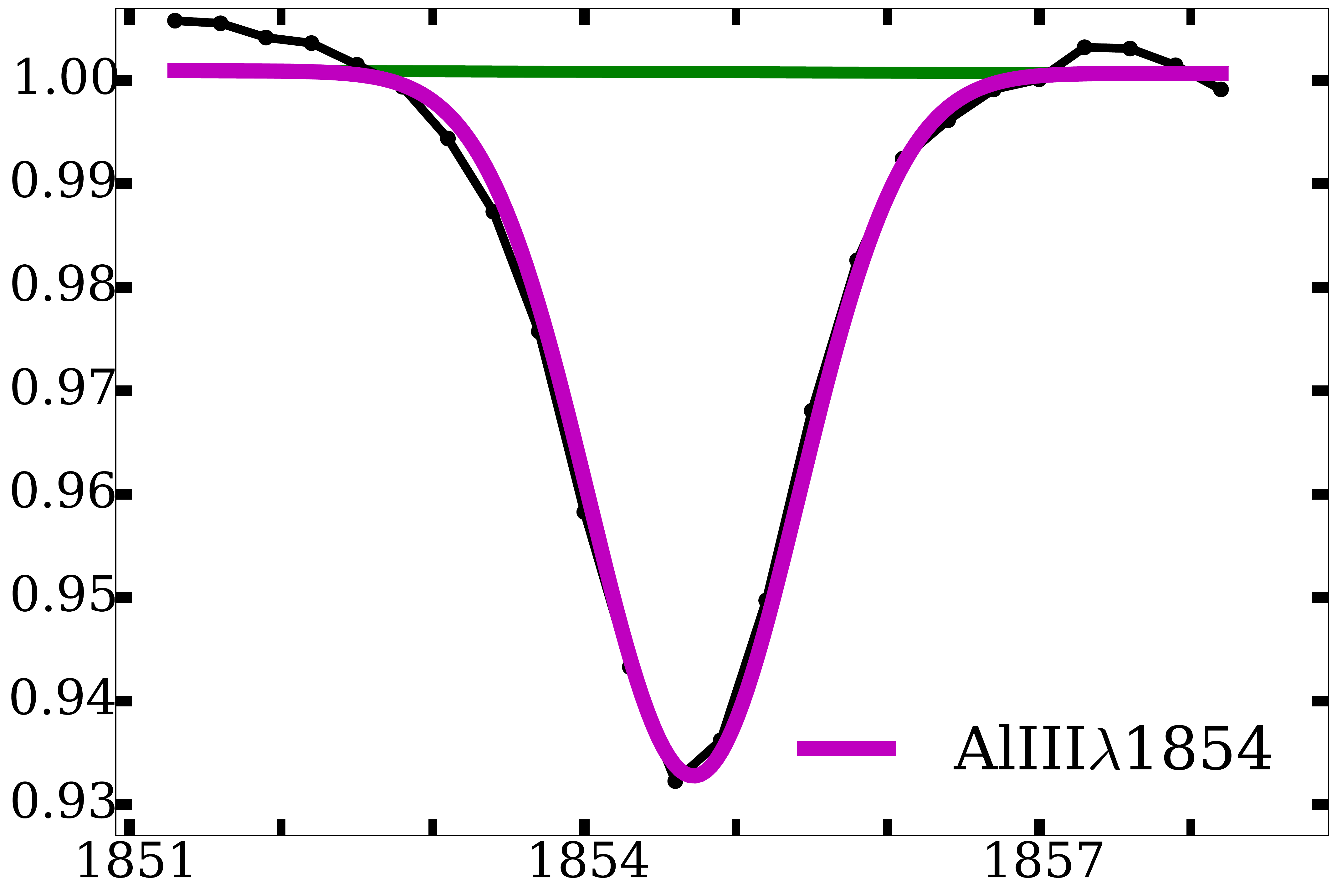}\includegraphics[width=0.25\textwidth]{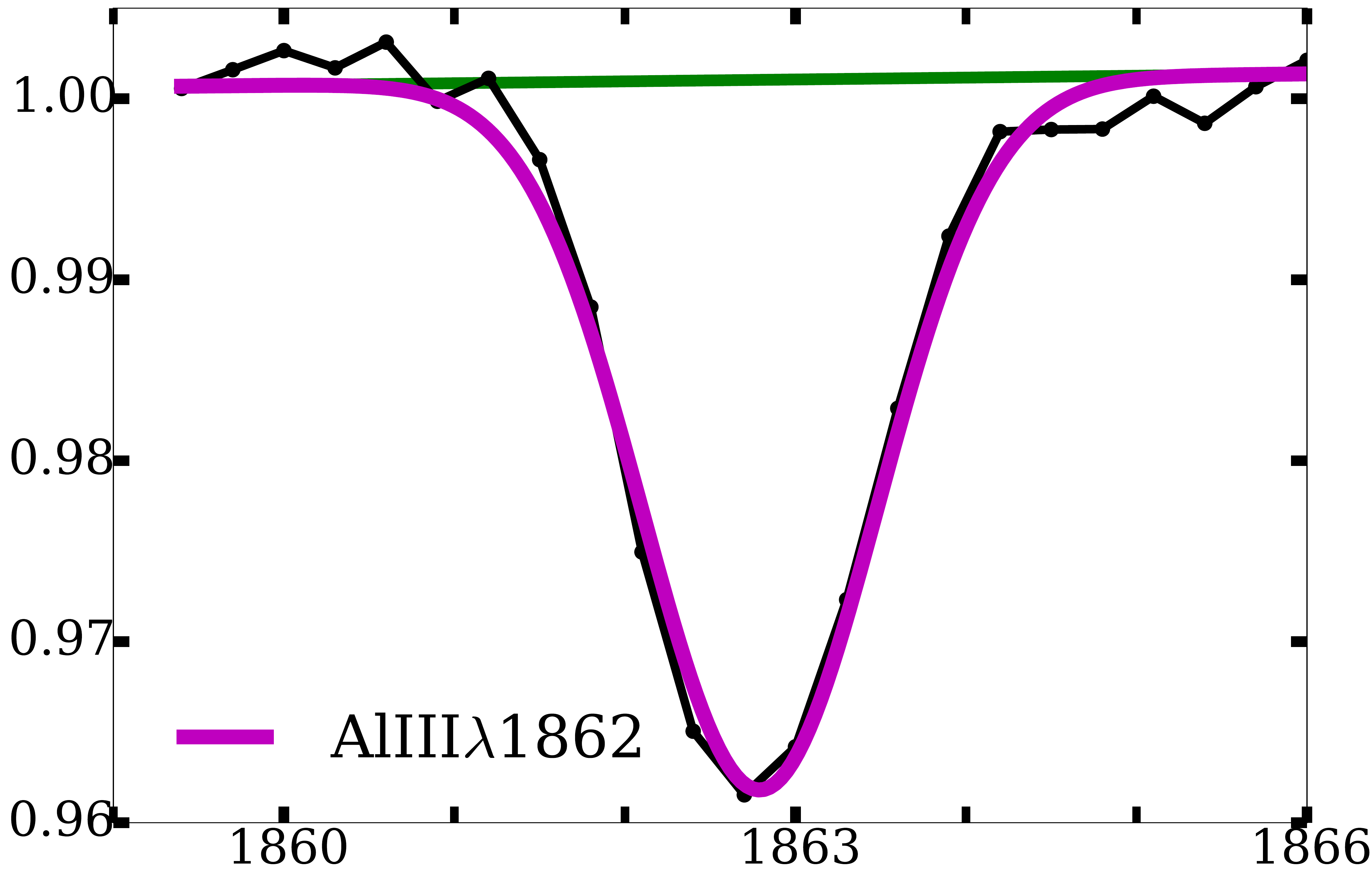}
\includegraphics[width=0.25\textwidth]{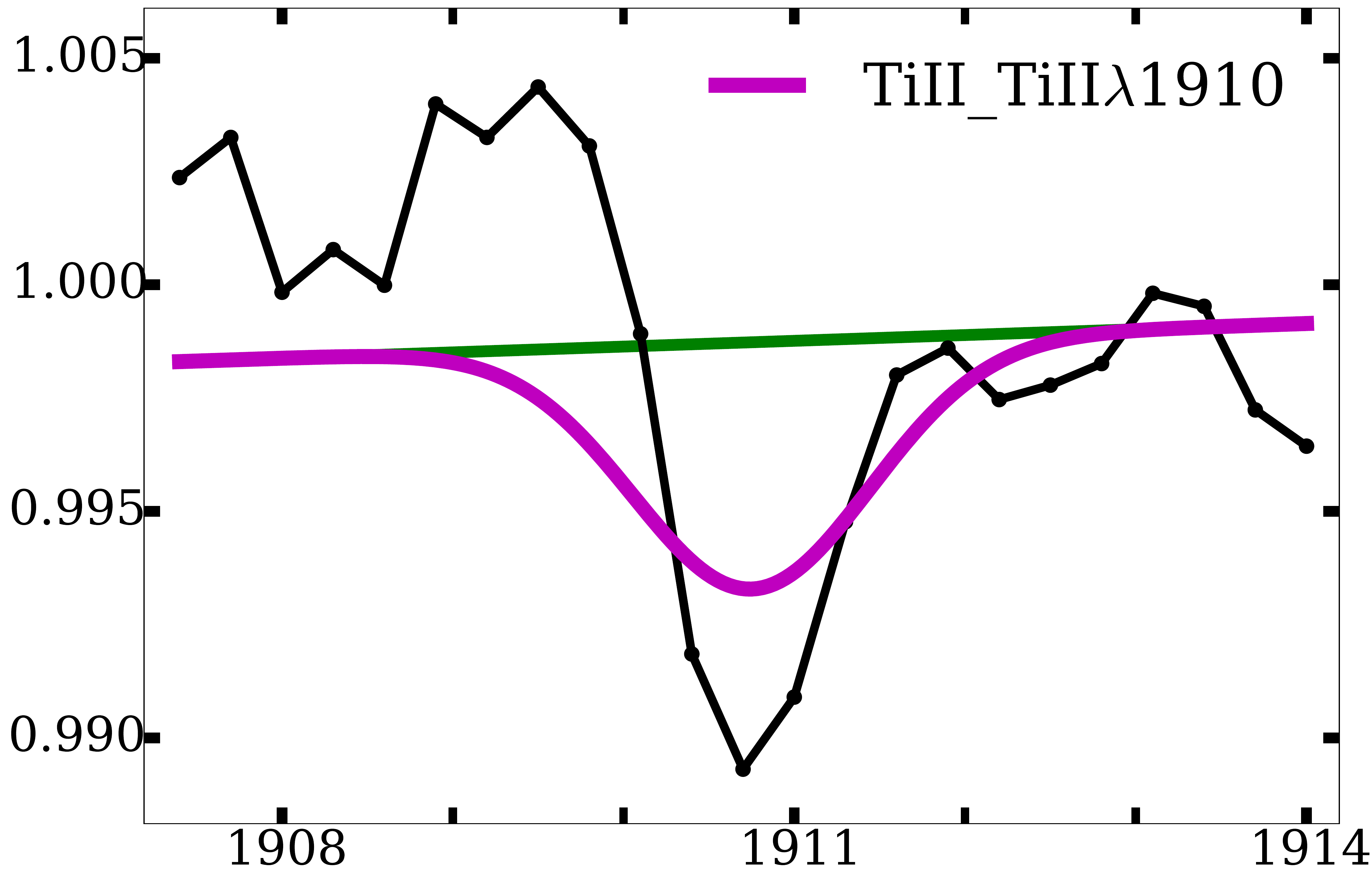}\includegraphics[width=0.25\textwidth]{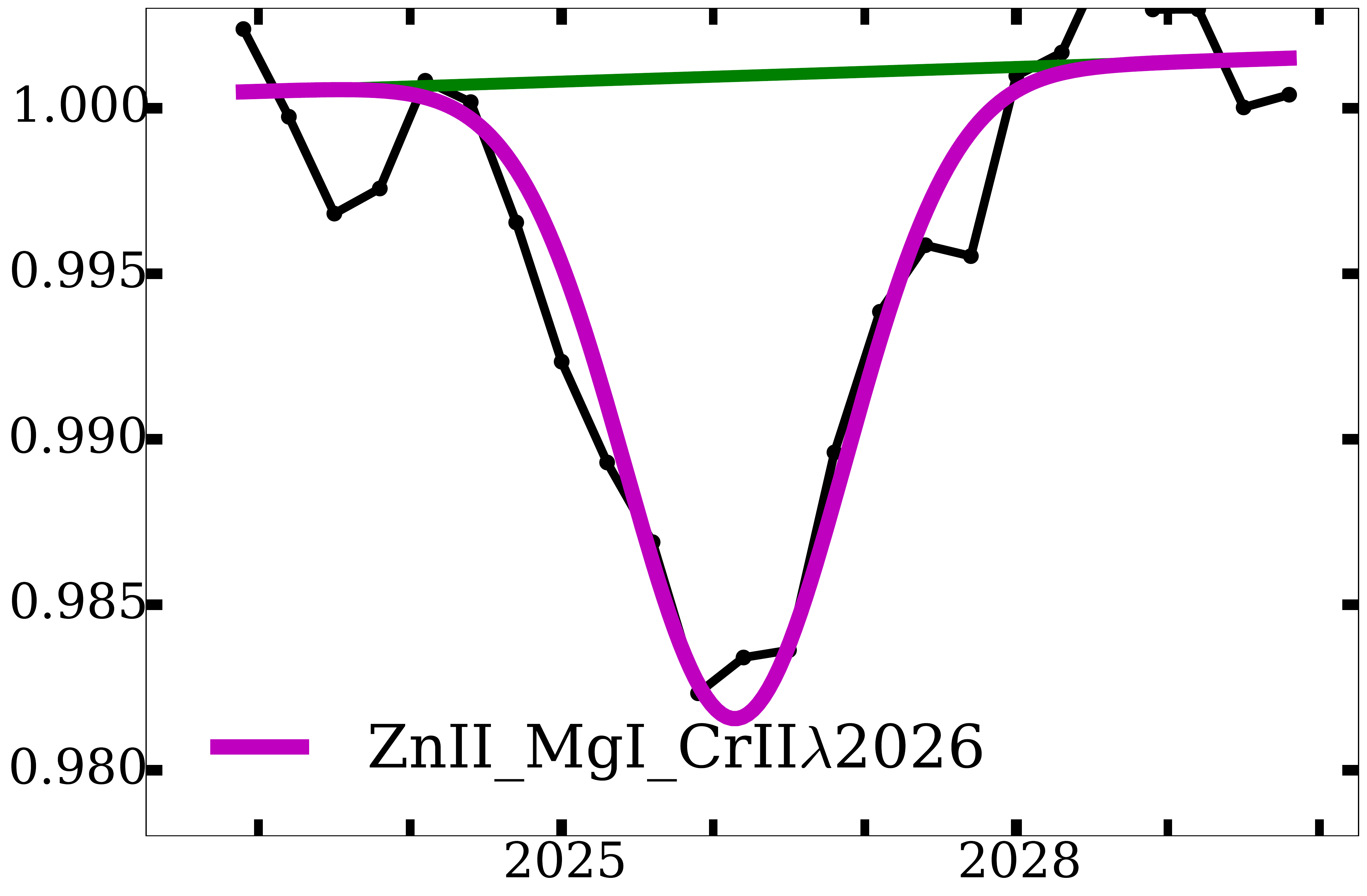}\includegraphics[width=0.25\textwidth]{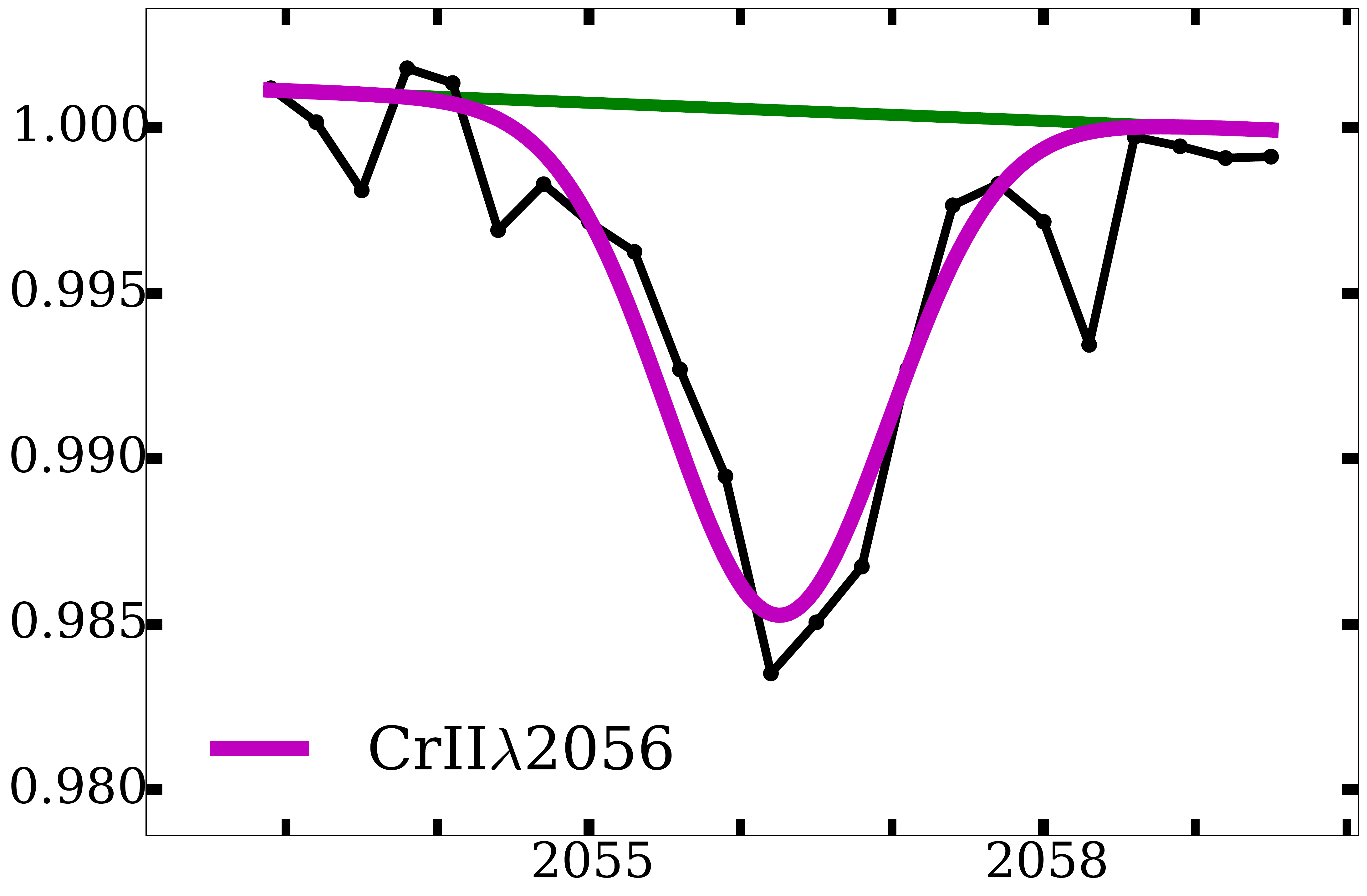}\includegraphics[width=0.25\textwidth]{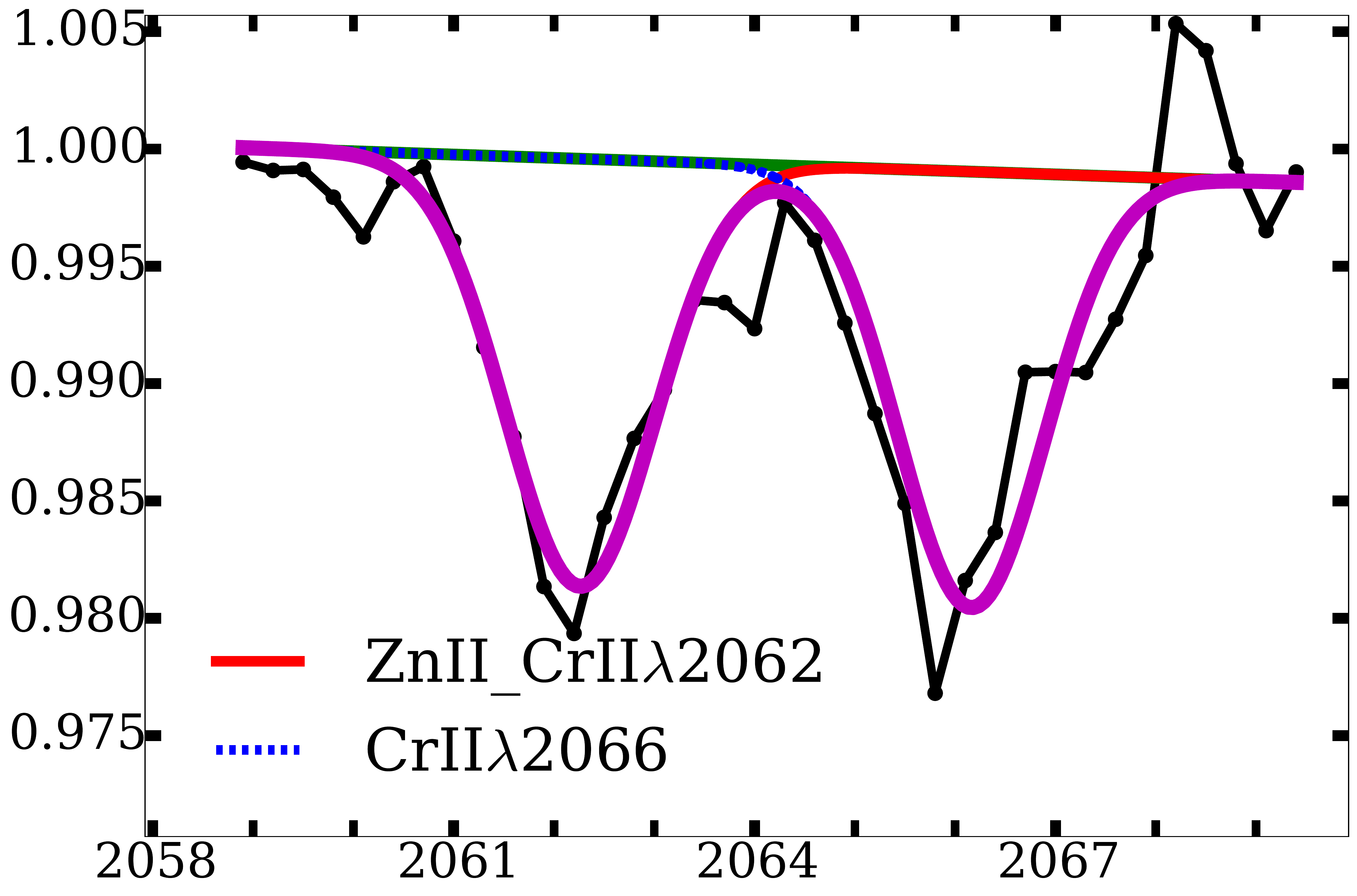}
\includegraphics[width=0.25\textwidth]{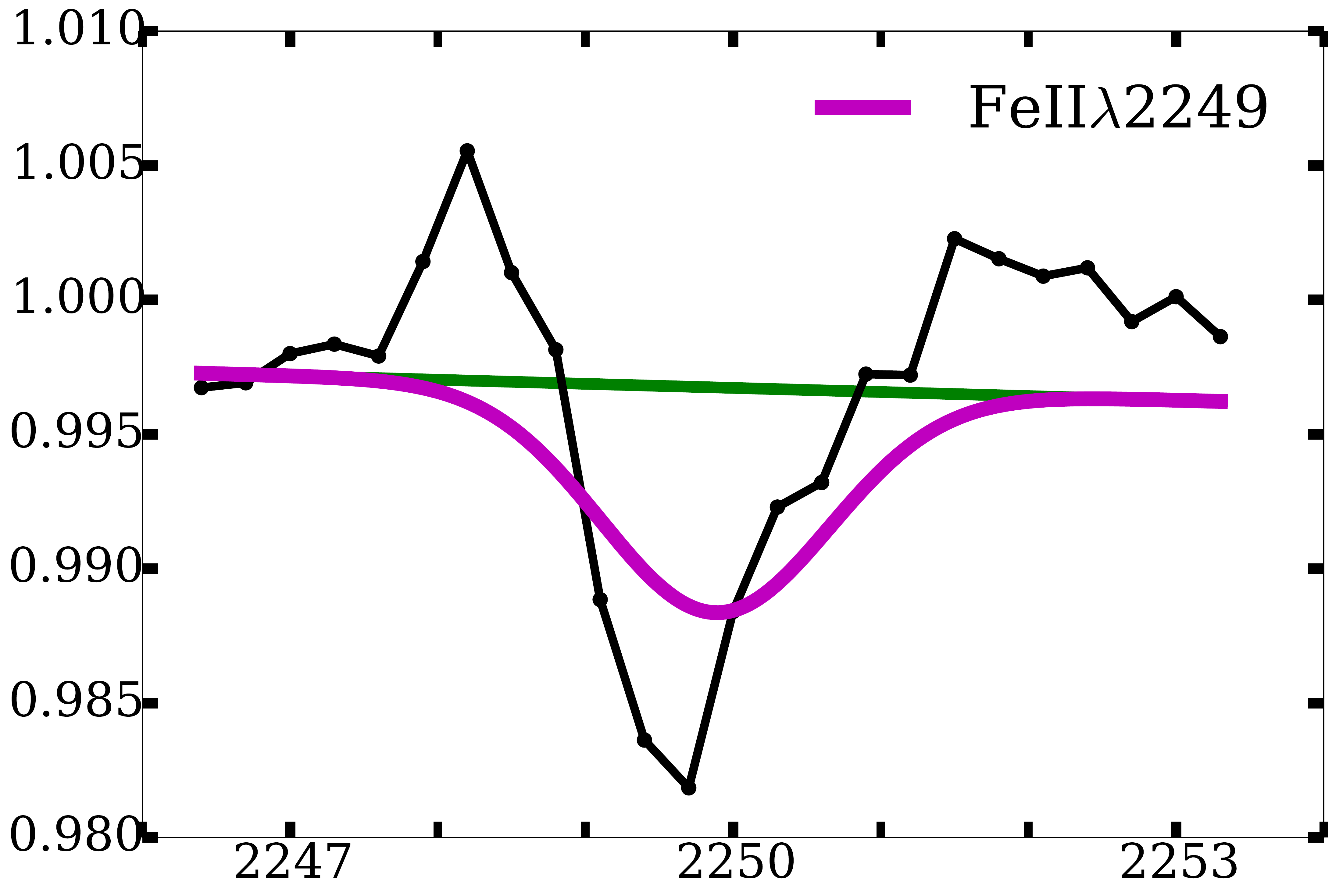}\includegraphics[width=0.25\textwidth]{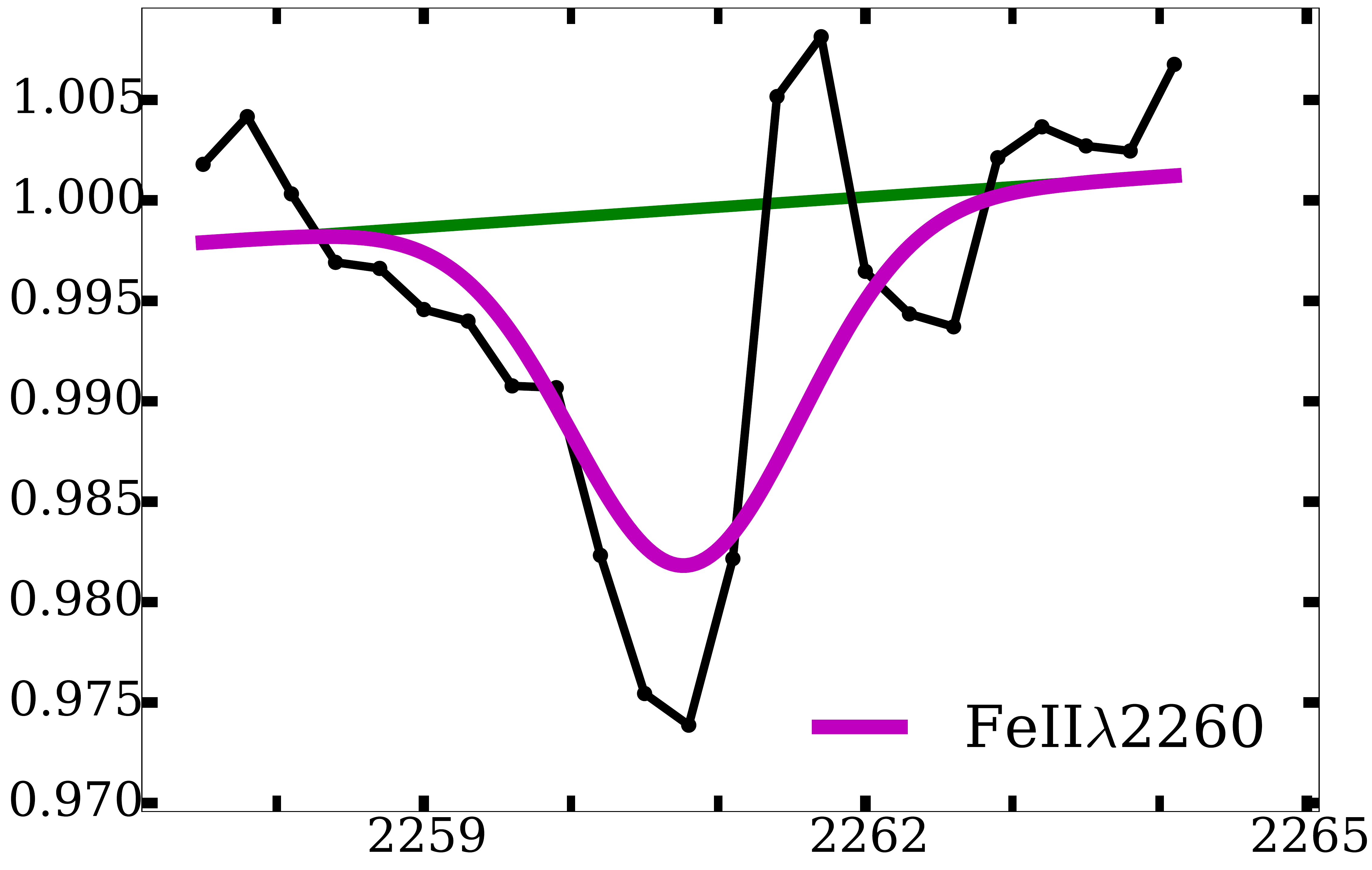}\includegraphics[width=0.25\textwidth]{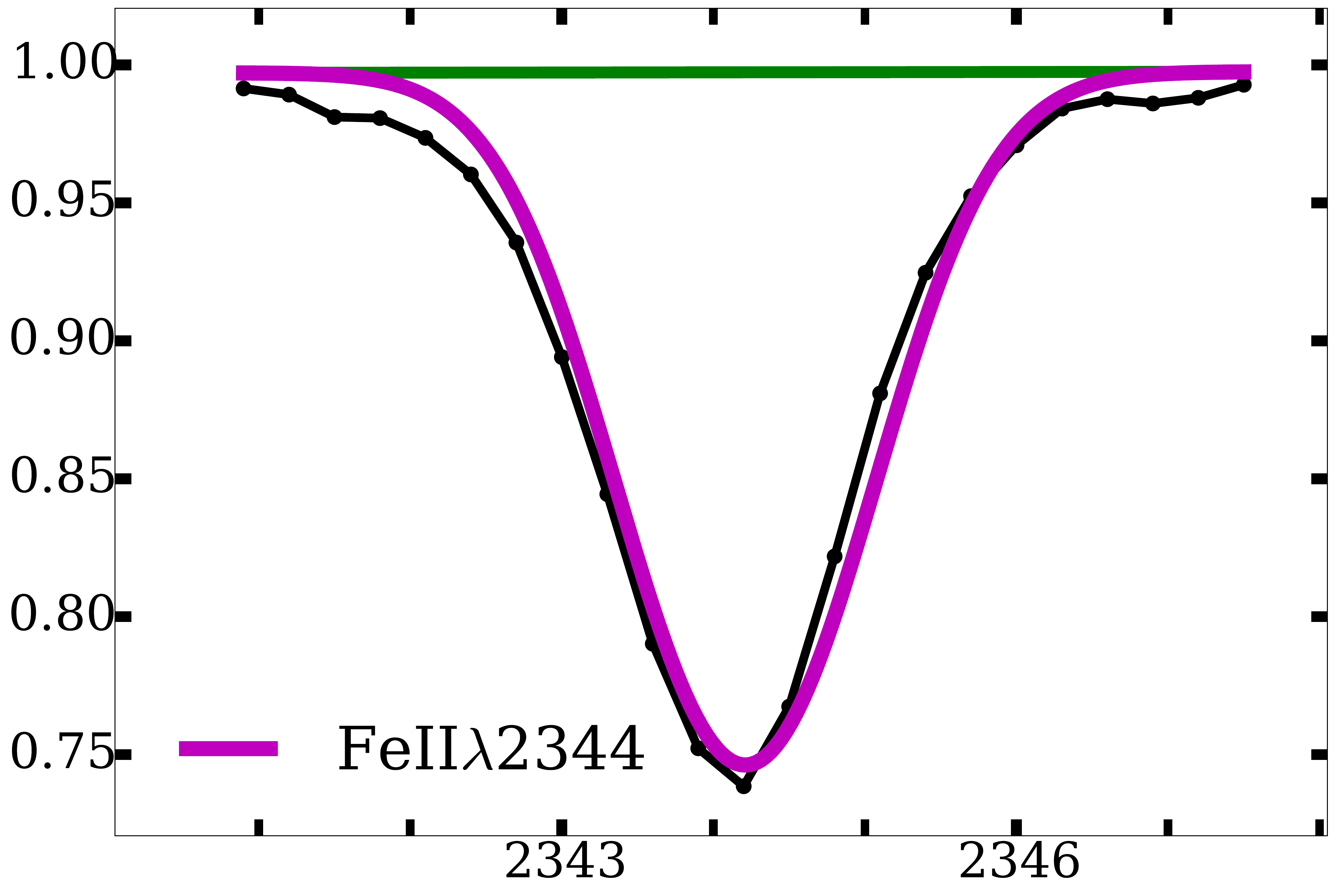}\includegraphics[width=0.25\textwidth]{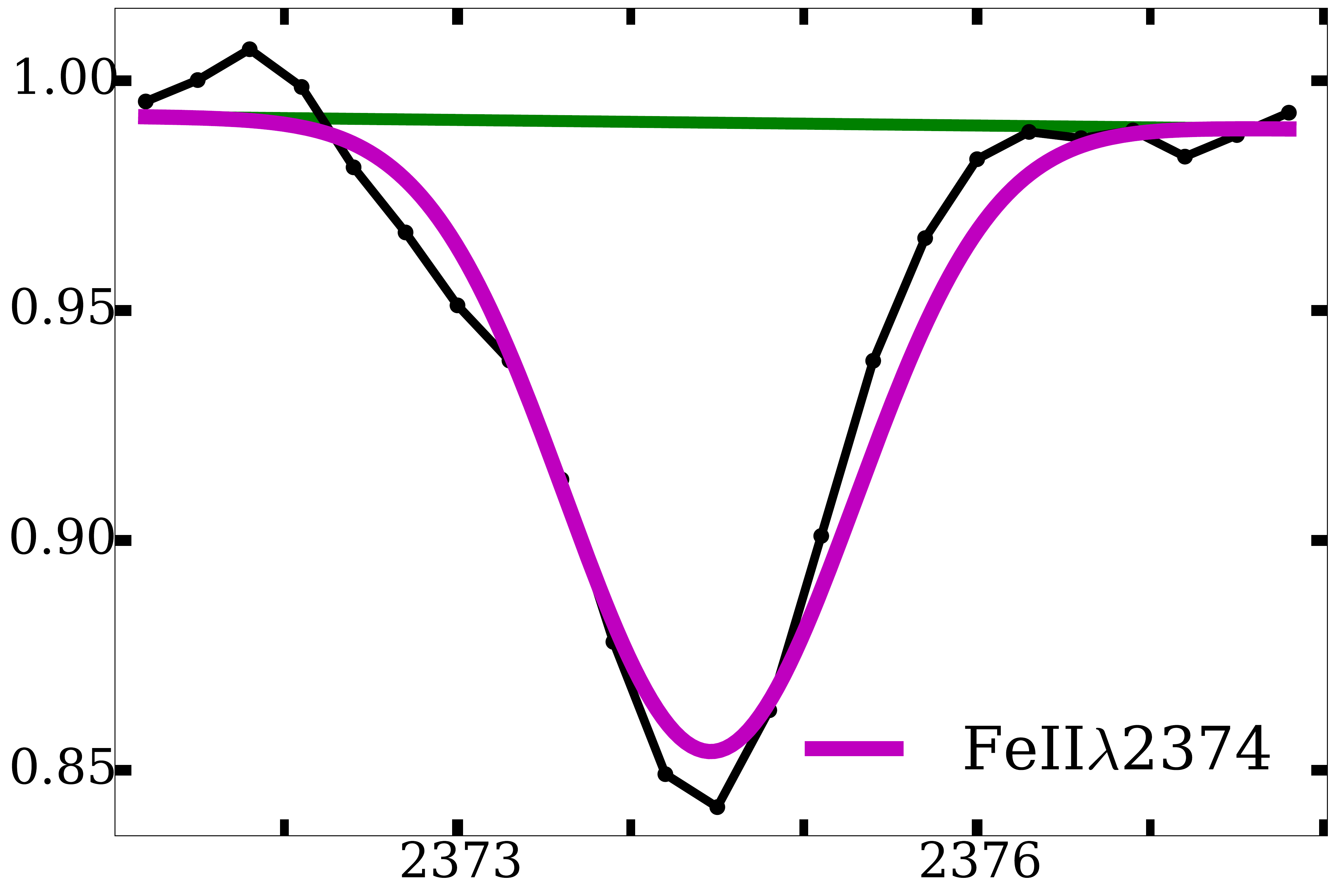}
\includegraphics[width=0.25\textwidth]{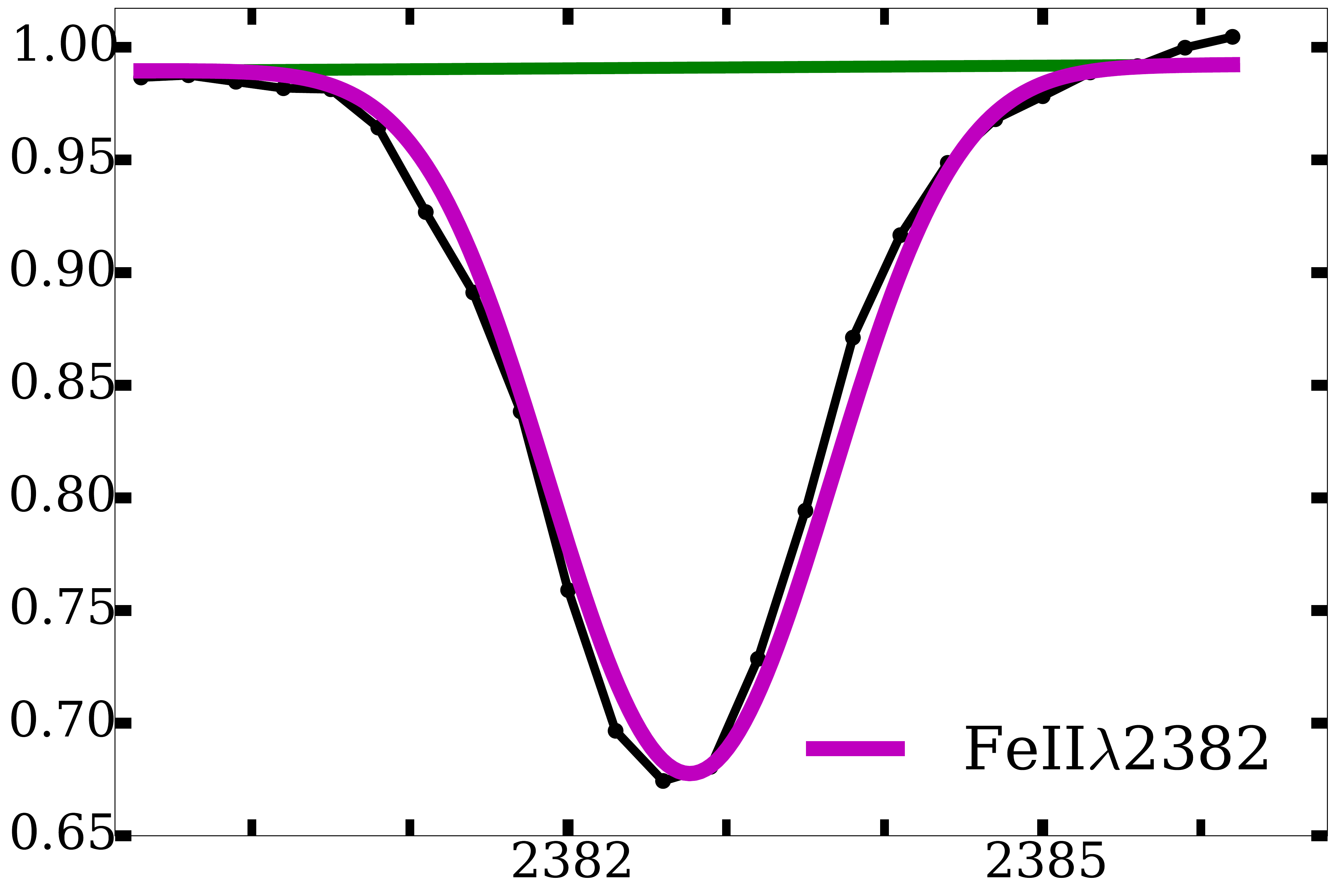}\includegraphics[width=0.25\textwidth]{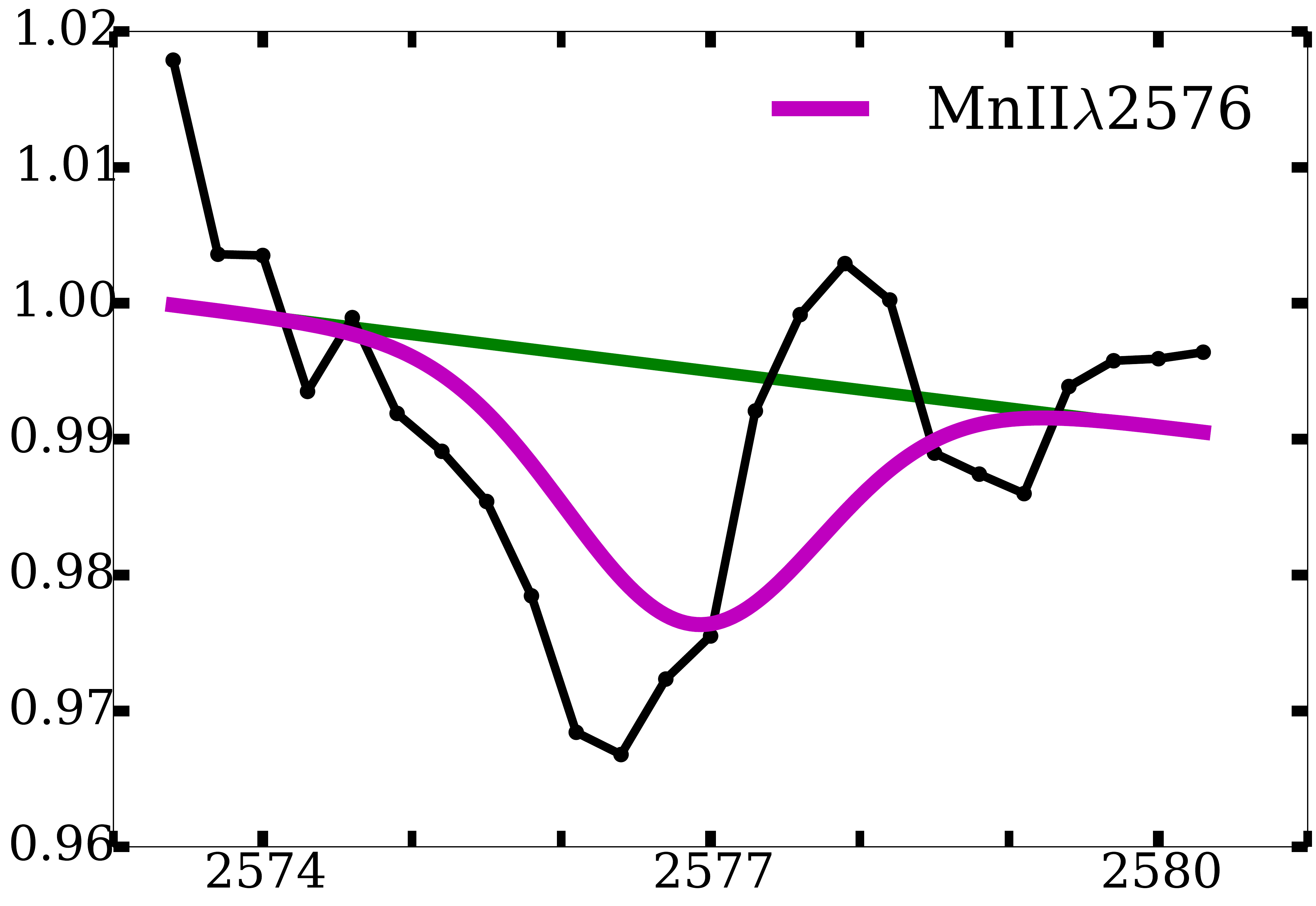}\includegraphics[width=0.25\textwidth]{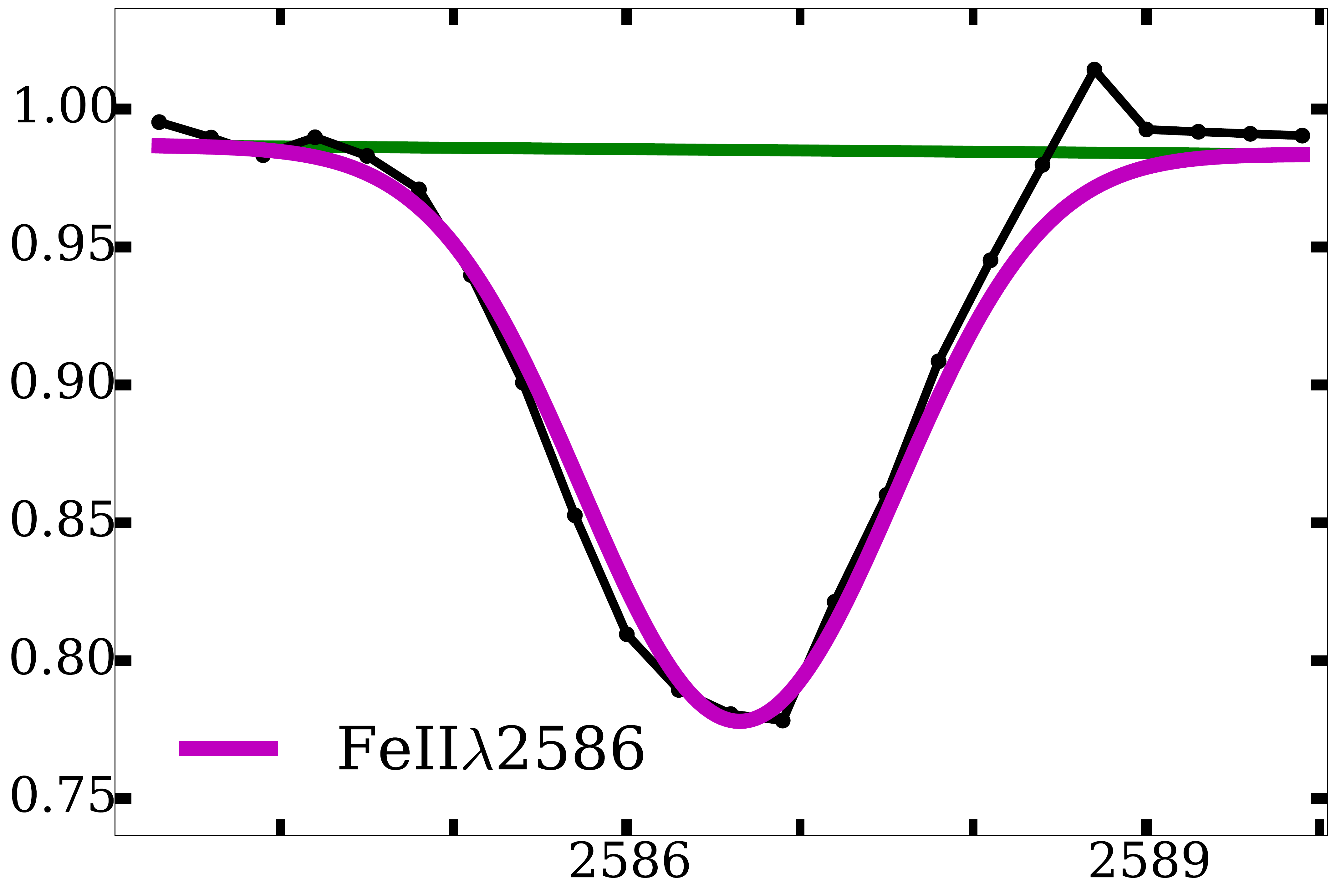}\includegraphics[width=0.25\textwidth]{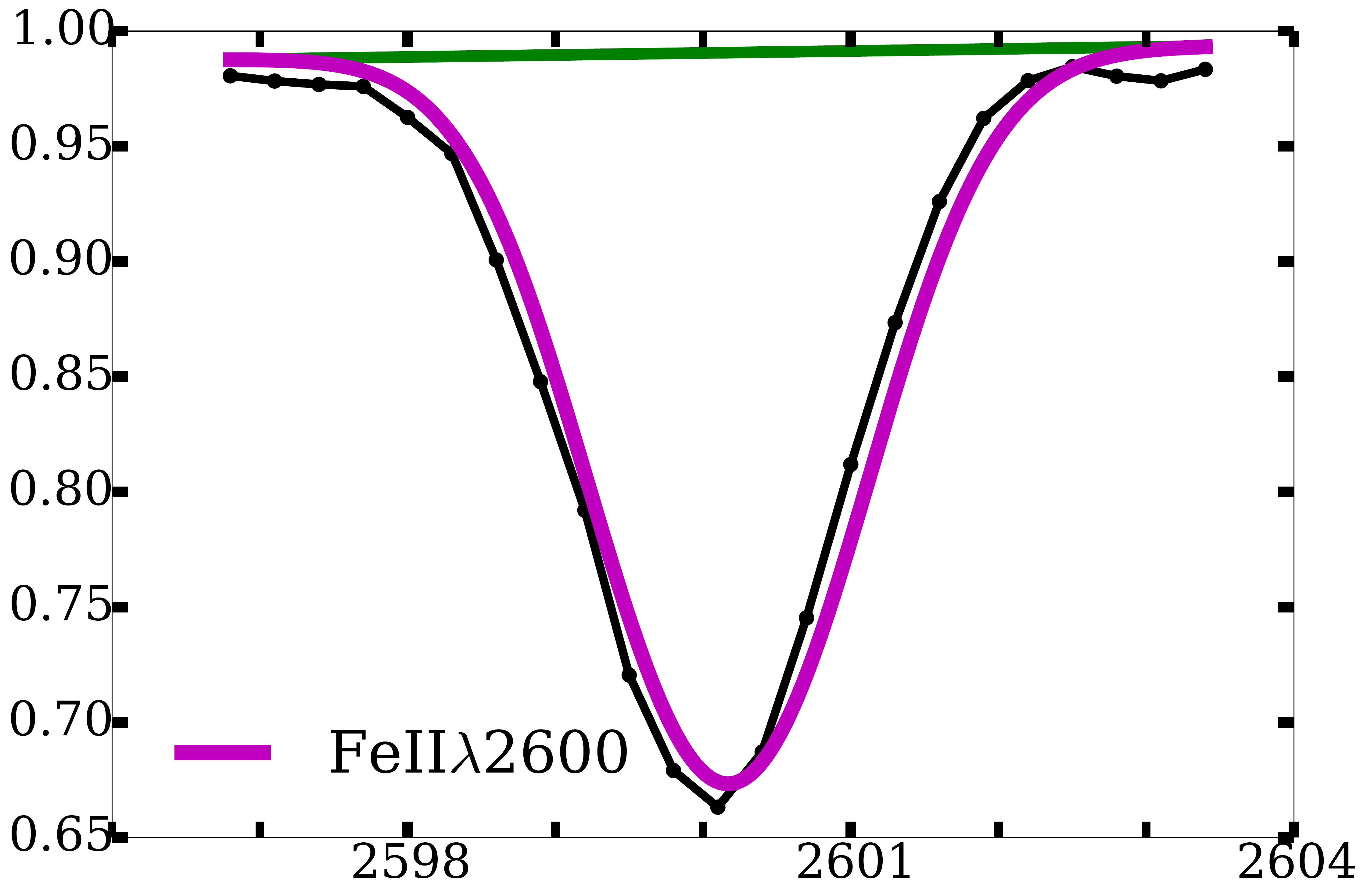}
\includegraphics[width=0.25\textwidth]{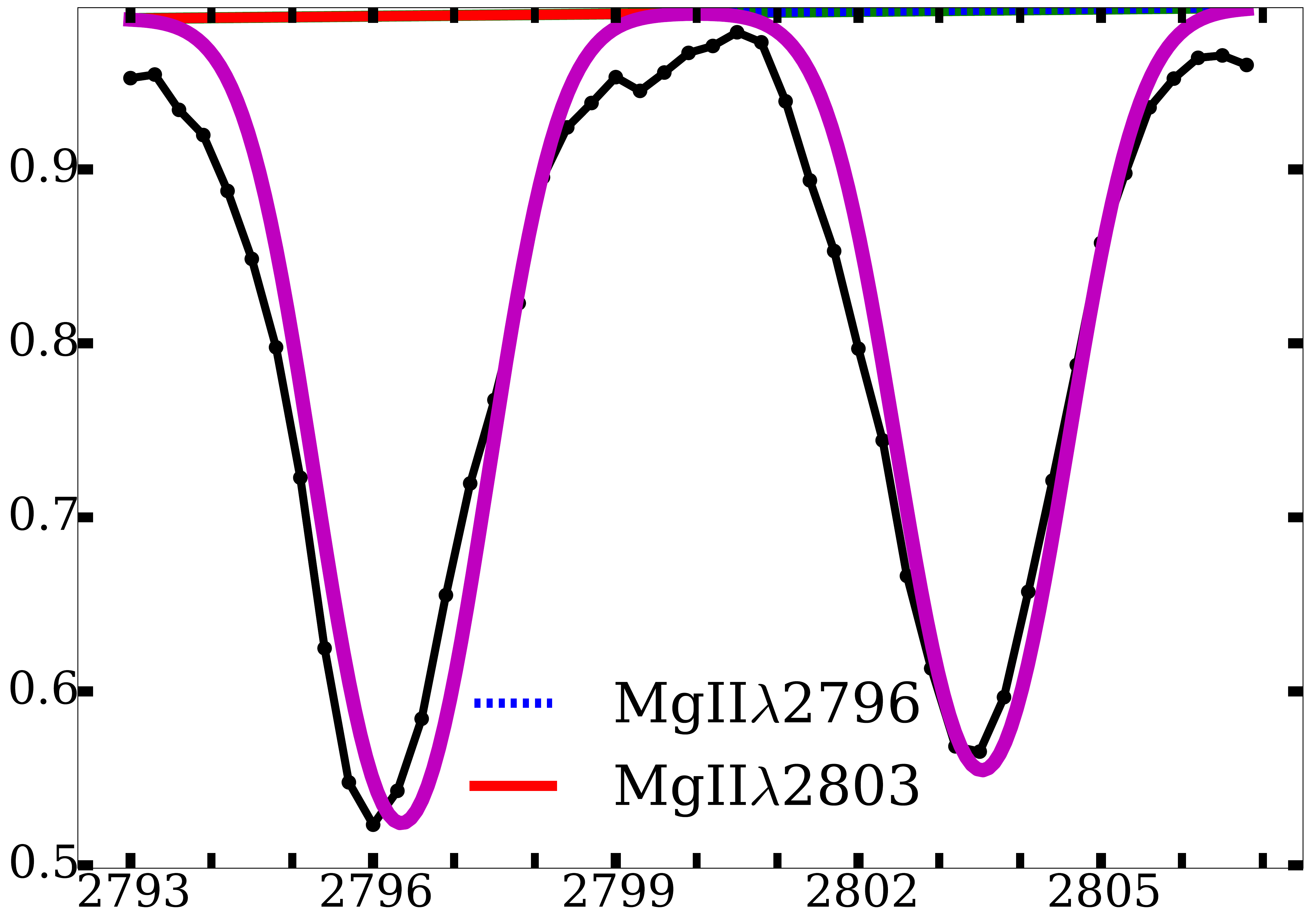}\includegraphics[width=0.25\textwidth]{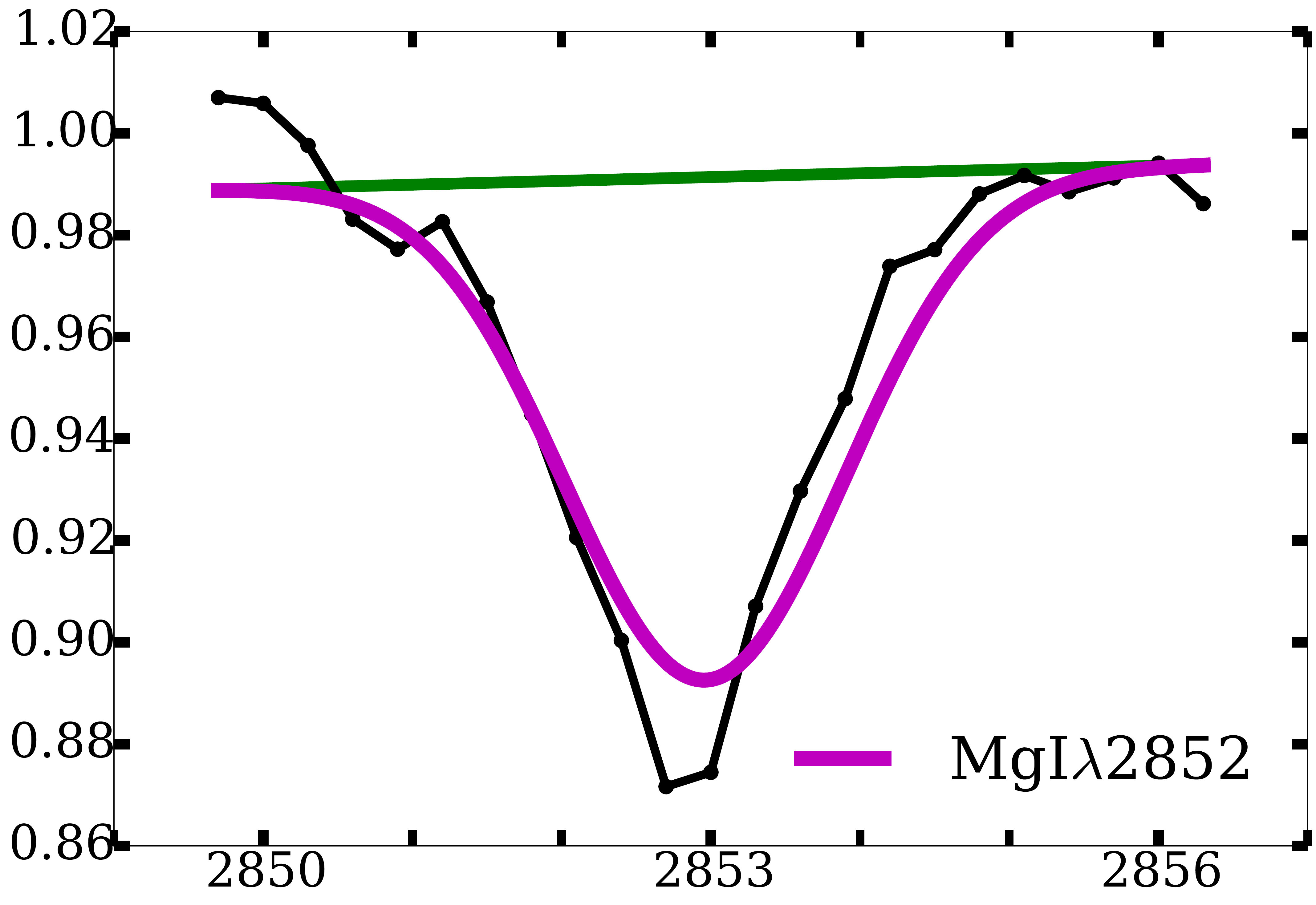}
\caption{Continuation of Figure \ref{fig:fit}}
\label{fig:fit2}
\end{figure*}

\clearpage
\bibliographystyle{apj}
\bibliography{dla_lluis}\label{References}

\begin{thebibliography}{}
\expandafter\ifx\csname natexlab\endcsname\relax\def\natexlab#1{#1}\fi

\bibitem[{{Akerman} {et~al.}(2005){Akerman}, {Ellison}, {Pettini}, \&
  {Steidel}}]{Akerman2005}
{Akerman}, C.~J., {Ellison}, S.~L., {Pettini}, M., \& {Steidel}, C.~C. 2005,
  \aap, 440, 499

\bibitem[{{Alam} {et~al.}(2015){Alam}, {Albareti}, {Allende Prieto}, {Anders},
  {Anderson}, {Anderton}, {Andrews}, {Armengaud}, {Aubourg}, {Bailey}, \&
  et~al.}]{Alam2015}
{Alam}, S., {Albareti}, F.~D., {Allende Prieto}, C., {et~al.} 2015, \apjs, 219,
  12

\bibitem[{{Asplund} {et~al.}(2009){Asplund}, {Grevesse}, {Sauval}, \&
  {Scott}}]{Asplund2009}
{Asplund}, M., {Grevesse}, N., {Sauval}, A.~J., \& {Scott}, P. 2009, \araa, 47,
  481

\bibitem[{{Baldwin}(1977)}]{Baldwin1977}
{Baldwin}, J.~A. 1977, \apj, 214, 679

\bibitem[{{Barnes} {et~al.}(2014){Barnes}, {Garel}, \& {Kacprzak}}]{Barnes2014}
{Barnes}, L.~A., {Garel}, T., \& {Kacprzak}, G.~G. 2014, \pasp, 126, 969

\bibitem[{{Barnes} \& {Haehnelt}(2014)}]{Barnes2014b}
{Barnes}, L.~A., \& {Haehnelt}, M.~G. 2014, \mnras, 440, 2313

\bibitem[{{Bautista} {et~al.}(2015){Bautista}, {Bailey}, {Font-Ribera},
  {Pieri}, {Busca}, {Miralda-Escud{\'e}}, {Palanque-Delabrouille}, {Rich},
  {Dawson}, {Feng}, {Ge}, {Gontcho}, {Ho}, {Le Goff}, {Noterdaeme},
  {P{\^a}ris}, {Rossi}, \& {Schlegel}}]{Bautista2015}
{Bautista}, J.~E., {Bailey}, S., {Font-Ribera}, A., {et~al.} 2015, \jcap, 5,
  060

\bibitem[{{Berg} {et~al.}(2015){Berg}, {Ellison}, {Prochaska}, {Venn}, \&
  {Dessauges-Zavadsky}}]{Berg2015}
{Berg}, T.~A.~M., {Ellison}, S.~L., {Prochaska}, J.~X., {Venn}, K.~A., \&
  {Dessauges-Zavadsky}, M. 2015, \mnras, 452, 4326

\bibitem[{{Berg} {et~al.}(2016){Berg}, {Ellison}, {S{\'a}nchez-Ram{\'{\i}}rez},
  {Prochaska}, {Lopez}, {D'Odorico}, {Becker}, {Christensen}, {Cupani},
  {Denney}, \& {Worsek}}]{Berg2016}
{Berg}, T.~A.~M., {Ellison}, S.~L., {S{\'a}nchez-Ram{\'{\i}}rez}, R., {et~al.}
  2016, \mnras, arXiv:1609.05968

\bibitem[{{Berry} {et~al.}(2014){Berry}, {Somerville}, {Haas}, {Gawiser},
  {Maller}, {Popping}, \& {Trager}}]{Berry2014}
{Berry}, M., {Somerville}, R.~S., {Haas}, M.~R., {et~al.} 2014, \mnras, 441,
  939

\bibitem[{{Bird} {et~al.}(2015){Bird}, {Haehnelt}, {Neeleman}, {Genel},
  {Vogelsberger}, \& {Hernquist}}]{Bird2015}
{Bird}, S., {Haehnelt}, M., {Neeleman}, M., {et~al.} 2015, \mnras, 447, 1834

\bibitem[{{Bland-Hawthorn} {et~al.}(2015){Bland-Hawthorn}, {Sutherland}, \&
  {Webster}}]{Bland2015}
{Bland-Hawthorn}, J., {Sutherland}, R., \& {Webster}, D. 2015, \apj, 807, 154

\bibitem[{{Boisse} {et~al.}(1998){Boisse}, {Le Brun}, {Bergeron}, \&
  {Deharveng}}]{Boisse1998}
{Boisse}, P., {Le Brun}, V., {Bergeron}, J., \& {Deharveng}, J.-M. 1998, \aap,
  333, 841

\bibitem[{{Calura} {et~al.}(2003){Calura}, {Matteucci}, \&
  {Vladilo}}]{Calura2003}
{Calura}, F., {Matteucci}, F., \& {Vladilo}, G. 2003, \mnras, 340, 59

\bibitem[{{Cen}(2012)}]{Cen2012}
{Cen}, R. 2012, \apj, 748, 121

\bibitem[{{Centuri{\'o}n} {et~al.}(2003){Centuri{\'o}n}, {Molaro}, {Vladilo},
  {P{\'e}roux}, {Levshakov}, \& {D'Odorico}}]{Centurion2003}
{Centuri{\'o}n}, M., {Molaro}, P., {Vladilo}, G., {et~al.} 2003, \aap, 403, 55

\bibitem[{{Cooke} {et~al.}(2013){Cooke}, {Pettini}, {Jorgenson}, {Murphy},
  {Rudie}, \& {Steidel}}]{Cooke2013}
{Cooke}, R., {Pettini}, M., {Jorgenson}, R.~A., {et~al.} 2013, \mnras, 431,
  1625

\bibitem[{{Cooke} {et~al.}(2011){Cooke}, {Pettini}, {Steidel}, {Rudie}, \&
  {Nissen}}]{Cooke2011}
{Cooke}, R., {Pettini}, M., {Steidel}, C.~C., {Rudie}, G.~C., \& {Nissen},
  P.~E. 2011, \mnras, 417, 1534

\bibitem[{{Cooke} {et~al.}(2015){Cooke}, {Pettini}, \& {Jorgenson}}]{Cooke2015}
{Cooke}, R.~J., {Pettini}, M., \& {Jorgenson}, R.~A. 2015, \apj, 800, 12

\bibitem[{{Crighton} {et~al.}(2015){Crighton}, {Murphy}, {Prochaska},
  {Worseck}, {Rafelski}, {Becker}, {Ellison}, {Fumagalli}, {Lopez}, {Meiksin},
  \& {O'Meara}}]{Crighton2015}
{Crighton}, N.~H.~M., {Murphy}, M.~T., {Prochaska}, J.~X., {et~al.} 2015,
  \mnras, 452, 217

\bibitem[{{Dawson} {et~al.}(2013){Dawson}, {Schlegel}, {Ahn}, {Anderson},
  {Aubourg}, {Bailey}, {Barkhouser}, {Bautista}, {Beifiori}, {Berlind},
  {Bhardwaj}, {Bizyaev}, {Blake}, {Blanton}, {Blomqvist}, {Bolton}, {Borde},
  {Bovy}, {Brandt}, {Brewington}, {Brinkmann}, {Brown}, {Brownstein}, {Bundy},
  {Busca}, {Carithers}, {Carnero}, {Carr}, {Chen}, {Comparat}, {Connolly},
  {Cope}, {Croft}, {Cuesta}, {da Costa}, {Davenport}, {Delubac}, {de Putter},
  {Dhital}, {Ealet}, {Ebelke}, {Eisenstein}, {Escoffier}, {Fan}, {Filiz Ak},
  {Finley}, {Font-Ribera}, {G{\'e}nova-Santos}, {Gunn}, {Guo}, {Haggard},
  {Hall}, {Hamilton}, {Harris}, {Harris}, {Ho}, {Hogg}, {Holder}, {Honscheid},
  {Huehnerhoff}, {Jordan}, {Jordan}, {Kauffmann}, {Kazin}, {Kirkby}, {Klaene},
  {Kneib}, {Le Goff}, {Lee}, {Long}, {Loomis}, {Lundgren}, {Lupton}, {Maia},
  {Makler}, {Malanushenko}, {Malanushenko}, {Mandelbaum}, {Manera}, {Maraston},
  {Margala}, {Masters}, {McBride}, {McDonald}, {McGreer}, {McMahon}, {Mena},
  {Miralda-Escud{\'e}}, {Montero-Dorta}, {Montesano}, {Muna}, {Myers},
  {Naugle}, {Nichol}, {Noterdaeme}, {Nuza}, {Olmstead}, {Oravetz}, {Oravetz},
  {Owen}, {Padmanabhan}, {Palanque-Delabrouille}, {Pan}, {Parejko},
  {P{\^a}ris}, {Percival}, {P{\'e}rez-Fournon}, {P{\'e}rez-R{\`a}fols},
  {Petitjean}, {Pfaffenberger}, {Pforr}, {Pieri}, {Prada}, {Price-Whelan},
  {Raddick}, {Rebolo}, {Rich}, {Richards}, {Rockosi}, {Roe}, {Ross}, {Ross},
  {Rossi}, {Rubi{\~n}o-Martin}, {Samushia}, {S{\'a}nchez}, {Sayres}, {Schmidt},
  {Schneider}, {Sc{\'o}ccola}, {Seo}, {Shelden}, {Sheldon}, {Shen}, {Shu},
  {Slosar}, {Smee}, {Snedden}, {Stauffer}, {Steele}, {Strauss}, {Streblyanska},
  {Suzuki}, {Swanson}, {Tal}, {Tanaka}, {Thomas}, {Tinker}, {Tojeiro},
  {Tremonti}, {Vargas Maga{\~n}a}, {Verde}, {Viel}, {Wake}, {Watson}, {Weaver},
  {Weinberg}, {Weiner}, {West}, {White}, {Wood-Vasey}, {Yeche}, {Zehavi},
  {Zhao}, \& {Zheng}}]{Dawson2013}
{Dawson}, K.~S., {Schlegel}, D.~J., {Ahn}, C.~P., {et~al.} 2013, \aj, 145, 10

\bibitem[{{Dessauges-Zavadsky} {et~al.}(2004){Dessauges-Zavadsky}, {Calura},
  {Prochaska}, {D'Odorico}, \& {Matteucci}}]{Dessauges2004}
{Dessauges-Zavadsky}, M., {Calura}, F., {Prochaska}, J.~X., {D'Odorico}, S., \&
  {Matteucci}, F. 2004, \aap, 416, 79

\bibitem[{{Dessauges-Zavadsky} {et~al.}(2003){Dessauges-Zavadsky},
  {P{\'e}roux}, {Kim}, {D'Odorico}, \& {McMahon}}]{Dessauges2003}
{Dessauges-Zavadsky}, M., {P{\'e}roux}, C., {Kim}, T.-S., {D'Odorico}, S., \&
  {McMahon}, R.~G. 2003, \mnras, 345, 447

\bibitem[{{Dessauges-Zavadsky} {et~al.}(2006){Dessauges-Zavadsky}, {Prochaska},
  {D'Odorico}, {Calura}, \& {Matteucci}}]{Dessauges2006}
{Dessauges-Zavadsky}, M., {Prochaska}, J.~X., {D'Odorico}, S., {Calura}, F., \&
  {Matteucci}, F. 2006, \aap, 445, 93

\bibitem[{{Dutta} {et~al.}(2014){Dutta}, {Srianand}, {Rahmani}, {Petitjean},
  {Noterdaeme}, \& {Ledoux}}]{Dutta2014}
{Dutta}, R., {Srianand}, R., {Rahmani}, H., {et~al.} 2014, \mnras, 440, 307

\bibitem[{{Eisenstein} {et~al.}(2011){Eisenstein}, {Weinberg}, {Agol},
  {Aihara}, {Allende Prieto}, {Anderson}, {Arns}, {Aubourg}, {Bailey},
  {Balbinot}, \& et~al.}]{Eisenstein2011}
{Eisenstein}, D.~J., {Weinberg}, D.~H., {Agol}, E., {et~al.} 2011, \aj, 142, 72

\bibitem[{{Ellison} {et~al.}(2001){Ellison}, {Yan}, {Hook}, {Pettini}, {Wall},
  \& {Shaver}}]{Ellison2001}
{Ellison}, S.~L., {Yan}, L., {Hook}, I.~M., {et~al.} 2001, \aap, 379, 393

\bibitem[{{Ellison} {et~al.}(2008){Ellison}, {York}, {Pettini}, \&
  {Kanekar}}]{Ellison2008}
{Ellison}, S.~L., {York}, B.~A., {Pettini}, M., \& {Kanekar}, N. 2008, \mnras,
  388, 1349

\bibitem[{{Fall} \& {Pei}(1993)}]{Fall1993}
{Fall}, S.~M., \& {Pei}, Y.~C. 1993, \apj, 402, 479

\bibitem[{{Faucher-Gigu{\`e}re} {et~al.}(2008){Faucher-Gigu{\`e}re},
  {Prochaska}, {Lidz}, {Hernquist}, \& {Zaldarriaga}}]{Faucher2008}
{Faucher-Gigu{\`e}re}, C.-A., {Prochaska}, J.~X., {Lidz}, A., {Hernquist}, L.,
  \& {Zaldarriaga}, M. 2008, \apj, 681, 831

\bibitem[{{Font-Ribera} {et~al.}(2012){Font-Ribera}, {Miralda-Escud{\'e}},
  {Arnau}, {Carithers}, {Lee}, {Noterdaeme}, {P{\^a}ris}, {Petitjean}, {Rich},
  {Rollinde}, {Ross}, {Schneider}, {White}, \& {York}}]{FontRibera2012}
{Font-Ribera}, A., {Miralda-Escud{\'e}}, J., {Arnau}, E., {et~al.} 2012, \jcap,
  11, 59

\bibitem[{{Foreman-Mackey} {et~al.}(2013){Foreman-Mackey}, {Hogg}, {Lang}, \&
  {Goodman}}]{emcee2013}
{Foreman-Mackey}, D., {Hogg}, D.~W., {Lang}, D., \& {Goodman}, J. 2013, \pasp,
  125, 306

\bibitem[{Foreman-Mackey {et~al.}(2014)Foreman-Mackey, Price-Whelan, Ryan,
  Emily, Smith, Barbary, Hogg, \& Brewer}]{triangle}
Foreman-Mackey, D., Price-Whelan, A., Ryan, G., {et~al.} 2014, triangle.py
  v0.1.1, doi:10.5281/zenodo.11020

\bibitem[{{Fox}(2011)}]{Fox2011}
{Fox}, A.~J. 2011, \apj, 730, 58

\bibitem[{{Fox} {et~al.}(2007{\natexlab{a}}){Fox}, {Ledoux}, {Petitjean}, \&
  {Srianand}}]{Fox2007b}
{Fox}, A.~J., {Ledoux}, C., {Petitjean}, P., \& {Srianand}, R.
  2007{\natexlab{a}}, \aap, 473, 791

\bibitem[{{Fox} {et~al.}(2007{\natexlab{b}}){Fox}, {Petitjean}, {Ledoux}, \&
  {Srianand}}]{Fox2007a}
{Fox}, A.~J., {Petitjean}, P., {Ledoux}, C., \& {Srianand}, R.
  2007{\natexlab{b}}, \aap, 465, 171

\bibitem[{{Fox} {et~al.}(2009){Fox}, {Prochaska}, {Ledoux}, {Petitjean},
  {Wolfe}, \& {Srianand}}]{Fox2009}
{Fox}, A.~J., {Prochaska}, J.~X., {Ledoux}, C., {et~al.} 2009, \aap, 503, 731

\bibitem[{{Fukugita} \& {M{\'e}nard}(2015)}]{Fukugita2015}
{Fukugita}, M., \& {M{\'e}nard}, B. 2015, \apj, 799, 195

\bibitem[{{Fumagalli} {et~al.}(2014){Fumagalli}, {O'Meara}, {Prochaska},
  {Kanekar}, \& {Wolfe}}]{Fumagalli2014}
{Fumagalli}, M., {O'Meara}, J.~M., {Prochaska}, J.~X., {Kanekar}, N., \&
  {Wolfe}, A.~M. 2014, \mnras, 444, 1282

\bibitem[{{Fumagalli} {et~al.}(2011){Fumagalli}, {Prochaska}, {Kasen}, {Dekel},
  {Ceverino}, \& {Primack}}]{Fumagalli2011}
{Fumagalli}, M., {Prochaska}, J.~X., {Kasen}, D., {et~al.} 2011, \mnras, 418,
  1796

\bibitem[{Goody \& Yung(1996)}]{Goody1964}
Goody, R., \& Yung, Y. 1996, Atmospheric Radiation: Theoretical Basis (OUP USA)

\bibitem[{{Gunn} {et~al.}(1998){Gunn}, {Carr}, {Rockosi}, {Sekiguchi}, {Berry},
  {Elms}, {de Haas}, {Ivezi{\'c}}, {Knapp}, {Lupton}, {Pauls}, {Simcoe},
  {Hirsch}, {Sanford}, {Wang}, {York}, {Harris}, {Annis}, {Bartozek},
  {Boroski}, {Bakken}, {Haldeman}, {Kent}, {Holm}, {Holmgren}, {Petravick},
  {Prosapio}, {Rechenmacher}, {Doi}, {Fukugita}, {Shimasaku}, {Okada}, {Hull},
  {Siegmund}, {Mannery}, {Blouke}, {Heidtman}, {Schneider}, {Lucinio}, \&
  {Brinkman}}]{Gunn1998}
{Gunn}, J.~E., {Carr}, M., {Rockosi}, C., {et~al.} 1998, \aj, 116, 3040

\bibitem[{{Gunn} {et~al.}(2006){Gunn}, {Siegmund}, {Mannery}, {Owen}, {Hull},
  {Leger}, {Carey}, {Knapp}, {York}, {Boroski}, {Kent}, {Lupton}, {Rockosi},
  {Evans}, {Waddell}, {Anderson}, {Annis}, {Barentine}, {Bartoszek}, {Bastian},
  {Bracker}, {Brewington}, {Briegel}, {Brinkmann}, {Brown}, {Carr},
  {Czarapata}, {Drennan}, {Dombeck}, {Federwitz}, {Gillespie}, {Gonzales},
  {Hansen}, {Harvanek}, {Hayes}, {Jordan}, {Kinney}, {Klaene}, {Kleinman},
  {Kron}, {Kresinski}, {Lee}, {Limmongkol}, {Lindenmeyer}, {Long}, {Loomis},
  {McGehee}, {Mantsch}, {Neilsen}, {Neswold}, {Newman}, {Nitta}, {Peoples},
  {Pier}, {Prieto}, {Prosapio}, {Rivetta}, {Schneider}, {Snedden}, \&
  {Wang}}]{Gunn2006}
{Gunn}, J.~E., {Siegmund}, W.~A., {Mannery}, E.~J., {et~al.} 2006, \aj, 131,
  2332

\bibitem[{{Haehnelt} {et~al.}(1998){Haehnelt}, {Steinmetz}, \&
  {Rauch}}]{Haehnelt1998}
{Haehnelt}, M.~G., {Steinmetz}, M., \& {Rauch}, M. 1998, \apj, 495, 647

\bibitem[{{Henry} \& {Prochaska}(2007)}]{Henry2007}
{Henry}, R.~B.~C., \& {Prochaska}, J.~X. 2007, \pasp, 119, 962

\bibitem[{{Howk} \& {Sembach}(1999)}]{Howk1999}
{Howk}, J.~C., \& {Sembach}, K.~R. 1999, \apjl, 523, L141

\bibitem[{{Jenkins}(2009)}]{Jenkins2009}
{Jenkins}, E.~B. 2009, \apj, 700, 1299

\bibitem[{{Jorgenson} {et~al.}(2013){Jorgenson}, {Murphy}, \&
  {Thompson}}]{Jorgenson2013}
{Jorgenson}, R.~A., {Murphy}, M.~T., \& {Thompson}, R. 2013, \mnras, 435, 482

\bibitem[{{Khare} {et~al.}(2004){Khare}, {Kulkarni}, {Lauroesch}, {York},
  {Crotts}, \& {Nakamura}}]{Khare2004}
{Khare}, P., {Kulkarni}, V.~P., {Lauroesch}, J.~T., {et~al.} 2004, \apj, 616,
  86

\bibitem[{{Khare} {et~al.}(2007){Khare}, {Kulkarni}, {P{\'e}roux}, {York},
  {Lauroesch}, \& {Meiring}}]{Khare2007}
{Khare}, P., {Kulkarni}, V.~P., {P{\'e}roux}, C., {et~al.} 2007, \aap, 464, 487

\bibitem[{{Khare} {et~al.}(2012){Khare}, {vanden Berk}, {York}, {Lundgren}, \&
  {Kulkarni}}]{Khare2012}
{Khare}, P., {vanden Berk}, D., {York}, D.~G., {Lundgren}, B., \& {Kulkarni},
  V.~P. 2012, \mnras, 419, 1028

\bibitem[{{Kisielius} {et~al.}(2015){Kisielius}, {Kulkarni}, {Ferland},
  {Bogdanovich}, {Som}, \& {Lykins}}]{Kisielius2015}
{Kisielius}, R., {Kulkarni}, V.~P., {Ferland}, G.~J., {et~al.} 2015, \apj, 804,
  76

\bibitem[{{Krogager} {et~al.}(2016){Krogager}, {Fynbo}, {Noterdaeme}, {Zafar},
  {M{\o}ller}, {Ledoux}, {Kr{\"u}hler}, \& {Stockton}}]{Krogager2016}
{Krogager}, J.-K., {Fynbo}, J.~P.~U., {Noterdaeme}, P., {et~al.} 2016, \mnras,
  455, 2698

\bibitem[{{Kulkarni} \& {Fall}(2002)}]{Kulkarni2002}
{Kulkarni}, V.~P., \& {Fall}, S.~M. 2002, \apj, 580, 732

\bibitem[{{Kulkarni} {et~al.}(2005){Kulkarni}, {Fall}, {Lauroesch}, {York},
  {Welty}, {Khare}, \& {Truran}}]{Kulkarni2005}
{Kulkarni}, V.~P., {Fall}, S.~M., {Lauroesch}, J.~T., {et~al.} 2005, \apj, 618,
  68

\bibitem[{{Kulkarni} {et~al.}(1997){Kulkarni}, {Fall}, \&
  {Truran}}]{Kulkarni1997}
{Kulkarni}, V.~P., {Fall}, S.~M., \& {Truran}, J.~W. 1997, \apjl, 484, L7

\bibitem[{{Kulkarni} {et~al.}(2007){Kulkarni}, {Khare}, {P{\'e}roux}, {York},
  {Lauroesch}, \& {Meiring}}]{Kulkarni2007}
{Kulkarni}, V.~P., {Khare}, P., {P{\'e}roux}, C., {et~al.} 2007, \apj, 661, 88

\bibitem[{{Kulkarni} {et~al.}(2015){Kulkarni}, {Som}, {Morrison}, {P{\'e}roux},
  {Quiret}, \& {York}}]{Kulkarni2015}
{Kulkarni}, V.~P., {Som}, D., {Morrison}, S., {et~al.} 2015, \apj, 815, 24

\bibitem[{{Ledoux} {et~al.}(2015){Ledoux}, {Noterdaeme}, {Petitjean}, \&
  {Srianand}}]{Ledoux2015}
{Ledoux}, C., {Noterdaeme}, P., {Petitjean}, P., \& {Srianand}, R. 2015, \aap,
  580, A8

\bibitem[{{Ledoux} {et~al.}(2006){Ledoux}, {Petitjean}, {Fynbo}, {M{\o}ller},
  \& {Srianand}}]{Ledoux2006}
{Ledoux}, C., {Petitjean}, P., {Fynbo}, J.~P.~U., {M{\o}ller}, P., \&
  {Srianand}, R. 2006, \aap, 457, 71

\bibitem[{{Lehner} {et~al.}(2008){Lehner}, {Howk}, {Prochaska}, \&
  {Wolfe}}]{Lehner2008}
{Lehner}, N., {Howk}, J.~C., {Prochaska}, J.~X., \& {Wolfe}, A.~M. 2008,
  \mnras, 390, 2

\bibitem[{{Lehner} {et~al.}(2014){Lehner}, {O'Meara}, {Fox}, {Howk},
  {Prochaska}, {Burns}, \& {Armstrong}}]{Lehner2014}
{Lehner}, N., {O'Meara}, J.~M., {Fox}, A.~J., {et~al.} 2014, \apj, 788, 119

\bibitem[{{McDonald} \& {Miralda-Escud{\'e}}(1999)}]{McDonald1999}
{McDonald}, P., \& {Miralda-Escud{\'e}}, J. 1999, \apj, 519, 486

\bibitem[{{Meiring} {et~al.}(2006){Meiring}, {Kulkarni}, {Khare}, {Bechtold},
  {York}, {Cui}, {Lauroesch}, {Crotts}, \& {Nakamura}}]{Meiring2006}
{Meiring}, J.~D., {Kulkarni}, V.~P., {Khare}, P., {et~al.} 2006, \mnras, 370,
  43

\bibitem[{{M{\o}ller} {et~al.}(2013){M{\o}ller}, {Fynbo}, {Ledoux}, \&
  {Nilsson}}]{Moller2013}
{M{\o}ller}, P., {Fynbo}, J.~P.~U., {Ledoux}, C., \& {Nilsson}, K.~K. 2013,
  \mnras, 430, 2680

\bibitem[{{Morton}(2003)}]{Morton2003}
{Morton}, D.~C. 2003, \apjs, 149, 205

\bibitem[{{Murphy} \& {Bernet}(2016)}]{Murphy2016}
{Murphy}, M.~T., \& {Bernet}, M.~L. 2016, \mnras, 455, 1043

\bibitem[{{Neeleman} {et~al.}(2015){Neeleman}, {Prochaska}, \&
  {Wolfe}}]{Neeleman2015}
{Neeleman}, M., {Prochaska}, J.~X., \& {Wolfe}, A.~M. 2015, \apj, 800, 7

\bibitem[{{Neeleman} {et~al.}(2013){Neeleman}, {Wolfe}, {Prochaska}, \&
  {Rafelski}}]{Neeleman2013}
{Neeleman}, M., {Wolfe}, A.~M., {Prochaska}, J.~X., \& {Rafelski}, M. 2013,
  \apj, 769, 54

\bibitem[{{Noterdaeme} {et~al.}(2009){Noterdaeme}, {Petitjean}, {Ledoux}, \&
  {Srianand}}]{Noterdaeme2009}
{Noterdaeme}, P., {Petitjean}, P., {Ledoux}, C., \& {Srianand}, R. 2009, \aap,
  505, 1087

\bibitem[{{Noterdaeme} {et~al.}(2014){Noterdaeme}, {Petitjean}, {P{\^a}ris},
  {Cai}, {Finley}, {Ge}, {Pieri}, \& {York}}]{Noterdaeme2014}
{Noterdaeme}, P., {Petitjean}, P., {P{\^a}ris}, I., {et~al.} 2014, \aap, 566,
  A24

\bibitem[{{Noterdaeme} {et~al.}(2015){Noterdaeme}, {Petitjean}, \&
  {Srianand}}]{Noterdaeme2015}
{Noterdaeme}, P., {Petitjean}, P., \& {Srianand}, R. 2015, \aap, 578, L5

\bibitem[{{Noterdaeme} {et~al.}(2012){Noterdaeme}, {Petitjean}, {Carithers},
  {P{\^a}ris}, {Font-Ribera}, {Bailey}, {Aubourg}, {Bizyaev}, {Ebelke},
  {Finley}, {Ge}, {Malanushenko}, {Malanushenko}, {Miralda-Escud{\'e}},
  {Myers}, {Oravetz}, {Pan}, {Pieri}, {Ross}, {Schneider}, {Simmons}, \&
  {York}}]{Noterdaeme2012}
{Noterdaeme}, P., {Petitjean}, P., {Carithers}, W.~C., {et~al.} 2012, \aap,
  547, L1

\bibitem[{{Padmanabhan} {et~al.}(2015){Padmanabhan}, {Choudhury}, \&
  {Refregier}}]{Padmanabhan2015}
{Padmanabhan}, H., {Choudhury}, T.~R., \& {Refregier}, A. 2015, ArXiv e-prints,
  arXiv:1505.00008

\bibitem[{{Palanque-Delabrouille} {et~al.}(2013){Palanque-Delabrouille},
  {Y{\`e}che}, {Borde}, {Le Goff}, {Rossi}, {Viel}, {Aubourg}, {Bailey},
  {Bautista}, {Blomqvist}, {Bolton}, {Bolton}, {Busca}, {Carithers}, {Croft},
  {Dawson}, {Delubac}, {Font-Ribera}, {Ho}, {Kirkby}, {Lee}, {Margala},
  {Miralda-Escud{\'e}}, {Muna}, {Myers}, {Noterdaeme}, {P{\^a}ris},
  {Petitjean}, {Pieri}, {Rich}, {Rollinde}, {Ross}, {Schlegel}, {Schneider},
  {Slosar}, \& {Weinberg}}]{Natalie2013}
{Palanque-Delabrouille}, N., {Y{\`e}che}, C., {Borde}, A., {et~al.} 2013, \aap,
  559, A85

\bibitem[{{P{\^a}ris} {et~al.}(2012){P{\^a}ris}, {Petitjean}, {Aubourg},
  {Bailey}, {Ross}, {Myers}, {Strauss}, {Anderson}, {Arnau}, {Bautista},
  {Bizyaev}, {Bolton}, {Bovy}, {Brandt}, {Brewington}, {Browstein}, {Busca},
  {Capellupo}, {Carithers}, {Croft}, {Dawson}, {Delubac}, {Ebelke},
  {Eisenstein}, {Engelke}, {Fan}, {Filiz Ak}, {Finley}, {Font-Ribera}, {Ge},
  {Gibson}, {Hall}, {Hamann}, {Hennawi}, {Ho}, {Hogg}, {Ivezi{\'c}}, {Jiang},
  {Kimball}, {Kirkby}, {Kirkpatrick}, {Lee}, {Le Goff}, {Lundgren}, {MacLeod},
  {Malanushenko}, {Malanushenko}, {Maraston}, {McGreer}, {McMahon},
  {Miralda-Escud{\'e}}, {Muna}, {Noterdaeme}, {Oravetz},
  {Palanque-Delabrouille}, {Pan}, {Perez-Fournon}, {Pieri}, {Richards},
  {Rollinde}, {Sheldon}, {Schlegel}, {Schneider}, {Slosar}, {Shelden}, {Shen},
  {Simmons}, {Snedden}, {Suzuki}, {Tinker}, {Viel}, {Weaver}, {Weinberg},
  {White}, {Wood-Vasey}, \& {Y{\`e}che}}]{Paris2012}
{P{\^a}ris}, I., {Petitjean}, P., {Aubourg}, {\'E}., {et~al.} 2012, \aap, 548,
  A66

\bibitem[{{P{\^a}ris} {et~al.}(2016){P{\^a}ris}, {Petitjean}, {Ross}, {Myers},
  {Aubourg}, {Streblyanska}, {Bailey}, {Armengaud}, {Palanque-Delabrouille},
  {Y{\`e}che}, {Hamann}, {Strauss}, {Albareti}, {Bovy}, {Bizyaev}, {Brandt},
  {Brusa}, {Buchner}, {Comparat}, {Croft}, {Dwelly}, {Fan}, {Font-Ribera},
  {Ge}, {Georgakakis}, {Hall}, {Jian}, {Kinemuchi}, {Malanushenko},
  {Malanushenko}, {McMahon}, {Menzel}, {Merloni}, {Nandra}, {Noterdaeme},
  {Oravetz}, {Pan}, {Pieri}, {Prada}, {Salvato}, {Schlegel}, {Schneider},
  {Simmons}, {Viel}, {Weinberg}, \& {Zhu}}]{Paris2016}
{P{\^a}ris}, I., {Petitjean}, P., {Ross}, N.~P., {et~al.} 2016, ArXiv e-prints,
  arXiv:1608.06483

\bibitem[{{P{\'e}rez-R{\`a}fols} {et~al.}(2014){P{\'e}rez-R{\`a}fols},
  {Miralda-Escud{\'e}}, {Lundgren}, {Ge}, {Petitjean}, {Schneider}, {York}, \&
  {Weaver}}]{PerezRafols2014}
{P{\'e}rez-R{\`a}fols}, I., {Miralda-Escud{\'e}}, J., {Lundgren}, B., {et~al.}
  2014, ArXiv e-prints, arXiv:1402.1342

\bibitem[{{P{\'e}roux} {et~al.}(2003{\natexlab{a}}){P{\'e}roux},
  {Dessauges-Zavadsky}, {D'Odorico}, {Kim}, \& {McMahon}}]{Peroux2003b}
{P{\'e}roux}, C., {Dessauges-Zavadsky}, M., {D'Odorico}, S., {Kim}, T.-S., \&
  {McMahon}, R.~G. 2003{\natexlab{a}}, \mnras, 345, 480

\bibitem[{{P{\'e}roux} {et~al.}(2005){P{\'e}roux}, {Dessauges-Zavadsky},
  {D'Odorico}, {Sun Kim}, \& {McMahon}}]{Peroux2005}
{P{\'e}roux}, C., {Dessauges-Zavadsky}, M., {D'Odorico}, S., {Sun Kim}, T., \&
  {McMahon}, R.~G. 2005, \mnras, 363, 479

\bibitem[{{P{\'e}roux} {et~al.}(2003{\natexlab{b}}){P{\'e}roux}, {McMahon},
  {Storrie-Lombardi}, \& {Irwin}}]{Peroux2003}
{P{\'e}roux}, C., {McMahon}, R.~G., {Storrie-Lombardi}, L.~J., \& {Irwin},
  M.~J. 2003{\natexlab{b}}, \mnras, 346, 1103

\bibitem[{{Petitjean} {et~al.}(2008){Petitjean}, {Ledoux}, \&
  {Srianand}}]{Petitjean2008}
{Petitjean}, P., {Ledoux}, C., \& {Srianand}, R. 2008, \aap, 480, 349

\bibitem[{{Pettini}(2004)}]{Pettini2004}
{Pettini}, M. 2004, in Cosmochemistry. The melting pot of the elements, ed.
  C.~{Esteban}, R.~{Garc{\'{\i}}a L{\'o}pez}, A.~{Herrero}, \&
  F.~{S{\'a}nchez}, 257--298

\bibitem[{{Pettini}(2006)}]{Pettini2006}
{Pettini}, M. 2006, in The Fabulous Destiny of Galaxies: Bridging Past and
  Present, ed. V.~{Le Brun}, A.~{Mazure}, S.~{Arnouts}, \& D.~{Burgarella}, 319

\bibitem[{{Pettini} {et~al.}(1990){Pettini}, {Boksenberg}, \&
  {Hunstead}}]{Pettini1990}
{Pettini}, M., {Boksenberg}, A., \& {Hunstead}, R.~W. 1990, \apj, 348, 48

\bibitem[{{Pettini} {et~al.}(2002){Pettini}, {Ellison}, {Bergeron}, \&
  {Petitjean}}]{Pettini2002}
{Pettini}, M., {Ellison}, S.~L., {Bergeron}, J., \& {Petitjean}, P. 2002, \aap,
  391, 21

\bibitem[{{Pettini} {et~al.}(1997){Pettini}, {King}, {Smith}, \&
  {Hunstead}}]{Pettini1997}
{Pettini}, M., {King}, D.~L., {Smith}, L.~J., \& {Hunstead}, R.~W. 1997, \apj,
  478, 536

\bibitem[{{Pettini} {et~al.}(1995){Pettini}, {Lipman}, \&
  {Hunstead}}]{Pettini1995}
{Pettini}, M., {Lipman}, K., \& {Hunstead}, R.~W. 1995, \apj, 451, 100

\bibitem[{{Pieri} {et~al.}(2014){Pieri}, {Mortonson}, {Frank}, {Crighton},
  {Weinberg}, {Lee}, {Noterdaeme}, {Bailey}, {Busca}, {Ge}, {Kirkby},
  {Lundgren}, {Mathur}, {P{\^a}ris}, {Palanque-Delabrouille}, {Petitjean},
  {Rich}, {Ross}, {Schneider}, \& {York}}]{Pieri2014}
{Pieri}, M.~M., {Mortonson}, M.~J., {Frank}, S., {et~al.} 2014, \mnras, 441,
  1718

\bibitem[{{Planck Collaboration}(2015)}]{Planck2015}
{Planck Collaboration}. 2015, ArXiv e-prints, arXiv:1502.01589

\bibitem[{{Pontzen} \& {Pettini}(2009)}]{Pontzen2009}
{Pontzen}, A., \& {Pettini}, M. 2009, \mnras, 393, 557

\bibitem[{{Prochaska}(2003)}]{Prochaska2003b}
{Prochaska}, J.~X. 2003, ArXiv Astrophysics e-prints, astro-ph/0310850

\bibitem[{{Prochaska} {et~al.}(2008){Prochaska}, {Chen}, {Wolfe},
  {Dessauges-Zavadsky}, \& {Bloom}}]{Prochaska2008}
{Prochaska}, J.~X., {Chen}, H.-W., {Wolfe}, A.~M., {Dessauges-Zavadsky}, M., \&
  {Bloom}, J.~S. 2008, \apj, 672, 59

\bibitem[{{Prochaska} {et~al.}(2003{\natexlab{a}}){Prochaska}, {Gawiser},
  {Wolfe}, {Castro}, \& {Djorgovski}}]{Prochaska2003}
{Prochaska}, J.~X., {Gawiser}, E., {Wolfe}, A.~M., {Castro}, S., \&
  {Djorgovski}, S.~G. 2003{\natexlab{a}}, \apjl, 595, L9

\bibitem[{{Prochaska} {et~al.}(2003{\natexlab{b}}){Prochaska}, {Gawiser},
  {Wolfe}, {Cooke}, \& {Gelino}}]{Prochaska2003c}
{Prochaska}, J.~X., {Gawiser}, E., {Wolfe}, A.~M., {Cooke}, J., \& {Gelino}, D.
  2003{\natexlab{b}}, \apjs, 147, 227

\bibitem[{{Prochaska} {et~al.}(2005){Prochaska}, {Herbert-Fort}, \&
  {Wolfe}}]{Prochaska2005}
{Prochaska}, J.~X., {Herbert-Fort}, S., \& {Wolfe}, A.~M. 2005, \apj, 635, 123

\bibitem[{{Prochaska} \& {Wolfe}(1997)}]{Prochaska1997}
{Prochaska}, J.~X., \& {Wolfe}, A.~M. 1997, \apj, 487, 73

\bibitem[{{Prochaska} \& {Wolfe}(2002)}]{Prochaska2002}
---. 2002, \apj, 566, 68

\bibitem[{{Prochaska} \& {Wolfe}(2009)}]{Prochaska2009}
---. 2009, \apj, 696, 1543

\bibitem[{{Prochaska} {et~al.}(2001){Prochaska}, {Wolfe}, {Tytler}, {Burles},
  {Cooke}, {Gawiser}, {Kirkman}, {O'Meara}, \&
  {Storrie-Lombardi}}]{Prochaska2001}
{Prochaska}, J.~X., {Wolfe}, A.~M., {Tytler}, D., {et~al.} 2001, \apjs, 137, 21

\bibitem[{{Quiret} {et~al.}(2016){Quiret}, {P{\'e}roux}, {Zafar}, {Kulkarni},
  {Jenkins}, {Milliard}, {Rahmani}, {Popping}, {Rao}, {Turnshek}, \&
  {Monier}}]{Quiret2016}
{Quiret}, S., {P{\'e}roux}, C., {Zafar}, T., {et~al.} 2016, \mnras, 458, 4074

\bibitem[{{Rafelski} {et~al.}(2014){Rafelski}, {Neeleman}, {Fumagalli},
  {Wolfe}, \& {Prochaska}}]{Rafelski2014}
{Rafelski}, M., {Neeleman}, M., {Fumagalli}, M., {Wolfe}, A.~M., \&
  {Prochaska}, J.~X. 2014, \apjl, 782, L29

\bibitem[{{Rafelski} {et~al.}(2012){Rafelski}, {Wolfe}, {Prochaska},
  {Neeleman}, \& {Mendez}}]{Rafelski2012}
{Rafelski}, M., {Wolfe}, A.~M., {Prochaska}, J.~X., {Neeleman}, M., \&
  {Mendez}, A.~J. 2012, \apj, 755, 89

\bibitem[{{Rahmani} {et~al.}(2010){Rahmani}, {Srianand}, {Noterdaeme}, \&
  {Petitjean}}]{Rahmani2010}
{Rahmani}, H., {Srianand}, R., {Noterdaeme}, P., \& {Petitjean}, P. 2010,
  \mnras, 409, L59

\bibitem[{{Rahmati} \& {Schaye}(2014)}]{Rahmati2014}
{Rahmati}, A., \& {Schaye}, J. 2014, \mnras, 438, 529

\bibitem[{{Ross} {et~al.}(2012){Ross}, {Myers}, {Sheldon}, {Y{\`e}che},
  {Strauss}, {Bovy}, {Kirkpatrick}, {Richards}, {Aubourg}, {Blanton}, {Brandt},
  {Carithers}, {Croft}, {da Silva}, {Dawson}, {Eisenstein}, {Hennawi}, {Ho},
  {Hogg}, {Lee}, {Lundgren}, {McMahon}, {Miralda-Escud{\'e}},
  {Palanque-Delabrouille}, {P{\^a}ris}, {Petitjean}, {Pieri}, {Rich}, {Roe},
  {Schiminovich}, {Schlegel}, {Schneider}, {Slosar}, {Suzuki}, {Tinker},
  {Weinberg}, {Weyant}, {White}, \& {Wood-Vasey}}]{Ross2012}
{Ross}, N.~P., {Myers}, A.~D., {Sheldon}, E.~S., {et~al.} 2012, \apjs, 199, 3

\bibitem[{{Rubin} {et~al.}(2015){Rubin}, {Hennawi}, {Prochaska}, {Simcoe},
  {Myers}, \& {Lau}}]{Rubin2015}
{Rubin}, K.~H.~R., {Hennawi}, J.~F., {Prochaska}, J.~X., {et~al.} 2015, \apj,
  808, 38

\bibitem[{{S{\'a}nchez-Ram{\'{\i}}rez}
  {et~al.}(2016){S{\'a}nchez-Ram{\'{\i}}rez}, {Ellison}, {Prochaska}, {Berg},
  {L{\'o}pez}, {D'Odorico}, {Becker}, {Christensen}, {Cupani}, {Denney},
  {P{\^a}ris}, {Worseck}, \& {Gorosabel}}]{Sanchez2015}
{S{\'a}nchez-Ram{\'{\i}}rez}, R., {Ellison}, S.~L., {Prochaska}, J.~X.,
  {et~al.} 2016, \mnras, 456, 4488

\bibitem[{Shull \& Thronson(2012)}]{Shull2012}
Shull, J., \& Thronson, H. 2012, The Environment and Evolution of Galaxies,
  Astrophysics and Space Science Library (Springer Netherlands)

\bibitem[{{Smee} {et~al.}(2013){Smee}, {Gunn}, {Uomoto}, {Roe}, {Schlegel},
  {Rockosi}, {Carr}, {Leger}, {Dawson}, {Olmstead}, {Brinkmann}, {Owen},
  {Barkhouser}, {Honscheid}, {Harding}, {Long}, {Lupton}, {Loomis}, {Anderson},
  {Annis}, {Bernardi}, {Bhardwaj}, {Bizyaev}, {Bolton}, {Brewington}, {Briggs},
  {Burles}, {Burns}, {Castander}, {Connolly}, {Davenport}, {Ebelke}, {Epps},
  {Feldman}, {Friedman}, {Frieman}, {Heckman}, {Hull}, {Knapp}, {Lawrence},
  {Loveday}, {Mannery}, {Malanushenko}, {Malanushenko}, {Merrelli}, {Muna},
  {Newman}, {Nichol}, {Oravetz}, {Pan}, {Pope}, {Ricketts}, {Shelden},
  {Sandford}, {Siegmund}, {Simmons}, {Smith}, {Snedden}, {Schneider},
  {SubbaRao}, {Tremonti}, {Waddell}, \& {York}}]{Smee2013}
{Smee}, S.~A., {Gunn}, J.~E., {Uomoto}, A., {et~al.} 2013, \aj, 146, 32

\bibitem[{{Smette} {et~al.}(2005){Smette}, {Wisotzki}, {Ledoux}, {Garcet},
  {Lopez}, \& {Reimers}}]{Smette2005}
{Smette}, A., {Wisotzki}, L., {Ledoux}, C., {et~al.} 2005, in IAU Colloq. 199:
  Probing Galaxies through Quasar Absorption Lines, ed. P.~{Williams}, C.-G.
  {Shu}, \& B.~{Menard}, 475--477

\bibitem[{{Som} {et~al.}(2013){Som}, {Kulkarni}, {Meiring}, {York},
  {P{\'e}roux}, {Khare}, \& {Lauroesch}}]{Som2013}
{Som}, D., {Kulkarni}, V.~P., {Meiring}, J., {et~al.} 2013, \mnras, 435, 1469

\bibitem[{{Som} {et~al.}(2015){Som}, {Kulkarni}, {Meiring}, {York},
  {P{\'e}roux}, {Lauroesch}, {Aller}, \& {Khare}}]{Som2015}
---. 2015, \apj, 806, 25

\bibitem[{{Srianand} {et~al.}(2005){Srianand}, {Petitjean}, {Ledoux},
  {Ferland}, \& {Shaw}}]{Srianand2005}
{Srianand}, R., {Petitjean}, P., {Ledoux}, C., {Ferland}, G., \& {Shaw}, G.
  2005, \mnras, 362, 549

\bibitem[{{Trenti} \& {Stiavelli}(2006)}]{Trenti2006}
{Trenti}, M., \& {Stiavelli}, M. 2006, \apj, 651, 51

\bibitem[{{Vladilo}(2002)}]{Vladilo2002}
{Vladilo}, G. 2002, \aap, 391, 407

\bibitem[{{Vladilo} {et~al.}(2011){Vladilo}, {Abate}, {Yin}, {Cescutti}, \&
  {Matteucci}}]{Vladilo2011}
{Vladilo}, G., {Abate}, C., {Yin}, J., {Cescutti}, G., \& {Matteucci}, F. 2011,
  \aap, 530, A33

\bibitem[{{Vladilo} {et~al.}(2001){Vladilo}, {Centuri{\'o}n}, {Bonifacio}, \&
  {Howk}}]{Vladilo2001}
{Vladilo}, G., {Centuri{\'o}n}, M., {Bonifacio}, P., \& {Howk}, J.~C. 2001,
  \apj, 557, 1007

\bibitem[{{Vladilo} {et~al.}(2008){Vladilo}, {Prochaska}, \&
  {Wolfe}}]{Vladilo2008}
{Vladilo}, G., {Prochaska}, J.~X., \& {Wolfe}, A.~M. 2008, \aap, 478, 701

\bibitem[{{Webster} {et~al.}(2015){Webster}, {Bland-Hawthorn}, \&
  {Sutherland}}]{Webster2015}
{Webster}, D., {Bland-Hawthorn}, J., \& {Sutherland}, R.~S. 2015, \apj, 804,
  110

\bibitem[{{Welty} {et~al.}(1999){Welty}, {Frisch}, {Sonneborn}, \&
  {York}}]{Welty1999}
{Welty}, D.~E., {Frisch}, P.~C., {Sonneborn}, G., \& {York}, D.~G. 1999, \apj,
  512, 636

\bibitem[{{Wolfe} {et~al.}(2003){Wolfe}, {Gawiser}, \& {Prochaska}}]{Wolfe2003}
{Wolfe}, A.~M., {Gawiser}, E., \& {Prochaska}, J.~X. 2003, \apj, 593, 235

\bibitem[{{Wolfe} {et~al.}(2005){Wolfe}, {Gawiser}, \& {Prochaska}}]{Wolfe2005}
---. 2005, \araa, 43, 861

\bibitem[{{Wolfe} {et~al.}(2004){Wolfe}, {Howk}, {Gawiser}, {Prochaska}, \&
  {Lopez}}]{Wolfe2004}
{Wolfe}, A.~M., {Howk}, J.~C., {Gawiser}, E., {Prochaska}, J.~X., \& {Lopez},
  S. 2004, \apj, 615, 625

\bibitem[{{Wolfe} \& {Prochaska}(1998)}]{Wolfe1998}
{Wolfe}, A.~M., \& {Prochaska}, J.~X. 1998, \apjl, 494, L15

\bibitem[{{Wolfe} \& {Prochaska}(2000)}]{WolfeProchaska2000}
---. 2000, \apj, 545, 591

\bibitem[{{Wolfe} {et~al.}(1986){Wolfe}, {Turnshek}, {Smith}, \&
  {Cohen}}]{Wolfe1986}
{Wolfe}, A.~M., {Turnshek}, D.~A., {Smith}, H.~E., \& {Cohen}, R.~D. 1986,
  \apjs, 61, 249

\bibitem[{{York} {et~al.}(1986){York}, {Dopita}, {Green}, \&
  {Bechtold}}]{York1986}
{York}, D.~G., {Dopita}, M., {Green}, R., \& {Bechtold}, J. 1986, \apj, 311,
  610

\bibitem[{{York} {et~al.}(2000){York}, {Adelman}, {Anderson}, {Anderson},
  {Annis}, {Bahcall}, {Bakken}, {Barkhouser}, {Bastian}, {Berman}, {Boroski},
  {Bracker}, {Briegel}, {Briggs}, {Brinkmann}, {Brunner}, {Burles}, {Carey},
  {Carr}, {Castander}, {Chen}, {Colestock}, {Connolly}, {Crocker}, {Csabai},
  {Czarapata}, {Davis}, {Doi}, {Dombeck}, {Eisenstein}, {Ellman}, {Elms},
  {Evans}, {Fan}, {Federwitz}, {Fiscelli}, {Friedman}, {Frieman}, {Fukugita},
  {Gillespie}, {Gunn}, {Gurbani}, {de Haas}, {Haldeman}, {Harris}, {Hayes},
  {Heckman}, {Hennessy}, {Hindsley}, {Holm}, {Holmgren}, {Huang}, {Hull},
  {Husby}, {Ichikawa}, {Ichikawa}, {Ivezi{\'c}}, {Kent}, {Kim}, {Kinney},
  {Klaene}, {Kleinman}, {Kleinman}, {Knapp}, {Korienek}, {Kron}, {Kunszt},
  {Lamb}, {Lee}, {Leger}, {Limmongkol}, {Lindenmeyer}, {Long}, {Loomis},
  {Loveday}, {Lucinio}, {Lupton}, {MacKinnon}, {Mannery}, {Mantsch}, {Margon},
  {McGehee}, {McKay}, {Meiksin}, {Merelli}, {Monet}, {Munn}, {Narayanan},
  {Nash}, {Neilsen}, {Neswold}, {Newberg}, {Nichol}, {Nicinski}, {Nonino},
  {Okada}, {Okamura}, {Ostriker}, {Owen}, {Pauls}, {Peoples}, {Peterson},
  {Petravick}, {Pier}, {Pope}, {Pordes}, {Prosapio}, {Rechenmacher}, {Quinn},
  {Richards}, {Richmond}, {Rivetta}, {Rockosi}, {Ruthmansdorfer}, {Sandford},
  {Schlegel}, {Schneider}, {Sekiguchi}, {Sergey}, {Shimasaku}, {Siegmund},
  {Smee}, {Smith}, {Snedden}, {Stone}, {Stoughton}, {Strauss}, {Stubbs},
  {SubbaRao}, {Szalay}, {Szapudi}, {Szokoly}, {Thakar}, {Tremonti}, {Tucker},
  {Uomoto}, {Vanden Berk}, {Vogeley}, {Waddell}, {Wang}, {Watanabe},
  {Weinberg}, {Yanny}, {Yasuda}, \& {SDSS Collaboration}}]{York2000}
{York}, D.~G., {Adelman}, J., {Anderson}, Jr., J.~E., {et~al.} 2000, \aj, 120,
  1579

\bibitem[{{York} {et~al.}(2006){York}, {Khare}, {Vanden Berk}, {Kulkarni},
  {Crotts}, {Lauroesch}, {Richards}, {Schneider}, {Welty}, {Alsayyad}, {Kumar},
  {Lundgren}, {Shanidze}, {Smith}, {Vanlandingham}, {Baugher}, {Hall},
  {Jenkins}, {Menard}, {Rao}, {Tumlinson}, {Turnshek}, {Yip}, \&
  {Brinkmann}}]{York2006}
{York}, D.~G., {Khare}, P., {Vanden Berk}, D., {et~al.} 2006, \mnras, 367, 945

\bibitem[{{Zafar} {et~al.}(2014){Zafar}, {Centuri{\'o}n}, {P{\'e}roux},
  {Molaro}, {D'Odorico}, {Vladilo}, \& {Popping}}]{Zafar2014}
{Zafar}, T., {Centuri{\'o}n}, M., {P{\'e}roux}, C., {et~al.} 2014, \mnras, 444,
  744

\bibitem[{{Zafar} {et~al.}(2013){Zafar}, {P{\'e}roux}, {Popping}, {Milliard},
  {Deharveng}, \& {Frank}}]{Zafar2013}
{Zafar}, T., {P{\'e}roux}, C., {Popping}, A., {et~al.} 2013, \aap, 556, A141

\bibitem[{{Zwaan} {et~al.}(2008){Zwaan}, {Walter}, {Ryan-Weber}, {Brinks}, {de
  Blok}, \& {Kennicutt}}]{Zwaan2008}
{Zwaan}, M., {Walter}, F., {Ryan-Weber}, E., {et~al.} 2008, \aj, 136, 2886

\end{thebibliography}

\end{document}